\documentclass[a4paper]{article}
\usepackage{graphicx}%
\usepackage{amsmath}%
\usepackage{amssymb}
\usepackage{bm}
\usepackage{color}
\usepackage{hhline}
\usepackage{cite}
\usepackage{pdfpages}

\usepackage{fancyhdr}

 \definecolor{colour}{rgb}{0,0.2,0.6}

\def\d{{\partial}}
\def\s{{\sigma}}
\def\e{{\epsilon}}
\def\k{{ {\bm k} }}
\def\p{{ {\bm p} }}
\def\q{{ {\bm q} }}
\def\Q{{ {\bm Q} }}

\def\0{{ {\bm 0} }}
\def\w{{\omega}}
\def\a{{\alpha}}
\def\b{{\beta}}

\allowdisplaybreaks[4]

\usepackage{fancyhdr}

 \definecolor{colour}{rgb}{0,0.2,0.6}

\begin{document}

\setcounter{page}{1}
\
\begin{center}

{DOI \color{colour}{https://doi.org/10.1080/00018732.2022.2144590}} 
\vspace{1mm} 

{\Large \bf{
Unconventional density waves and superconductivities in
Fe-based superconductors
and other strongly correlated electron systems}}

\vspace{2mm} 

\emph{
Hiroshi Kontani$^{1}$,
Rina Tazai$^{2}$,
Youichi Yamakawa$^{1}$, 
Seiichiro Onari$^{1}$
}

$^1$ Department of Physics, Nagoya University,
Furo-cho, Nagoya 464-8602, Japan. 
 \\
$^2$ Yukawa Institute for Theoretical Physics, 
Kyoto University, Kyoto 606-8502, Japan
\\
\vspace{2mm} 

{kon@slab.phys.nagoya-u.ac.jp}
 
\end{center}

\noindent
{\bf {Abstract}}
In this article, 
we review the recent significant progress in the theoretical studies of the 
electronic states by mainly focusing on Fe-based and cuprate superconductors.
These superconductors are ``unconventional'' in that 
strong electron-electron correlation mediates the pairing;
\color{black}
they are different from conventional phonon-mediated 
BCS superconductors.
\color{black}
To seek the high-$T_{\rm c}$ pairing mechanism,
many scientists have focused on the 
mysterious spontaneous rotational symmetry breaking above $T_{\rm c}$,
such as nematic order at $\q={\bm0}$ and smectic order at $\q\ne{\bm0}$.
Such exotic correlation-driven symmetry breaking in metals 
has become a central issue in condensed matter physics.
We demonstrate the emergence of the nematic and smectic orders due to 
orbital polarization ($n_{xz}\ne n_{yz}$) and the symmetry breaking
in the correlated intersite hopping (= bond order $\delta t_{i,j}$)
in Fe-based and cuprate superconductors.
In addition, we discuss exotic spontaneous loop current orders 
driven by the pure imaginary $\delta t_{i,j}$.
These interesting ``unconventional density-waves'' 
originate from the quantum interference between different spin fluctuations
that is described by the vertex correction (VC) in the field theory. 
\color{black}
In the next stage, we discuss electron-correlation driven superconductivity
due to the fluctuations of unconventional density-waves.
\color{black}
For this purpose, we suggest the beyond-Migdal-Eliashberg gap equation
by including the VCs into the equation.
In Fe-based superconductors, high-$T_{\rm c}$ $s$-wave superconductivity 
can be mediated by nematic and smectic fluctuations 
because the pairing interaction is magnified by the VCs.
We also discuss the multipolar fluctuation pairing mechanism in 
heavy fermion systems, owing to the cooperation between
the strong spin-orbit interaction and the strong electron correlation.
To summarize, we suggest that 
the quantum interference mechanism described by the VCs 
plays essential roles in not only various 
unconventional density-waves,
but also exotic superconducting states in many strongly correlated metals.
We finally discuss some interesting future issues
with respect to the quantum interference mechanism.

\noindent
{\bf {Keywords:}} unconventional superconductivity, electronic nematic order, current order, Fe-based superconductors, cuprate superconductors

\setcounter{tocdepth}{3}
\tableofcontents

\section{Introduction
\label{sec:Kontani1}
}

\subsection{Unconventional superconductivity due to electron correlation 
\label{sec:Kontani1-1}
}

In conventional BCS-type superconductors,
the electron-phonon interaction mediates the pairing interaction.
Since the phonon-mediated interaction is attractive,
$s$-wave superconductivity is realized.
Fundamental superconducting (SC) electronic properties, such as
transition temperature $T_{\rm c}$ and thermodynamic quantities,
are well explained based on the BCS theory
\cite{Schrieffer}.
\color{black}
In general, the realized $T_{\rm c}$ is low since the phonon-mediated
pairing interaction is weak and the Debye temperature is 
only on the order of $100$K.
However, there are several exceptional high-$T_c$ 
phonon-mediated superconductors,
such as MgB$_2$ and hydrogen-based compounds.
The $T_{\rm c}$ of the latter compounds
exceeds 200K under ultrahigh pressure 
(over $200$GPa) thanks to high Debye temperature of H-ion oscillation
\cite{Hydrogen}.
\color{black}

In contrast, in unconventional superconductors,
the pairing interactions originate from 
strong electron-electron correlations.
Since the correlation-driven pairing interaction 
strongly depends on systems,
unconventional superconductivity exhibits amazing variety
in various strongly correlated electron systems.
For example, nodal $d$-wave superconductivity
is realized in cuprate high-$T_{\rm c}$ superconductors
and heavy fermion compound Ce$M$In$_5$ ($M$=Co,Rh,Ir).
In contrast, fully-gapped or nodal $s$-wave superconductivity 
is realized by electron-electron correlation
in Fe-based superconductors.
Also, spin-triplet superconductivity 
is expected to be realized in several U-based heavy fermions. 

The gap function is obtained by solving the gap equation.
A simplified expression of the linearized gap equation is
\cite{Schrieffer,Eliashberg,Scalapino-rev},
\begin{eqnarray}
\lambda\Delta_\k = -\frac1N \sum_\p V^{\rm SC}(\k-\p)\delta(\e_\p-\mu)\Delta_\p \ln(\omega_c/T)
\label{eqn:gapeq-simplest}
\end{eqnarray}
where $T$ is the temperature, $\mu$ is the chemical potential,
and $\e_\p$ is the conduction electron energy.
$V^{\rm SC}(\k-\p)$ is the pairing interaction 
due to the collective bosonic fluctuations, and
$\omega_c$ is the fluctuation energy-scale.
The gap function $\Delta_\k$ is uniquely determined 
as the eigenfunction of the largest eigenvalue $\lambda$
in Eq. (\ref{eqn:gapeq-simplest}), and $T_{\rm c}$ 
is given by the condition $\lambda=1$.
Therefore, to reveal the unconventional SC state,
we have to know accurate $V^{\rm SC}(\q)$ in the normal state.
For this purpose, we study the normal state electronic properties
before analyzing the SC state.
One of the main aim of this article is to explain the 
``phase diagram in the normal state''
that is the parent state of the superconductivity.

\begin{figure}[t]
\centering
\includegraphics[width=.9\linewidth]{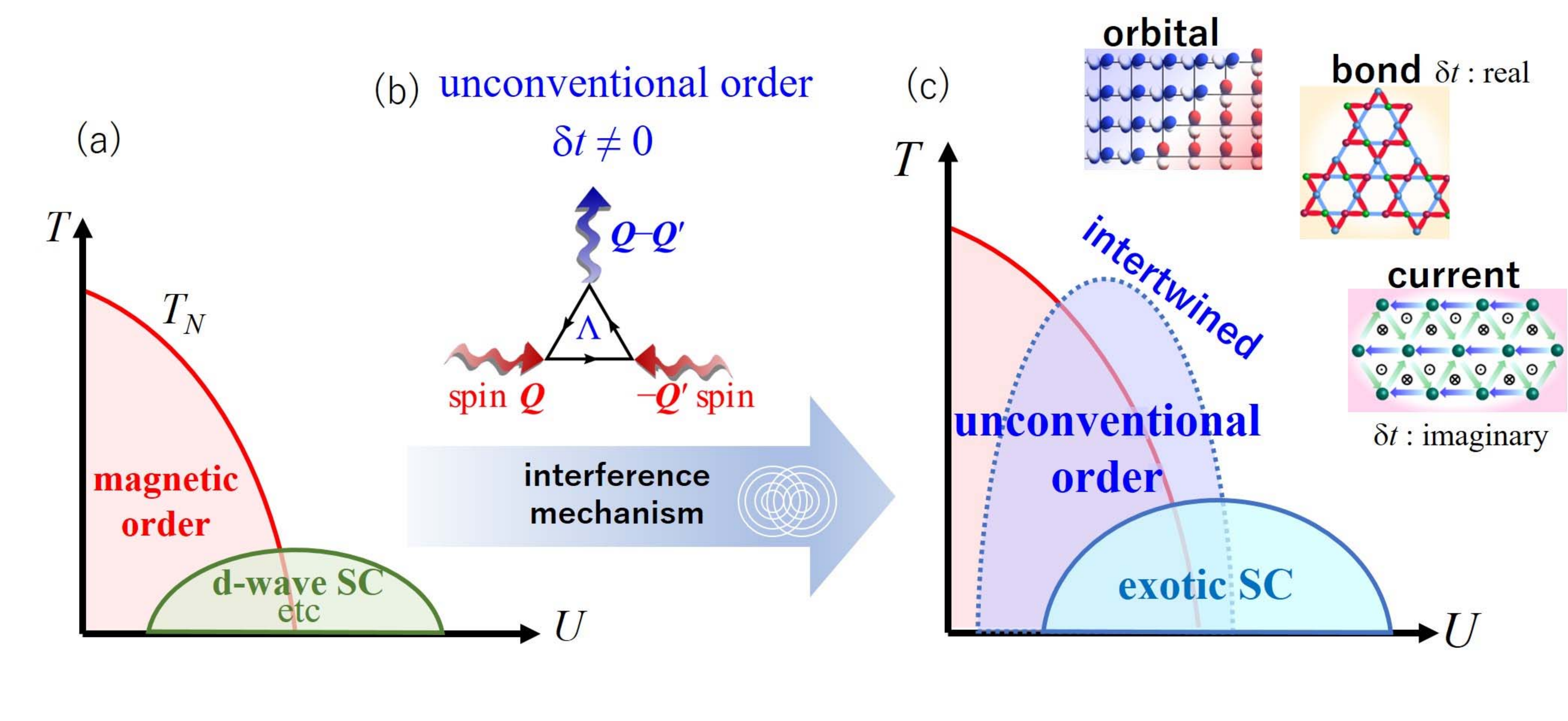}
\caption{
(a) A schematic phase diagram due to the spin fluctuation mechanism
with antiferromagnetic (AFM) state and $d$-wave superconductivity.
Strong spin fluctuations appear near the AFM quantum critical point (QCP).
(b) Paramagnon interference mechanism
that induces various nonmagnetic nematic or bond orders in metals.
(c) The expected phase diagram with exotic bond/orbital order
that is the unconventional (= non-$A_{1g}$ symmetry) order parameters.
It originates from the paramagnon interference mechanism.
Exotic superconducting state is expected to emerge 
near the bond/orbital order QCP
}
\label{fig:phase-AFM}
\end{figure}

The most famous correlation-driven superconductivity 
would be the spin-fluctuation-mediated singlet pairing state.
A schematic phase diagram of the spin fluctuation scenario
is shown in Fig. \ref{fig:phase-AFM} (a).
Here, the antiferromagnetic (AFM) second-order 
transition temperature $T_N$ decreases with $x$, 
and the AFM order disappears ($T_N=0$) at 
the quantum critical point (QCP) $x=x_c$.
Near the QCP, strong spin fluctuations develop at the nesting vector $\Q_s$,
and the spin susceptibility $\chi^s(\q)$ shows large peak at $\q=\Q_s$.
For the spin-singlet Cooper pair,
spin fluctuations give repulsive (=positive) interaction 
as $V^{\rm SC}(\q) \propto \chi^s(\q)$.
Therefore, in general,
the $d$-wave gap function $\Delta(\k)\propto \cos k_x - \cos k_y$
is mediated by the
staggered spin fluctuations with $\Q_s\approx (\pi,\pi)$,
which is expected to be realized in Ce$M$In$_5$ ($M$=Co,Rh,Ir).
Near the ferromagnetic ($\Q_c={\bm0}$) QCP,
the spin triplet superconductivity is expected to appear.
This ``spin-fluctuation pairing mechanism''
has been established by many theorists.
\cite{Scalapino-rev}. 

Here, we consider strongly correlated metals
with large on-site Coulomb interaction 
\begin{eqnarray}
H_U= \sum_i U n_{i\uparrow}n_{i\downarrow}=
\sum_i U(n_i^2-m_i^2)/4,
\end{eqnarray}
where $i$ is the site index,
$n_{i\s}=c_{i\s}^\dagger c_{i\s}$,
$n_i=n_{i\uparrow}+n_{i\downarrow}$, and 
$m_i=n_{i\uparrow}-n_{i\downarrow}$.
In strongly correlated metals, strong AFM fluctuations frequently
appear since $U$ tends to induce on-site spin polarization 
$\langle m_i \rangle\ne0$,
while on-site charge polarization is suppressed by $U$.
This consideration leads to the phase diagram 
in Fig. \ref{fig:phase-AFM} (a).
Nonetheless of this commonsense, 
the phase diagrams of Fe-based and cuprate
high-$T_{\rm c}$ superconductors are very different 
from Fig. \ref{fig:phase-AFM} (a) because of the {\it presence
of bond-order and orbital-order}.
These orders are called nematic at wavevector $\q={\bm0}$,
and smectic at $\q\ne{\bm0}$.
The main origin of these nonmagnetic orders,
which cannot be explained by mean-field-level approximations,
is the quantum interference between spin fluctuations
exhibited in Fig. \ref{fig:phase-AFM} (b)
\cite{RTazai-JPSJ-rev}.
This {\it paramagnon interference mechanism}
totally modifies the phase diagram in strongly correlated metals:
Figure \ref{fig:phase-AFM} (c) is an example of expected phase diagrams
due to the paramagnon interference mechanism.
The fluctuations near the nematic or smectic QCP
would induce non $d$-wave SC state,
which will be explained in this article.

\color{black}
In the last decade, interesting quantum phase transitions 
have been discovered in many strongly-correlated superconductors. 
In Fe-based superconductors, 
we encounter the electronic nematic states, which are the uniform ($\q={\bm 0}$)
rotational symmetry breaking states due to electron correlations.
In $R$1111 families ($R$=La,Ce,Pr,Nd,Sm)
\cite{1111-Buechner,1111-Buechner2},
A122 families ($A$=Ba,Ca,Sr)  
\cite{122-phase-diagram}
and FeSe families
\cite{phase-11-PT,phase-11-PT-2,FeSeTe}, 
$B_{1g}$ symmetry nematic order
with sizable orbital polarization ($n_{xz}\ne n_{yz}$) appears
at the structural transition temperature $T_S$.
\color{black}
Interestingly, $B_{2g}$ symmetry nematic order emerges in 
heavily $e$-doped system AFe$_2$As$_2$ (A=Cs,Rb)
\cite{B2g-NMR,B2g-Ishisa,B2g-STM}.

\color{black}
In cuprate superconductors, 
stripe order formation was reported by
x-ray \cite{Bianconi1996}
and neutron \cite{Tranquada1996}
scattering studies.
Recently, the emergence of charge-density-wave (CDW) order 
with finite wavevector $\Q = (\pi/2, 0)$ or $(0, \pi/2)$ 
has been confirmed by resonant X-ray scattering measurements
\cite{Y-Xray1,Bi-Xray1,Keimer-rev,cuprate-3D,cuprate-3D2,cuprate-3D3},
the Scanning tunneling spectroscopy/Scanning unneling microscopy
(STM/STS) measurements
\cite{STM-Kohsaka,STM-Fujita,Davis:2013ce},
and the Nuclear magnetic resonance (NMR) studies 
\cite{Zheng-1,Julien-NMR}.
\color{black}
These experimental findings manifest that the
spin, charge, and orbital degrees of freedom are all active and 
strongly coupled in these high-$T_{\rm c}$ superconductors. 
Then, strong fluctuations of orbital/charge degrees of freedoms
may be significant for high-$T_{\rm c}$ superconductivity.
\color{black}
Note that the CDW order below $T_{\rm CDW}\sim200$K 
is ``short range'' with the correlation length $\xi \ll 10$nm.
It becomes the true long-range order under high magnetic field ($\sim20$T)
\cite{CDW-long-range-H1,CDW-long-range-H2}
or applying the uniaxial strain ($|\e_{xx}|\lesssim0.01$)
\cite{CDW-long-range-strain}
below $\sim100$K.
\color{black}

In this article, we review the recent progress on the study of the 
mechanisms of unconventional (= non-$A_{1g}$ symmetry) 
order parameter and superconductivity,
mainly in Fe-based and cuprate superconductors and related compounds.
We explain that the strong coupling between
spin, charge, and orbital degrees of freedom are induced by 
the inter-mode coupling due to vertex corrections (VCs),
which are neglected in previous mean-field-type approximations.
The VCs are important key ingredients to understand the nematicity, CDW 
and superconductivity in many strongly correlated electron systems.

In later subsections of Sect. 1, we review the 
phase diagrams in Fe-based and cuprate superconductors.
In Sect. 2 - Sect. 5, 
we discuss the origin of unconventional ordered states,
such as the orbital order and $d$-wave bond order,
by considering the spin-fluctuation-driven VCs.
In Sect. 6, we discuss the pairing mechanism in Fe-based superconductors,
by focusing on both orbital and spin fluctuations.
In Sect. 7, we review the multipole order and 
multipole fluctuation pairing mechanism in $f$-electron systems
with strong spin-orbit interaction (SOI).
The discussions of this article are summarized in Sect. 8.

\subsection{Fe-based superconductors
\label{sec:Kontani1-2}
}

\begin{figure}[htb]
\centering
\includegraphics[width=.7\linewidth]{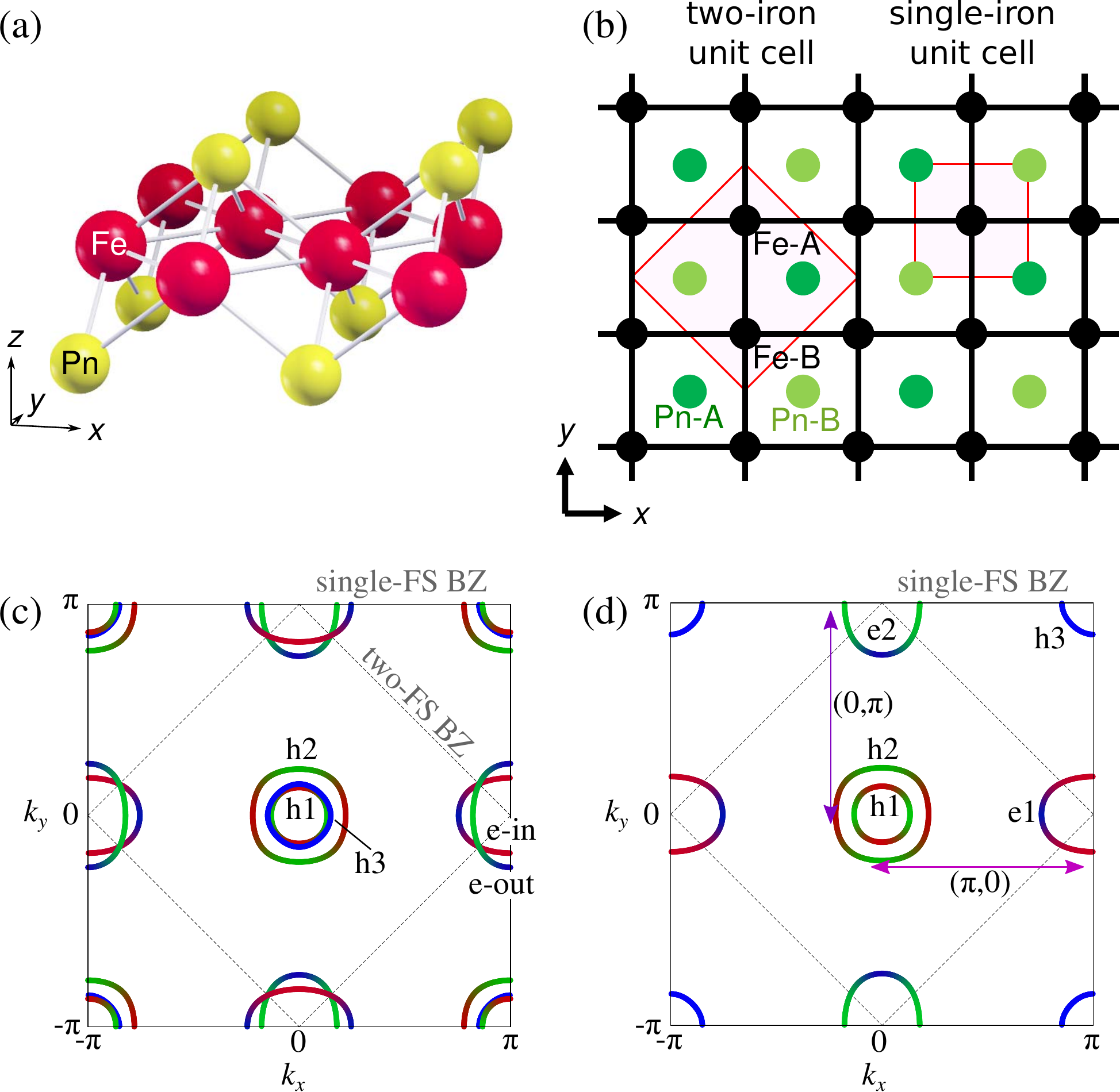}
\caption{
(a)(b) Crystal structure of FePn-layer (Pn=As,Se), in
which the unit cell is given by the two-iron unit cell composed
of Fe-A, Fe-B, Pn-A, and Pn-B. The single-iron unit cell is
realized by applying the ``unfold-gauge transformation''.
(c) Original folded and (d) unfolded FSs 
in 1111 and 122 families with $n_d=6$.
}
\label{fig:FeAs-layer}
\end{figure}

The discovery of Fe-based high-$T_{\rm c}$ superconductors by Kamihara and Hosono 
is an epoch-making event of in condensed matter physics.
$T_{\rm c}=26$K in the first discovered 1111 compound LaFeAsO
\cite{Hosono-1111}
had immediately increased to 56K by replacing La with other rare-earth elements
\cite{1111-rare,Hosono-1111-2}.
Also, $T_{\rm c}=30\sim50$K in 122 compounds $A$Fe$_2$As$_2$
($A$=Ba,Ca,Sr, etc)
\cite{Ba122-38K,Saffat-22K,Hosono-phase1111}.
In 11 compounds, $T_{\rm c}\sim40$K is realized in Li-intercalated FeSe 
\cite{Lu_LiFeOHFeSe_crystal,Du2016_LiFeOHFeSe_STM,Noji2014_FeSe_intercalate,Yan2016_LiFeOHFeSe_QPI,Niu2015_LiFeOHFeSe_ARPES,Ren2017_LiFeOHFeSe_ARPES,Gu2018_LiFeOHFeSe_QPI}, 
and $T_{\rm c}>60$K is reported
in single layer FeSe grown on SrTiO$_3$ substrates 
\cite{Wang2012_FeSeSTO_1st, Lee2014_FeSeSTO_ARPES, Fan2015_FeSe_STM, Zhang2016_FeSeSTO_ARPES, Shi2017_FeSeSTO_phase}.
In all families,
the metallicity and superconductivity occur in two-dimensional 
FePn (Pn=As,Se) plane, which is shown in Figs. \ref{fig:FeAs-layer} (a) and (b).
In usual compounds, the $d$-electron filling of each Fe ion is $n_d\sim6$.

In this article, we set $x$ and $y$ axes parallel to the nearest Fe-Fe
bonds, and represent the $z^2$, $xz$, $yz$, $xy$, and $x^2-y^2$
$d$-orbitals as 1,2,3,4, and 5, respectively.
The Fermi surfaces (FSs) are mainly composed of $t_{2g}$ orbitals
($l = 2-4$), although $z^2$ orbitals hole-pocket exists in 122 families.
Since the Pn-A (Pn-B) ions form the upper- (lower-) plane, 
the unit cell contains Fe-A and Fe-B.
The FSs of this original ten-orbital tight-binding model are
shown in Fig. \ref{fig:FeAs-layer} (c).
The orbital character is shown by green ($d_{xz}$),
red ($d_{yz}$), and blue ($d_{xy}$) colors.
The good nesting between hole-pockets and electron-pockets 
with $\Q\sim(\pi,0),(0,\pi)$ causes the stripe AFM order.
In Refs. \cite{KKuroki-PRL2008,TMiyake-JPSJ2010}, 
the authors introduced the ``unfold-gauge transformation''
$|l\rangle \rightarrow -|l\rangle$ ($l=1,4,5$) only for Fe-B sites.
Due to this gauge transformation, the unit cell is
halved to become the single-iron unit cell.
The FSs of the unfolded five-orbital tight-binding model are
shown in Fig. \ref{fig:FeAs-layer} (d).
This unfolded model is very convenient for theoretical analyses.

In almost all Fe-based superconductors, 
the $s$-wave superconductivity is realized.
However, basic properties of the gap function, 
such as the momentum and orbital dependences and
the presence or absence of the sign-reversal in the gap function,
are still under debate in many compounds.
To uncover the pairing mechanism,
we should first understand the basic many-body electronic properties
in the normal state.

\begin{figure}[htb]
\centering
\includegraphics[width=.9\linewidth]{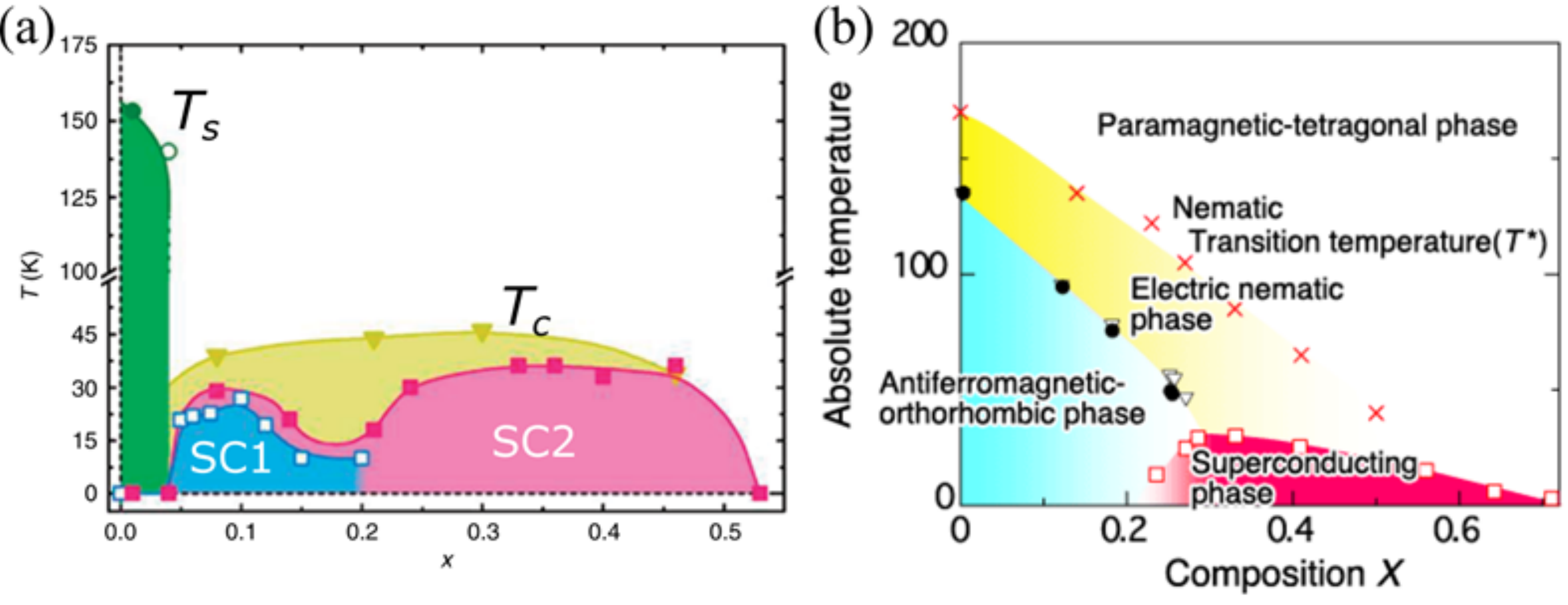}
\caption{
Experimental phase diagrams of FeAs families with $n_d\sim6$:
(a) LaFeAsO$_{1-x}$H$_x$ at ambient pressure ($P=0$) and at $P=2$GPa.
Reprinted from Ref. \cite{Hosono-phase1111}.
(b) BaFe$_2$(As$_{1-x}$P$_x$)$_2$
\cite{Kasahara-hidden}.
The electronic nematic transition occurs at $T=T_S$,
below which small orthorhombic structural transition 
($\delta=(a-b)/(a+b)\sim 3\times10^{-3}$) is accompanied.
In (b), hidden nematicity with tiny 
otrhorhombicity ($\delta \sim 10^{-4}$) appears at $T=T^*$, 
which will be discussed in Sect. \ref{sec:Kontani4}.
Reprinted by permission from Springer Nature:
Nature \cite{Kasahara-hidden}, copyright 2012.
}
\label{fig:phase-1111}
\end{figure}

First, we explain the diverse phase diagrams of Fe-based superconductors.
Figure \ref{fig:phase-1111} shows the phase diagrams
for (a) 1111 family
\cite{Hosono-phase1111}
and (b) 122 family 
\cite{Kasahara-hidden}
with $n_d\sim6$.
$T_S$ is the second-order structural transition temperature
from tetragonal state ($C_4$, $a=b$) to orthorhombic ($C_2$, $a>b$) state.
The ratio $(a-b)/(a+b)$ below $T_S$ is at most $0.3\%$.
As decreasing the temperature,
stripe magnetic order with
$\Q_s\approx(\pi,0)$ or $(0,\pi)$ appears at $T_N \ (\le T_S)$
in both 1111 and 122 compounds.
These phase diagrams indicate the
close relationship between the structural transition and magnetism.

The origin of the orthorhombic transition is the spontaneous
rotational symmetry breaking due to electron correlation,
so it is called the ``electronic nematic state''.
The nematic states in Figs. \ref{fig:phase-1111} (a) and (b) 
possess $B_{1g}$ ($d_{x^2-y^2}$) symmetry,
where $x$ and $y$ axes are along the nearest Fe-Fe direction shown in 
Fig. \ref{fig:FeAs-layer} (b).
To explain ubiquitous electronic nematicity,
up to now, the spin nematic order, the orbital order,
and the bond order have been proposed as the real order parameters.
The microscopic mechanism for each order parameter
has been developed intensively in the last decade.
The origin and the mechanism of nematicity is one of the 
main topics of this review article.

In addition, almost hidden nematic transition with tiny nematicity 
($(a-b)/(a+b)\sim10^{-4}$) occurs at $T=T^* \ (>T_S)$ 
in several FeAs families.
For example, 
it is recognized in the phase diagram of BaFe$_2$(As$_{1-x}$P$_x$)$_2$
in Fig. \ref{fig:phase-1111} (b).
At present, the origin of this ``slight nematic state'' is unknown.
Some extrinsic origins (such as surface induced nematicity) 
have been discussed so far.
As the intrinsic order,
we propose the smectic orbital order at $\q=(\pi,0)$ 
based on a microscopic theory,
as we will explain in Sect. \ref{sec:Kontani4-6}.
We stress that similar smectic order has been recently
observed by the Angle-resolved photoemission spectroscopy (ARPES) 
studies in (Ba,K)Fe$_2$As$_2$ \cite{Shimojima-hidden},
CaKFe$_4$As$_4$ and KCa$_2$Fe$_4$As$_4$F$_2$ \cite{Zhou-hidden},
and electron-doped thin layer FeSe 
\cite{smectic-FeSe-STM}.

\begin{figure}[htb]
\centering
\includegraphics[width=.99\linewidth]{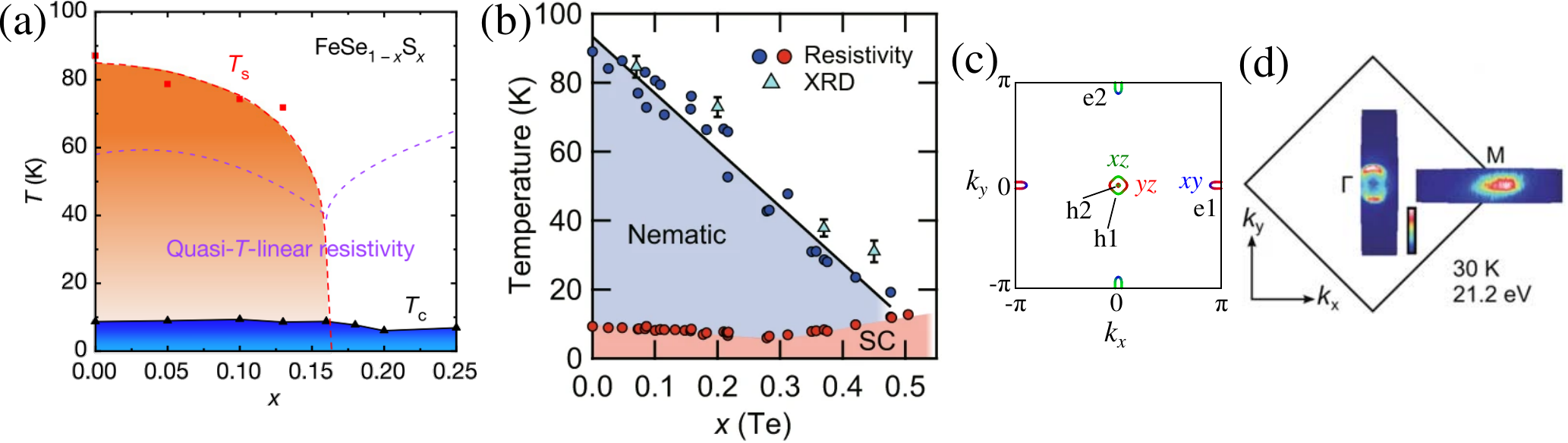}
\caption{
(a) Experimental phase diagram of FeSe$_{1-x}$S$_{x}$.
Reprinted by permission from Springer Nature:
Nature \cite{FeSeS}, copyright 2019.
(b) Experimental phase diagram of FeSe$_{1-x}$Te$_{x}$.
Reprinted from Ref. \cite{FeSeTe}.
(c) FS of the unfolded FeSe model in the tetragonal state.
(d) FS of the nematic state in FeSe observed by ARPES.
Reprinted with permission from \cite{FeSe-Suzuki}.
Copyright (2015) by the American Physical Society.
}
\label{fig:phase-11}
\end{figure}

Next, we discuss the FeSe family without carrier doping ($n_d=6$):
The phase diagram of FeSe$_{1-x}$S$_{x}$ \cite{FeSeS}
and that of FeSe$_{1-x}$Te$_{x}$ \cite{FeSeTe}
are shown in Fig. \ref{fig:phase-11} (a).
In both compounds,
no magnetism appears in the nematic phase below $T_S$,
and superconductivity emerges in the nematic phase.
The symmetry of the nematic state in FeSe$_{1-x}$Te$_{x}$ and FeSe$_{1-x}$S$_{x}$ 
is $B_{1g}$ ($d_{x^2-y^2}$).
Similar phase diagram is realized in FeSe$_{1-x}$S$_{x}$.
Above $T_S$, observed low-energy spin fluctuations are very small
in both S- and Te-doped FeSe.
The discovery of ``nematicity without magnetism in FeSe families''
gives us significant information on the origin of nematicity.
The unfolded FSs in the tetragonal state ($T>T_S$)
is presented in Fig. \ref{fig:phase-11} (c).
The $C_4$ symmetry of the FS shape is spontaneous violated below $T_S$.
Figure \ref{fig:phase-11} (d) shows the
FS of detwined FeSe below $T_S$ observed by ARPES 
\cite{FeSe-Suzuki,FeSe-ARPES-noY}.
Due to the smallness of the Fermi energy $E_F$,
the SC coherence length $\xi$ 
becomes comparable to the Fermi momentum $k_{\rm F}$,
so the BCS-BEC crossover is expected to be realized \cite{Kasahara-BEC}.

\begin{figure}[htb]
\centering
\includegraphics[width=.5\linewidth]{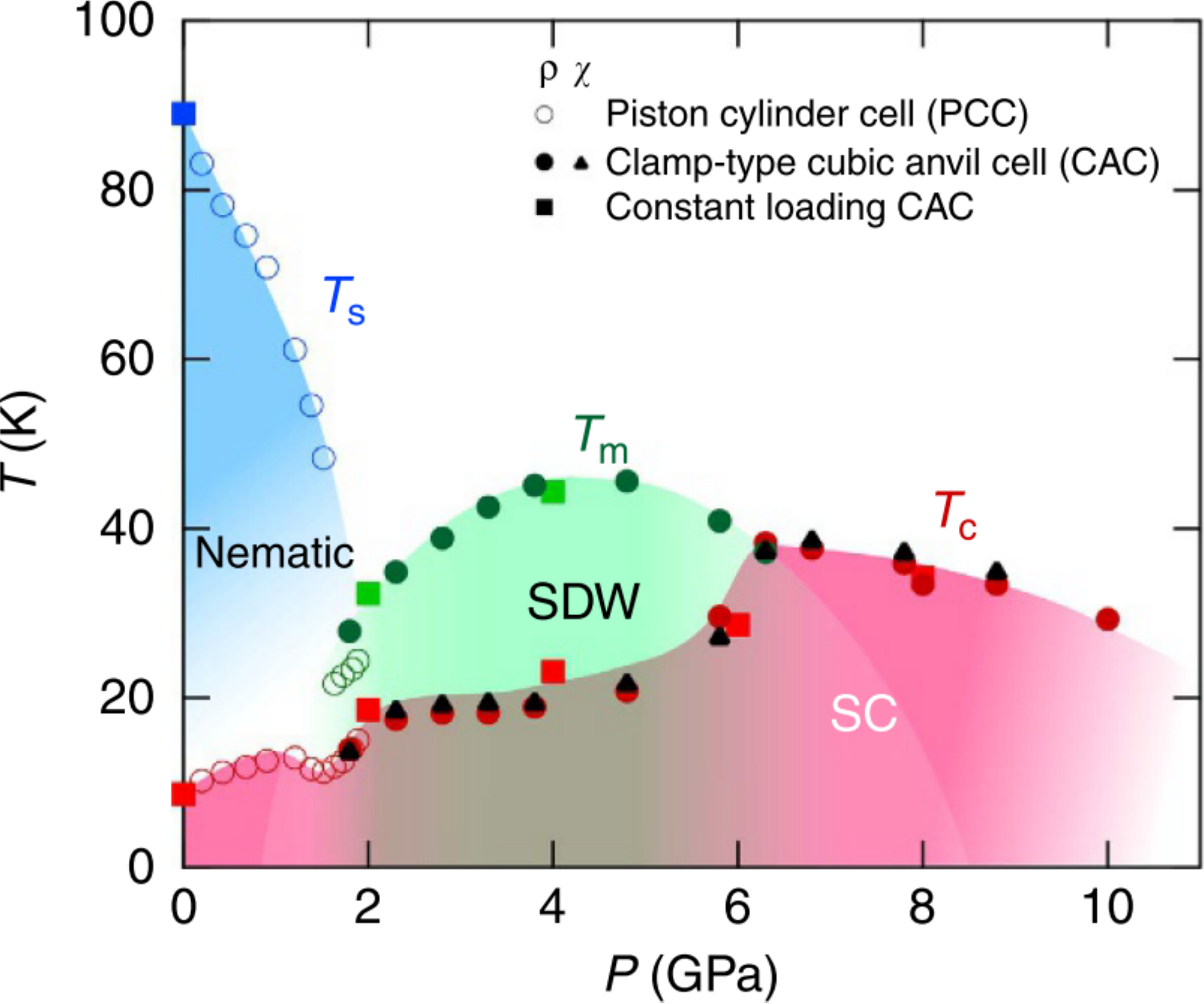}
\caption{
$T$-$P$ phase diagram of bulk FeSe.
Reprinted from Ref. \cite{phase-11-PT-2}.
}
\label{fig:phase-11-PT}
\end{figure}

The pressure-temperature ($P$-$T$) phase diagram of FeSe
is also remarkably interesting:
As shown in Fig. \ref{fig:phase-11-PT},
the nematic order is quickly suppressed under pressure,
and instead, the magnetic order appears
\cite{phase-11-PT,phase-11-PT-2}.
$T_{\rm c}$ gradually increases with $P$, and
the maximum $T_{\rm c}\sim40$K is realized at $P\sim6$GPa.
In the $P$-$T$ phase diagram of FeSe$_{1-x}$S$_x$ with $x\sim0.1$, 
the nematic phase at $P\sim0$GPa is completely separated from
the pressure induced magnetic order phase,
which appears just around $P\sim5$GPa
\cite{phase-11-PT-2}.
We will explain the $P$-$T$ phase diagram of FeSe
in terms of the quantum interference mechanism in later section.

The thin atomic-layer FeSe attracts great attention 
because this system exhibits the highest $T_{\rm c}$ ($\gtrsim70$K)
among Fe-based superconductors
\cite{Wang2012_FeSeSTO_1st,Lee2014_FeSeSTO_ARPES,Fan2015_FeSe_STM,Zhang2016_FeSeSTO_ARPES,Shi2017_FeSeSTO_phase}.
In mono atomic-layer FeSe, all hole-pockets disappear because
$10\sim20$\% electron carrier is naturally introduced.
In this e-doped FeSe,
electronic nematic order disappears, and instead
high-$T_{\rm c}$ SC state emerges on two electron-pockets.
Similar e-doped high-$T_{\rm c}$ SC state in FeSe is also realized
by $Li$-intercalation
\cite{Lu_LiFeOHFeSe_crystal, Du2016_LiFeOHFeSe_STM,Noji2014_FeSe_intercalate,Yan2016_LiFeOHFeSe_QPI,Niu2015_LiFeOHFeSe_ARPES,Ren2017_LiFeOHFeSe_ARPES,Gu2018_LiFeOHFeSe_QPI}, 
$K$-deposition
\cite{Miyata2015_KdoseFeSe, Wen2016_KdosedSTO_phase}, 
and ionic-liquid gating
\cite{FeSe-ion1,FeSe-ion2}.
In e-doped FeSe, the observed low-energy spin fluctuations are tiny 
since the hole-pocket is absent.
Thus, the Fermi pocket nesting is inessential
for realizing high-$T_{\rm c}$ state in electron-doped FeSe.

Recently, nonmagnetic smectic order at $\q\sim(\pi/4,0)$ is discovered
in atomic layer FeSe \cite{smectic-FeSe-STM},
in the vicinity of the high-$T_{\rm c}$ SC phase.
This experimental finding indicates that the high-$T_{\rm c}$ SC state is 
mediated by the smectic order fluctuations.
In this article, 
we explain the emergence of the smectic order in e-doped FeSe,
and propose the mechanism of the smectic fluctuation mediated high-$T_{\rm c}$ $s$-wave SC state.

Finally, we discuss RbFe$_2$As$_2$ and CsFe$_2$As$_2$,
which are heavily hole-doped ($n_d=5.5$) 122 compounds.
Here, the symmetry of nematic state is $B_{2g}$ ($d_{xy}$),
which is 45 degree rotated from the $B_{1g}$ nematicity in usual compounds.
This $B_{2g}$ nematicity in these compounds
has been reported by the NMR \cite{B2g-NMR} 
and STM \cite{B2g-STM},
and the nematic susceptibility 
\cite{B2g-Ishisa}
measurements.
The discovered $B_{2g}$ nematicity provides 
a very severe constraint on the mechanism of nematicity.
In this article, we try to explain rich variety of nematicity 
in Fe-based superconductors 
in terms of the paramagnon interference mechanism.

In summary, the nematic states universally appear
in Fe-based superconductors, while the symmetry of nematicity 
and the relation between nematicity and magnetism
exhibit amazing system dependence.
These facts strongly indicate that the electronic properties,
especially the SC states,
are strongly compound-dependent.
On the other hand, one may expect that the rich variety of nematicity 
can be understood based on the same mechanism,
from the perspective of universality.
Thus, diverse nematicity in Fe-based superconductors
is a significant test for constructing the reliable theory of nematicity.
The knowledge of nematicity is significant 
to establish the mechanism of superconductivity,
which is another significant open problem in Fe-based superconductors.
It is widely believed that the nematic fluctuations 
mediate the superconductivity, since high-$T_{\rm c}$ state 
is frequently realized next to the nematic phase.
The aim of the present study is to discuss the
origin of nematicity and the pairing mechanism 
from a unified viewpoint.

\subsection{cuprate superconductors}
\label{sec:Kontani1-3}

\begin{figure}[htb]
\centering
\includegraphics[width=.9\linewidth]{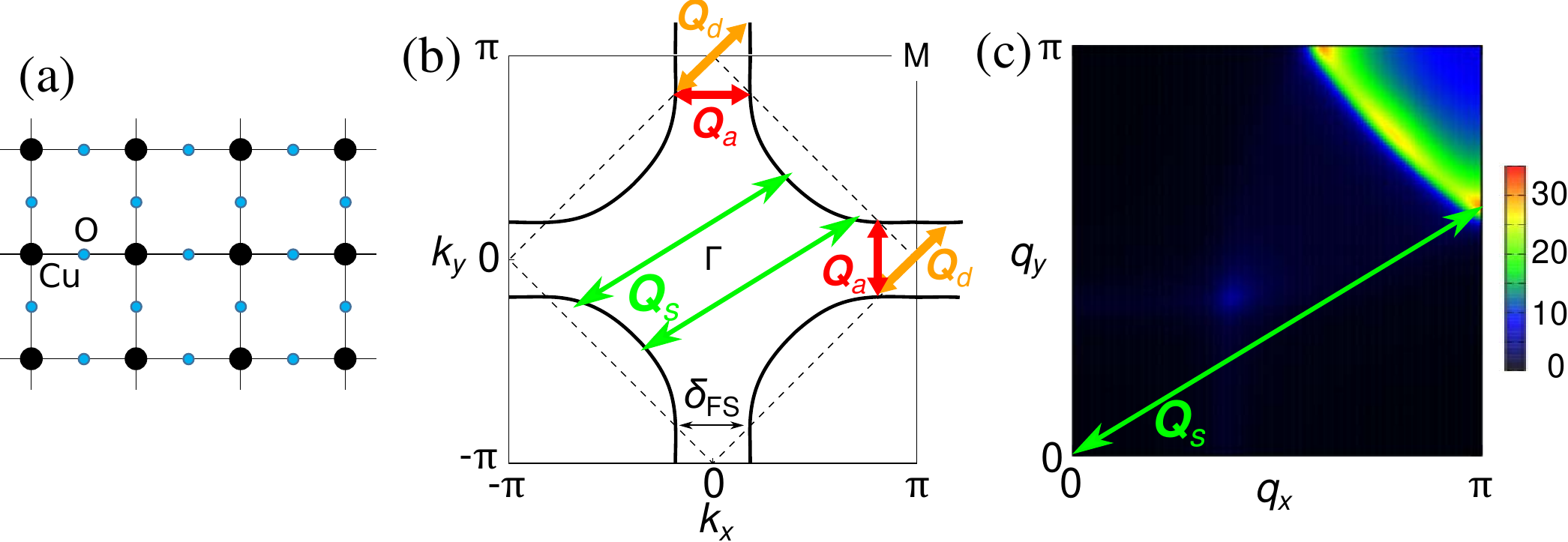}
\caption{
(a) Schematic CuO$_2$ plane of cuprate superconductors.
(b) FSs of hole-doped cuprate (YBCO) and main nesting vectors.
(c) Spin susceptibility $\chi^s(\q)$ obtained by the RPA.
}
\label{fig:YBCO-FS}
\end{figure}

The discovery of cuprate high-$T_{\rm c}$ superconductors in 1986
has triggered significant progress 
in the study of strongly correlated electron systems
\cite{Moriya-review1,Moriya-review2,Yamada-text,Nagaosa-rev,Ogata-rev}.
The superconductivity is realized by introducing hole-carrier
or electron-carrier into the parent
antiferromagnetic ($\Q=(\pi,\pi)$) insulators.
In the hole-doping systems, the maximum $T_{\rm c}$ 
at ambient pressure is about $40$K in La-compounds 
(e.g.,  La$_{2-x}$Sr$_x$CuO$_4$, La$_{2-x}$Nd$_x$CuO$_4$ ),
and around $100$K in Y-compounds (e.g.,  Ba$_2$Cu$_3$O$_{7-x}$ ),
and Bi, Ta, and Hg-compounds.
In the electron-doping compounds, 
the maximum $T_{\rm c}$ is around $30$K
in $R_{2-x}$Ce$_x$CuO$_4$ ($R$=Nd,Pr).
In all compounds, the metallicity and superconductivity occur 
in two-dimensional CuO$_2$ plane,
which is shown in Fig. \ref{fig:YBCO-FS} (a).
It is widely believed that the spin-fluctuation-mediated
$d$-wave superconductivity is realized.
The FS and $\chi^s(\q)$ in the RPA for YBCO model
are shown in Fig. \ref{fig:YBCO-FS} (b) and (c), respectively.

In real cuprate superconductors, however,
exotic orders composed of charge and spin degrees of freedom 
emerge in a cooperative fashion
\cite{Tranquada-rev}.
These unconventional density-wave states
cannot be explained by simple mean-field-type approximations.
For example, in Y-based, Bi-based, and Hg-based cuprates,
the charge-density-wave (CDW) order 
with finite wavevector $\Q = (\pi/2, 0)$ or $(0, \pi/2)$ appears
inside the pseudogap region in the presence of spin fluctuations
\cite{Y-Xray1,Bi-Xray1,Keimer-rev,cuprate-3D,cuprate-3D2,cuprate-3D3,STM-Kohsaka,STM-Fujita,Davis:2013ce,Zheng-1,Julien-NMR}.
Thus, spin and charge degrees of freedom are 
strongly coupled in high-$T_{\rm c}$ cuprates.
Then, strong fluctuations of charge degrees of freedoms
may be significant for high-$T_{\rm c}$ superconductivity.

\begin{figure}[htb]
\centering
\includegraphics[width=.5\linewidth]{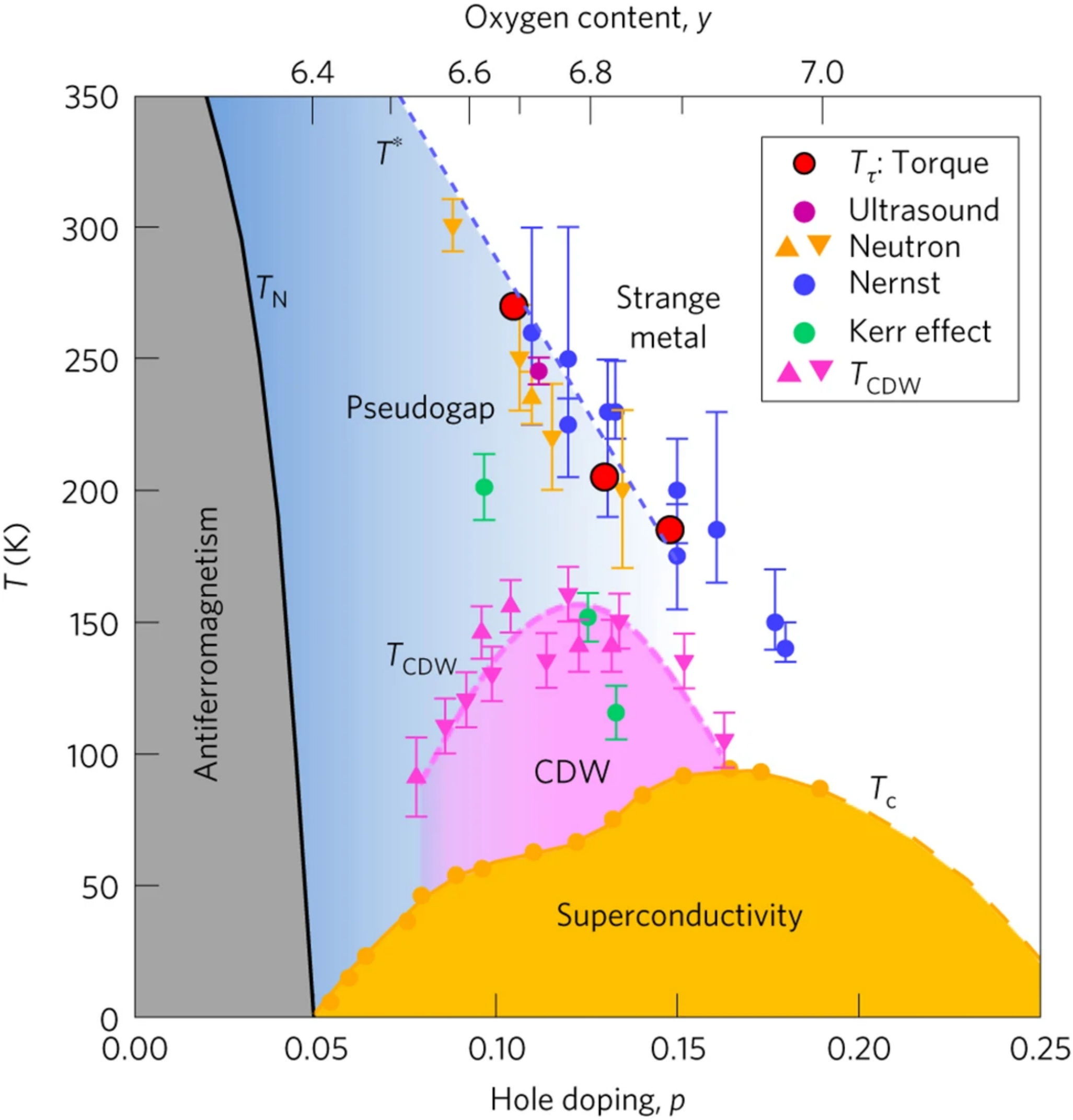}
\caption{
Experimental phase diagram of cuprate 
high-$T_{\rm c}$ superconductor YBCO.
Reprinted by permission from Springer Nature:
Nat. Phys. \cite{Sato-Matsuda2017}, copyright 2017.
}
\label{fig:YBCO-phase}
\end{figure}

Figure \ref{fig:YBCO-phase} shows the phase diagram of YBCO compound
\cite{Sato-Matsuda2017}.
Above the pseudogap temperature $T^*$,
the antiferromagnetic fluctuations develop monotonically
as decreasing the temperature.
The observed non-Fermi liquid behaviors,
such as $1/T_1T\propto T^{-1}$, and $\rho\propto T$,
and $R_H\propto T^{-1}$,
are satisfactorily explained based on the 
spin fluctuation theories.
Below $T^*$, the pseudogap appears in the electron density-of-states (DOS),
$N(0)$, so the Knight shift starts to decrease.
Below $T^*$, the increment of spin fluctuations becomes moderate,
so Fermi liquid behavior tends to be recovered. 
Until now, a number of theoretical studies have been performed
\cite{George-rev,Yokoyama,Varma2,various1,various2,Tohyama-rev,Nagaosa-rev,Dagotto-rev,Imada-rev,Ogata-rev}.
However, the origin and nature of the pseudogap have been 
significant open problems.

The CDW phase around $x\sim0.1\sim0,15$ 
in Fig. \ref{fig:YBCO-phase} has been discovered
in the last decade by $X$-ray measurements
\cite{Y-Xray1,Bi-Xray1,Zheng-1,Keimer-rev,cuprate-3D,cuprate-3D2,cuprate-3D3}
STM/STS measurements
\cite{STM-Kohsaka,STM-Fujita,Davis:2013ce},
and NMR studies
\cite{Zheng-1,Julien-NMR}.
The wavevector of the CDW state is $\q\approx(0.5\pi,0),(0,0.5\pi)$,
which corresponds to the nesting vector between the adjacent hot spots;
see Fig. \ref{fig:YBCO-FS}.
The CDW phase emerges universally in all hole-doped compounds,
and its high transition temperature ($T_{\rm CDW}\sim200$K)
indicates that the CDW is driven by the electron correlation.
However, it is highly nontrivial why the CDW is driven by the 
electron correlation.
It cannot be explained in the mean-field approximations.
To explain the CDW phase, various order parameters have been proposed, such as
the pair-density-wave (PDW) 
\cite{Lee:2014ka,Agterberg}
and the bond order
\cite{Kivelson-NJP,Halboth:2000tt,Davis:2013ce,Yamakawa-CDW,Kawaguchi-CDW,Tsuchiizu2018,Orth:2017,Fradkin:2015co,Yamase}.
The latter is the modulation of the hopping integrals
given by the symmetry-breaking in the self-energy.

Recently, 
experimental evidences for the phase transition at $T^*$
have been accumulated.
The nematic transition has been observed at $T^*$
in Y-, Hg-, and Bi-based compounds
\cite{Sato-Matsuda2017,Hg-Murayama,Fjimori-nematic}.
As the candidates of the pseudogap order parameters,
the uniform ($\q=\bm{0}$) bond order
\cite{Kawaguchi-CDW,Tsuchiizu2018},  
charge loop current
\cite{Varma}, 
spin loop current
\cite{HKontani-sLC},
and various composite orders
\cite{Chubukov-cuprate2014}
have been discussed intensively.

In this article,
we discuss the origin of the unconventional density waves
by considering the electron correlation effects.
We explain the importance of the quantum interference 
as the origin of the unconventional density waves.
It is important to uncover the origin of the CDW and the 
pseudogap order parameter, since the QCPs of these orders 
might be significant for the pairing mechanism,
as indicated by the phase diagram in Fig. \ref{fig:YBCO-phase}.

\subsection{Other exotic strongly correlated electron systems
\label{sec:Kontani1-4}
}
In previous subsections, we explained the phase diagrams of
two high-$T_{\rm c}$ superconductors.
Other candidates of superconductors due to electron correlation 
would be Ru-based superconductor Sr$_2$RuO$_4$
\cite{Maeno},
Cr-based superconductors 
$A_2$Cr$_3$As$_3$ and $A$Cr$_3$As$_3$ ($A$=K, Rb, Cs, Na)
\cite{Cr-SC},
V-based superconductors
\cite{Kagome,Kagome-Tazai},
and organic superconductors
$\kappa$-(BEDT-TTF)$_2$X
\cite{Kanoda-rev,Kino-Kontani}
and (TMTSF)$_2$PF$_6$.
\cite{Jerome}.

In addition, there are many exotic superconductors
in $f$-electron systems.
The $d$-wave superconductivity in Ce$M$In$_5$ ($M=$Rh, Co, Ir)
has been established by many experiments
\cite{HF-rev}.
Also, spin triplet superconductivity is expected 
in several U-based compounds;
UPt$_3$, UGe$_2$, UCoGe, UTe$_2$, etc
\cite{HF-rev}.
Among these compounds, 
CeCu$_2$Si$_2$ is one of the most famous $f$-electron superconductor,
since it is the first discovered unconventional superconductor in 1979
\cite{Steglich}.
For long time, CeCu$_2$Si$_2$ is believed to be a $d$-wave superconductor
since strong AFM fluctuations are observed above $T_{\rm c}$.
However, recent experimental study has revealed the realization of  
the fully-gapped $s$-wave gap state \cite{Yama-122,Kit1-122,Kit2-122},
based on the penetration depth, impurity effect on $T_{\rm c}$,
and specific heat measurements.
These surprising results mean that very
strong (charge-channel) attractive pairing interaction does exist
in heavy fermion systems with large on-site $U$.

\begin{figure}[htb]
\centering
\includegraphics[width=.55\linewidth]{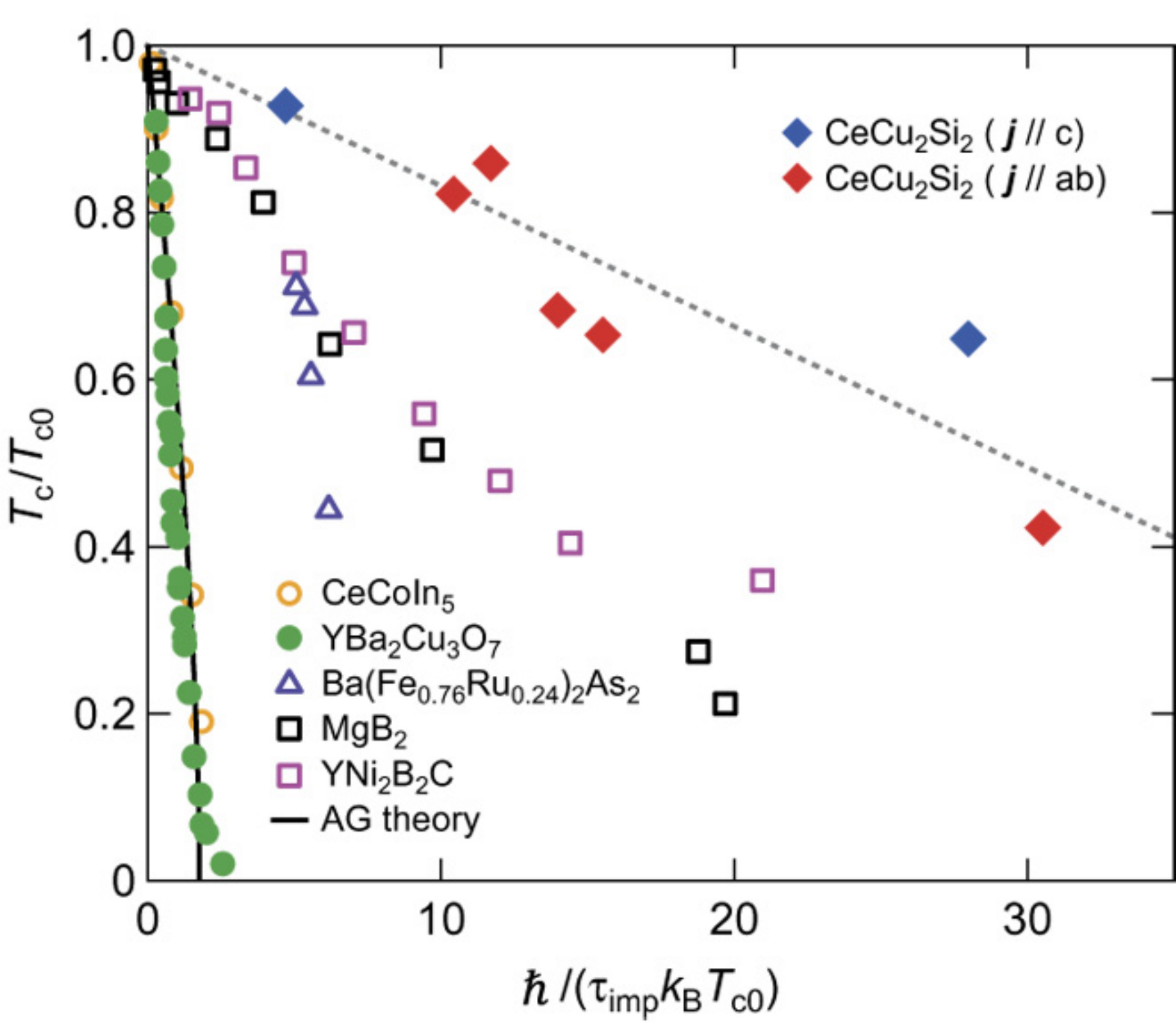}
\caption{
Impurity effect on $T_{\rm c}$ in CeCu$_2$Si$_2$
measured by electron irradiation study.
Reprinted from Ref. \cite{Yama-122}.
}
\label{fig:CeCu2Si2-impurity}
\end{figure}

Figure \ref{fig:CeCu2Si2-impurity}
shows the impurity effect on $T_{\rm c}$ in CeCu$_2$Si$_2$ 
and other superconductors obtained by the irradiation experiments
\cite{Yama-122}.
The horizontal axis represents the normalized impurity scattering strength.
The $T_{\rm c}$ of $d$-wave superconductors, YBCO and CeCoIn$_5$,
are quickly suppressed by impurity scattering,
by following the Abrikosov-Gorkov (AG) theory.
This is because the gap function with sign reversal is 
easily destroyed by the impurity scattering when the mean-free-path 
of electrons $l_{\rm m.f.p}$ is comparable to the coherence length.
In sharp contrast, $T_{\rm c}$ in CeCu$_2$Si$_2$ is extremely 
strong against impurity scattering, much stronger than
the conventional (anisotropic) $s$-wave superconductor
MgB$_2$ and YNi$_2$B$_2$C.
This experiment clearly indicates that CeCu$_2$Si$_2$ 
exhibits the $s$-wave gap function without sign reversal.

To understand the origin of $s$-wave pairing glue in CeCu$_2$Si$_2$,
in Sect. \ref{sec:Tazai},
we focus on the higher-rank multipole operators,
which is active thanks to the strong spin-orbit interaction (SOI)
in heavy fermion systems.
We find that charge-channel (quadrupole or hexadecapole) 
multipole fluctuations are induced by the quantum interference 
among magnetic fluctuations near the magnetic QCP.
In addition, the obtained multipole fluctuations  
are magnified with the aid of moderate electron-phonon interaction.
We reveal that the derived strong charge-channel multipole fluctuations 
give sizable attractive pairing interaction,
so the $s$-wave superconductivity in CeCu$_2$Si$_2$
is naturally understood in terms of the multipole fluctuations theory.

The present theory 
is also useful for understanding the mechanism of 
rich multipolar order physics in heavy fermion systems, 
many of which have been unsolved at present.
We successfully applied the present theory
to the multipole order in CeB$_6$ \cite{tazai-last}, 
which is a famous multipolar heavy fermion system.

\section{Form factors of unconventional orders}
\label{sec:Kontani2}

In strongly correlated metals, various symmetry breakings occur
in the normal state.
In this section, we discuss the variety of symmetry breaking states
and their microscopic origin based on the recently developed theories.

\subsection{Form factors}
\label{sec:Kontani2-1}

Here, we introduce the unconventional density orders
in a square lattice single orbital tight-binding Hubbard
Hamiltonian, $H=H_0+H_I$.
Here, $H_I$ is the Coulomb interaction term,
and $H_0$ is the kinetic term given as
\begin{eqnarray}
H_0 = \sum_{i,j,\s} t_{i,j} c_{i\s}^\dagger c_{j\s}
= \sum_{\k,\s} \e_\k c_{\k\s}^\dagger c_{\k\s}
\end{eqnarray}
where $i,j$ are the site indices in real space,
$\k$ is the momentum, and $\s \ (=\pm1)$ is the spin index.
Both $t_{i,j}=t_{i-j}$ and its Fourier transformation $\e_\k$
belong to $A_{1g}$ representation of the square lattice.

Due to the Coulomb interaction $H_I$,
the kinetic term is modified by the self-energy correction $\Sigma$.
In this subsection, we neglect its energy-dependence 
to simplify the discussion.
The self-energy without symmetry breaking,
$\Sigma_{i,j}^0$, belongs $A_{1g}$ representation.
In the field theory, the quantum phase transition
is given by the spontaneous symmetry breaking in the self-energy.
The self-energy after the phase transition is expressed as
$\Sigma_{i,j}^\s=\Sigma_{i,j}^0+\Delta\Sigma_{i,j}^\s$,
where $\Delta\Sigma_{i,j}^\s$ is the order parameter 
that belongs to a non-$A_{1g}$ representation.

Its Fourier transformation is given as
\begin{eqnarray}
\Delta\Sigma_{\q}^\s(\k)
= \sum_{i,j}\Delta \Sigma_{i,j}^\s 
e^{i\k\cdot({\bm r}_i-{\bm r}_j)}e^{-i\q\cdot{\bm r}_j}
\label{eqn:Fourier}
\end{eqnarray}
where $\q$ is the wavenumber of the order parameter.
Hereafter, we put 
$\Delta \Sigma_{\q}^\s(\k)\equiv \Delta E \cdot f_{\q}^\s(\k)$,
where $f_{\q}^s(\k)$ is the normalized 
($\frac1N \sum_\k|f_{\q}^\s(\k)|^2=1$)
dimensionless order parameter.
We also introduce the 
charge (spin channel form factor 
$f_{\q}^{c(s)}(\k) \equiv f_{\q}^\uparrow(\k) +(-) f_{\q}^\downarrow(\k)$.

In the present article, we call 
$f_{\q}^{x}(\k)$ ($x=s,c$) or its Fourier transformation $f_{i,j}^{x}$
the ``form factor'' of the density wave.
The form factor is a central and essential concept
of the unconventional density-wave state.
We stress that the Hermite condition
$f_{i,j}^{x}= (f_{j,i}^{x})^*$ should be satisfied
because we consider the thermal equilibrium phase transitions.

\begin{figure}[htb]
\centering
\includegraphics[width=.99\linewidth]{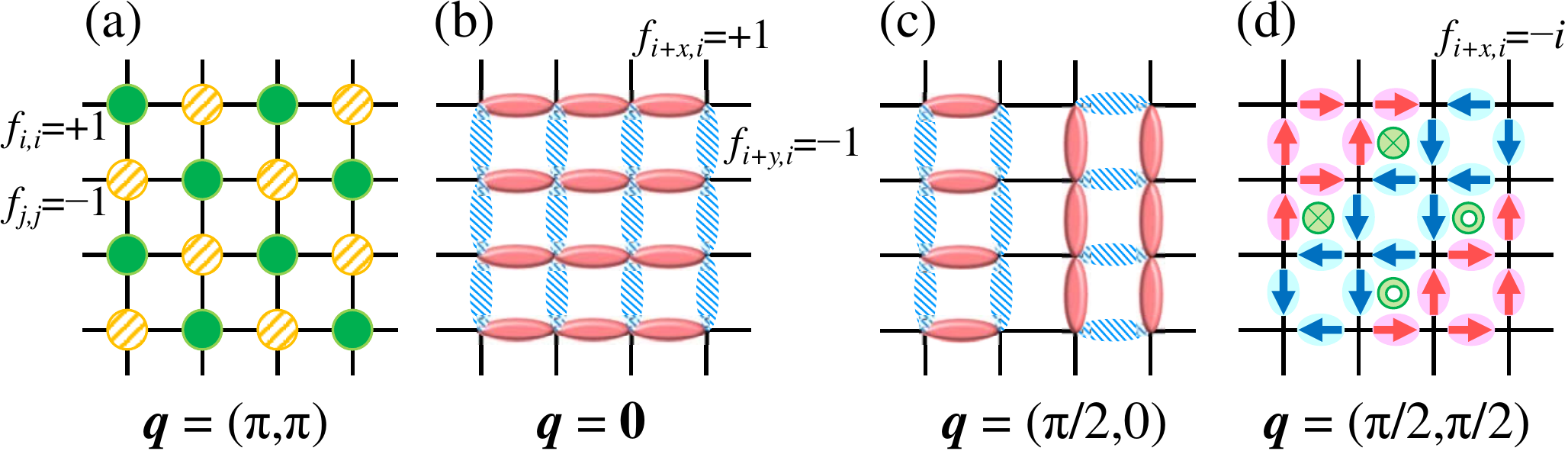}
\caption{
(a) Local density order (CDW or SDW) with the wavevector $\q=(\pi,pi)$.
The form factor is $f=1$.
(b)(c) Non-local $d$-wave bond order with the wavevector
$\q={\bm 0}$ and $\q=(\pi/2,0)$.
The form factor is $f=\cos k_x - \cos k_y$.
(d) Non-local current order with the wavevector $\q=(\pi/2,\pi/2)$.
The form factor is $f=\sin k_x + \sin k_y$.
}
\label{fig:schematic-OP}
\end{figure}

Here, we introduce typical examples of the density-wave states,
which are actually expected to emerge in several 
strongly correlated metals.
\\

{\bf (i) Conventional local order $f_{i,i}^x={\rm real}$:} \ 
Figure \ref{fig:schematic-OP} (a) 
exhibits the conventional CDW or SDW at wavevector $\q=(\pi,\pi)$.
Here, $f_{i,i}^{x}= e^{i{\bm r}_i\cdot\q}$ for general $\q$.
(Note that the ferro-CDW is prohibited by the charge conservation law.)
\\

{\bf (ii) Nonlocal bond order $f_{i,j}^x=f_{j,i}^x={\rm real}$:} \ 
Figures \ref{fig:schematic-OP} (b) and (c)
exhibits the $d$-wave bond orders
at wavevectors $\q={\bm 0}$ and $\q=(\pi/2,0)$, respectively.
Its form factor at $\q={\bm0}$ is $f_{x^2-y^2}^x(\k) = \cos k_x - \cos k_y$ in $\k$-space,
and its real-space expression is
$f_{i,j}^x= (\delta_{i_x,j_x\pm1}\delta_{i_y,j_y}-(x\rightarrow y) )$,
where $(i_x,i_y)$ is the coordinate of site $i$.
Here, the relation $f_{i,j}^x=f_{j,i}^x$ holds.
The $d$-wave bond order is observed in the CDW phase in cuprates
and in the nematic state in FeSe (within the $d_{xy}$ orbital).
\\

{\bf (iii) Nonlocal current order $f_{i,j}^x=-f_{j,i}^x={\rm imaginary}$:} \
Figure \ref{fig:schematic-OP} (d)
exhibits the current order at wavevector $\q=(\pi/2,\pi/2)$.
Its form factor at $\q={\bm0}$ is $f_{x+y}^x(\k) = \sin k_x + \sin k_y$ in $\k$-space,
and its real-space expression is
$f_{i,j}^x= i(\delta_{i_x,j_x+1}-\delta_{i_x,j_x-1})\delta_{i_y,j_y}
+(x\rightarrow y) ) e^{i({\bm r}_i+{\bm r}_j)\cdot\q/2}$.
Here, the relation $f_{i,j}^x=-f_{j,i}^x$ holds.
$f^{c(s)}$ induces the loop charge (spin) current order,
and its existence is hotly discussed in cuprates and irritates.
(Note that the macroscopic current is prohibited by the 
translational symmetry even in ferro-current orders
\cite{HKontani-sLC}.
\\

Both bond and current orders are difficult to 
be explained within the mean-field-level approximation,
as we will explain in Sect. \ref{sec:Kontani2-2}.
However, these unconventional orders are caused by the 
non-local vertex corrections (VCs),
which is the main issue of the present article.

\begin{figure}[htb]
\centering
\includegraphics[width=.99\linewidth]{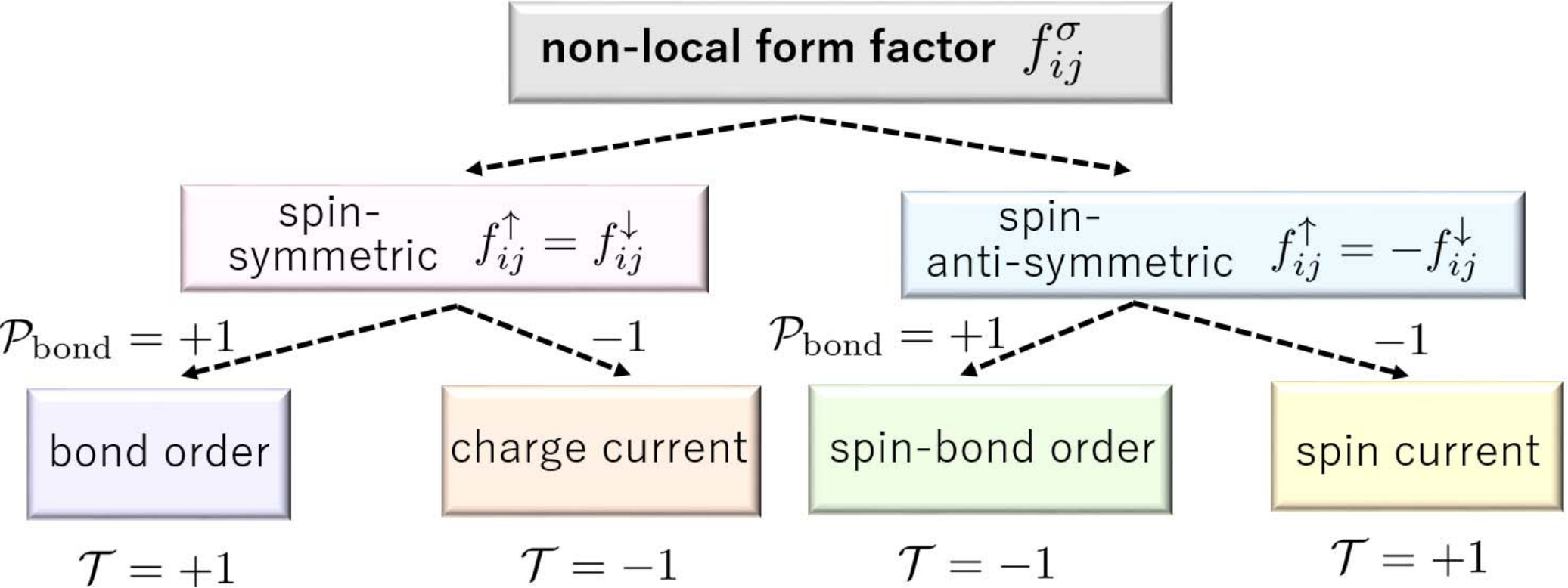}
\caption{
Classification of the non-local form factor $f_{i,j}^\s$.
${\cal P}_{\rm bond}$ is the parity of the site exchange 
($i\leftrightarrow j$),
and ${\cal T}$ represents the time-reversal symmetry.
The hermitian condition of form factor is 
${\cal S}\cdot{\cal T}\cdot{\cal P}_{\rm bond}=1$,
where ${\cal T}$ is the parity of spin exchange.
}
\label{fig:Tazai-Tree}
\end{figure}

In Fig. \ref{fig:Tazai-Tree},
we summarize the classification of the non-local form factor $f_{i,j}^\s$.
Here, ${\cal P}_{\rm bond}$ 
is the parity (=eigenvalue $\pm1$) of the site exchange operator 
${\hat P}_{\rm bond}\{f_{i,j}^\s\} \equiv f_{j,i}^\s$.
${\cal P}_{\rm spin}$ is the parity of the spin-flip operator 
${\hat P}_{\rm spin}\{f_{i,j}^\s\} \equiv f_{i,j}^{-\s}$.
${\cal T}$ is the parity of the time-reversing operator 
${\hat T}\{f_{i,j}^\s\} \equiv (f_{i,j}^{-\s})^*$.

Finally, we introduce the orbital order in
a square lattice two-orbital tight-binding Hubbard model.
Its kinetic term is 
\begin{eqnarray}
H_0=\sum_{\s,i,j,l,m} 
t_{il,jm} c_{i l \s}^\dagger c_{j m \s},
\label{eqn:H0-two-orbital}
\end{eqnarray}
where $l,m=1,2$ represents the $d$-orbital ($1=xz$, $2=yz$).
$t_{il,jm}$ is the hopping integral 
between the orbital $l$ at site $i$ and
the orbital $m$ at site $j$.

\begin{figure}[htb]
\centering
\includegraphics[width=.9\linewidth]{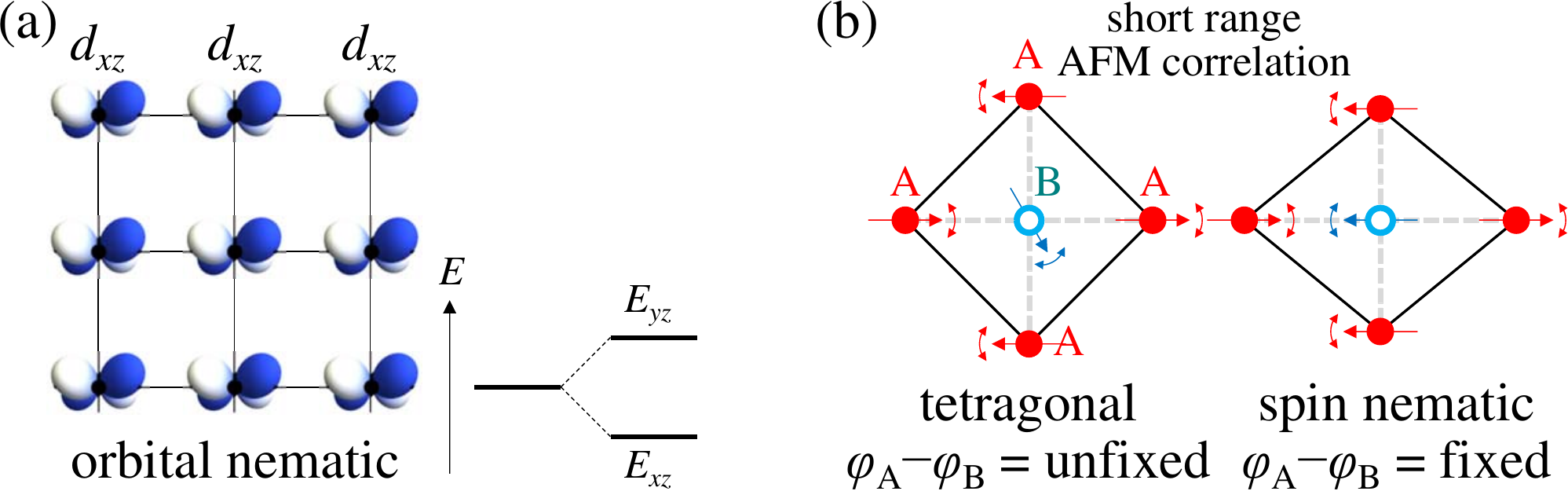}
\caption{
(a) Orbital nematic ($\q={\bm 0}$) order 
due to the orbital polarization $n_{i,xz}> n_{i,yz}$.
The form factor is ${\hat f}={\hat \tau}_z$ in the 
($d_{xz}$, $d_{yz}$) orbital basis.
The corresponding energy splitting is shown.
(b) Tetragonal ($C_4$) state and (d) nematic ($C_2$) state
in square lattice model with spin degrees of freedom.
$\phi_A$ ($\phi_B$) is the spin direction of sublattice A (B).
In the paramagnetic state, $\langle\phi_A\rangle=\langle\phi_B\rangle=0$.
The spin nematic ($C_2$) state is realized 
when $\langle\phi_A\phi_B\rangle\ne0$, depicted in (d).
}
\label{fig:schematic-OP2}
\end{figure}

In $2$-orbital Hubbard models,
the form factor is expressed as the $2\times 2$ matrix form,
${\hat f}_{\q}^{c,s}(\k)$.
In this subsection, we consider a simple
$\k$-independent form factor ${\hat f}^c={\hat \tau}_z$,
where ${\hat \tau}_\mu$ is the Pauli matrix in the $d$-orbital basis.
The induced orbital polarization $n_{i,xy}\ne n_{i,yz}$
at $\q={\bm 0}$ is depicted in Fig. \ref{fig:schematic-OP2} (a).
The orbital order is realized in the nematic phase in
many Fe-based superconductors.
The orbital order is ``non-local'' and ``non-$A_{1g}$'' in the orbital space,
and therefore ``unconventional''.
This unconventional order is difficult to 
be explained within the mean-field-level approximation.
However, it is caused by the VCs,
which is the main issue of the present article.

Finally, we explain the scenario of spin nematic order
for the nematic transition proposed in Fe-based superconductors
\cite{Fernandes1,Fernandes2,Fernandes-Nature,Chubukov-nematic-rev}.
Figure \ref{fig:schematic-OP2} (b) shows the 
square lattice quantum spin model.
When the magnetic frustration is large,
the paramagnetic state 
$\langle\phi_A\rangle=\langle\phi_B\rangle=0$ is realized,
where $\phi_A$ ($\phi_B$) is the spin direction of sublattice A (B).
Then, nonmagnetic spin nematic ($C_2$) state can be established
by emerging the order parameter $\langle\phi_A-\phi_B\rangle\ne0$,
which is depicted in Fig. \ref{fig:schematic-OP2} (b).
(Mathematically, this is the non-local spin quadrupole order.)
In the spin nematic scenario,
only the rotational symmetry breaks $(C_4\rightarrow C_2)$ at $T=T_S$,
and the time-reversal symmetry breaks at $T=T_{N}$ successively.
In this sense, the spin nematic order is 
sometimes called the ``vestigial order''.
The expected phase diagram with $T_S>T_N (>0)$ would be consistent 
with 1111 and 122 phase diagrams in Figs. \ref{fig:phase-1111} (a) and (b), 
while the origin of the nematicity without magnetism in 
Fe(Se,Te) and Fe(Se,S), shown in Figs. \ref{fig:phase-11} (a) and (b) 
respectively, may not be trivial.

\subsection{Difficulties in the mean-field approximation}
\label{sec:Kontani2-2}

Here, we discuss the microscopic mechanism of 
the unconventional orders introduced in the previous subsections.

\begin{figure}[htb]
\centering
\includegraphics[width=.75\linewidth]{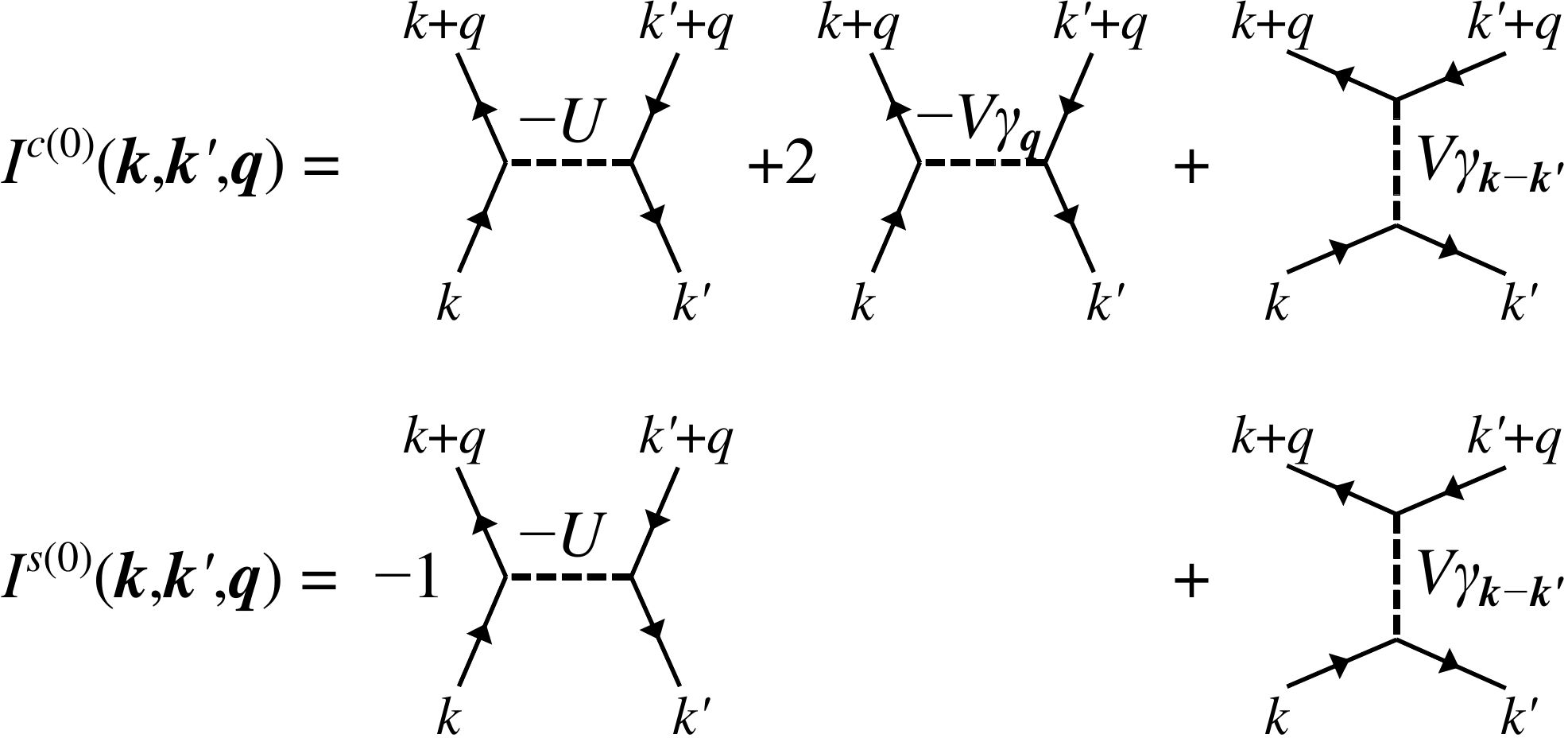}
\caption{
Bare charge- and spin-channel interactions.
$I^{c(0)}=I^{(0)}_{\s,\s}+I^{(0)}_{\s,{\bar \s}}$ and 
$I^{s(0)}=I^{(0)}_{\s,\s}-I^{(0)}_{\s,{\bar \s}}$.
Here, the Fock term $\gamma_\k= 2(\cos k_x + \cos k_y)$
originates from non-local interaction $V$.
}
\label{fig:Gamma0}
\end{figure}

As the first step, we perform the mean-field analysis
for the single orbital Hubbard model.
We introduce the on-site Coulomb interaction $U$ 
in addition to the {\it off-site Coulomb interaction $V_{i,j}$}
in the interaction term:
\begin{eqnarray}
H_I&=&\frac U2 \sum_i n_{i\s}n_{i-\s}
+ \sum_{i,\ne j}V_{i-j} n_{i}n_{j}
\label{eqn:H1r}
\end{eqnarray}
where $n_{i\s}=c_{i\s}^\dagger c_{i\s}$ and $n_{i}=\sum_\s n_{i\s}$,
and $\s=\pm1$ is the spin index.
For simplicity, here we consider $V_{i-j}$ only for the nearest sites.
As a consequence of $SU(2)$ symmetry in spin space,
$H_I$ is written as
\begin{eqnarray}
H_I&=&\frac1N \sum_{\k\k'\q,\s\s'} I_{\s\s'}^{(0)}(\k,\k',\q)
c_{\k+\q,\s}^\dagger c_{\k',\s'}^\dagger c_{\k'+\q,\s'} c_{\k,\s}
\\
I_{\s\s'}^{(0)}(\k,\k',\q)&=&(I^{c(0)}(\k,\k',\q)+I^{s(0)}(\k,\k',\q)\s\s')/2
\end{eqnarray}
and $I^{c(0)}$ and $I^{s(0)}$ are explicitly given as
\begin{eqnarray}
I^{c(0)}(\k,\k',\q)&=&-U-2V\gamma_\q+V\gamma_{\k-\k'}
\label{eqn:H1c} \\
I^{s(0)}(\k,\k',\q)&=&U+V\gamma_{\k-\k'}
\label{eqn:H1s} 
\end{eqnarray}
where $\displaystyle \gamma_{\k}=2(\cos k_x+\cos k_y)$ originates from the 
Fourier transformation of $V_{i,j}$.
The diagrammatic expressions of $I^{c(0)}$ and $I^{s(0)}$ 
are shown in Fig. \ref{fig:Gamma0}.
The following relationship is important:
$\displaystyle \gamma_{\k-\k'}=\sum_\Gamma^{x\pm y,x^2\pm y^2} f_{\Gamma}(\k)f_{\Gamma}(\k')$,
where $\displaystyle f_{x^2\pm y^2}(\k)=\cos k_x \pm \cos k_y$ and 
$\displaystyle  f_{x\pm y}(\k)=\sin k_x \pm \sin k_y$.
Note that $f_\Gamma$ is normalized as
$\displaystyle \frac1N \sum_\k |f_\Gamma(\k)|^2=1$.

In the mean-field theory,
the Stoner criterion determines the
condition of the phase transition at $\q$.
The spin ($x=s$) or charge ($x=c$) channel Stoner factor at $\q$
with respect to the form factor $f_\Gamma$ is
$\a_x^\Gamma(\q) =  g_x^\Gamma(\q)\chi^{(0)}_{\Gamma}(\q)$, where
\begin{eqnarray}
g_x^\Gamma(\q)&\equiv&\frac1{N^2}\sum_{\k,\k'}I^{x(0)}(\k,\k',\q)f_{\Gamma}(\k+\q/2) f_{\Gamma}(\k'+\q/2) 
\\
\chi^{(0)}_{\Gamma}(\q)
&=& \frac1N \sum_\k \frac{n(\e_{\k+\q})-n(\e_\k)}{\e_\k-\e_{\k+\q}} 
|f_{\Gamma}(\k+\q/2)|^2 
\label{eqn:chi0-Q0}
\end{eqnarray}
and we denote $\a_x^\Gamma \equiv \max_\q \a_x^\Gamma(\q)$.
Note that $\chi^{(0)}_{\Gamma=1}(\q)$ is equal to the  
density-of-states (DOS) $N(0)$ at $\q={\bm0}$.
Here, $g_x^\Gamma(\q)$ for each form factor is given as
\\

{\bf (i) Local order ($\Gamma=1$):}
\begin{eqnarray}
g_c^1(\q)=-U-2V\gamma_\q, \ g_s^1(\q)=U, 
\end{eqnarray}

{\bf (ii) $d$-wave bond order ($\Gamma=x^2-y^2$):}
\begin{eqnarray}
g_c^{x^2-y^2}(\q)= g_s^{x^2-y^2}(\q)=V,
\end{eqnarray}

{\bf (iii) Current order ($\Gamma=x\pm y$):}
\begin{eqnarray}
g_c^{x\pm y}(\q)= g_s^{x\pm y}(\q)=V,
\end{eqnarray}
We note that only the Fock term ($=V \gamma_{\k-\k'}$)
in Eqs. (\ref{eqn:H1c}) and (\ref{eqn:H1s})
contribute to the non-local orders (ii) and (iii).
In contrast, the Hartree term ($= U, 2V\gamma_\q$)
contribute to the local order (i).
Note that we dropped the Fock term for the local order (i) 
because it is very small.

As for the local order (i), at $\q=(\pi,\pi)$,
the SDW Stoner factor $\a_S^1$ is larger than 
the CDW Stoner factor $\a_C^1$ for $U>4V$. 
Therefore, the SDW (CDW) order occurs in the case of 
$U>4V$ ($U<4V$) when the interaction is strong.

As for the non-local orders (ii) and (iii),
$\a_x^{\Gamma}$ ($\Gamma=x^2-y^2, x\pm y$) 
will exceeds SDW Stoner factor $\a_S^{\Gamma=1}$ under the 
condition $4V\gtrsim U$,
because $\chi^{(0)}_{\Gamma}(\q)<4\chi^{(0)}_{\Gamma=1}(\q)$
due to the relation $|f_\Gamma(\k)|\le2$.
Thus, in the present $U$-$V$ Hubbard model,
$\max\{\a_{s}^1,\a_C^1\}$ is larger than $\a_{s,c}^{\Gamma\ne1}$,
so SDW/CDW orders always dominates over the bond and current orders.
Therefore, additional non-local interactions,
such as the RKKY interaction,
must be introduced to realize the bond/current order
\cite{Nersesyan}
within the mean-field approximation.
This difficulty will be resolved by developing the 
many-body theory beyond the RPA, as we will explain 
in Sect. \ref{sec:Kontani3}

Finally, we briefly discuss the Stoner factors
in the two-orbital Hubbard model $H=H_0+H_I$.
The local multiorbital Coulomb interaction $H_I$
is composed of intra- and inter-orbital repulsions, $U$ and $U'$,
in addition to the Hund's coupling, $J$.
\color{black}
The relation $U=U'+2J$ holds with respect to 
$t_{2g}$ (=$xz,yz,xy$) orbitals.
\color{black}
In the mean-field approximation,
the spin-channel Stoner factor is always larger (smaller) 
than the charge channel Stoner factor when $J$ is positive (negative). 
\cite{Takimoto,Kontani-RPA}.
Therefore, under the realistic condition $J>0$,
the SDW always dominates over the orbital order
within the mean-field approximation.
In later sections,
we will explain that the orbital order can emerge even for $J>0$
by introducing the VCs into the theory.

\subsection{Spin fluctuation theory 
\label{sec:Kontani2-3}
}

In the previous subsection, we see that
the non-local (bond or current) order
can be induced by off-site interactions.
However, its realization condition is very severe
within the mean-field approximation.
Hereafter, we study the non-local orders
based on the Hubbard model with local Coulomb interaction $U$
($V_{i-j}\equiv 0$ in Eq. (\ref{eqn:H1r}))
by constructing ``beyond mean-field approximation''.
In other words, we develop the mechanism of non-local orders
driven by the off-site many-body interaction described by 
higher-order VCs.

To prepare for that, we briefly review the 
random-phase-approximation (RPA)
and fluctuation-exchange (FLEX) approximation.
In these approximations,
the irreducible susceptibility without VC is given as
\begin{eqnarray}
\chi^{(0)}(q)&=& -T\sum_k G(k+q)G(k),
\label{eqn:chi0}
\end{eqnarray}
where $q=(\q,\w_l=2\pi Tl)$ and $k=(\k,\e_n=\pi T(2n+1))$.
$G$ is the electron Green function given as
\begin{eqnarray}
G(k)&=&\frac1{i\e_n+\mu-\e_\k-\Sigma(k)}
\end{eqnarray}
where $\Sigma$ is the self-energy.
Note that Eq. (\ref{eqn:chi0}) at $\w_l=0$
becomes $\chi_{\Gamma=1}^{(0)}(\q)$ in Eq. (\ref{eqn:chi0-Q0}) in the case of $\Sigma=0$.

Below, we consider the on-site Coulomb Hubbard model
in Eqs. (\ref{eqn:H1r})-(\ref{eqn:H1s}) with $V=0$.
As we discussed in the previous subsection,
the spin (charge) susceptibility is magnified by 
the factor $(1-\a_{s(c)}^{\Gamma=1}(\q))^{-1}$,
and therefore
%
\begin{eqnarray}
\chi^{s(c)}(q)= \chi^{(0)}(q)/(1-U^{s(c)}\chi^{(0)}(q))
\label{eqn:chi-RPA}
\end{eqnarray}
where $U^{s}=U$ and $U^{c}=-U$.
Thus, $\chi^{s}(\Q,0)$ diverges when $\a_S$ approaches to unity.
In cuprate superconductors, $\chi^{(0)}(\Q,0)\sim N(0)\sim1 {\rm [eV^{-1}]}$,
whereas $\chi^{s}(\Q,0)$ exceeds $100{\rm [eV^{-1}]}$
in optimally or slightly under-doped cuprates above $T_{\rm c}$
\cite{neutron}.
This fact indicates that $(1-\a_S)^{-1}\gg100$ in cuprate superconductors.

The self-energy due to the spin fluctuations is given as
\begin{eqnarray}
\Sigma(k)&=& T\sum_q V^{\rm FLEX}(q)G(k+q),
\\
V^{\Sigma}(q)&=& \frac32 (U^{s})^2\chi^s(q)
+\frac12 (U^{c})^2\chi^c(q)-(U^{s})^2\chi^{(0)}(q)
\end{eqnarray}
which becomes important near the magnetic crucial point; 
$\a_S^{\Gamma=1}\lesssim1$.
In the FLEX approximation, both the susceptibilities and 
the self-energy are calculated self-consistently.

For convenience, we introduce a phenomenological form of 
spin susceptibility
 \cite{SCR,Monthoux1,Monthoux2,MMP,Stojkovic}: 
\begin{eqnarray}
\chi^{s}(\q,\w+i0) = \sum_{\Q}
 \frac{\chi_Q}{1+\xi_{s}^2 (\q-\Q)^2 - i \w/\w_{s}} ,
 \label{eqn:kai_qw}
\end{eqnarray}
where $\Q$ is the AF wavevector
and $\xi_{s}$ is the AF correlation length.
In cuprates for $T>T^*$,
both $\chi_Q$ and $1/\w_{s}$ are scaled by $\xi_{s}^2$ 
as follows \cite{Mag_scaling}:
\begin{eqnarray}
& &\xi_{s}^2 \approx c_0/( T-\theta_s ), 
 \label{eqn:parameters1} \\
& &\chi_Q \approx c_1\cdot \xi_{s}^2, \ \ \
 1/\w_{s} \approx c_2\cdot \xi_{s}^2,
 \label{eqn:parameters2}
\end{eqnarray}
where $\theta_S$, $c_0$, $c_1$ and $c_2$ are constants.
Since $\chi_Q\w_{s}\propto \xi_{s}^0$ 
in eq. (\ref{eqn:parameters2}), the dynamical exponent is $z=2$.
The relationship $\w_{s} \gtrsim T$ is satisfied in optimally-doped cuprates.

The relationships in Eqs. (\ref{eqn:parameters1}) and 
(\ref{eqn:parameters2}) are explained by the 
self-consistent-renormalization (SCR) theory, 
in which the ``mode-mode coupling effect'' 
is calculated self-consistently\cite{Moriya-text}.
Relationships (\ref{eqn:parameters1}) and (\ref{eqn:parameters2})
are also satisfied by the FLEX approximation 
\cite{Bickers,Dahm,Takimoto,d-p-model,Trellis,Wermbter,Manske-PRB}.
In both theories, long-range magnetic order does not occur
in pure 2D systems, because both theories satisfy
the Mermin-Wagner theorem \cite{GVI}.


We stress that the charge susceptibility 
in Eq. (\ref{eqn:chi-RPA}) remains small.
Therefore, theoretical methods beyond the FLEX approximation
should be established to explain the strong charge-channel fluctuations 
observed in Fe-based and cuprate superconductors.

\subsection{Vertex corrections for susceptibilities: Self-consisting vertex correction (SC-VC) theory}
\label{sec:Kontani2-4}

Now, we analyze SDW and CDW susceptibilities with constant form factor 
($f_{\Gamma=1}=1$) by including the VCs:
\begin{eqnarray}
\chi^x(q)&=& (1-U^{x}\Phi^x(q))^{-1}\Phi^x(q),
\label{eqn:suscep-SCVC}
\\
\Phi^x(q)&=& \chi^{(0)}(q)+X^x(q),
\\
X^x(q)&=&T^2\sum_{kk'} G(k+q)G(k)\Gamma^{x}_{U-{\rm irr}}(k,k';q)G(k'+q)G(k')
\label{eqn:suscep-SCVC-X}
\end{eqnarray}
where $x=c,s$.
Here, $\Phi^x(q)$ and $\Gamma^x_{U-{\rm irr}}(k,k';q)$ 
are irreducible with respect to single $U^x$.
$\Gamma^x$ is expressed by using the 
irreducible vertex with respect to the particle-hole (p-h) channel $I^{x}$
by using the following Bethe-Salpeter equation.
\begin{eqnarray}
\Gamma^x(k,k';q)= I^{x}(k,p;q)
-T\sum_p I^{x}(k,p;q)G(p+q)G(p)\Gamma^{x}(p,k';q)
\label{eqn:BS1}
\end{eqnarray}
Note that 
$\Gamma^{x}_{U-{\rm irr}}$ is given by Eq. (\ref{eqn:BS1})
by replacing $I^{x}$ with $I^{x}-U^x$

\begin{figure}[htb]
\centering
\includegraphics[width=.99\linewidth]{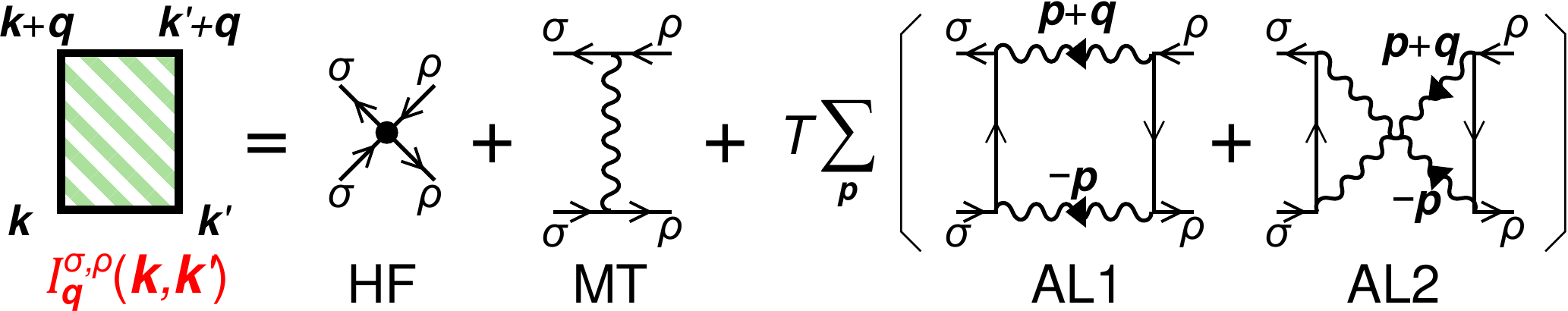}
\caption{
Hartree-Fock (HF) term, MT term, and two AL terms 
included in the irreducible four-point vertex $I$.
Only the HF term is included in the RPA.
The nonlocal interaction by MT and AL terms cause
exotic bond and current orders
\cite{HKontani-PRB2011,SOnari-PRL2012,HKontani-sLC,RTazai-PRB2021}.
}
\label{fig:fig-MTAL}
\end{figure}

To avoid unphysical results,
it is important to satisfy the following Ward identity:
%
\begin{eqnarray}
\Sigma_{\s}(x_i,x_j)&=&
\frac{\delta\Phi_{\rm LW}}{\delta G(x_i,x_j)}
\\
I_{\s,\s;\s'\s'}(x_i,x_j;x_i',x_j')&=&
-\frac{\delta^2\Phi_{\rm LW}}{\delta G(x_i,x_j)\delta G(x_i',x_j')}
\end{eqnarray}
where $x_i=({\bm r}_i,\e_n)$, and
$\Phi_{\rm LW}$ is the Luttinger-Ward function.
In the $\k$-representation,
the Luttinger-Ward function in the FLEX scheme is given as
\begin{eqnarray}
\Phi_{\rm LW}= {\rm Tr} [ \frac32 {\rm ln}(1-U\chi^0(q))
+\frac12 {\rm ln}(1+U\chi^0(q)) +\frac{U^2}{2}\{\chi^0(q)\}^2
+U\chi^0(q)]
\end{eqnarray}
Then, 
$\Gamma_{\s,\s;\s'\s'}^I$ is given by three terms:
one Maki-Thompson (MT) and two Aslamazov-Larkin (AL) terms.
\begin{eqnarray}
I^{x}&=&I^{x,{\rm MT}}+I^{x,{\rm AL1}}+I^{x,{\rm AL2}}
\label{eqn:MTAL}
\end{eqnarray}
which are expressed in Fig. \ref{fig:fig-MTAL},
and their analytic expressions are given in 
Ref. \cite{HKontani-sLC}.

For general wavevector $\q$ of the density-wave,
the kernel in Eq. (\ref{eqn:MTAL}) in the momentum representation is given as
\begin{eqnarray}
&& I^{c,{\rm MT}}(\q;k,k')=
-\left(\frac32 V_0^s(k-k')+\frac12 V_0^c(k-k')\right)
 \\
&& I^{c,{\rm AL}}(\q;k,k')=
T\sum_p \left( \frac32 V_0^s(p+{\q})V_0^s(p) 
+\frac12 V_0^c(p+{\q})V_0^c(p) \right)
G(k-p)[G(k'-p)+G(k'-p+\q)]
\end{eqnarray}
where double-counting terms in the $U^2$-order should be subtracted.
%
%
Near the magnetic QCP, the relationship 
$\Phi^c(\q) \gg \chi^{(0)}(\q)$ is realized for $|\q|\ll 1$
since $I^{c,{\rm AL1,2}}$ is proportional to the convolution, 
$C_q\equiv \sum_\k \chi^s(\k+\q)\chi^s(\k)\propto \xi_s^2$.
(In contrast, $\Phi^s(\q) \sim \chi^{(0)}(\q)$ even for $\xi_s\gg1$
\cite{YYamakawa-PRX2016}.)
In a single orbital Hubbard model,
the charge susceptibility
$\chi^c(\q)= (1+U\Phi^c(\q))^{-1}\Phi^c(\q)$
is always smaller than $U^{-1}$ due to the suppression by the Hartree term.
In multiorbital Hubbard models, in contrast, the orbital order
is realized when $\Phi^c(\q)\gg U^{-1}$, as we will explain 
in the next section.

Hereafter, we calculate the susceptibilities on the basis of
Eqs. (\ref{eqn:suscep-SCVC})-(\ref{eqn:suscep-SCVC-X}) self-consistently
by setting $\Gamma^c_{U-{\rm irr}}=I^{c,{\rm MT}}+I^{c,{\rm AL}}$ for simplicity.
We call this theoretical framework the
self-consistent vertex correction (SC-VC) method
\cite{SOnari-PRL2012}.
The nematic order in Fe-based superconductors
is satisfactorily derived from the SC-VC method.

\section{Why vertex corrections cause unconventional orders?
\label{sec:Kontani3}
}

\subsection{Orbital order due to vertex corrections
\label{sec:Kontani3-1}
}

\begin{figure}[htb]
\centering
\includegraphics[width=.8\linewidth]{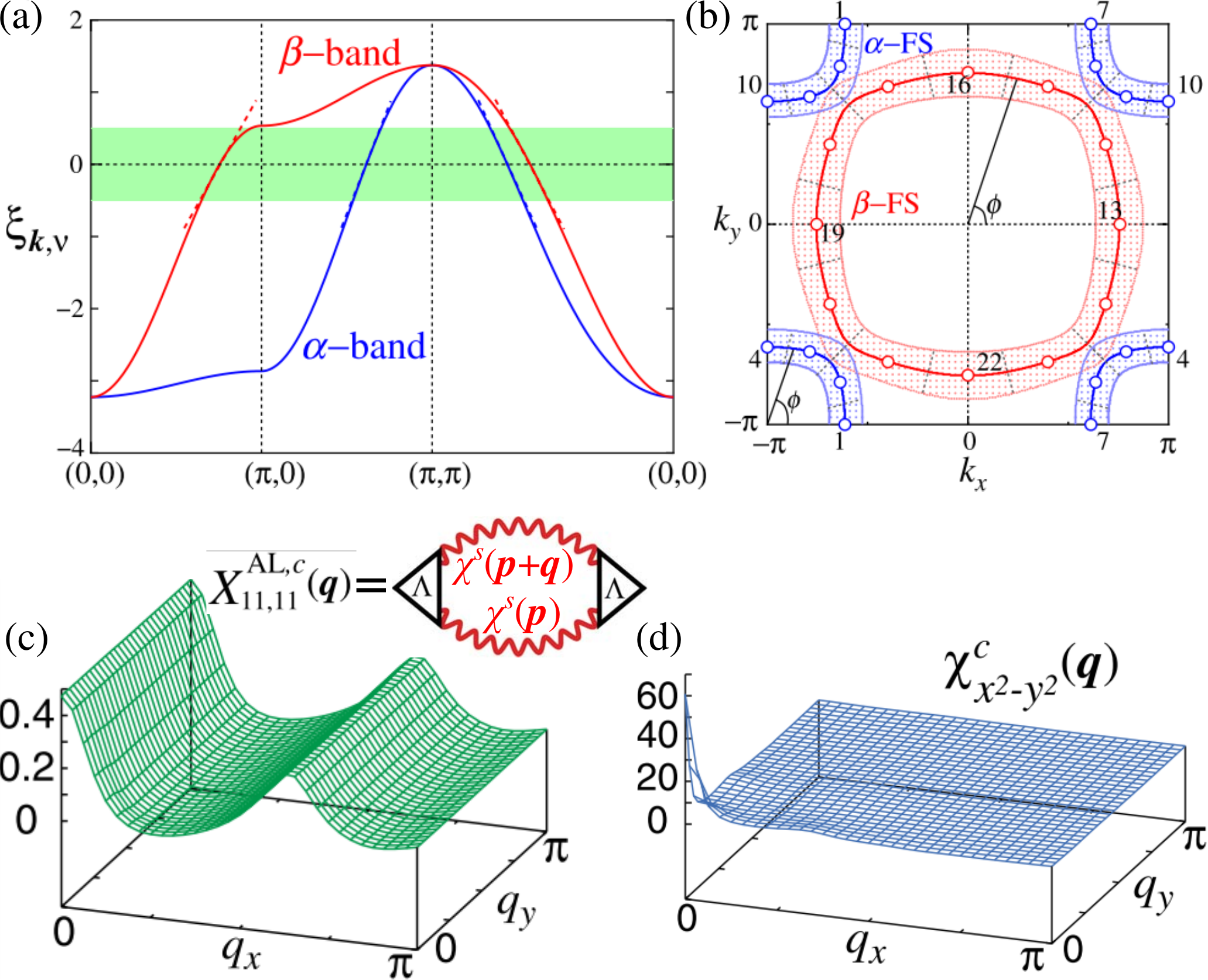}
\caption{
(a) Band structure of the two-orbital model.
The low-energy excitations of electrons 
($|\xi_{\bm k,\nu}^\mathrm{linear}| \le \Lambda_0$)
are denoted by the shaded area.
(b)  FSs of the two-orbital model for $n=2.7$.
The colors correspond to $1 = xz$ (red) and $2 = yz$ (blue).
The arrow represents the major nesting vector.
The patch index ($1\sim 24$) on the FSs is shown.
(c) AL term $X^c_{11,11}(\q)$ and (d) orbital susceptibility
$\chi^c_{x^2-y^2}(\q)$ at $n=3.3$ obtained by the SC-VC method.
Strong ferro-orbital susceptibility is induced by
large $X^c_{11,11}({\bm 0})$.
Cited from Ref. \cite{Ohno:2013hc}.
}
\label{fig:FS-2orbital}
\end{figure}

In order to understand why exotic density waves are caused by the VCs,
we first explain the orbital order in a simple two-orbital Hubbard model.
We demonstrate that $\Phi^c({\bm 0})\gg \chi^{(0)}({\bm 0})$
due to AL terms give orbital order in multiorbital Hubbard models with $U$.
Here, we consider the following simple two-orbital model $H=H_0+H_I$:
\begin{eqnarray}
H_0&=&\sum_{\k;\s=\uparrow,\downarrow;l,m=1,2} 
\xi_\k^{lm} c_{\k l \s}^\dagger c_{\k m \s},
\label{eqn:Ham0-2orb}
\\
H_I&=& U\sum_{i,l} n_{il\uparrow}n_{il\downarrow}
+U'\sum_{i,l< m} n_{il}n_{im}
\nonumber \\
& &
+J\sum_{i,l<m,\s\s'} c_{im\s}^\dagger c_{il\s'}^\dagger 
c_{im\s'}c_{il\s}
+J'\sum_{i,l\ne m} c_{il\uparrow}^\dagger c_{il\downarrow}^\dagger 
c_{im\downarrow}c_{im\uparrow}
\label{eqn:HamI-2orb}
\end{eqnarray}
where $n_{il\s}= c_{il\s}^\dagger c_{il\s}$ and $n_{il}= \sum_\s n_{il\s}$, and
$l,m=1,2$ represents the $d$-orbital; $1=xz$ and $2=yz$.
This model describes the $\a$-FS and $\b$-FS
of Ru-oxides, and
it has been used in the study of 
anomalous and spin Hall effects \cite{Kontani-AHE}.
Here, we introduce the nearest intra-orbital ($t$) and the 
next-nearest inter-orbital ($t'$) hopping integrals;
$\xi_\k^{11}=-2t\cos k_x$, $\xi_\k^{22}=-2t\cos k_y$, 
and $\xi_\k^{12}= 4t'\sin k_x \sin k_y$.
The bandstructure and the FSs for the electron filling $n=3.3$ 
are shown in Figs. \ref{fig:FS-2orbital} (a) and (b), respectively.
$H_I$ represents the multiorbital Coulomb
interaction composed of intra (inter) orbital interaction  
$U$ ($U'$) and the exchange interaction $J$.

Hereafter, we calculate the susceptibilities by using the RPA.
The local quadruple (order) order parameters in the present model is
\cite{Kontani-RPA,Ohno:2013hc}:
\begin{eqnarray}
{\hat O}_{x^2-y^2}^j 
\!\! &=& \!\!\! 
\sum_\sigma
(c_{j,1,\sigma}^\dagger c_{j,1,\sigma}-c_{j,2,\sigma}^\dagger c_{j,2,\sigma})
= n_{j,1} - n_{j,2},  \,\,\,
\label{eq:Ox2-y2}
\\
{\hat O}_{xy}^j 
\!\! &=& \!\!\!
\sum_\sigma 
(c_{j,1,\sigma}^\dagger c_{j,2,\sigma}+c_{j,2,\sigma}^\dagger c_{j,1,\sigma}),
\end{eqnarray}
where $j$ is the site index.
The quadrupole susceptibility per spin is given by
($\gamma=x^2-y^2$ or $xy$)
$\chi_\gamma^c(\bm q) = \frac{1}{2}\int_0^{\beta} d\tau 
\left\langle T_\tau \, {\hat O}_{\gamma}(\bm q,\tau) \, 
{\hat O}_{\gamma}(-\bm q,0)
\right\rangle
$,
where $\tau$ is the imaginary time and $\beta = 1/(k_B T)$.
The divergence of $\chi^c_{x^2-y^2}(\bm q=\bm 0)$ reflects the
emergence of the orbital nematic state ($\langle n_{xz}\rangle \neq
\langle n_{yz}\rangle$).
In addition, we analyze the charge susceptibility
$\chi^{\mathrm{c}}_{\Gamma=1}(\bm q)= \frac{1}{2}
  \int_0^\beta d\tau  \langle T_\tau n(\bm q,\tau) n(-\bm q,0)\rangle$
where $n(\bm q,\tau)$ is the charge density operator.

In this model,
the $2\times2$ matrix Green function in the $d$-orbital basis is
\begin{eqnarray}
{\hat G}(k)&=&((i\e_n+\mu){\hat 1} - \xi_\k-\hat\Sigma)^{-1}
\end{eqnarray}
which is approximately diagonal when $|t'/t|\ll1$.
Then, the irreducible susceptibility is also diagonal 
in the orbital basis approximately:
%
\begin{eqnarray}
{\hat \Phi}^x(q)&\approx&
\left(
    \begin{array}{cc}
      \Phi_1^x(q) & 0 \\
      0 & \Phi_2^x(q)
    \end{array}
  \right)
\end{eqnarray}
Here, we consider a simplified Coulomb interaction 
with $J=0$ and $U\ne U'$ in Eq. (\ref{eqn:HamI-2orb}).
According to Fig. \ref{fig:Gamma0},
the spin-channel and charge-channel Hartree terms are given as
${\hat U}^{s}=U{\hat \tau_0}$ and 
${\hat U}^{c}=-U{\hat \tau_0}+2U'{\hat \tau}_x$, respectively.
By following Eq. (\ref{eqn:suscep-SCVC}), we obtain
${\hat \chi}^{x}(q)= (1-{\hat U}^{x}{\hat \Phi}^{x}(q))^{-1}{\hat \Phi}^{x}(q)$.
for $x=s,c$.

Then, the SDW susceptibility at $q=0$ is
$\chi^{s}(0) = \Phi^s_1(1-\a_S)^{-1}$, where $\a_S= U\Phi^s_1$.
Also, the orbital susceptibility with respect to ${\hat O}_{x^2-y^2}$ at $q=0$ is
$\chi_{x^2-y^2}^c (0)= \Phi^c_1(1-\a_C)^{-1}$, where $\a_C=(2U'-U)\Phi^c_1$.
In the RPA, $\Phi^c_1=\chi^{(0)}(0)$, and therefore $\a_S>\a_C$ for $U>U'$.
Since the relation $U>U'$ holds in transition metals,
the SDW instability dominates over the orbital order instability within the RPA.
However, when $\Phi^c_1$ is larger than $\Phi^s_1$ owing to the VCs,
orbital order can emerge in the paramagnetic phase.
This novel orbital order mechanism is important for the nematic state of
Fe-based superconductors.
The divergence of $\chi_{x^2-y^2}^c$ immediately leads to the  
ferro-quadrupole order $n_1 \ne n_2$,
resulting in the ``nematic'' deformation of the FSs.


Finally, we perform the numerical study for the two-orbital model.
Here, we set $J/U=0.1$ under the relation $U=U'+2J$ \cite{Ohno:2013hc}.
In the RPA, $\chi^s(\q)$ have a peak at the nesting vector $\q\sim(0.3\pi,0.3\pi)$.
In contrast, $\chi^c_{x^2-y^2}(\q)\sim O(1)$ in energy unit $|t|$, 
so the orbital order is not realized in the RPA
\cite{Ohno:2013hc}.
Now, we calculate the VCs on the charge-channel susceptibilities
based on the SC-VC theory, which was introduced in Sect. \ref{sec:Kontani2-4}.
Figure \ref{fig:FS-2orbital} (c) exhibits the charge-channel AL-type VC
for the irreducible susceptibility, $X^c_1(q)$, on orbital 1.
(Note that $X^c_2(q_x,q_y,\w_l)=X^c_1(q_y,q_y,\w_l)$.)
We find that $X_1^c(q) \sim 0.5$ is comparable to the DOS per orbital $N(0)$,
while $X_1^s(q)$ exhibits small negative value \cite{Ohno:2013hc}.
In this case, the relation $\Phi^c_1 \sim 2 \Phi^s_1$ holds, 
and therefore the orbital order appears in the paramagnetic state.
Figure \ref{fig:FS-2orbital} (d) shows the obtained $\chi^c_{x^2-y^2}(\q)$,
derived from the self-consistent calculation of $\chi^{s,c}(q)$ and $X^{s,c}(q)$ 
\cite{Ohno:2013hc}.
In this numerical study, the nematic orbital order is realized when $\a_S\sim0.9$.
To summarize, the interference between spin fluctuations, which is described
by the AL diagram, gives rise to the ferro-orbital order.

\subsection{Orbital order owing to VCs: fRG theory
\label{sec:Kontani3-2}
}

We have shown that the spin-fluctuation-driven VCs induces the 
orbital order in a two-orbital Hubbard model
with on-site Coulomb interaction.
However, limited numbers of VCs were considered in the previous subsection.
Thus, the importance of VCs should be clarified  by other unbiased 
theoretical techniques.
For this purpose, the functional-renormalization-group (fRG) theory 
is quite suitable because 
the RG method enable us to perform the systematic calculations of VCs.
The energy band structure and 
the FSs  obtained from $H_0$ [Eq.\ (\ref{eqn:Ham0-2orb})] 
are shown in Figs.\ \ref{fig:FS-2orbital} (a) and (b).
The $\alpha$ band forms a hole-like 
FS centered at $\k=(\pi,\pi)$ while the $\beta$ band forms an electron-like 
 FS centered at $\k=(0,0)$.

\begin{figure}[htb]
\centering
\includegraphics[width=.9\linewidth]{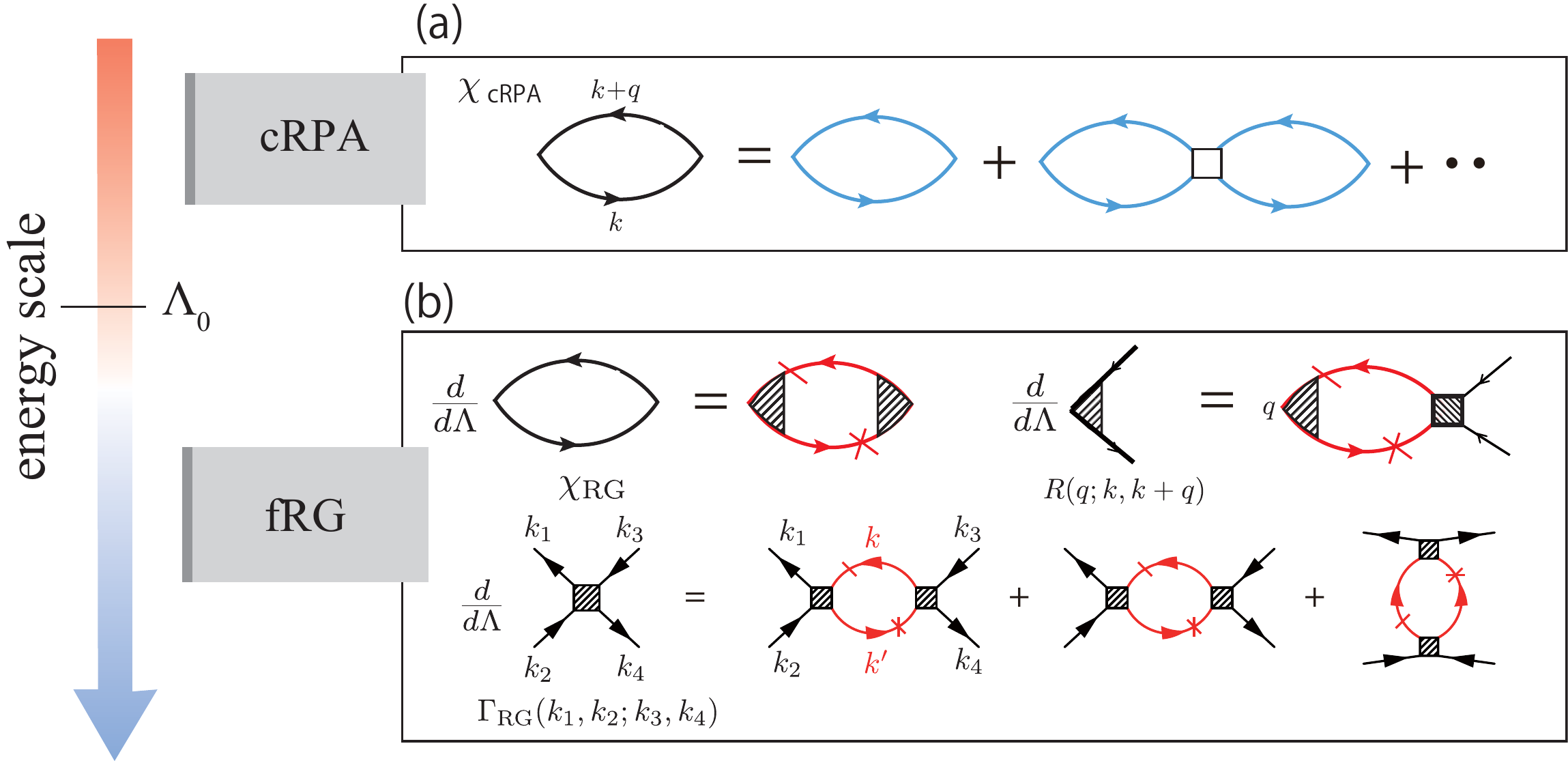}
\caption{
Schematic explanation for the RG+cRPA method \cite{Tsuchiizu-PRL}.
The higher-energy process ($|\xi_{\k,\nu}|>\Lambda_{0}$) 
is calculated by the constrained RPA.
Next, we calculate the lower-energy process
($|\xi_{\k,\nu}|<\Lambda_{0}$) using the fRG method.
The slashed (crossed) line represents on-shell 
Green function with $\Lambda_{l+dl}<|\xi_{\k,\nu}|<\Lambda_{l}$
($\xi_{\k,\nu}|<\Lambda_{l}$), where $\Lambda_l=\Lambda_0 e^-l$.
The RG equations for four-point vertex, three-point vertex,
and susceptibility are shown.
}
\label{fig:RGpluscRPA}
\end{figure}

We apply the one-loop RG method,
by which the VCs given by the parquet equations are calculated.
The RG equations are shown in Fig.\ \ref{fig:RGpluscRPA}, where 
$\chi(\bm q)$, $R(\bm q;k_1,k_2)$, and $\Gamma(k_1,k_2;k_3,k_4)$
are the susceptibility, the three-point and four-point vertices, respectively
\cite{Bourbonnais2003}.
The scattering processes of electrons having energies less than a 
 cutoff $\Lambda_0$ are integrated within the RG scheme.
Here, the band dispersion near the Fermi level is divided into 
$N_p \ (=24)$ patches, as shown in Fig.\ \ref{fig:FS-2orbital} (b).
Here, the vertex functions $R$ and $\Gamma$
are functions of the patch indices $k_i$,
and their frequency dependences are ignored
\cite{Zanchi:1998ua,Zanchi:2000ua,Halboth:2000tt}.
In contrast to conventional patch fRG \cite{Metzner:2012jv},
in Ref. \cite{Tsuchiizu-PRL},
we introduced the initial cutoff $\Lambda_0$ ($T\ll\Lambda_0\gg W_{\rm band}$)
shown in Fig.\ \ref{fig:FS-2orbital} (a):
The VCs due to lower energy contributions ($<\Lambda_0$) 
are calculated accurately by solving the RG equation,
while higher-energy contributions ($>\Lambda_0$) are 
calculated within the RPA since the VCs are less important
\cite{Hirsch:1985wz}.

In a conventional patch RG scheme,
the higher-energy contributions are treated less accurately 
because of the projection of momenta on the FS.
In the RG+cRPA method \cite{Tsuchiizu-PRL}, in contrast,
the higher-energy contributions are accurately 
calculated numerically by using fine $\bm k$ meshes.
This ``RG+cRPA method''
is a natural combination of the merits of the RG (for lower energy)
and RPA (for higher energy), and enables us to obtain accurate results.
In the RG+cRPA theory, the initial values of 
$\Gamma$, $R$, and $\chi$ in the RG equations are given by 
the ``constrained'' RPA (cRPA), shown in Fig.\ \ref{fig:RGpluscRPA}.
The cRPA susceptibility
$\chi_\mathrm{cRPA}(\bm q)$
is given by omitting the low-energy particle-hole (p-h) contributions
($|E|<\Lambda_0$) from the RPA susceptibility.
Then, the contribution from the RG and that from the cRPA are not over-counted.
We have verified that the numerical results 
depend on $\Lambda_0$ only weakly.

\begin{figure}[htb]
\centering
\includegraphics[width=.9\linewidth]{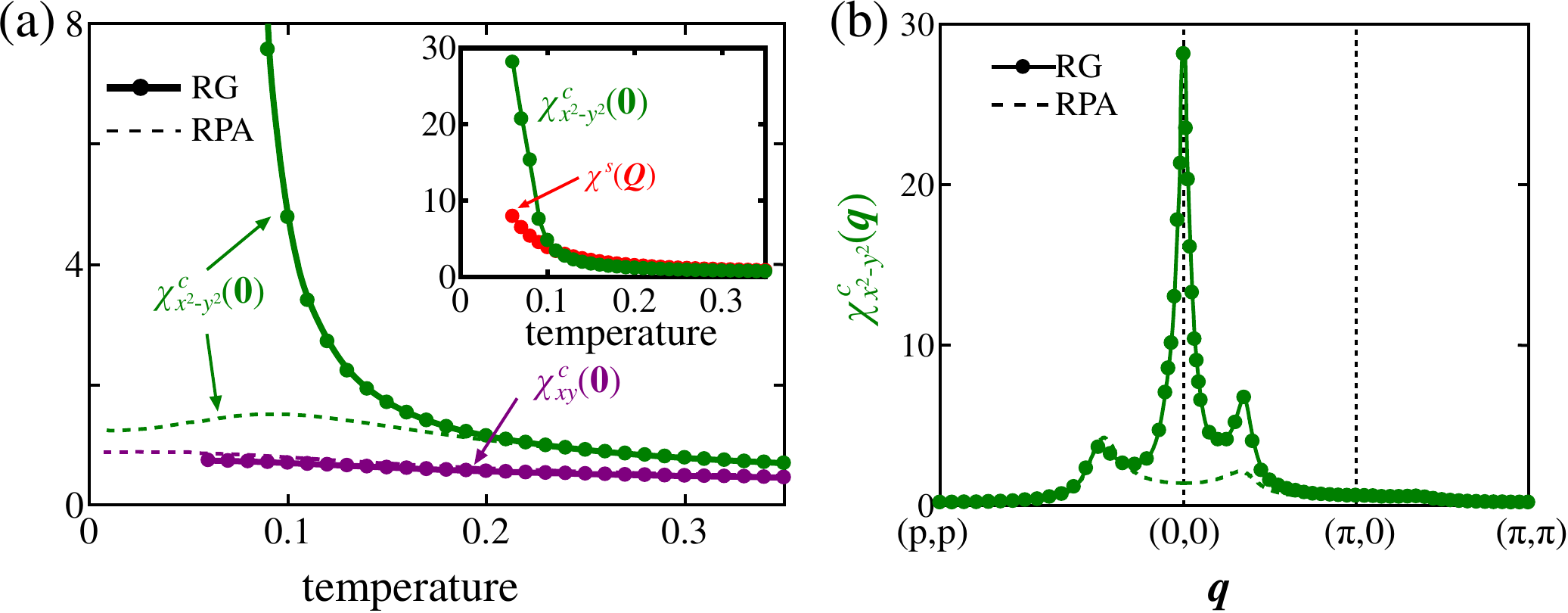}
\caption{
(a) Temperature dependences of $\chi^c_{x^2-y^2}(\bm 0)$ and 
$\chi^c_{xy}(\bm 0)$ for $n=3.3$, $U=2.13$, and $U'/U=0.9$.
The solid (dashed) lines represent the RG (RPA) results.
In the inset, the same data of 
 $\chi^\mathrm{s}(\bm Q)$ and $\chi^{c}_{x^2-y^2}(\bm 0)$
are plotted on a different vertical scale.
(b) Momentum dependences of 
$\chi^c_{x^2-y^2}(q)$ for $T = 0.06$.
The solid (dashed) lines represent the RG (RPA) results.
Cited from Ref. \cite{Tsuchiizu-PRL}. 
}
\label{fig:fRG-2orb-chi}
\end{figure}

As we already explained,
only $\chi^\mathrm{s}(\bm q)$  
is strongly enhanced for $U> U'$
in the RPA without VCs \cite{Takimoto,Ohno:2013hc}.
That is, orbital fluctuations remain small in the RPA.

Next, we perform the RG+cRPA study \cite{Tsuchiizu-PRL}.
The obtained $T$ dependences of the spin and quadrupole
susceptibilities are shown in Fig.\ \ref{fig:fRG-2orb-chi} (a).
In the high temperature ($T\gtrsim 0.3$) region,
all the susceptibilities
exhibit similar behavior to the RPA results \cite{Hirsch:1985wz}.
Even at low temperatures, $\chi^c_{xy}(\bm 0)$
show the same $T$ dependences as in RPA.
The effect of VCs suppresses $\chi^\mathrm{s}(\bm Q)$ at low temperatures.
The most striking feature of Figs.\ \ref{fig:fRG-2orb-chi} (a) and (b)
are the critical enhancement of 
$\chi_{x^2-y^2}^c(\q)$ around $\q\approx{\bm0}$ 
at low temperatures, which cannot be derived from RPA.
Thus, spin-fluctuation-driven ferro-orbital fluctuations
due to the AL process shown in the previous subsection
is well confirmed by the present fRG study.
Thus, the importance of the VCs 
on the charge-channel fluctuations has been confirmed by the fRG study.
The present mechanisms of
the orbital nematic phase would be realized 
in various multiorbital systems.

\subsection{Density-wave equation: Optimization of $\k$-dependent form factors}
\label{sec:Kontani3-3}

In previous subsections,
we studied the mechanism of the orbital order
in multiorbital Hubbard models,
by assuming that orbital order parameters are local.
However, as we discussed in Sect. \ref{sec:Kontani2},
nonlocal p-h interaction (due to VCs)
can induce various non-local order parameters described by the form factor,
such as the bond order and current order.
In order to study symmetry breaking phenomena 
with non-local order parameters,
we have to develop the theory of unconventional density waves
with nontrivial form factors.
For this purpose, we introduce generalized density-wave (DW) states
for both charge- and spin-channels.
They are expressed in real space as follows
\cite{Nersesyan}:
\begin{eqnarray}
D_{i,j}^{\s,\rho} &\equiv& (1-{\cal P}_{\rm A_{1g}}) \langle c_{i\s}^\dagger c_{j\rho} \rangle
\nonumber \\
&=& d^c_{i,j} \delta_{\s,\rho} + \bm{d}^s_{i,j} \cdot \bm{\s}_{\s,\rho},
\label{eqn:D1S}
\end{eqnarray}
where $D_{i,j}^{\s,\rho}=\{D_{j,i}^{\rho,\s}\}^*$,
and $d^c_{i,j}$ (${\bm d}^s_{i,j}$) is 
the charge (spin) channel order parameter.
${\cal P}_{\rm A_{1g}}$ is the projection operator on the $A_{1g}$ symmetry space.
General DW states with various symmetries
(such as SDW, CDW, orbital order, bond order, current order)
can be expressed by Eq. (\ref{eqn:D1S}).

The DW formation given by Eq. (\ref{eqn:D1S})
directly drives the ``symmetry breaking in the self-energy'':
$\Sigma_0(i-j) \rightarrow \Sigma_0(i-j)+\Delta\Sigma(i,j)$.
Here, $\Sigma_0(i-j)$ is the without symmetry breaking, 
and it possesses $A_{1g}$ symmetry.
$\Delta\Sigma(i,j)$ is the symmetry-breaking component,
which breaks the $A_{1g}$ symmetry.
It is expressed as
%
\begin{eqnarray}
\Delta \Sigma_{i,j}^{\s\rho}
&=& f_{i,j} \delta_{\s,\rho} + \bm{g}_{i,j} \cdot \bm{\s}_{\s,\rho}
\end{eqnarray}
which we call the form factors in this article.
For example, the bond order 
is given by real even-parity function $f_{i,j}=f_{j,i}$, and
the spin current order is given by pure imaginary odd-parity vector 
${\bm g}_{i,j}=-{\bm g}_{j,i}$,
as we will discuss below.
After the Fourier transformation in Eq. (\ref{eqn:Fourier}),
\begin{eqnarray}
\Delta \Sigma_\q^{\s\rho}(\k)
&=& f_\q(\k) \delta_{\s,\rho} + \bm{g}_{\q}(\k) \cdot \bm{\s}_{\s,\rho}
\end{eqnarray}
which we call the form factors in this paper.
Below, we assume $\bm{g}_{\q}(\k)= g_{\q}(\k) {\bm e}_z$ 
without losing generality.
The DW is interpreted as the electron-hole pairing
condensation
\cite{Nersesyan}.

Now, we consider the symmetry breaking in the self-energy by following 
Ref. \cite{RTazai-arXiv2021}.
The self-energy is functional of the Green function: $\Sigma^{[G]}_{i,j}$.
It is given as $\Sigma^{[G]}_{i,j}= \delta \Phi[G]/\delta G_{i,j}$,
where $\Phi[G]$ is Luttinger Ward function composed of $G$ and $U$.
Here, we denote the self-energy and Green function without symmetry 
breaking as $\Sigma^0=\Sigma^{[G^0]}$ and 
$G^0=((G^{\rm fr})^{-1}-\Sigma^0)^{-1}$.
Then, the symmetry breaking in the self-energy is given as
%
\begin{eqnarray}
\Delta \Sigma_{i,j}&=&(1-{\cal P}_{\rm A_{1g}}) \Sigma^{[G]}_{i,j}
\label{eqn:DSigma}
\end{eqnarray}
where $i=({\bm r},\e_n,\s)$ and $G=((G^0)^{-1}-\Delta\Sigma)^{-1}$.
Here, $\Sigma^{[G]}$ in the right-hand-side is composed of the full Green function
with $\Delta\Sigma\ne0$: $G=((G^0)^{-1}-\Delta\Sigma)^{-1}$.
Based on this ``full DW equation'' in Eq. (\ref{eqn:DSigma}),
we can obtain the self-consistent solution of $\Delta\Sigma$.

Next, we expand the right-hand-side of Eq. (\ref{eqn:DSigma})
with respect to $\Delta \Sigma$ 
by using the Dyson equation $G= G^0+G^0\Delta\Sigma G$.
The obtained linearized self-consistent equation is
\begin{eqnarray}
\lambda \Delta \Sigma_{i,j}&=& T \sum_{l,m} I(i,j;l,m)
[G^0\cdot \Delta\Sigma \cdot G^0]_{l,m}
\label{eqn:DSigma2}
\end{eqnarray}
where $\lambda$ is the eigenvalue and 
$\Delta \Sigma$ is the eigenfunction.
$I(i,j;l,m)= \delta \Sigma_{i,j}^{[G^0]}/\delta G^0_{l,m}$
is the four-point vertex which is irreducible with respect to
the p-h channel.

After the Fourier transformation of Eq. (\ref{eqn:DSigma2}),
we obtain the ``linearized DW equation''  
for both spin/charge channels \cite{Kawaguchi-CDW}:
%
\begin{eqnarray}
& & \!\!\!\! 
\lambda_{\q}f_\q(k)= -\frac{T}{N}\sum_{p}I_\q^c(k,p)G(p)G(p+\q)f_\q(p) ,
\label{eqn:DWeq3} \\
& & \!\!\!\! 
\eta_{\q}g_\q(k)= -\frac{T}{N}\sum_{p}I_\q^s(k,p)G(p)G(p+\q)g_\q(p) ,
\label{eqn:DWeq4}
\end{eqnarray}
where $\lambda_{\q}$ ($\eta_\q$) is the eigenvalue 
that represents the charge (spin) channel DW instability,
$k\equiv (\k,\e_n)$, $p\equiv (\p,\e_m)$, and
$\e_n$, $\e_m$ are fermion Matsubara frequencies.
The charge (spin) channel kernel function is 
$I_\q^{c(s)} = I_\q^{\uparrow,\uparrow}+(-)I_\q^{\uparrow,\downarrow}$.
These DW equations are interpreted as the 
``spin/charge channel electron-hole pairing equations''. 
This DW equations is easily extended to 
the multiorbital Hubbard models.

In the RPA, 
$I_q$ is given by the first HF term in Fig. \ref{fig:fig-MTAL}:
$I_\q^{c} =-U$ and $I_\q^{s} =U$.
Then, the form factors $f_\q$ and $g_\q$ become constants,
which represents conventional CDW and SDW order parameters.
Their eigenvalues are $\lambda_q=-U\chi^{(0)}(q)$
and $\eta_q=U\chi^{(0)}(q)$, respectively.

However, $I_\q^{c}(k,p)$ by MT and AL terms possesses $k,p$-dependence.
In Sect. \ref{sec:Kontani4} and Sect. \ref{sec:Kontani5},
we discuss that various non-local DW orders in 
strongly correlated electron systems can be understood 
using the irreducible four-point vertices in Eq. (\ref{eqn:MTAL}).
They are composed of 
one single-magnon exchange (MT) term 
and two double-magnon exchange (AL) terms:
The significance of these VCs have been
revealed by the functional-renormalization-group (fRG) study,
in which higher-order VCs are produced in an unbiased way
\cite{Tsuchiizu2018,Tsuchiizu-PRL,Tazai-FRG,Tazai-kappa}.
The AL terms frequently cause the $\q={\bm0}$ CDW instability 
since its functional form $\propto \sum_\k\chi^s(\k+\q)\chi^s(\k)$ 
is large for $\q\approx{\bm0}$ in two-dimensional systems
\cite{SOnari-PRL2012}.

\begin{figure}[htb]
\centering
\includegraphics[width=0.7\linewidth]{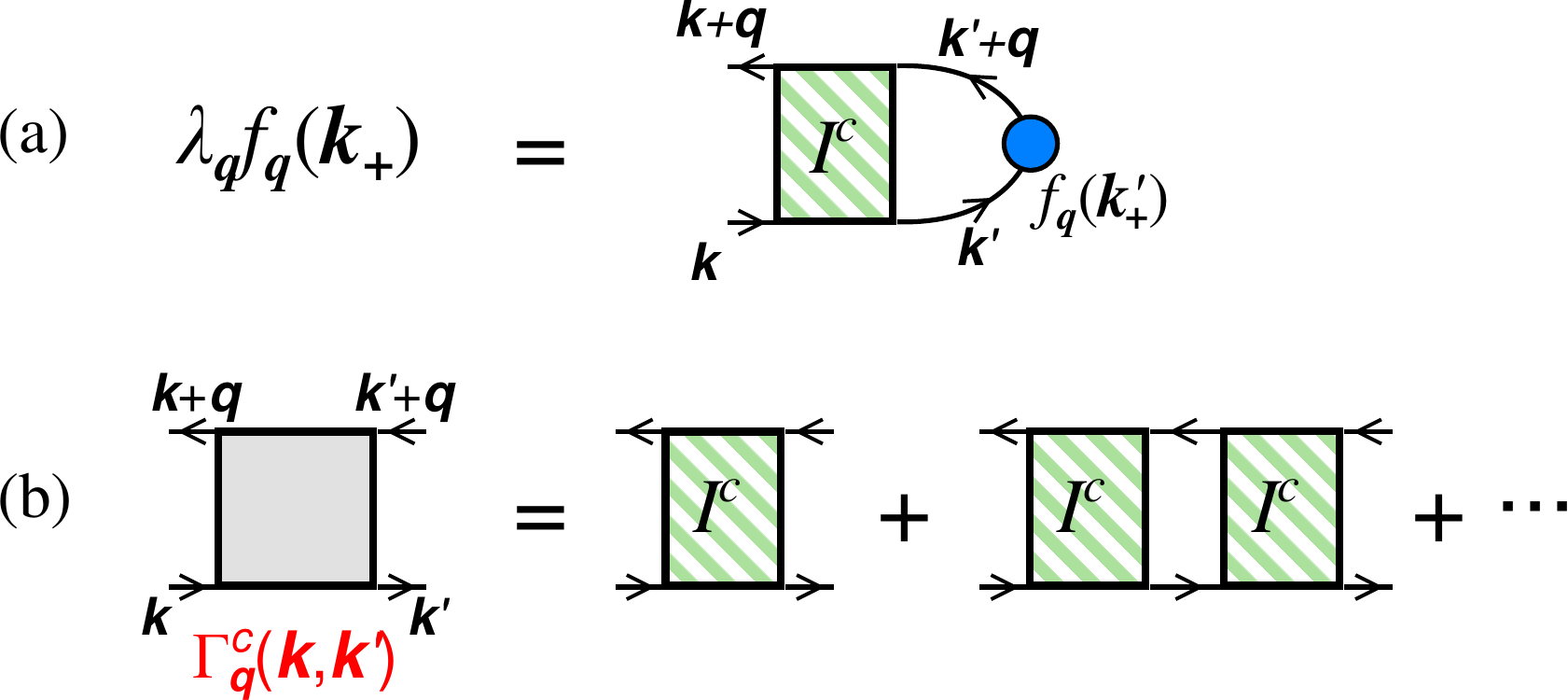}
\caption{
(a) Schematic linearized DW equation for general wavenumber $\q$
\cite{Kawaguchi-CDW,HKontani-sLC}. 
$f_\q(\k)$ is the charge-channel form factor.
The irreducible vertex $I^c$ includes the
HF, MT, AL1, AL2 terms shown in Fig. \ref{fig:fig-MTAL}.
(b) Full vertex $\Gamma^c$ composed of infinite series of $I^c$'s.
($\Gamma^c$ is the solution of the Bethe-Salpeter equation (\ref{eqn:BS1}).)
$\Gamma^c$ is also generated by solving the DW equation.
}
\label{fig:DW-eq-diagram}
\end{figure}

In principle, the DW order parameter is given as the 
``symmetry breaking in the self-energy $\Delta\Sigma$'',
similarly to the superconductivity given as the
symmetry-breaking in the anomalous self-energy.
The solution $\Delta\Sigma$ obtained by the DW equation 
naturally explains the nematic symmetry breaking 
in Fe-based \cite{Onari-FeSe} and cuprate \cite{Kawaguchi-CDW}
superconductors.
By solving the linearized equation,
the higher-order diagrams with respect to these terms are generated.
Examples of generated higher-order AL+MT processes 
are given in Fig. \ref{fig:DW-eq-diagram} (b).

Here, we stressed the importance of the AL terms for various DW states.
In contrast, the MT term is significant for the
non-Fermi liquid transport phenomena
\cite{Kontani-ROP} and the charge current order
\cite{RTazai-PRB2021}.
The MT term is also important in the SC gap equation 
that represents the particle-particle (p-p) condensation.
Mathematically, 
the gap equation is given by replacing the p-h propagator [$G(p+\q)G(p)$]
in the DW equation with the p-h propagator at $\q={\bm 0}$ [$G(p)G(-p)$].
In this replacement, the AL terms disappears because of the 
particle conservation laws in the three-point vertex.
Since the AL terms are larger than the MT term near the magnetic QCP,
the DW order transition can occur above $T_{\rm c}$.

\section{Unconventional orders in single-orbital models: 
with a focus on cuprate superconductors
\label{sec:Kontani5}
}


\subsection{Motivation}
\label{sec:Kontani5-0}

In cuprate superconductors, 
the normal electronic states exhibit amazing variety
as we explained in Sect. \ref{sec:Kontani1-3}.
Exotic orders composed of charge and spin degrees of freedom 
emerge in a cooperative fashion; see Ref. \cite{Tranquada-rev}.
In Y-based, Bi-based, and Hg-based cuprates,
the charge-density-wave (CDW) order 
with finite wavevector $\Q = (\pi/2, 0)$ or $(0, \pi/2)$ appears
in the presence of strong spin fluctuations
\cite{Y-Xray1,Bi-Xray1,Zheng-1,Keimer-rev,cuprate-3D,cuprate-3D2,cuprate-3D3,STM-Kohsaka,STM-Fujita,Davis:2013ce,Zheng-1,Julien-NMR}.
The discovery of these unconventional density-wave (DW) states
in high-$T_{\rm c}$ cuprates have triggered significant
progress in the field of strongly correlated electron systems.
This significant open issue sets a very severe constraint on the theory,
and the theoretical study will serves to understand 
the high-$T_{\rm c}$ pairing mechanism.
The rich variety of correlation-driven spontaneous symmetry breaking in metals 
becomes a central issue in condensed matter physics.

The aim of this section is to discuss
the mechanisms of unconventional DW states in 
cuprate superconductors near around the optimum doping,
with the electron filling $n\sim0.85$.
For this purpose,
we analyze a simple single-orbital square-lattice Hubbard model,
which describes an effective model of cuprate superconductors.
It is expressed as
\begin{eqnarray}
H=\sum_{\k,\s}\e_\k c_{\k\s}^\dagger c_{\k\s}
+U\sum_i n_{i\uparrow}n_{i\downarrow}.
\label{eqn:Hubbard-cuprate}
\end{eqnarray}
We denote the hopping integrals $(t_1,t_2,t_3)=(-1,1/6,-1/5)$,
where $t_l$ is the $l$-th nearest hopping integral
\cite{Kontani-ROP,Springer}.
Hereafter, we set the unit of energy as $|t_1|=1$,
which corresponds to $\sim 4000$ [K] in cuprates,
and fix the temperature $T=0.05 \ (\sim 200{\rm K})$.
The FS at filling $n=0.85$ is given in Fig. \ref{fig:YBCO-FS} (b).

\color{black}
As for the value of $U$,
early constrained LDA (cLDA) studies lead $U=7\sim10$ eV 
in La$_2$CuO$_4$ \cite{cLDA-cuprates1,cLDA-cuprates2,cLDA-cuprates3}.
On the other hand, recent cRPA analysis predicts 
$U\approx3.2$ eV ($U/|t|\approx6.5$) in La$_2$CuO$_4$,
and $U\approx2.2$ eV ($U/|t|\approx4.5$) 
in Hg-based cuprates \cite{cRPA-cuprates}.
Hereafter, we use smaller $U$ when the self-energy is neglected
in electron Green functions
in order to satisfy the paramagnetic condition $\a_S<1$.
\color{black}

The spin susceptibility
in the random-phase-approximation (RPA) is
$\chi^s(q)= \chi^0(q)/(1-U\chi^0(q))$, where $\chi^0(q)$ 
is the irreducible susceptibility without $U$ and $q\equiv(\q,\w_l)$.
The spin Stoner factor is defined as 
$\a_S\equiv \max_q\{U\chi^0(q)\}= U\chi^0(\Q_s,0)$.
Figure  \ref{fig:YBCO-FS} (c) shows the obtained $\chi^s(q)$
at $\a_S=0.99$ ($U=3.27$).
Here, $\chi^s(\Q_s,0)\sim30$ [$1/{t_1}$]
$ \sim80$ [$\mu_{\rm B}^2/{\rm eV}$],
which is still smaller than 
Im$\chi^s(\Q_s,E=31{\rm meV})\sim 200$ [$\mu_{\rm B}^2/{\rm eV}$]
at $T\sim200$K in 60K YBCO
\cite{neutron}.
Thus, $\a_S>0.99$ in real compounds.
\color{black}
In the FLEX approximation,
the relation $\a_S\lesssim 1$ is satisfied 
for $U/|t|\gg 3.3$ without fine tuning of $U$,
because of the large negative feedback of 
spin-fluctuation-induced self-energy on $\a_S$
in two-dimensional systems.
In Ref. \cite{Kontani-ROP},
non-Fermi liquid transport phenomena are 
satisfactorily for $U/|t|=5\sim8$ 
based on the FLEX + current VC study.
It is notable that the FLEX approximation satisfies 
the Mermin-Wagner theorem \cite{GVI},
so $\a_S$ never exceeds unity in two-dimensional systems.
\color{black}

In strongly correlated metals,
the ``correlation-driven density-wave (DW)''
has been studied intensively 
\cite{SOnari-PRL2012,Tsuchiizu-PRL,Sachdev-CDW1,Sachdev-CDW2,Husemann,Chubukov-cuprate2014,Davis:2013ce,Kivelson-NJP}.
Various beyond-mean-field approximations have been developed to
explain the nematic and smectic orders in cuprates.
Here, we focus on the impact of th vertex corrections (VCs)
that describe the paramagnon interference process in Fig. \ref{fig:phase-AFM} (b).
In the Fermi liquid theory,
The irreducible VC $I$ is derived from the Ward-identity ($\delta{\hat \Sigma}/\delta {\hat G}$).
In the one-loop approximation, $I$ is composed of
the Maki-Thompson (MT) and Aslamazov-Larkin (AL) VCs
as we depicted in Fig. \ref{fig:fig-MTAL}.
As studied in Ref. \cite{Sachdev-CDW1},
the higher-order MT processes give the diagonal bond order with 
$\q=\Q_d=(\delta,\delta)$.
However, this wavevector is inconsistent with experiments.
The axial bond order is given by the lowest-order AL process 
if small inter-site Coulomb interaction exists \cite{Yamakawa-CDW},
while the uniform nematic order that 
is observed at $T=T^*$ \cite{Sato-Matsuda2017} is bot explained.
Therefore, new theoretical method should be developed.

Hereafter, we analyze the DW instabilities 
based on the DW equation as well as 
the functional renormalization group (fRG) theory.
In both theories, we obtain the uniform ($\q={\bm 0}$) 
$d$-wave bond order, which is
schematically shown in Fig. \ref{fig:schematic-OP} (b).
This uniform bond order strongly enlarges the 
axial nematic bond order instability at $\q=\Q_a$
shown in Fig. \ref{fig:schematic-OP} (c).
These studies lead to the prediction that 
the uniform bond order occurs at 
$T^*$, and axial $\q=\Q_a$ CDW is induced at $T_{\rm CDW}<T^*$.
The higher-order AL processes are significant 
for the rich variety of bond orders.

\subsection{Analysis by DW equation for cuprates
\label{sec:Kontani5-1}
}

\begin{figure}[!htb]
\centering
\includegraphics[width=.7\linewidth]{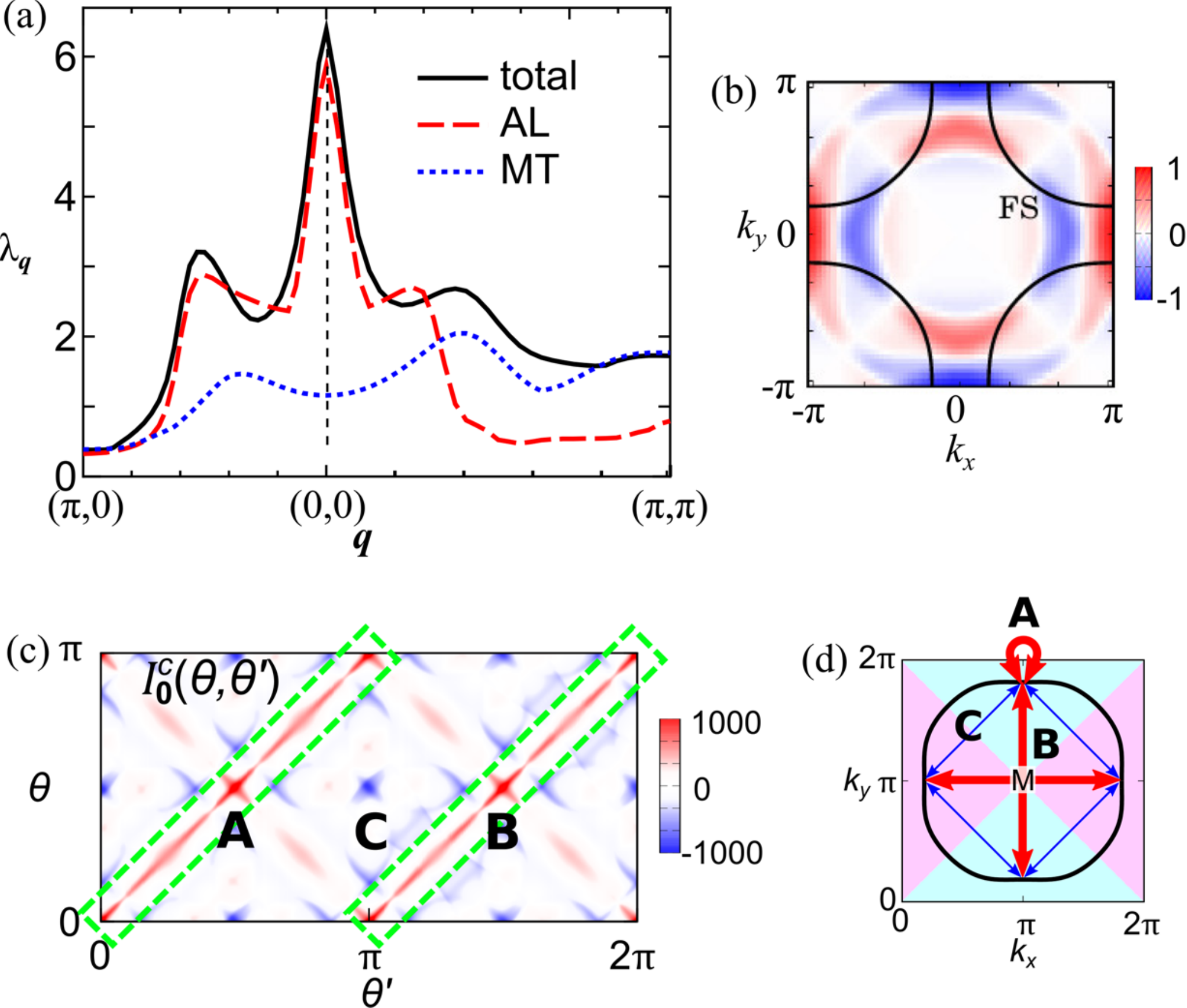}
\caption{
(a) Charge-channel eigenvalue $\lambda_\q$ 
of DW equation due to the MT+AL processes.
Cited from \cite{Kawaguchi-CDW}.
The eigenvalue due to AL processes are also shown.
(b) Form factor for $\q={\bm 0}$ ($d$-wave), $f_\q(\k)$,
normalized as $\max_\k\{f_\q(\k)\}=1$.
(c)Charge-channel
kernel functions on the FS, $I_{\q={\bm 0}}^{c}(\theta,\theta')$,
where $\theta$ represents the position of $\k$.
We see that $I_{{\bm 0}}^{c}$ has large positive values
due to the AL1 and AL2 terms shown in Fig. \ref{fig:fig-MTAL} (a).
(d) Origin of $d$-wave bond order.
Red (blue) color arrows represent the attractive (repulsive) interaction.
}
\label{fig:DW-YBCO-result}
\end{figure}

Here, we analyze the DW instabilities 
in a simple Hubbard model (\ref{eqn:Hubbard-cuprate})
by using the DW equation,
without assuming any $\q$-dependence and the form factor.
%
Figure \ref{fig:DW-YBCO-result} (a) shows the 
$\q$-dependence of the eigenvalue $\lambda_\q$ obtained for 
$\a_S=0.995$ at $T=50$meV.
(In FIg. \ref{fig:DW-YBCO-result},
we introduce the imaginary part of the self-energy $\gamma=0.3$ eV.)
Here, $\lambda_\q$ is the largest at $\q={\bm 0}$,
meaning that the uniform DW emerges at the highest temperature.
As shown in Fig. \ref{fig:DW-YBCO-result} (b), the corresponding form factor
$f_{{\bm 0}}(\k)$ has the $d$-wave symmetry.
The second largest peak is Fig. \ref{fig:DW-YBCO-result} (a)
locates $\q=\Q_a=(\delta,0)$, which corresponds to the axial CDW.
Since these form factors have sign reversal in $\k$-space,
the contribution from the Hartree term ($U$) is absent.
To summarize, the present theory predicts the emergence of the
nematic ($\q={\bm0}$) and smectic $\Q=\Q_a$ bond orders shown in 
Figs. \ref{fig:schematic-OP} (b) and (c), respectively.
Thus, the present theory predicts that uniform bond order transition emerges 
at a high temperature,
and it stabilize the axial $\q=\Q_a$ bond order at $T=T_{\rm CDW}$.

To find the origin of the DW instability,
we solve the linearized DW equation by including only AL terms.
The obtained ``charge-channel'' eigenvalue $\lambda^{\rm AL}_\q$ 
is shown in Fig. \ref{fig:DW-YBCO-result} (a).
The similarity between $\lambda_\q$ and $\lambda^{\rm AL}_\q$ 
means that the DW instabilities at $\q={\bm0}$ and $\Q_a$
originate from the AL processes, whereas the instability at $\q=\Q_d$ 
is mainly derived from the MT processes.
Thus, it is confirmed that the higher-order AL processes
cause the multi bond order transition at both $\q={\bm 0}$ and $\Q_a$.

We stress that the large eigenvalues 
in Fig. \ref{fig:DW-YBCO-result}  
are strongly suppressed to $O(1)$ by considering
the small quasiparticle weight $z = m/m^* \sim O(10^{-1})$ 
and large quasiparticle damping $\gamma \gg T$ in cuprates
\cite{Kawaguchi-CDW,SOnari-PRB2019}.

We discuss the reason why $d$-wave bond order is obtained
based on a simplified DW equation (\ref{eqn:DWeq2-dwave}):
The charge-channel ``electron-hole pairing interaction''
$I^c_{\q={\bm 0}}(\k,\k')$ on the FS is shown 
in Fig. \ref{fig:DW-YBCO-result} (c).
Here, $\theta$ represents the position of $\k$ 
shown in Fig. \ref{fig:DW-YBCO-result} (d).
$I_\q^c(\k,\k')$ in Fig. \ref{fig:DW-YBCO-result} (c)
gives large attractive interaction for 
{\bf A}: $\k\approx\k'$ and 
{\bf B}: $\k\approx-\k'$,
and weak repulsive interaction for 
{\bf C}.
Then, we naturally obtain the $d$-wave form factor $g_\q(\k)$ 
shown in Fig. \ref{fig:DW-YBCO-result} (d).
Here, red (blue) arrows represent the attractive (repulsive) 
interaction by {\bf A} and {\bf B} ({\bf C}).
The large positive $I_{{\bm 0}}^{c}$ 
around {\bf A} originates from the p-h channel in AL1 term,
and that around {\bf B} originates from the p-p channel in AL2 term
 \cite{HKontani-sLC}.
The moderate repulsive $I_{{\bm 0}}^{c}$ 
around {\bf C} originates from the MT term.

\begin{figure}[htb]
\centering
\includegraphics[width=.8\linewidth]{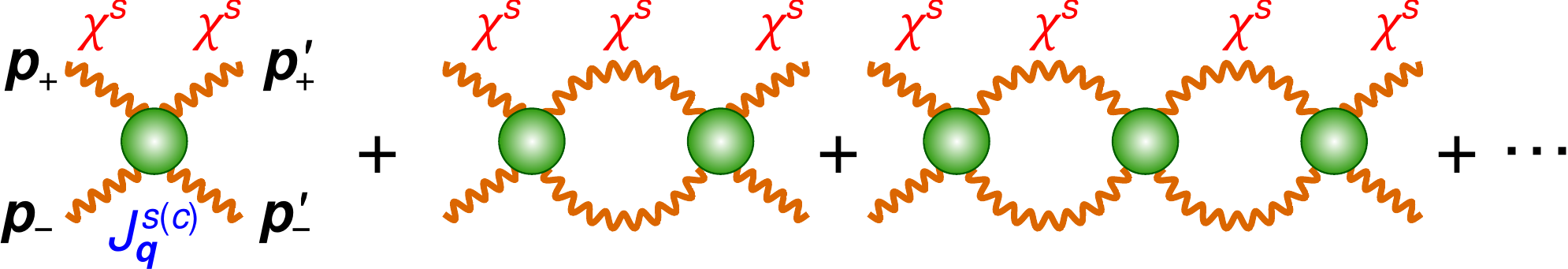}
\caption{
Diagrammatic expression for the even/odd parity magnon pair
condensation, which is the physical origin of the sLC and bond order.
Cited from Ref. \cite{HKontani-sLC}.
}
\label{fig:YBCO-DW-phase}
\end{figure}


In the DW equation formalism, the order parameter 
of bond order is the form factor that represents the 
electron-hole pairing.
Another physical interpretation of the bond order is the 
``condensation of even (odd) parity magnon-pairs'',
which is the origin of the nematic order in quantum spin systems
\cite{Andreev,Coleman,Shannon}.
In fact, the two-magnon propagator shown in Fig. \ref{fig:YBCO-DW-phase} 
diverges when the eigenvalue of DW equation reaches unity
\cite{HKontani-sLC}. 
The condensation of magnon-pairs and the p-h pair condensation 
occur simultaneously.
Thus, the bond order discussed here and the spin nematic order 
in quantum spin systems are the same phenomenon.
The predicted multistage symmetry breaking  
will be a key ingredient in understanding pseudogap phase 
and electronic nematicity in cuprates and other strongly correlated
electron systems.


\subsection{Renormalization group study for cuprates
\label{sec:Kontani5-2}
}

In this subsection,
we analyze the unconventional DW formation
based on the functional renormalization group (fRG) theory.
One of the great merit of the fRG method is that 
huge numbers of higher-order VCs are calculated in unbiased way.

Here, we study a standard three-orbital $d$-$p$ Hubbard model
\cite{Yamakawa-CDW,Tsuchiizu2016,Hansmann2014,Tsuchiizu2018}
expressed as
$H=\sum_{\bm k, \sigma} \bm c_{\bm k, \sigma}^\dagger \,
\hat h_0(\bm k)  \, \bm c_{\bm k, \sigma}^{}
+U\sum_{\bm j} n_{d, \bm j,\uparrow} n_{d, \bm j,\downarrow}$, 
where $\bm c_{\bm k, \sigma}^\dagger=
(d_{\bm k,\sigma}^\dagger, p_{x,\bm k,\sigma}^\dagger, 
 p_{y,\bm k,\sigma}^\dagger)$
is the creation operator for the electron 
on $d$, $p_x$, and $p_y$ orbitals,
and $\hat h_0(\bm k)$ is the kinetic term.
$U$ is the Hubbard-type on-site Coulomb interaction for the $d$ orbital,
and $n_{d,\bm j,\sigma}=d^\dagger_{\bm j,\sigma}d_{\bm j,\sigma}$ at site $\bm j$.
Hereafter, we study the $10 \%$ hole doping case.

\begin{figure}[htb]
\centering
\includegraphics[width=.99\linewidth]{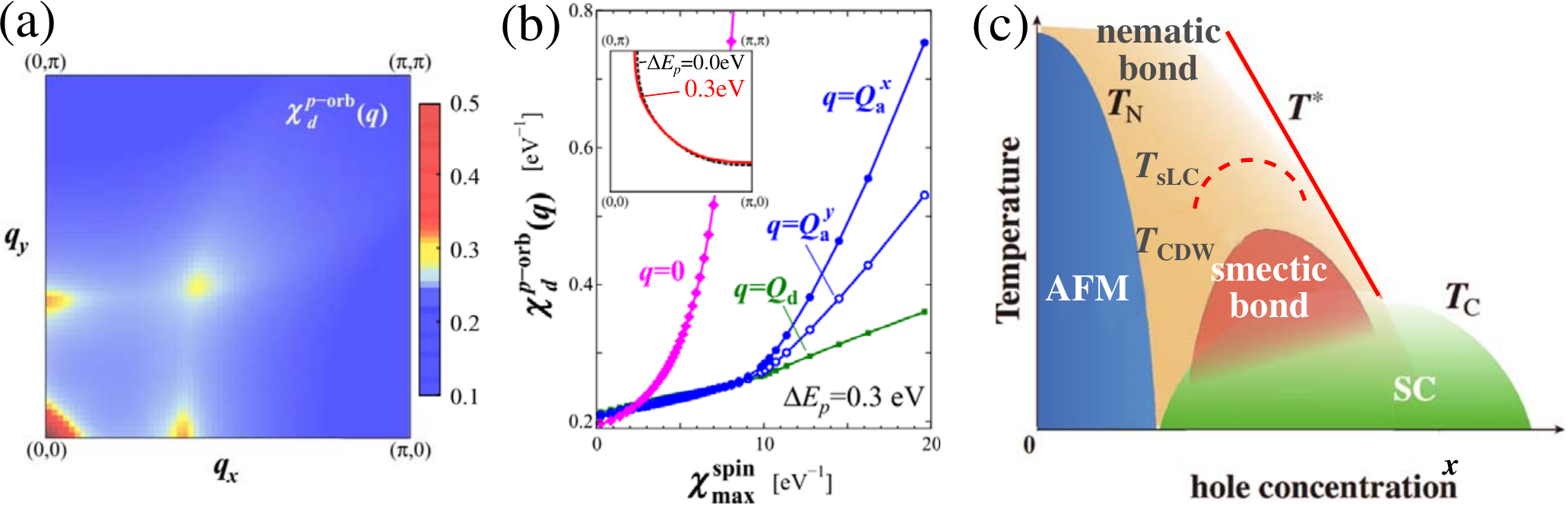}
\caption{
(a) RG+cRPA result of the $p$O-CDW susceptibility 
$\chi^{p\mbox{-}\mathrm{orb}}_d(\bm q)$ obtained for $U=4.32$ eV.  
The axial wavevector is $\bm Q_\mathrm{a}\approx  (0.37\pi, 0)$ 
and the diagonal wavevector is
$\bm Q_\mathrm{d}\approx  (0.40\pi,0.40\pi)$.  
Both $\bm Q_\mathrm{a}$ and $\bm Q_\mathrm{d}$ 
correspond to the wavevector connecting the hot spots.
(b) RG+cRPA result of 
$\chi^{p\mbox{-}\mathrm{orb}}_d(\bm q)$ 
at $\bm q={\bm0}$, $\bm Q_\mathrm{a}^{x,y}$ and $\Q_\mathrm{d}$
as a function of $\chi^\mathrm{spin}_\mathrm{max}$ for $\Delta E_{p}=0.3$ eV.
The inset shows the FS.
(c) Schematic phase diagram of cuprate superconductor
predicted by the DW-equation and RG+cRPA theory.
Nematic and smectic bond order are shown in 
Figs. \ref{fig:schematic-OP} (b) and (c).
Cited from Ref. \cite{Tsuchiizu2016}.
}
\label{fig:YBCO-RG}
\end{figure}

By using the RG+cRPA theory in Ref. \cite{Tsuchiizu2016},
we find that the spin susceptibility for $d$-electrons,
\begin{equation}
\chi^{\mathrm{spin}}(\bm q)
=
\frac{1}{2}
\int_0^{1/T} d\tau \,
\left\langle 
S_d(\bm q,\tau)
S_d(-\bm q,0)
\right\rangle,
\end{equation}
and the $B_{1g}$-symmetry ($d$-symmetry) charge-channel susceptibility 
for $p$-electrons,
\begin{equation}
\chi^{p\mbox{-}\mathrm{orb}}_{d}(\bm q)
=
\frac{1}{2}
\int_0^{1/T} \! d\tau \,
\left\langle 
n^{p\mbox{-}\mathrm{orb}}_d(\bm q,\tau)
n^{p\mbox{-}\mathrm{orb}}_d(-\bm q,0)
\right\rangle , 
\end{equation}
are the most enhanced susceptibilities 
\cite{Tsuchiizu2018}.
Here, 
$S_d(\bm q,\tau)$ is the $d$-electron spin operator, and
$n^{p\mbox{-}\mathrm{orb}}_d(\bm{q}) \equiv  n_{x}(\bm{q}) - n_{y}(\bm{q})$
($n_{x(y)} (\bm q ) =  \sum_{\bm k, \sigma}
 p_{x(y),\bm{k},\sigma}^\dagger p_{x(y),\bm{k+q},\sigma}$) is the 
$p$-orbital charge-density-wave ($p$O-CDW) operator with $B_{1g}$ symmetry.
If $\chi^{p\mbox{-}\mathrm{orb}}_d(\bm q)$ diverges at $\bm q=\bm Q_{\rm a}$
[$\bm q=\bm 0$], the bond order similar to Fig.\ \ref{fig:schematic-OP} (c) 
[Fig.\ \ref{fig:schematic-OP}(b)] is realized.

Hereafter, we perform the RG+cRPA analysis by putting 
$N_p=128$ and $\Lambda_0=0.5$ eV.
In the RG+cRPA method, the numerical accuracy of the susceptibilities
is drastically improved by applying the cRPA for the higher-energy processes,
because the $N_p$-patch RG scheme is less accurate.
We verified that the numerical results are essentially 
independent of the choice of $\Lambda_0$
when  $E_{\rm F}\gtrsim \Lambda_0\gg T$.
In Fig.\ \ref{fig:YBCO-RG}(a), we show the 
$p$O-CDW susceptibility $\chi^{p\mbox{-}\mathrm{orb}}_{d}(\bm q)$
given by the RG+cRPA method for $U=4.32$ eV at $T=0.1$ eV.  
The obtained large peaks at 
$\bm q=\bm 0$, $\bm Q_\mathrm{a}$, and $\bm Q_\mathrm{d}$
originate from the VCs, since 
the RPA result is less singular.
The obtained highest peak at $\bm q=\bm 0$
is consistent with the experimental uniform
nematic transition at $T^* \ (>T_{\rm CDW})$
\cite{Sato-Matsuda2017}.
We also obtain the 
peak structures at $\bm q=\bm Q_\mathrm{a}$ and $\bm Q_\mathrm{d}$.
Note that the temperature $T=0.1$ eV is comparable to $T^*\sim300$ K
if the mass-enhancement factor $m^*/m_{\rm band}\sim 3$ is considered.

In order to discuss the CDW instabilities \textit{inside} the 
nematic phase, we perform the RG+cRPA analysis 
in the presence of the uniform $p$O-CDW order
$H'=-\frac{1}{2}\Delta E_{p} [n_x(\bm 0)-n_y(\bm 0)]$.
In Fig.\ \ref{fig:YBCO-RG} (b), we plot 
the peak values of $\chi^{p\mbox{-}\mathrm{orb}}_d(\bm q)$
in the uniform $p$O-CDW state with $\Delta E_{p}=0.3$ eV.
Due to small $\Delta E_{p}>0$,
$\chi^{p\mbox{-}\mathrm{orb}}_d(\bm q)$ at 
$\bm q =\bm Q_\mathrm{a}^x =(\delta,0)$ strongly increases 
whereas that at $\bm q =\bm Q_\mathrm{a}^y=(0,\delta)$ decreases.
Thus, the $p$O-CDW at $\bm q=\bm Q_\mathrm{a}^x$ is expected to emerge 
below $T_\mathrm{\rm CDW}$, consistently with the phase diagram in 
Fig.\ \ref{fig:YBCO-phase}.

Because $p$-electrons are non-interacting in this $d$-$p$ Hubbard model, 
the enhancement of $\chi^{p\mbox{-}\mathrm{orb}}_d(\bm q)$
originates from the $d$-orbital Coulomb interaction.
The obtained $p$O-CDW susceptibility is equivalent to
the $d$-wave bond order between $d$-orbitals,
which is obtained in Ref. \cite{Tsuchiizu2018}
based on the DW equation.
Figure \ \ref{fig:YBCO-RG}(c) represents
the schematic phase diagram of cuprate superconductor
predicted by the DW-equation and RG+cRPA theory.
Nematic and smectic bond orders are shown in 
Figs. \ref{fig:schematic-OP} (b) and (c), respectively.

\subsection{Odd parity current orders: spin loop current}
\label{sec:Kontani5-3}

In previous subsections,
we discussed the spin-fluctuation-mediated
$d$-wave orbital/bond formation in Fe-based and cuprate superconductors.
They are expressed by the ``charge-channel, even-parity 
(${\cal P}_{\rm bond}=+1$)'' form factor $f^c_\k=f^c_{-\k}$.
Here, we explain that the ``odd-parity (${\cal P}_{\rm bond}=-1$)''
form factor can be caused by the same spin fluctuation mechanism.
We derive the $p$-wave spin-channel form factor 
$f^s\propto (\sin k_x, \sin k_y)$ 
that accompanies the spontaneous spin loop current (sLC).
The obtained sLC may be the origin of the pseudogap behaviors
in cuprate superconductors.
In the sLC state, the time reversal symmetry is preserved $({\cal T}=+1)$.

\begin{figure}[!htb]
\centering
\includegraphics[width=.7\linewidth]{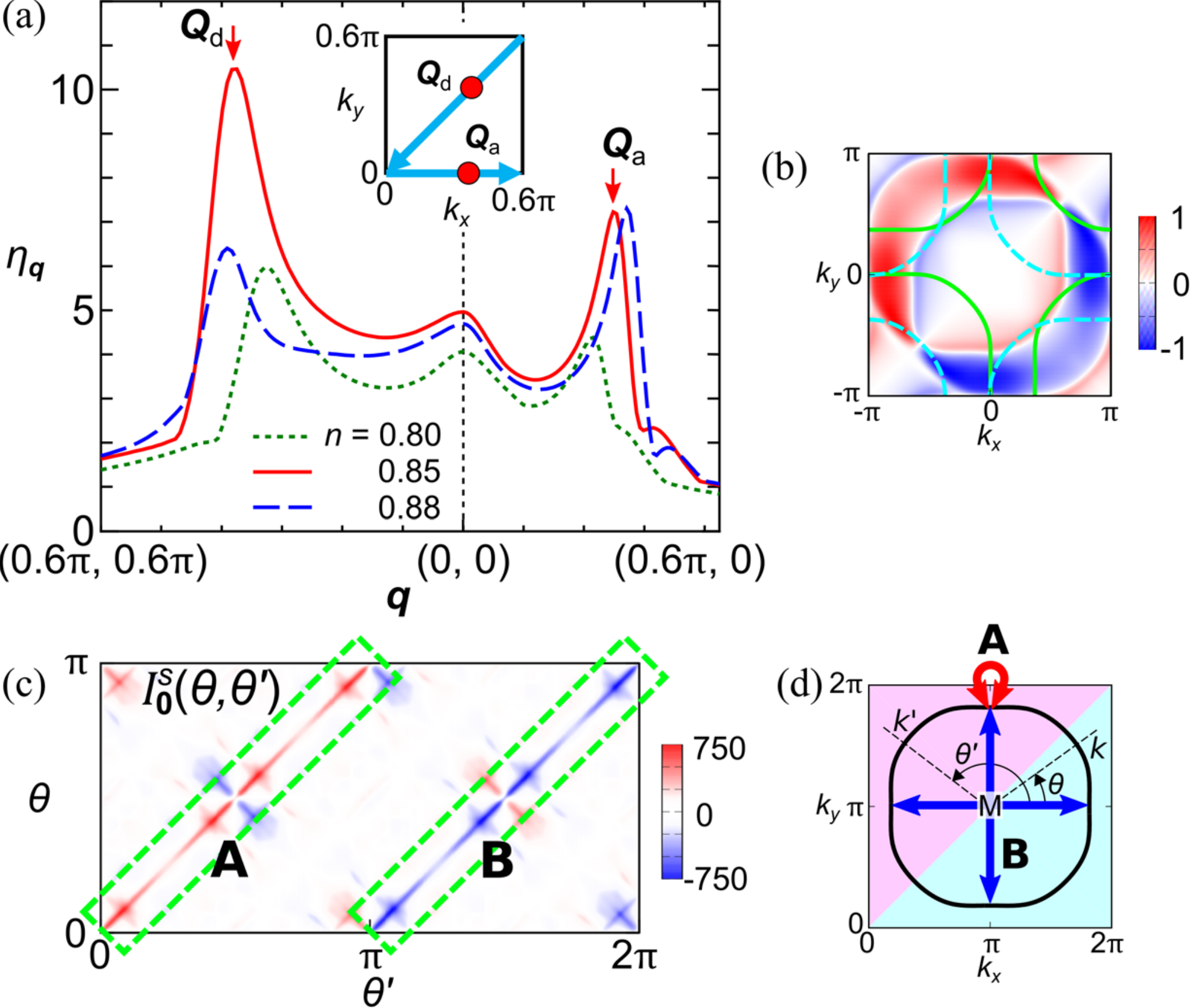}
\caption{
(a) Spin-channel eigenvalue $\eta_\q$ of DW equation
discussed in Ref. \cite{HKontani-sLC}.
(b) Form factor for $\q={\bm 0}$ ($p$-wave)
normalized by its maximum value.
(c) Spin-channel
kernel functions on the FS, $I_{\q={\bm 0}}^{s}(\theta,\theta')$,
where $\theta$ represents the position of $\k$.
We see that $I_{{\bm 0}}^{s}$ has large positive and negative values
for {\bf A} $\theta'\approx \theta$ due to AL1
and {\bf B} $\theta'\approx \theta+\pi$  due to AL2, respectively.
AL1 and AL2 are shown in Fig. \ref{fig:fig-MTAL} (a).
(d) Origin of $d$-wave bond order.
Red (blue) color arrows represent the 
attractive (repulsive) interaction due to {\bf A} ({\bf B}).
Cited from Ref. \cite{HKontani-sLC}. 
}
\label{fig:DW-current}
\end{figure}

Here, we discuss the spin-fluctuation-driven sLC order.
Figure \ref{fig:DW-current} (a) exhibits the 
spin-channel eigenvalue $\eta_{\q}$
derived from the DW eq. (\ref{eqn:DWeq4}).
Peaks of $\eta_{\q}$ are located at the nesting vectors 
$\q=\Q_{\rm d}$ (diagonal) and $\q=\Q_{\rm a}$ (axial).
The obtained form factor $g_\q(\k)$ at
$\q=\Q_{\rm d}$ (diagonal sLC) is shown in Fig. \ref{fig:DW-current} (b).
The obtained odd-parity solution $g_\q(\k)=-g_\q(-\k)$
corresponds to the sLC order.


To understand why sLC state is obtained,
we simplify Eq. (\ref{eqn:DWeq4}) by taking 
the Matsubara summation analytically 
by approximating that $I_\q^{s}$ and $g_\q(k)$ are static:
\begin{eqnarray}
\eta_{\q}g_\q(\k)= \frac{1}{N}\sum_{\p}I_\q^s(\k,\p)F_\q(\p)g_\q(\p) ,
\label{eqn:DWeq2}
\end{eqnarray}
where 
$\displaystyle F_\q(\p)\equiv -T\sum_m G(p+\q)G(p)= \frac{n(\e_{\p+\q})-n(\e_{\p})}{\e_{\p}-\e_{\p+\q}}$ is a positive function.
We exhibit the spin-channel ``electron-hole pairing interaction''
$I^s_{\q={\bm 0}}(\k,\k')$ on the FS in Fig. \ref{fig:DW-current} (c), where
$\theta$ is the position of $\k$ shown in Figs. \ref{fig:DW-current} (d).
We see that $I_\q^s(\k,\k')$ 
gives large attractive and repulsive interactions
at {\bf A} ($\k\approx\k'$) and {\bf B} ($\k\approx-\k'$), respectively.
Thus, the $p$-wave form factor $g_\q(\k)$ is naturally obtained as
we explain in Fig. \ref{fig:DW-current} (d).
Here, red (blue) arrows represent the attractive (repulsive) interaction.

As shown in Ref. \cite{HKontani-sLC}, large $I^s$
originates from the convolution of transverse spin fluctuations 
$C_\q^{\rm tr} = \sum_\k \chi^{s}_{\pm}(\k+\q)\chi^{s}_{\mp}(\k)$.
As we found in Ref. \cite{HKontani-sLC},
$I^s \approx \mbox{[AL1]}-\mbox{[AL2]}$, and
the expressions of AL1 and AL2 are shown in Fig. \ref{fig:fig-MTAL}.
Since $\mbox{[AL1]}$ and $\mbox{[AL2]}$ take large positive value 
for $\k\approx\k'$ and $\k\approx-\k'$, respectively,
the numerical results in Fig. \ref{fig:DW-current} (c)
is naturally understood.
In contrast, the charge channel kernel
$I_\q^c(\k,\k')$ gives an attractive interaction for both
$\k\approx\pm \k'$ as shown in Fig. \ref{fig:DW-YBCO-result} (c),
because $I^c = 3(\mbox{[AL1]}+\mbox{[AL2]})/2$.
Therefore, the $d$-wave change bond order is obtained
\cite{HKontani-sLC}.



In the present sLC mechanism due to the spin-flipping AL process $C_\q^{\rm tr}$
\cite{HKontani-sLC},
the $\bm{g}$-vector will be parallel to $z$-direction
when $\chi^s_{x(y)}(\Q_s)>\chi^s_z(\Q_s)$ (XY-anisotropy)
due to the spin-orbit interaction (SOI). 
When the XY-anisotropy of $\chi^s_{\mu}(\Q_s)$ is very large,
$I^c$ due to AL terms is multiplied by $2/3$
whereas $I^s$ is unchanged,
so it is suitable condition for the sLC order.

Next, we investigate the spin current in real space,
which is driven by a fictitious Peierls phase 
due to the ``spin-dependent self-energy'' $\delta t_{i,j}^\s = \s g_{i,j}$.
In the current order, $\delta t_{i,j}$ is purely imaginary
and odd with respect to $i \leftrightarrow j$.
The conservation law $\dot{n}_i^\s = \sum_j j_{i,j}^\s$
directly leads to the definition the spin current operator 
from site $j$ to site $i$ as
$j_{i,j}^\s =-i\sum_{\s}\s (h_{i,j}^\s c_{i\s}^\dagger c_{j\s}-(i \leftrightarrow j))$,
where $h_{i,j}^\s = t_{i,j}+\delta t_{i,j}^\s$.
Then, the spontaneous spin current from $j$ to $i$ is
$J_{i,j}^s=\langle j_{i,j}^s \rangle_{\hat{h}^\s}$.
The spin current for the
commensurate sLC order at $\q_{\rm sLC}=(\pi/2,\pi/2)$
is shown in Fig. \ref{fig:schematic-OP} (d)
\cite{HKontani-sLC}.

\subsection{Odd parity current orders: charge loop current in quasi 1D systems}
\label{sec:Kontani5-4}

Spontaneous current orders due to odd-parity order parameters
attract increasing attention in various strongly correlated metals.
Here, we propose a novel spin-fluctuation-driven charge loop current (cLC)
mechanism based on the functional renormalization group (fRG) theory.
By analyzing a simple frustrated Hubbard model,
we find that the ferro cLC appears 
between the antiferromagnetic (AFM) and $d$-wave SC ($d$SC) phases.
The key ingredients of the present cLC mechanism are the
geometrical frustration and magnetic criticality.
The cLC is expressed as the $p$-wave charge-channel form factor
without time-reversal symmetry (${\cal T}=-1$).
The present study indicates that the cLC is ubiquitous in metals near the 
magnetic criticality with geometrical frustration.

\begin{figure}[!htb]
\centering
\includegraphics[width=.7\linewidth]{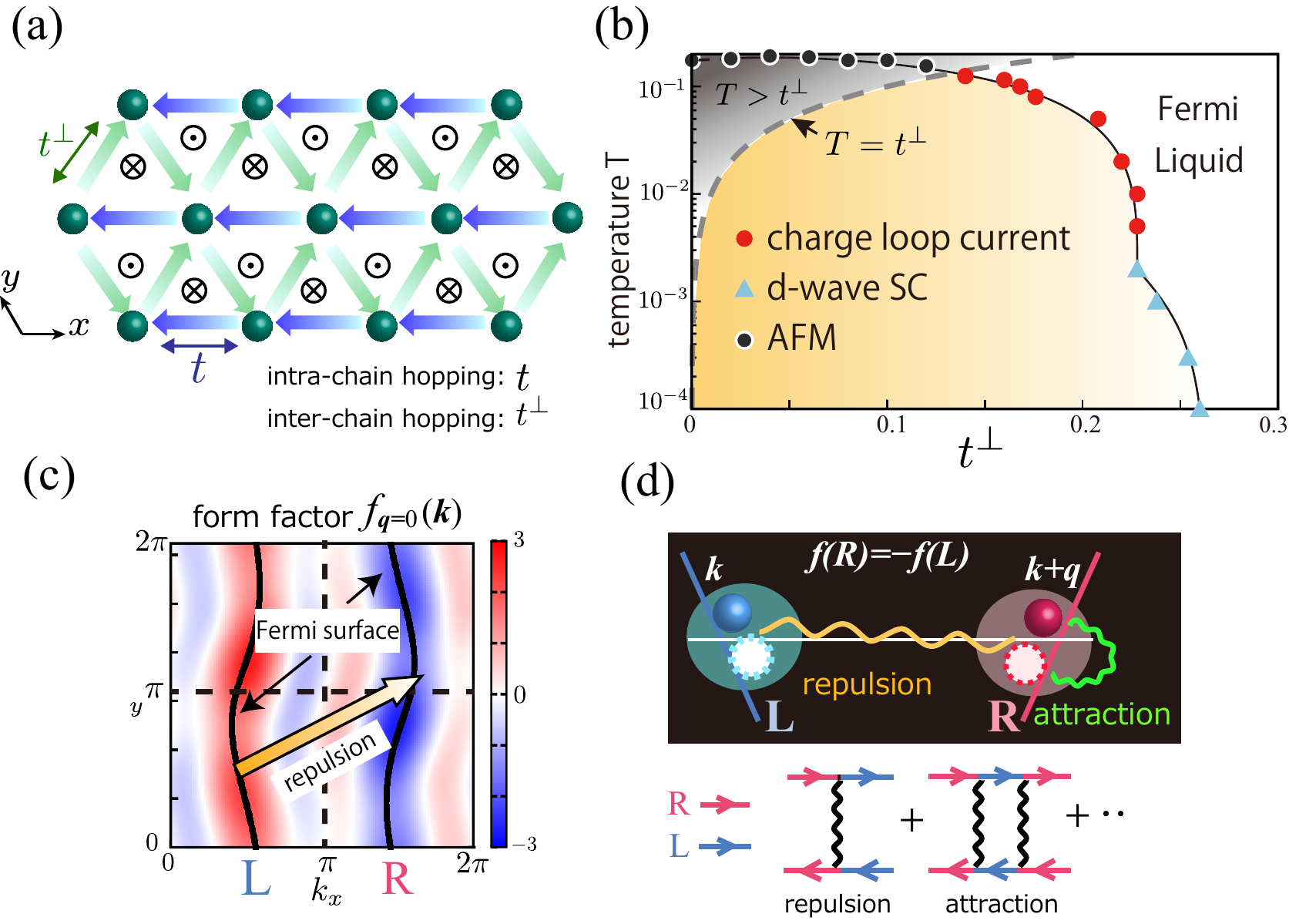}
\caption{
(a) Geometrically frustrated coupled chain model.
$t$ and $t^\perp$ represent the intra- and inter-chain hopping integrals, respectively.
Intra-unit-cell cLC pattern obtained in Ref. \cite{RTazai-PRB2021}
is illustrated.
(b) Obtained transition temperatures as function of $t^\perp$.
The cLC state is realized for $t^\perp=0.1\sim0.2$.
The AFM and d-wave SC appear for $t^\perp\lesssim0.1$ and
$t^\perp\gtrsim0.2$, respectively.
The cLC appears in the Fermi liquid (FL) regime. 
(c) Obtained charge-channel odd-parity form factor
$f^{\q={\bm0}}(\k)$ ($\propto \sin k_x + b \sin 3k_x$).
(d) (upper) The cLC order induced by 
inter-branch repulsion and intra-branch attraction.
(lower) Inter-branch repulsion [intra-branch attraction]
due to odd-number [even-number] spin fluctuation exchange processes.
Cited from Ref. \cite{RTazai-PRB2021}.
}
\label{fig:fRG-current}
\end{figure}

Here, we calculate the spin and charge susceptibilities
with nonlocal form factor $\chi_f^{s,c}(\k)$
in the geometrically frustrated coupled chain model
shown in Fig. \ref{fig:fRG-current} (a).
The kinetic energy is given as
$\e_\k= 2t\cos k_x + 2t_\perp(\cos k_y + \cos(k_x+k_y))$.
Here, we optimize the form factor $f^{\q}(\k)$ 
so as to maximize $\chi_f^{s,c}(\k)$ in the RG theory
by following Ref. \cite{RTazai-PRB2021},
under the constraint 
$\sum_{\k} |f^{\q}_{\k}|^2=1$ at each $\q$-point. 
For this purpose, we introduce the Fourier expansion form of $f^{\q}(\k)$ as
\begin{eqnarray}
f^{\q}(\k)=\sum_{n,m=1}^{7}a_{n m}^{\q} h_{n}(k_x) 
h_m(k_y),\end{eqnarray}
where $\frac{h_{n}(k)}{\sqrt{2}}=\{ \frac{1}{\sqrt{2}}, \cos k, \cos 2k, 
\cos 3k, \sin k, \sin 2k, \sin 3k \}$ for $n=1,2,3,4,5,6,7$, respectively.
More detailed explanations are presented in 
Ref. \cite{SOnari-PRR2020}.
Hereafter, we study the half-filling case at $U=2.0$.


Figure \ref{fig:fRG-current} (b) shows the 
transition temperatures obtained by the fRG method as function of $t^\perp$.
The charge-channel odd-parity solution at wavevector $\q={\bm0}$, 
which corresponds to the cLC state, is obtained for $t^\perp=0.1\sim0.2$.
The AFM and d-wave SC appear for $t^\perp\lesssim0.1$ and
$t^\perp\gtrsim0.2$, respectively.
The cLC appears in the Fermi liquid (FL) regime,
where inter-chain coherence is established.
The optimized charge-channel form factor at $\q=\bm{0}$
is depicted in Fig. \ref{fig:fRG-current} (c).
The relation 
$f^{\bm{0}}(\k)\simeq -f^{\bm{0}}(-\k) (\propto \sin k_x+b\sin 3k_x )$ holds.
Then, the real-space order parameter is 
$\delta t_{ij}=-\delta t_{ji}$ that leads to the emergence of ferro-cLC order. 
The schematic picture of the cLC in real space is depicted in 
Fig. \ref{fig:fRG-current} (a), which is a magnetic-octupole-toroidal order. 
Thus, the intra-unit-cell cLC order is obtained without any bias
in a simple frustrated chain Hubbard model.

Here,
the $p$-wave form factor $f^{\bm{0}}(\k_L)=-f^{\bm{0}}(\k_R)$
is induced by the inter-branch repulsion and the intra-branch attraction
in Fig. \ref{fig:fRG-current} (d).
By means of the $g$-ology theory,
it originates from the enhancement of the forward scatterings ($g_2$,$g_4$)
\cite{RTazai-PRB2021}.
The origin of the $p$-wave form factor is also understood 
based on the ''particle-hole (ph) gap equation'':
Inter-branch repulsion [intra-branch attraction]
due to odd-number [even-number] spin fluctuation exchange processes,
as shown in Fig. \ref{fig:fRG-current} (d).
Thus, the cLC is explained by the spin-fluctuation-driven mechanism 
based on 2D FL concept
\cite{Kontani-ROP,Moriya-review1,Moriya-review2,Yamada-text}.
Therefore, the cLC order emerges next to the AFM phase 
in Fig. \ref{fig:fRG-current} (b).
We stress that the cLC phase in the FL regions is replaced with the AFM phase
if we remove the geometrical frustration.
The present theory provides important guidelines for generating the cLC in materials; 
the geometrical frustration and the magnetic criticality.
It is an important issue to understand the cLC order
recently observed in ladder \cite{cLC-2leg} 
and square lattice \cite{TRSB-neutron1,TRSB-neutron2,SHG-cuprate,TRSB-iridate,SHG-iridate} 
systems based on the present cLC mechanism.

\section{Unconventional orders in Fe-based superconductors}
\label{sec:Kontani4}

\subsection{Motivation}
\label{sec:Kontani4-0}

The normal state electronic states in 
Fe-based superconductors exhibit amazing variety
as we explained in Sect. \ref{sec:Kontani1-2}.
The discovery of the ``electronic nematic states'' 
in Fe-based superconductors has triggered significant
progress in the field of strongly correlated electron systems.
For example, the $B_{1g}$ nematic states with (without)
magnetization emerge in BaFe$_2$As$_2$ (FeSe) families, 
and $B_{2g}$ nematicity appears in RbFe$_2$As$_2$.
Such rich variety of the electronic states
set a very severe constraint on the theory of Fe-based superconductors,
and the theoretical study will serve to understand the 
pairing mechanism of high-$T_{\rm c}$ superconductivity.
Now, it is widely accepted that 
the correlation-driven spontaneous symmetry breaking in metals 
exhibits richer variety than we had expected before the discovery 
of LaFeAsO in 2008 \cite{Hosono-1111},
and this issue has become a central open problem
in condensed matter physics.

In strongly correlated metals,
the ``correlation-driven density-wave (DW)''
has been studied intensively 
\cite{SOnari-PRL2012,Tsuchiizu-PRL,Sachdev-CDW1,Sachdev-CDW2,Husemann,Chubukov-cuprate2014,Davis:2013ce,Kivelson-NJP}.
Various beyond-mean-field approximations have been developed to 
explain the electronic nematic phases in Fe-based superconductors,
such as spin nematic scenarios 
\cite{Kivelson-spin-nematic,Fernandes1,DHLee-spin-nematic,QSi-spin-nematic,Valenti-spin-nematic,Fernandes-Nature}
and the orbital/charge-order scenarios 
\cite{HKontani-PRB2011,SOnari-PRL2012,Onari-SCVCS,Onari-FeSe,YYamakawa-PRX2016,Text-SCVC,JP,Fanfarillo,Chubukov-nematic-rev}
These scenarios were successfully applied to the nematicity in 
BaTi$_2$Sb$_2$O \cite{Nakaoka-Ti}
and 1T-TaS$_2$  \cite{Hirata}.

In this section, we try to explain various key experiments 
of many Fe-based superconductors based on the itinerant picture,
by focusing on the  ``paramagnon interference mechanism''
shown in Fig. \ref{fig:phase-AFM} (b).
First, we introduce the unfolded five-orbital Hubbard model
for Fe-based superconductors, $H=H_0+H_I$.
The kinetic term is given as
\begin{eqnarray}
H_0&=&\sum_{\k;\s=;l,m} 
h_0^{lm}(\k) c_{\k,l,\s}^\dagger c_{\k,m,\s},
\label{eqn:Ham}
\end{eqnarray}
where $l,m$ are the $d$-orbital indices.
Hereafter, we denote
$d_{3z^2-r^2}$, $d_{xz}$, $d_{yz}$, $d_{xy}$, $d_{x^2-y^2}$ as $l=1,2,3,4,5$.
The unfolded Fermi surfaces (FSs) for Ba122 and FeSe are shown in
Fig. \ref{fig:FeAs-layer} (c) and \ref{fig:phase-11} (d), respectively.
In both compounds, the 
FSs are mainly composed of $xz$, $yz$, and $xy$ orbitals.
The multiorbital Coulomb interaction term is given in Eq. (\ref{eqn:HamI-2orb}).
%
In transition metals, the relations $U=U'+2J$ and $J=J'$ holds approximately.
In addition, $J/U=0.1\sim0.2$, which means that $U'/U=0.8\sim0.6$
according to first principles study
\cite{TMiyake-JPSJ2010}.
$H_I$ in Eq. (\ref{eqn:HamI-2orb}) is compactly expressed as
\begin{eqnarray}
H_I= 
\frac14 \sum_{i,1\sim 4} ({\hat U})_{1,2,3,4} c_{1}^\dagger c_2 c_3 c_{4}^\dagger ,
\label{eqn:HU2} 
\end{eqnarray}
where $1\equiv(l_1,\s_1)$ and so on.
Reflecting the $SU(2)$ symmetry,
${\hat U}$ in Eq. (\ref{eqn:HU2}) is expressed as
\begin{eqnarray}
{\hat U}= \frac12 {\hat U}^c\delta_{\s_1,\s_2}\delta_{\s_4,\s_3}
+\frac12 {\hat U}^s{\bm \s}_{\s_1,\s_2}\cdot {\bm \s}_{\s_4,\s_3}
\label{eqn:HU3} 
\end{eqnarray}
where ${\hat U}^{c(s)}$ is the charge (spin) channel
Coulomb interaction in $5^2\times 5^2$ matrix form;
see Refs. \cite{HKontani-PRB2011,YYamakawa-PRX2016}.
(Note that $U^s=U$ and $U^c=-U$ in single-orbital models.)

\color{black}
According to the constrained RPA (cRPA) study
\cite{TMiyake-JPSJ2010},
the averaged intra-orbital Coulomb interaction on Fe-ion
${\bar U}$ in the $d$-$p$ orbital (8 orbital) model
ranges from $4$eV to $7$eV for 1111, 122, 111, and 11 compounds.
In contrast, ${\bar U}$ is reduced to $2.5$eV to $4$eV 
in the $d$ orbital (5 orbital) model,
due to the screening effect by $p$-orbitals.
In later sections, we set the Coulomb interaction in Eq. (\ref{eqn:HU2}) 
as ${\hat U}\equiv r{\hat U}_{\rm X}^{d-p}$
(X=LaFeAsO, FeSe, and BeFe$_2$As$_2$),
which is given by the cRPA result for the $d$-$p$ orbital model
\cite{TMiyake-JPSJ2010}
multiplied by the reduction factor $r$.
Here, $r$ is the solo model parameter
in the majority of the present study.
Here, we set $U\sim1.7$eV 
($r\sim0.4$ in LaFeAsO and $r\sim0.25$ in FeSe)
when the self-energy is neglected
to satisfy the paramagnetic condition $\a_S<1$.
If FLEX self-energy is included in the electron Green functions,
we use larger $U$ ($U\sim2.8$eV and $r\sim0.4$) for FeSe model
because $\alpha_S$ is reduced by the self-energy. 
\color{black}

\subsection{Nematic order and fluctuations due to paramagnon-interference: the SC-VC theory}  
\label{sec:Kontani4-1}

In this subsection, 
we analyze the orbital fluctuations driven by the VCs
based on the 
self-consistent vertex correction (SC-VC) method in 
Sect. \ref{sec:Kontani2-4}.
The formalism in Sect. \ref{sec:Kontani2-4}
can be naturally extended to multiorbital Hubbard models
\cite{SOnari-PRL2012}.

\begin{figure}[htb]
\centering
\includegraphics[width=.99\linewidth]{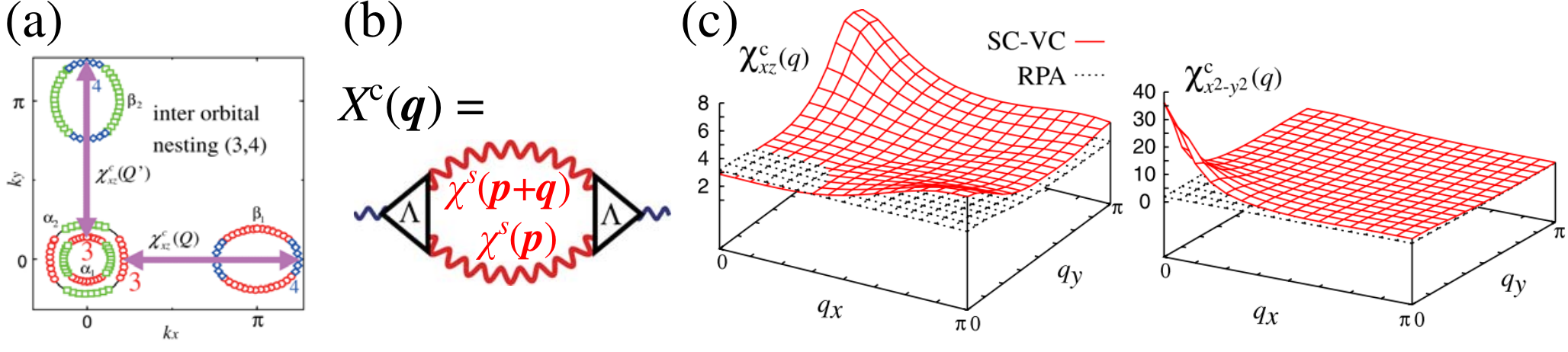}
\caption{
(a) FSs of La1111 model at $n=6.1$ \cite{SOnari-PRL2012}.
(b) VC for the charge-channel susceptibility in the SC-VC theory.
In this theory, $I_\q^{\s,\rho}$ is given in Fig. \ref{fig:fig-MTAL}.
(c) Obtained enhancements of orbital susceptibilities
$\chi_{xz}(\q)$ and $\chi_{x^2-y^2}(\q)$ in La1111 Hubbard model
\cite{SOnari-PRL2012}. 
They exhibit peaks at $\q\approx(\pi,0),(0,\pi)$ and at $\q\approx{\bm0}$,
respectively.
Both susceptibilities in the RPA shown by broken lines are very small.
Cited from Ref. \cite{SOnari-PRL2012}.
}
\label{fig:SCVC-1111}
\end{figure}

Here, study the five-orbital Hubbard model 
derived from the Density-functional theory (DFT) band calculation 
for La1111 compound
\cite{KKuroki-PRL2008,SOnari-PRL2012}.
Its unfolded FS is similar to Fig. \ref{fig:FeAs-layer} (d),
whereas the $xy$-orbital hole pocket around $(\pi,\pi)$ point is absent;
see Fig. \ref{fig:SCVC-1111} (a).
In the RPA, in which any VCs are neglected,
strong spin fluctuations develop at $\q\approx (\pi,0),(0,\pi)$.
In contrast, no charge-channel fluctuations develop at all.
Thus, the nematic phase transition without magnetism at $T_S \ (>T_N)$
cannot be explained by the RPA.

For convenience,
we introduce the matrix expressions of the quadrupolar operators 
with respect to orbitals 2-4 as:
\begin{eqnarray}
&&({\hat O}_{x^2-y^2})_{lm}= (-1)^l \ \ {\rm for} \ (lm)=(22),(33), 
\label{eqn:O-x2y2} \\
&&({\hat O}_{xz})_{lm}= +1 \ \ {\rm for} \ (lm)=(3,4),(4,3),
\label{eqn:O-xz} 
\end{eqnarray}
and other elements are zero.
Note that $({\hat O}_{\mu\nu})_{lm} \propto \langle l |(l_\mu l_\nu+ l_\nu l_\mu)| m \rangle$ 
with $\mu,\nu=x,y,z$.
Then, $\Gamma$-channel susceptibility is given as
$\chi_\Gamma^x(q)= {\rm Tr} \{ {\hat O}_\Gamma {\hat \chi}^x(q){\hat O}_\Gamma \}
\ (= \sum_{ll'mm'}({\hat O}_\Gamma)_{ll'} \chi_{ll',mm'}^x(q)({\hat O}_\Gamma)_{m'm})$

In the five-orbital model, 
the Green function ${\hat G}(k)$ is expressed as $5\times5$ matrix,
and the $25\times25$ matrix of the irreducible susceptibility in the RPA is
\begin{eqnarray}
[{\hat \chi}^{(0)}(q)]_{ll',mm'}=-T\sum_k G_{lm}(k+q)G_{m'l'}(k).
\end{eqnarray}
The bare Coulomb interaction
${\hat U}^{x}$ ($x=c,s$) is expressed as $25\times25$ matrix.
Note that $U_{ll,ll}^s=-U_{ll,ll}^c=U$.
In the present model,
$\chi^s_{22,22}(\q)$ [$\chi^s_{33,33}(\q)$] possesses broad peak around $\q=(0,\pi)$ [$\q=(\pi,0)$]
due to the intra-orbital nesting between electron- and hole-pockets; see Fig. \ref{fig:SCVC-1111} (a).
In contrast, $\chi^s_{44,44}(\q)$ exhibits small peak around $\q=(\pi,\pi/2),(\pi/2,\pi)$
due to the weak nesting between electron-pockets.
In Ba122 model, the peak of $\chi^s_{44,44}(\q)$ shifts to $\q=(0,\pi),(\pi,0)$
because of the additional $d_{xy}$-orbital hole-pocket around M point; 
h3 in Fig. \ref{fig:FeAs-layer} (d).

Next, we calculate the charge-channel susceptibility 
${\hat \chi}^c(q)$ by including both the MT
and AL terms in Fig. \ref{fig:fig-MTAL},
by applying the SC-VC theory 
given as Eqs. (\ref{eqn:suscep-SCVC})-(\ref{eqn:suscep-SCVC-X}).
In the presence of moderate spin fluctuations,
${\hat \chi}^{c}(q)$ is strongly enhanced by the charge-channel 
AL-VC $X^{{\rm AL},c}\sim T\sum \chi^s\chi^s$,
which is shown in Fig. \ref{fig:SCVC-1111} (b)
 \cite{SOnari-PRL2012}.
On the other hand, it is verified that 
${\hat X}^{{\rm AL},s}\sim T\sum \chi^s\chi^c$ is less important
in Fe-based superconductors
 \cite{YYamakawa-PRX2016}.
Figure \ref{fig:SCVC-1111} (c)
shows the charge-channel susceptibilities
given by the self-consistent calculation of the AL and MT terms
(SC-VC method) for $n=6.1$, $J/U=0.088$ and $U=1.53$eV.
Both orbital susceptibilities
$\chi^c_{x^2-y^2}(\q)= 2(\chi^c_{22,22}- \chi^c_{22,33})$ and
$\chi^c_{xz}(\q)= 2(\chi^c_{34,34}+\chi^c_{34,43})$ 
are strongly enhanced by the charge AL term, ${\hat X}^{{\rm AL},c}$.
(The obtained results are essentially unchanged even if MT term is dropped.)
The enhancements of other orbital susceptibilities are small.
We well discuss in Sect. \ref{sec:Kontani4-6}  
that the relation $\chi^c_{xz}(\Q)>\chi^c_{x^2-y^2}({\bm 0})$
is realized in Ba122 model,
in which large $xy$-orbital hole pockets exists.

In summary, we developed the SC-VC method 
for multiorbital Hubbard models, and obtained strong
nematic and smectic orbital fluctuations in Fe-based superconductors
due to the AL process that describes the paramagnon interference effect
\cite{HKontani-PRB2011}.
The AL process is dropped in the RPA.
In the SC-VC theory,
the structure transition ($\a^c\approx1$)
occurs prior to the magnetic transition ($\a^s\approx1$)
for $J/U\lesssim 0.2$, consistently with experiments.
(In the DW equation explained in Sect. \ref{sec:Kontani3-3}
nonmagnetic orbital order is realized even for $J/U> 0.2$.)
When $\a^s\sim \a^c$, both $s_{++}$- and $s_\pm$-states could be realized,
depending on model parameters like the impurity concentration
\cite{Kontani-RPA,Hirsch-cross}.
Thus, the orbital-fluctuation-mediated 
superconductivity is expected near the nematic QCP.

\subsection{
Comparison between 1111 and 11 systems: the SC-VC theory  
}
\label{sec:Kontani4-2}


Near the quantum critical point (QCP) of the nematic order,
prominent nematic criticality has been observed in various compounds
by many experimental techniques,
such as the shear modulus measurement \cite{Yoshizawa,Bohmer},
in-plane resistivity anisotropy \cite{Fisher-review,B2g-Ishisa},
Raman spectroscopy \cite{Gallais,Gallais2,Raman-spectroscopy},
and the phonon softening measurement by x-ray scattering
\cite{Egami-softening}.
In these measurements, one can derive 
the ``electronic nematic susceptibility'' 
driven by the electron correlation,
free from the electron-phonon interaction.
For example, Raman nematic spectroscopy is free from
the acoustic phonon contribution because of the relationship 
$(\w_k/k)_{\rm photon} \gg (\w_k/k)_{\rm phonon}$
\cite{Kontani-Raman}.
It has been established that the nematicity 
of Fe-based superconductors originates from the 
electron correlation, while its transition temperature 
is raised by just several tens of Kelvins,
which we call the Jahn-Teller energy $T_{\rm JT}$.
That is, the relationship $T_{\rm S}= T_{\rm nem}+T_{\rm JT}$ holds,
where $T_{\rm S}$ is the observed structural transition temperature,
and $T_{\rm nem}$ is the correlation-driven nematic transition temperature.
In many Fe-based superconductors, $T_{\rm JT}$ is just $20\sim30$K.
For simplicity,
we set $T_S=T_{nem}$ by neglecting the Jahn-Teller energy.
in later discussions.


Experimentally, both AFM susceptibility and 
the electronic nematic susceptibility follows the 
Curie-Weiss behaviors in the ``tetragonal phase above $T_S$'': 
$\chi^s(\Q)\propto (T-T_{\rm N})^{-1}$
and $\chi_{\rm nem}\propto (T-T_{\rm S})^{-1}$, respectively.
Interestingly,
the relationship between $T_{\rm N}$ and $T_{\rm S}$
is strongly depends on compounds.
In many Ba122 families,
the relation $T_{\rm S}\gtrsim T_{\rm N}$ holds,
which is naively expected as the spin-fluctuation-driven nematicity.
In 1111 families and NaFeAs ($T_S=60$K), in contrast,
$T_{\rm S}$ is clearly larger than $T_{\rm N}$.
The AFM order appears inside the nematic phase
because the AFM correlation is magnified by the nematic order.
In FeSe with $T_{S}=90$K,
no AFM order appears down to zero temperature.
In fact, $T_{\rm N}$ derived from the Weiss temperature of $\chi^s(\Q)$ above $T_S$
takes large negative value.
Therefore, $\chi_{\rm nem}$ is strongly enlarged 
in the $C_4$ phase near the nematic phase,
whereas $\chi^s(\q)$ at $T=T_S$ drastically depends on materials.
Such drastic material dependences of $\chi^s(\Q)$ and $\chi_{\rm nem}$
put a strong constraint on the theory of nematicity.

\begin{figure}[htb]
\centering
\includegraphics[width=.8\linewidth]{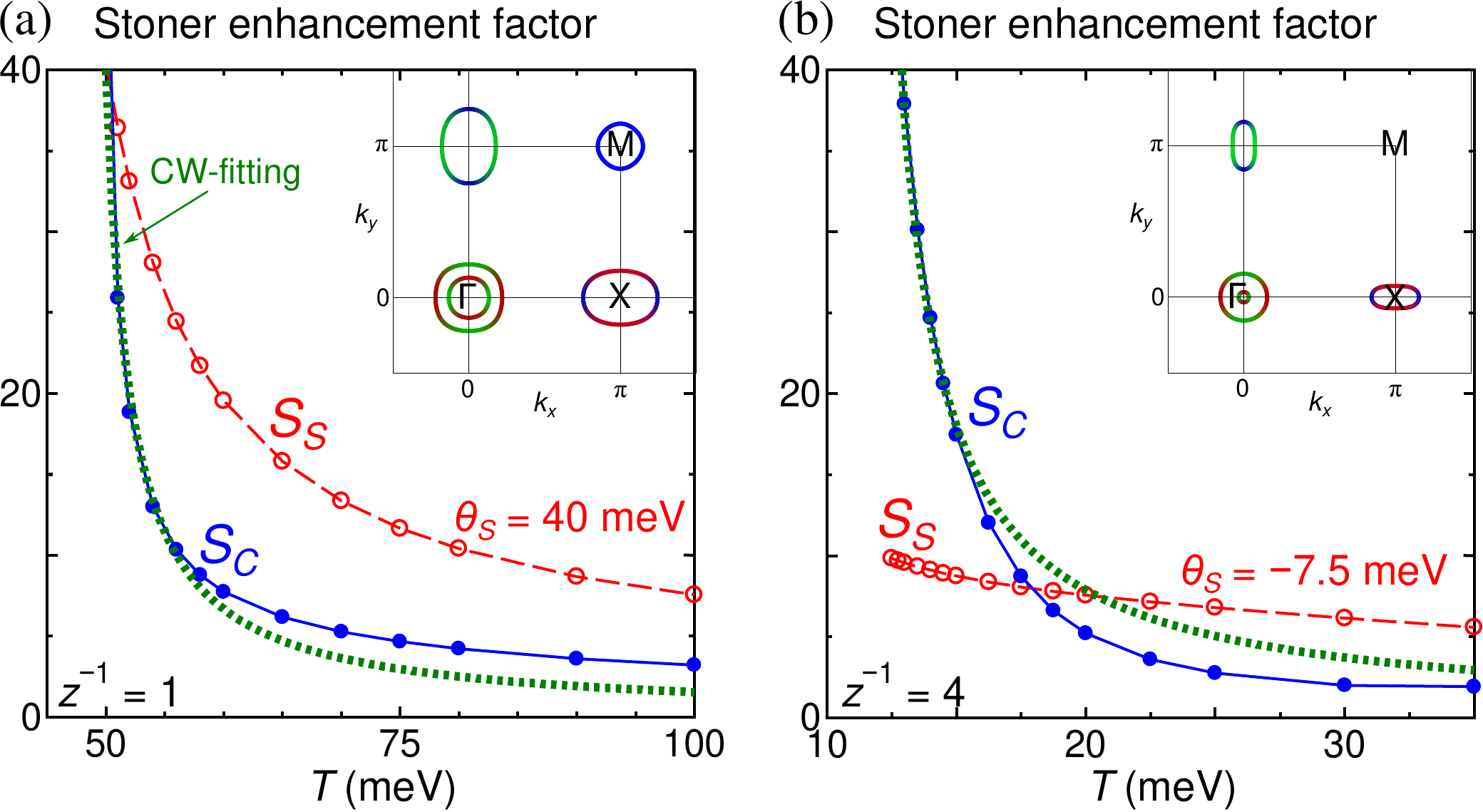}
\caption{
Enhancement factor of the nematic susceptibility ($S_c= 1/(1-\a_C)$)
and that of the spin susceptibility ($S_s= 1/(1-\a_S)$)
obtaind for (a) LaFeAsO model and (b) FeSe model
\cite{YYamakawa-PRX2016}.
In FeSe model, $S_c$ stronlgy increases due to the AL-process, 
while $S_s$ is nearly constant.
Cited from Ref. \cite{YYamakawa-PRX2016}.
}
\label{fig:FeSC-nematic}
\end{figure}

Now, we analyze the LaFeAsO model based on the SC-VC theory.
Figure  \ref{fig:FeSC-nematic} (a) shows the 
spin-channel Stoner enhancement factor $S_s\equiv (1-\a_S)^{-1}$ 
in LaFeAsO model with $U=1.74$ eV.
We also show the charge-channel Stoner enhancement factor $S_c\equiv (1-\a_C)^{-1}$
that is proportional to the nematic orbital susceptibility.
Both $S_{c}$ and $S_{s}$ follow the Curie-Weiss behaviors with 
$\theta_C=T_{\rm S}=48$ meV and $\theta_S =T_{\rm N}=40$ meV, respectively.
Similar results are obtained by analyzing Ba122 Hubbard model
\cite{YYamakawa-PRX2016}.
The obtained relation $T_{\rm S}\gtrsim T_{\rm N}$,
which is naively expected as the spin-fluctuation-driven mechanism,
is consistent with experimental reports in Ba122 families
\cite{Yoshizawa,Bohmer,Raman-spectroscopy,Gallais,Kontani-Raman,Fisher-review}.

Next, we analyze the FeSe model,
in which the particle-hole asymmetry of the bandstructure
is very large, and the ratio $\bar{J}/\bar{U}$ is just $\sim0.1$
\cite{TMiyake-JPSJ2010}.
These situations are favorable for the nematic ordering
due to paramagnon interference
\cite{YYamakawa-PRX2016}.
Figure  \ref{fig:FeSC-nematic} (b) shows the enhancement factors in 
FeSe model, in which the renormalized Coulomb is
$U^*=zU=1.75$eV and $z=0.25$ (i.e., $U=7$eV).
We stress that $S_{C}$ approximately follow the Curie-Weiss 
behavior with the Weiss temperature $\theta_C=12$ meV, which is
consistent with the experimental positive $\theta_C$ in FeSe \cite{Ishida-NMR}.
In contrast, the spin Weiss temperature
takes a large negative value ($\theta_S\sim-7.5$ meV),
so the nematicity without magnetization in FeSe families
is naturally explained by the paramagnon interference mechanism.

Here, we discuss the origin of the relation $S_C\gg S_S$ in FeSe.
The nematic susceptibility is enhanced not only by $\chi^s(\Q)$,
but also by the ``significant $T$-dependence of 
the spin-charge coupling term $\Lambda \sim T\sum_k (G(k))^2 G(k+\Q)$''
in Fig. \ref{fig:phase-AFM} (b)
\cite{Chubukov-nematic-rev,Kontani-Raman}.
According to Ref. \cite{YYamakawa-PRX2016},
the relation $\Lambda^2 \propto T^{-a}$ 
with $a\approx 1$ is realized in FeSe models,
by reflecting prominent particle-hole asymmetric bandstructure.
Since ${\hat X}^{{\rm AL},c} \sim |\Lambda|^2 (T \xi_s^2)$
given in Fig. \ref{fig:SCVC-1111} (b) increases at low temperatures,
and therefore $\chi^c_{x^2-y^2}({\bm0})$ is enlarged 
by following the Curie-Weiss law.
Another favorable condition for the nonmagnetic nematicity in FeSe 
is the absence of h3:
Its existence magnifies the $d_{xy}$-orbital spin fluctuations 
while they are unimportant for the AL-process driven orbital order.
In this case, spin fluctuations develop only in ($d_{xz},d_{yz}$)-orbitals.
This ``orbital selective spin fluctuations'' is favorable 
for the orbital polarization $n_{xz}\ne n_{yz}$
\cite{YYamakawa-PRX2016}.

To summarize, we studied the origin of the nematicity 
in Fe-based superconductors, 
by paying the special attention to the  {\it nonmagnetic nematic order} in FeSe.
Based on the SC-VC theory,
we succeeded in explaining the rich variety of the phase diagrams
in Fe-based superconductors, such as the 
nonmagnetic/magnetic nematic order in FeSe/LaFeAsO. 
In the present theory,
the ratio $\theta_S/\theta_C$ decreases 
when the size of the FS and the ratio $J/U$ are small.
The present results are verified by the 
DW equation analysis 
\cite{Onari-FeSe,SOnari-PRB2019,SOnari-PRR2020,Onari-Kontani-Front2022}
as we explain in the next subsection.

\subsection{Nematic electronic states in FeSe below $T_S$: DW equation analysis
\label{sec:Kontani4-3}
}

FeSe provides a very suitable platform to understand the 
central issues of Fe-based superconductors, i.e., 
the relationship between nematicity and magnetism,
and the origin of high-$T_{\rm c}$ superconductivity above 60K.
As we discussed in previous subsections,
the nonmagnetic nematicity in bulk FeSe at ambient pressure 
is naturally explained by the paramagnon interference mechanism.

Recently, the nematic electronic states below $T_S=90$K have been 
precisely determined by experimentally.
The two main characteristics of the nematic state in FeSe would be 
(i) Sign reversing $xz,yz$ orbital order
\cite{FeSe-Suzuki,FeSe-ARPES-noY}, and
(ii) Disappearance of the e-pocket around Y point
\cite{FeSe-Shen-singleE,FeSe-ARPES-Luke2020,FeSe-ARPES-Huh2019}.
Here, 
we reproduce these two characteristics theoretically,
by calculating the momentum and orbital dependences of the form factor
self-consistently based on the DW equation (\ref{eqn:DWeq3})
introduced in Sect. \ref{sec:Kontani3-3}.

\begin{figure}[htb]
\centering
\includegraphics[width=.60\linewidth]{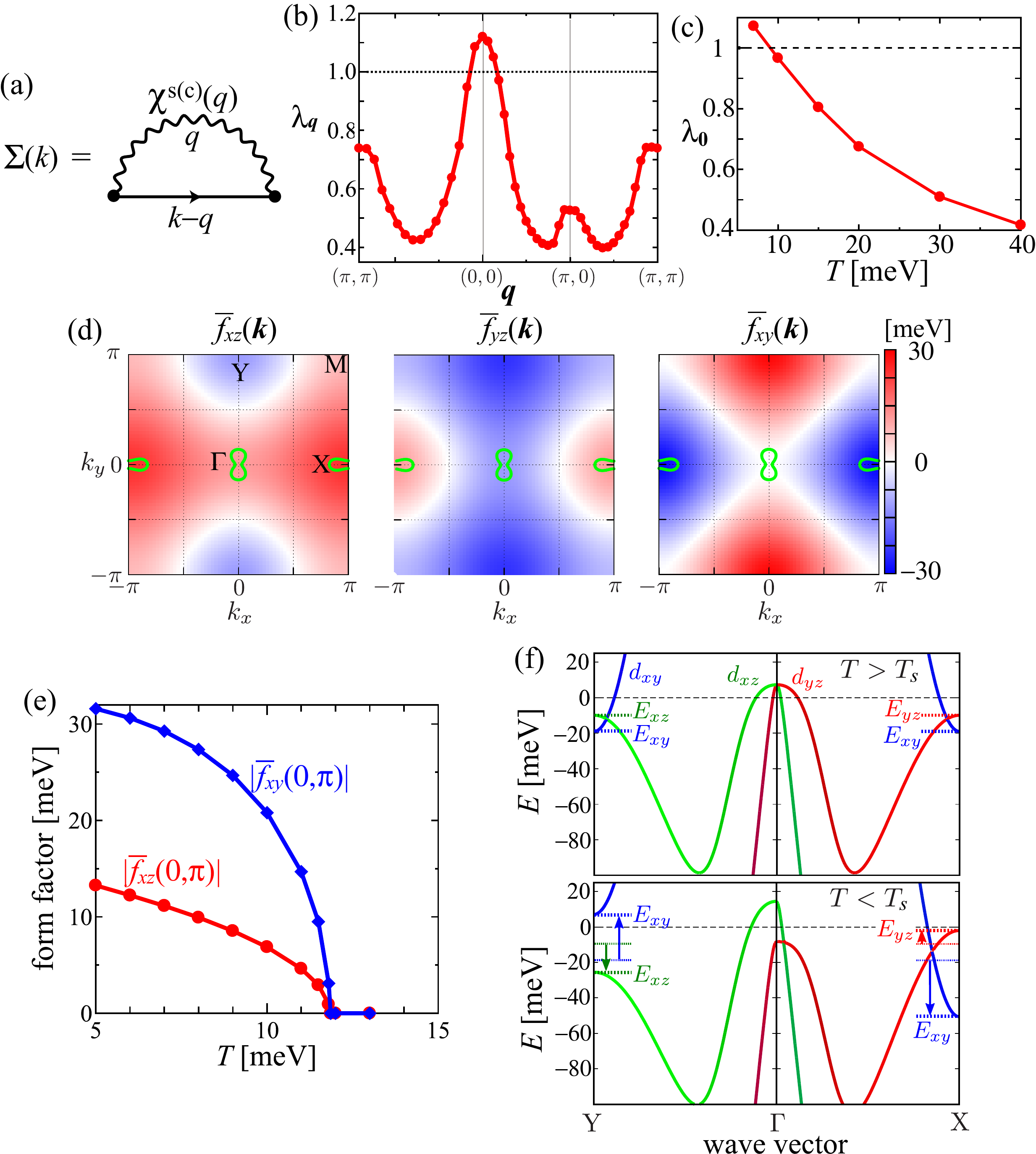}
\caption{
(a) Self-energy $\Sigma(k) =\delta \Phi_{\rm LW}/\delta G(k)$
in the one-loop approximation.
(b) Obtained $\q$-dependence of $\lambda_\q$ and
(c) $T$-dependence of $\lambda_{\q={\bm0}}$ 
given by the ``linearized DW equation'' with the self-energy
in the framework of the conserving approximation
\cite{RTazai-arXiv2021}.
(d) Renormalized form factors 
${\bar f}_{l,l}^{\q={\bm0}}(\k)$ for $l=xz$ ,$yz$, and $xy$,
derived from the full DW equation'' in the nematic state
\cite{RTazai-arXiv2021}.
The e-pocket around Y point disappears under the nematic order parameters.
The obtained ``one e-pocket + one h-pocket'' in the nematic phase 
of FeSe is consistent with recent experimental reports
\cite{FeSe-ARPES-noY,FeSe-Shen-singleE,FeSe-ARPES-Luke2020,FeSe-ARPES-Huh2019}.
Both ${\bar f}_{2,2}^{\bm0}(\k)$ along $k_y$-axis and 
${\bar f}_{3,3}^{\bm0}(\k)$ along $k_x$-axis exhibit sign reversal.
(e) Renormalized form factors as functions of $T$, and
(f) Bandstructure in the nematic phase
\cite{RTazai-arXiv2021}.
Here, two characteristics (i) and (ii) in the nematic state
in FeSe are well reproduced.
Sign reversal in the bandshift along $\Gamma$-X,Y axis 
is observed experimentally.
Cited from Ref. \cite{RTazai-arXiv2021}.
}
\label{fig:SCVC-11}
\end{figure}

In the DW equation method, macroscopic conservation laws are rigorously
satisfied if the kernel of the DW equation and the self-energy 
are derived from the same Luttinger-Ward function $\Phi_{\rm LW}$ 
\cite{Baym-Kadanoff}.
To satisfy the conservation laws,
here we calculate the $5\times5$ self-energy ${\hat \Sigma}(k)$
in the FLEX approximation shown in Fig. \ref{fig:SCVC-11} (a).
The obtained mass-enhancement factor for orbital $l$ is
$z_l^{-1}=1-{\rm Re}\d\Sigma_{l,l}/\d\e|_{\e=0}$.
In the present study, 
we obtain $z_{xz}^{-1}\approx3.5$ and $z_{xy}^{-1}\approx6.5$,
which are consistent with experimental values.
Both band-dispersion and form factor 
are renormalized by the factor $z$.
Here, we set ${\hat U}$ in Eq. (\ref{eqn:HU2}) 
as $r{\hat U}_{\rm FeSe}^{d-p}$, 
which is the cRPA Coulomb interaction for $d$-$p$ model of FeSe 
\cite{TMiyake-JPSJ2010}
multiplied by the reduction factor $r$.
Now, we set $r=0.37$ that corresponds to $U\sim2.6$eV.

Now, we solve the linearized DW equation (\ref{eqn:DWeq3})
by using the full Green function with ${\hat \Sigma}(k)$
in order to satisfy the conservation laws.
Figures \ref{fig:SCVC-11} (b) and (c) represents the 
obtained eigenvalue $\lambda_\q$
at $T=5$me and $r=0.40$.
The ferro ($\q={\rm 0}$) DW instability strongly develops
due to the interference between small spin fluctuations ($\a_S\lesssim 0.9$).
In addition, the eigenvalue $\lambda_{\q={\bm0}}$ reaches unity at 
$T\approx10$meV for $r\ge0.36$,
which is consistent with the experimental transition temperature $T_S=90$K.
The obtained form factor ${\hat f}^{\q={\bm0}}(\k)$
induces $B_{1g}$ nematic orbital order ($n_{xz}\ne n_{yz}$),
consistently with the enlarged orbital 
susceptibility $\chi^c_{x^2-y^2}(\q)$ at $\q={\bm 0}$
obtained by the SC-VC theory in Fig. \ref{fig:SCVC-1111} (c).

Next, we solve the ``full DW equation'' in Eq. (\ref{eqn:DSigma}) 
self-consistently, 
in order to analyze the electronic nematic states under $T_S$
\cite{RTazai-arXiv2021,Onari-Kontani-Front2022}.
Figure \ref{fig:SCVC-11} (d) 
shows the renormalized form factors 
${\bar f}_{l}(\k) \equiv z_l {f}_{l,l}^{{\bm0}}(\k)$ 
for $l=3 \ (yz)$ and $l=4 \ (xy)$.
In the obtained $B_{1g}$ symmetry solution satisfies 
the relations ${\bar f}_{xz}(k_x,k_y)=-{\bar f}_{yz}(k_y,k_x)$ 
and ${\bf f}_{xy} \sim \cos k_x - \cos k_y$.
The obtained nematic FS is shown in each panel.
Figure \ref{fig:SCVC-11} (e) exhibits the
$T$-dependence of the form factor, 
and the band dispersion in the nematic state
is shown in Fig. \ref{fig:SCVC-11} (f).
We see that the e-pocket around Y-point is lifted by 
the $d_{x^2-y^2}$-wave bond order ${\bar f}_{xy}(0,\pi) >0$. 
Thus, the obtained coexistence of
the bond order on $xy$-orbital 
the orbital order on $xz,yz$-orbital with sign reversal
are consistent with the two characteristics (i) and (ii)
in the nematic states in FeSe
\cite{FeSe-Suzuki,FeSe-ARPES-noY,FeSe-ARPES-noY,FeSe-Shen-singleE,FeSe-ARPES-Luke2020,FeSe-ARPES-Huh2019}.

In Figs. \ref{fig:SCVC-11} (b)-(g),
we performed the conserving approximation by 
taking account of the FLEX self-energy $\Sigma_{\rm FLEX}$.
It is found that the effect of $\Sigma_{\rm FLEX}$ is just to 
reduce the DW instability,
while the $q$-dependence of $\lambda_\q$ and the form factor
are essentially unchanged \cite{Onari-FeSe}.
Therefore, in later subsections,
we sometimes drop $\Sigma_{\rm FLEX}$
in order to simplify the analysis.


\subsection{FeSe phase diagram under pressure: DW equation analysis
\label{sec:Kontani4-4}
}

\begin{figure}[htb]
\centering
\includegraphics[width=.7\linewidth]{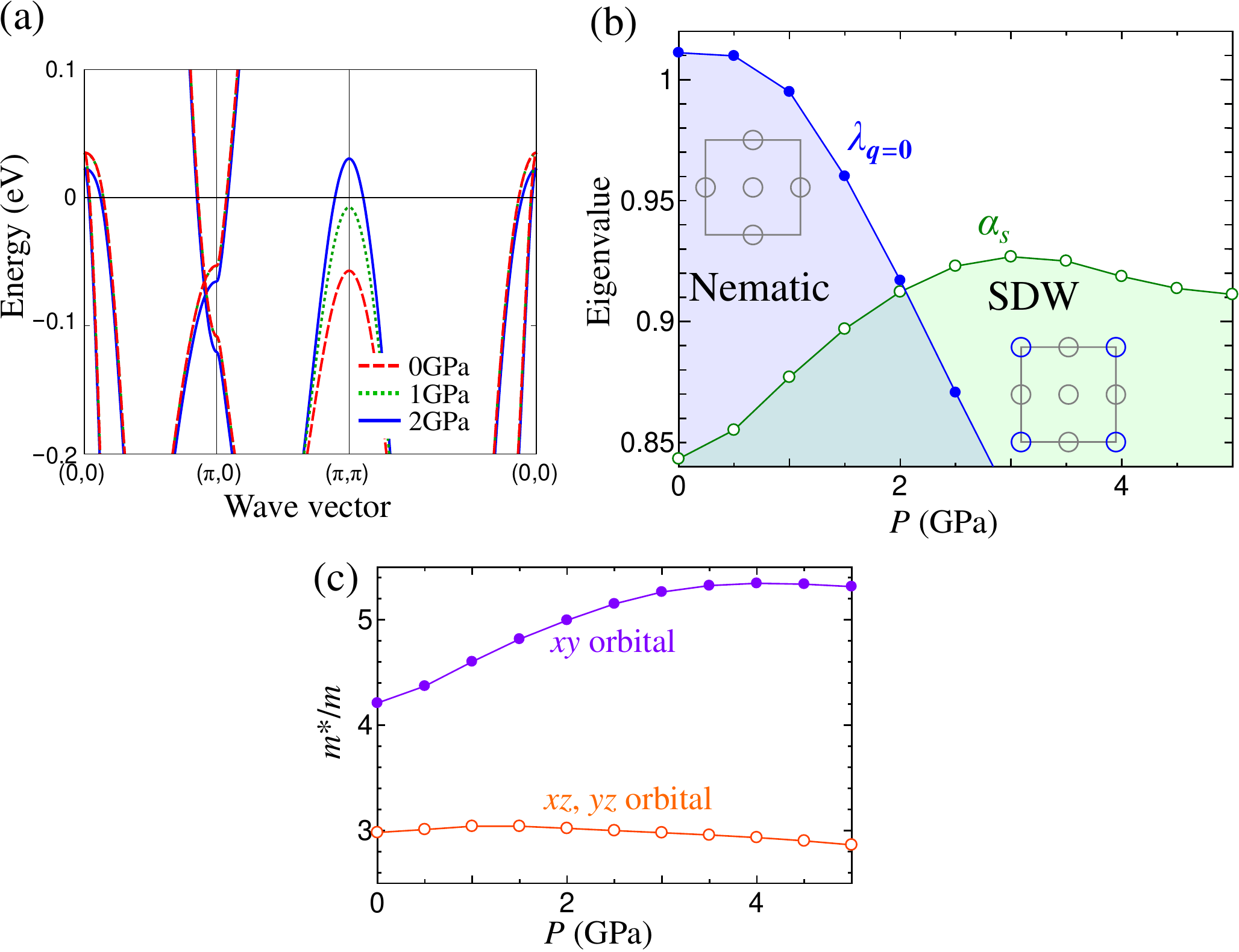}
\caption{
(a) Pressure-induced change in the bandstructure in FeSe.
(b) Obtained eigenvalue of the nematic order $\lambda_{\bm 0}$
and the spin Stoner factor $\a_S$, and
(c) Obtained mass-enhancement factor,
as function of $P$ at $T=5{\rm meV}$.
(Y. Yamakawa {\it et al.}, unpublished.)
}
\label{fig:SCVC-11-P}
\end{figure}

Next, we discuss the rich $P$-$T$ phase diagram in FeSe 
exhibited in Fig. \ref{fig:phase-11-PT},
which vividly demonstrates the strong interplay between the 
nematicity, magnetism and superconductivity.
For this purpose, we construct the first principles model 
for FeSe as function of $P$, and discovered theoretically the
pressure-induced emergence of $d_{xy}$-orbital hole-pocket.
The obtained $P$-dependent bandstructure is
shown in Figs. \ref{fig:SCVC-11-P} (a).
The $xy$-orbital level around M point prominently shifts upwards
because the Se-height increases under pressure.
For this reason, $xy$-orbital pocket appears for $P\gtrsim2$GPa.
This pressure-induced Lifshitz transition 
has been confirmed by the sign change in $R_{\rm H}$ \cite{FeSe-RH}
and the abrupt increment of the knight shift \cite{FeSe-Knight-shift}.

Now, we analyze this FeSe model using the linearized DW equation (\ref{eqn:DWeq3})
\cite{YYamakawa-unpublished2021}. 
We calculate the self-energy ${\hat \Sigma}(k)$ based on the FLEX approximation,
and incorporate it into the DW equation.
Figure \ref{fig:SCVC-11-P} (b) shows the 
obtained eigenvalue of the nematic order $\lambda$
and the spin Stoner factor $\a_S$ as function of $P$ at $T=5{\rm meV}$.
With increasing $P$, $\lambda$ monotonically decreases 
because of the slight increment of the bandwidth.
Around $P=2$GPa, $xy$-orbital e-pocket emerges.
Due to this pressure-induced Lifshitz transition,
the spin susceptibility on the $d_{xy}$ orbital is enhanced,
whereas that on $d_{xz},d_{yz}$ orbitals is gradually reduced.
Since the nematicity is mainly driven by the spin fluctuations on $d_{xz},d_{yz}$ orbitals 
through the intra-orbital VCs, 
the nematic eigenvalue $\lambda$ in Fig. \ref{fig:SCVC-11-P} (b) remains small for $P>2$GPa.
Thus, rich $T$-$P$ phase diagram in the normal state of FeSe 
is naturally understood based on the paramagnon interference mechanism.

Figure \ref{fig:SCVC-11-P} (c) show the 
mass-enhancement factors $z_l^{-1}=(m^*/m)_l$ for the orbitals 
$l=xz(yz)$ and $l=xy$ obtained by the FLEX approximation.
The obtained relations $z_{xy}^{-1}\approx5$ and $z_{xz}^{-1}\approx3$ 
are consistent with the LDA+DMFT analysis.


\subsection{$B_{2g}$ nematic order in heavily hole-doped $A$Fe$_2$As$_2$ ($A$=Cs,Rb)
\label{sec:Kontani4-5}
}

In previous subsections, we explain the $B_{1g}$ nematic order in
typical Fe-based superconductors 
based on the paramagnon interference mechanism.
Surprisingly,
a new type of nematic order was recently discovered
in heavily hole-doped ($n_d=5.5$) compound AFe$_2$As$_2$ (A=Cs, Rb)
\cite{B2g-Ishisa,B2g-NMR,B2g-STM}.
The discovered nematicity has $B_{2g}$ (=$d_{xy}$) symmetry, 
rotated by $45^\circ$ from the $B_{1g}$ (=$d_{x^2-y^2}$) nematicity 
in usual compounds with $n_d\approx6$.
The discovery of $B_{2g}$ nematicity provides a very useful information
to figure out the unique mechanism of nematicity 
for all Fe-based superconductors.
To reveal the origin of the $B_{2g}$ nematicity,
spin nematic (or vestigial order) scenario has been proposed
in Refs. \cite{Fernandes-B2g,QSi-B2g}.
In this article,
we investigate the symmetry-breaking in the self-energy
based on the DW equation (\ref{eqn:DWeq3}).

\begin{figure}[htb]
\centering
\includegraphics[width=.8\linewidth]{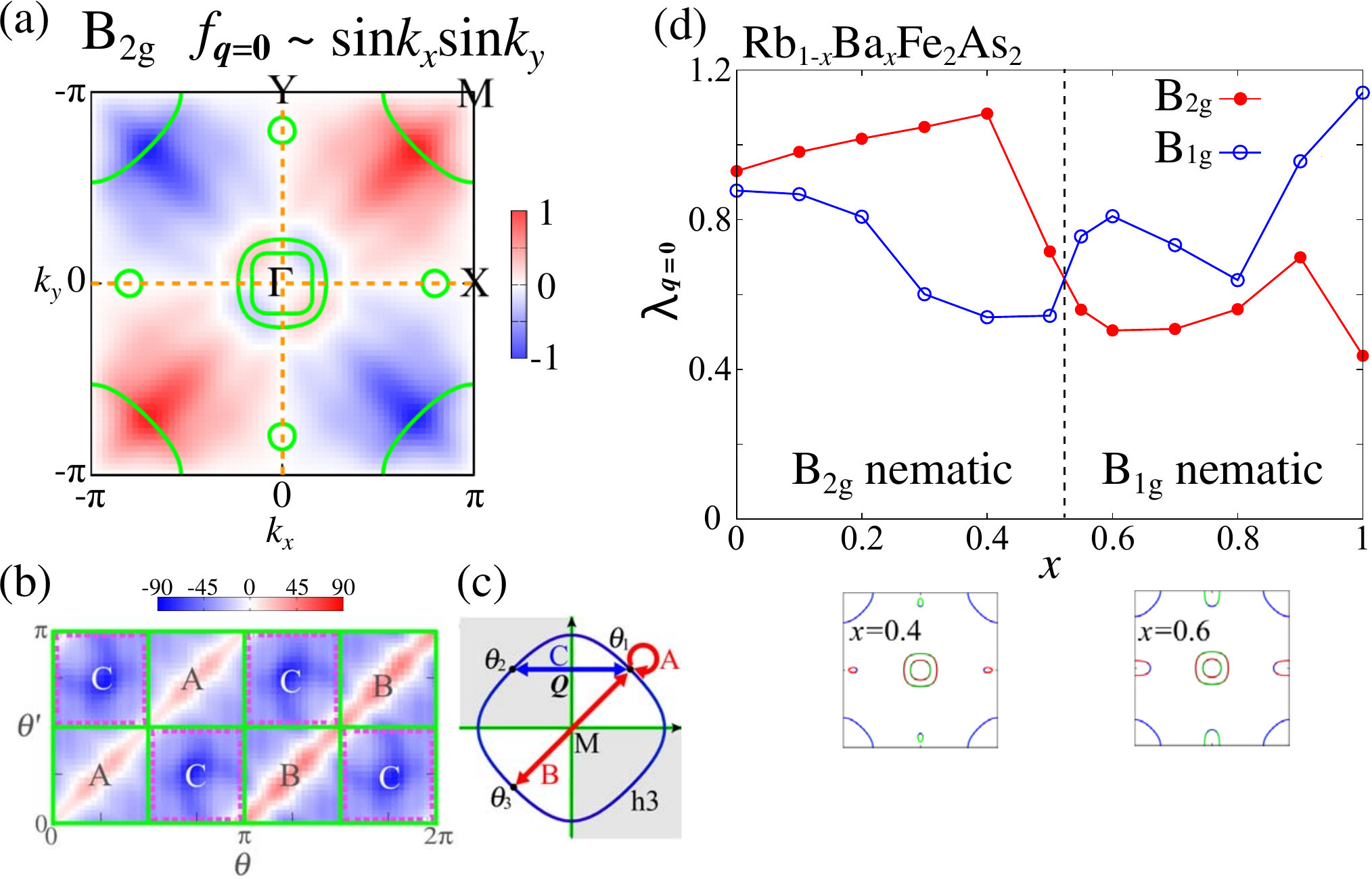}
\caption{
(a) The FSs of RbFe$_2$As$_2$ model, together with
the obtained form factor $f_{l,l}^{\q={\bm0}}(\k)$ for $l=4$ $(=d_{xy})$.
The $B_{\rm 2g}$ symmetry ($=d_{xy}$-wave) bond order is derived
\cite{SOnari-PRB2019}.
(b) $I_\q^c(\k,\k')$ on the FS h3.
$\theta$ and $\theta'$ are positions of $\k$ and $\k'$ on FS h3, respectively.
(c) $d_{xy}$ wave form factor due to attractive (repulsive)
interaction A and B (C).
(d) $B_{1g}$ and $B_{2g}$ eigenvalues in Rb$_{1-x}$K$_x$Fe$_2$As$_2$.
The symmetry of dominant nematic fluctuations 
changes at the Lifshitz transition $x\sim0.5$
\cite{SOnari-PRB2019},
consistently with recent experimental report
\cite{B2g-Ishisa}.
Cited from Ref. \cite{SOnari-PRB2019}. 
}
\label{fig:Cs122}
\end{figure}

The FSs of heavily hole-doped system AFe$_2$As$_2$
($n_d=5.5$) derived from the first principles study are
shown in Fig. \ref{fig:Cs122} (a).
The hole FS around M point composed of $d_{xy}$-orbital is large,
while the Dirac pockets near X and Y points are small.
The arrows $\Q$ denote the most important 
intra-$d_{xy}$-orbital nesting vector.
In the RPA or FLEX approximation, 
strong spin fluctuations at $\q\approx \Q$ is obtained,
consistently with inelastic neutron scattering study
\cite{neutron-Lee2011,Suzuki-KFe2As2-2011}.

Now, we perform the linearized DW equation analysis
for RbFe$_2$As$_2$ model.
Figure \ref{fig:Cs122}(a) presents
the dominant form factor at $\q={\bm 0}$, $f^{\q=\bm{0}}_{4,4}(\k)$,
for the largest eigenvalue $\lambda=0.93$.
The obtained solution has $B_{2g}$-symmetry
since the relation
$f^{\bm{0}}_{4}(k_x,k_y)\propto \sin k_x \sin k_y$ holds.
Thus, the primary nematic order is the 
``next-nearest-neighbor bond order''.
The obtained $B_{2g}$ bond order is consistent with the
experimental $d_{xy}$-wave nematicity in AFe$_2$As$_2$
\cite{B2g-NMR,B2g-Ishisa,B2g-STM}.

To understand why $B_{2g}$ bond order state is obtained,
we simplify Eq. (\ref{eqn:DWeq3}) by taking 
the Matsubara summation analytically 
by approximating that $I_\q^{c}$ and $g_\q(k)$ are static:
\begin{eqnarray}
\lambda_{\q}f_\q(\k)= \frac{1}{N}\sum_{\k'}I_\q^c(\k,\k')F_\q(\k')f_\q(\k') ,
\label{eqn:DWeq2-dwave}
\end{eqnarray}
where $\displaystyle F_\q(\p)\equiv -T\sum_m G(p+\q)G(p)
= \frac{n(\e_{\p+\q})-n(\e_{\p})}{\e_{\p}-\e_{\p+\q}}$
is a positive function,
and $n(\e)$ is Fermi distribution function.
In general, the peak positions of $\lambda_{\q}$
in Eq. (\ref{eqn:DWeq2-dwave})
are located at $\q={\bm 0}$ and/or nesting vectors with small wavelength
($\q=\Q_{\rm a}, \Q_{\rm d}$ in the present model).
The reason is that $I_\q^c \sim T\sum_{p}\chi^s(\p+\q)\chi^s(\p)$ by AL terms
is large for small $|\q|$,
and $F_\q(\p)$ is large for wide area of $\p$
when $\q$ is a nesting vector.

Figure \ref{fig:Cs122}(b) presents $I_\q^c(\k,\k')$ on h3,
where $\theta$ and $\theta'$ are positions of $\k$ and $\k'$ on h3,
respectively.
The large positive value around A originates from AL1 term in $I_\q^c$.
Also, the large positive value around B originates from AL2 term.
They give large attractive interaction 
for an even-parity form factor $f(\k)=f(-\k))$
in the p-h gap equation (\ref{eqn:DWeq2-dwave})
\cite{SOnari-PRB2019}.
On the other hand, $I_\q^c$ is negative at C, where $\k-\k'\sim\Q$,
due to the AL and MT terms.
As a result, 
the $d_{xy}$-wave form factor $f(\k)\sim \sin k_x \sin k_y$ is obtained, 
as we summarize in Fig. \ref{fig:Cs122}(c).

We comment that simple $s$-wave solution $f(\k)\sim {\rm const}$ 
is not obtained in the DW equation
because it is prohibited by the Hartree term
\cite{SOnari-PRB2019}.


Next, we discuss the doping-dependence of the nematicity:
We introduce reliable model Hamiltonian for
A$_{1-x}$Ba$_x$Fe$_2$As$_2$, by interpolating between 
CsFe$_2$As$_2$ model and BaFe$_2$As$_2$ model with the ratio $1-x:x$.
Here, the FSs with four Dirac pockets change to the FSs with 
two electron pockets at the Lifshitz transition point $x_c\sim0.5$. 
Figure \ref{fig:Cs122} (d) shows the $x$ dependence of 
$\lambda_{\bm{q}=\bm{0}}$ for $B_{2g}$ and $B_{1g}$ symmetries 
\cite{SOnari-PRB2019}.
For $x<x_c$, the $B_{2g}$ nematic instability is dominant 
since it is mainly driven by strong spin fluctuations 
due to the $d_{xy}$ orbital nesting at $\Q$.
For $x>x_c$, the $B_{1g}$ nematicity becomes dominant,
because it is mainly caused by the spin fluctuations in $d_{xz,yz}$ orbitals
 \cite{SOnari-PRL2012,Onari-SCVCS,YYamakawa-PRX2016}.
Thus, the present theory presents a unified explanation for 
both the $B_{1g}$ nematicity in non-doped $(n_d\approx 6)$
systems and $B_{2g}$ nematicity in heavily hole-doped compounds,
by focusing on the impact of the Lifshitz transition.

Recently, field-angle dependent specific heat measurement has
been performed for RbFe$_2$As$_2$
\cite{Mizukami-B2g}.
The observed field-dependence of $H_{2c}$
indicates that the $B_{2g}$ nematicity is established well above $T_{\rm c}$.
On the other hand,
small $B_{2g}$ nematic susceptibility was recently reported 
by means of a piezoelectric-based strain cell
\cite{B2g-Boehmer1,B2g-Boehmer2}, 
so further studies are necessary to clarify the $B_{2g}$ nematicity.
We note that the absence of the specific heat jump at $T_{S} = 40K$
in RbFe$_2$As$_2$
\cite{B2g-Ishisa}
is naturally understood based on the recent theoretical scaling relation
$\Delta C/T_S \propto T^b$ ($b\sim3$)
derived in Ref. \cite{RTazai-arXiv2021}.

To summarize, both $B_{1g}$ and $B_{2g}$ nematicity
in A$_{1-x}$Ba$_x$Fe$_2$As$_2$ are naturally induced by 
the paramagnon interference mechanism.
The present study demonstrates that
the nature of nematicity is sensitively controlled by modifying
the orbital character and the topology of the FS.

\subsection{Hidden nematic order in Ba122 family
\label{sec:Kontani4-6}
}

In several Fe-based superconductors, 
slight $C_4$ symmetry breaking occurs at $T^*$,
which is tens of Kelvin higher than the structural transition 
temperature $T_S$; see Fig. \ref{fig:phase-1111} (b).
This ``slight nematic state'' at $T_S<T<T^*$
with tiny orthorhombicity [$\phi=(a-b)/(a+b) \ll 0.1$\%]
has accumulated great interest
\cite{Kasahara-hidden}.
Similar ``slight nematic state'' above $T_S$
has been also discovered in NaFeAs
\cite{Zheng-hidden}.

To explain this long-standing mystery,
the emergence of the smectic bond order
at $T=T^* >T_{S}$ has been proposed
 \cite{SOnari-PRR2020,Onari-Kontani-Front2022}.
Based on this smectic order scenario, we can understand
characteristic phenomena below $T^*$, such as 
the pseudogap in the DOS and the small nematicity $\psi\propto T^*-T$.
The smectic order at $T^*$ does not interrupt the ferro-orbital order at
$T_S$ thanks to the prominent orbital selectivity of nematicity.

As we explained in Fig. \ref{fig:SCVC-1111} (c),
the smectic fluctuations develop in La1111 Hubbard model
using the SC-VC theory, although they are slightly smaller than 
the nematic fluctuations.
The former fluctuations become larger in models with 
larger $xy$-orbital hole-pocket at $n_d\sim6$,
and this hole pocket actually exists in both BaFe$_2$As$_2$ and NaFeAs.
The emergence of smectic + nematic orbital order
in BaFe$_2$As$_2$ model is verified in Ref. \cite{SOnari-PRR2020}
by analyzing the DW equation.

\begin{figure}[htb]
\centering
\includegraphics[width=.99\linewidth]{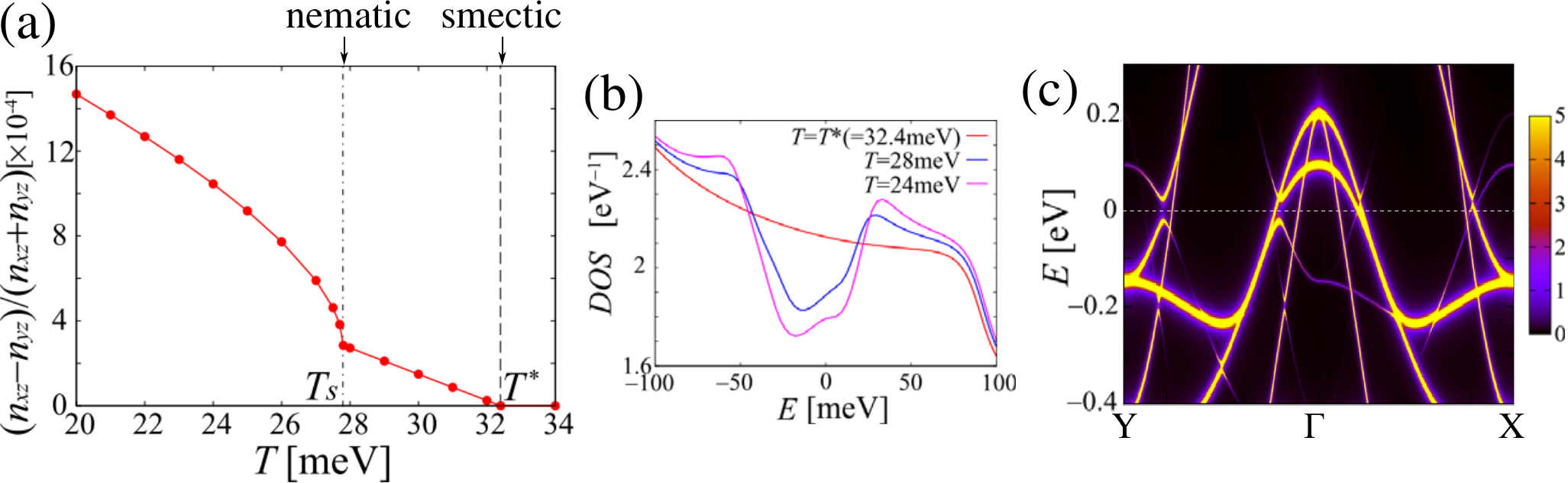}
\caption{
(a) Nematic orbital order $\Delta n =n_{xz}-n_{yz}$ as function of $T$.
$T^*$ ($T_S$) is the nematic (smectic) transition temperature.
$\Delta n$ is linear in $(T^*-T)$ below $T^*$,
and $\Delta n \propto \sqrt{T_S-T}$ below $T_S$
\cite{SOnari-PRR2020}.
Obtained (b) pseudogap in the DOS and
(c) unfolded bandstructure with Dirac dispersion
\cite{SOnari-PRR2020}.
Cited from Ref. \cite{SOnari-PRR2020}. 
}
\label{fig:Ba122-AFBO}
\end{figure}

Figure \ref{fig:Ba122-AFBO} (a) shows the obtained
nematic orbital order $\Delta n =n_{xz}-n_{yz}$ as function of $T$.
The obtained smectic transition temperature $T^*$
is slightly higher than the ferro-nematic transition temperature $T_S$.
The smectic (nematic) order parameter is proportional to 
$\sqrt{T_0-T}$ with $T_0=T^*$ ($T_0=T_S$).
Below $T^*$, $\Delta n$ is proportional to the square of 
smectic order parameter,
and therefore $\Delta n \propto (T^*-T)$.
Below $T_S$, the relation $\Delta n \propto \sqrt{T_S-T}$ holds.
The smectic order parameter is given by the combination of 
intra-orbital and inter-orbital bond orders \cite{SOnari-PRR2020}.
The smectic order originates from the interference 
between AFM and FM fluctuations.
(The FM fluctuations are induced in the small $d_{xy}$-orbital hole-pocket.)
In addition, the spin-charge coupling term $\Lambda_{\Q-\Q'}$
in Fig. \ref{fig:phase-AFM} (b) is large when 
$\q\equiv \Q-\Q'$ is the nesting vector $\q\approx(\pi,0)$.
The interference between AFM and FM fluctuations also
causes the smectic bond order in Ba122 and NaFeAs
\cite{SOnari-PRR2020}.

The obtained inter-orbital smectic bond order
naturally explains the pseudogap and the band-folding, 
as illustrated in Figs. \ref{fig:Ba122-AFBO} (b) and (c) respectively.
The hidden smectic order explains key experiments in both 
BaFe$_2$As$_2$ and NaFeAs, 
but it is not expected to occur in FeSe because of the absence of 
the $d_{xy}$-orbital hole-pocket.  
Theoretically, the smectic fluctuations significantly
contribute to the pairing mechanism \cite{Kontani-RPA,SOnari-PRR2020}.
It is noteworthy that 
the smectic order without magnetization has been recently
observed by ARPES studies in (Ba,K)Fe$_2$As$_2$ \cite{Shimojima-hidden},
CaKFe$_4$As$_4$ and KCa$_2$Fe$_4$As$_4$F$_2$ \cite{Zhou-hidden}.

\section{Superconductivity in Fe-based superconductors}
\label{sec:Onari}

\subsection{Spin and orbital fluctuation theories}

\subsubsection{Basic idea}

Study of unconventional superconductivity in strongly correlated electron systems
has long history.
As a typical example, the nodal $d$-wave state is realized in high-$T_{\rm c}$ cuprates and heavy
fermion compound Ce$M$In$_5$ ($M=$Co,Ru,Ir). The $d$-wave state was
confirmed by the phase sensitive experiments such as 
impurity effect\cite{impurity-cuprate1,impurity-cuprate2,impurity-cuprate3}, resonance peak in neutron scattering\cite{resonance-cuprate1,resonance-cuprate2,resonance-cuprate3}, and $\pi$ junction\cite{Wollman}.
It is believed that the $d$-wave superconductivity is explained by
the spin-fluctuation mechanism. In the spin-fluctuation mechanism, the AF spin
fluctuations with peak at $\Q=(\pi,\pi)$ act as repulsive pairing
interaction between singlet Cooper pairs $(\k,-\k)$ and $(\k+\Q,-\k-\Q)$, so the signs of
gap functions $\Delta(\k)$ and $\Delta(\k+\Q)$ are opposite. Thus, the spin-singlet $d_{x^2-y^2}$-wave state is realized.
On the other hand, spin-triplet superconductivity has been observed in several
U-based heavy fermion compounds such as UGe$_2$, URhGe, and
UCoGe\cite{U-based-review}. As the mechanism of these spin-triplet
superconductivity, the ferro spin-fluctuation mechanism has been proposed, but the
mechanism of the spin-triplet superconductivity is still open question.

High-$T_{\rm c}$ superconducting states in Fe-based superconductors 
have been the most intensively studied theme in the last decade.
The $s$-wave (=$A_{1g}$ symmetry) gap state is realized in almost all compounds.
In some typical compounds, simple fully gapped $s$-wave state is expected to appear
\cite{ARPES1,ARPES4,ARPES-Shen}.
However, the momentum and orbital dependences of gap function exhibit
remarkable substance dependences.
Rich diversity of the SC states is a remarkable 
characteristic of Fe-based superconductors.
At present, the mechanism of superconductivity is an open question.
Just after the discovery of La1111 compound,
spin-fluctuation pairing mechanism has been proposed by focusing on 
the adjacent stripe AFM phase 
\cite{KKuroki-PRL2008,Mazin},
which is essentially similar to that of $d$-wave superconductors
such as cuprates and Ce$M$In$_5$.  

On the other hand, in many Fe-based superconductors,
high-$T_{\rm c}$ superconducting phase is next to the nematic phase,
and sizable nematic fluctuations are observed
by Raman scattering\cite{Yoshizawa,Simayi,Bohmer,Goto} 
and shear modulus\cite{Gallais,Gallais2} measurements.
Based on these observations,
novel charge-channel fluctuation pairing mechanism has been proposed 
\cite{Kontani-RPA}.
Recently, it was revealed that orbital order/fluctuations are
derived from the paramagnon interference mechanism,
as we discussed in Sect. \ref{sec:Kontani4},
and theories of orbital fluctuation mediated pairing have been developed.

\begin{figure}[!htb]
\centering
\includegraphics[width=0.8\linewidth]{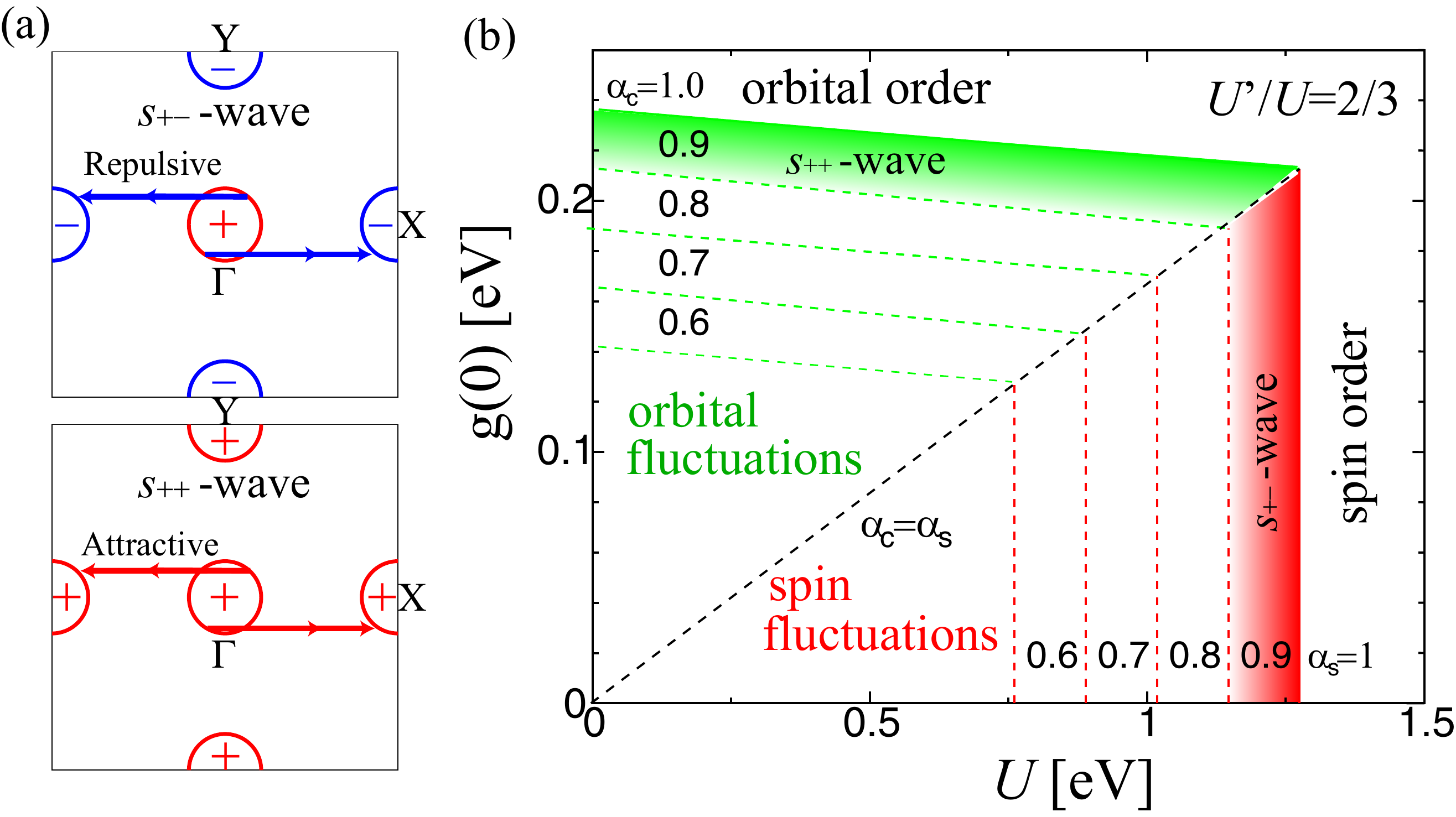}
\caption{
(a) Schematic pictures of $s_{\pm}$-wave state and $s_{++}$-wave state.
(b) Obtained $U$-$g(0)$ phase diagram for $n=6.1$
\cite{Kontani-RPA}.
Near the orbital-density-wave boundary, $s_{++}$-wave SC state
is realized by orbital fluctuations.
Cited from Ref. \cite{Kontani-RPA}. 
}
\label{fig:S+-S++}
\end{figure}

Hereafter, we analyze SC states in Fe-based superconductors by using the gap equation.
The spin-singlet gap equation is given by
\begin{eqnarray}
\lambda_{\rm SC}\hat{\Delta}(k)=-T\sum_{k'}\hat{V}^{\rm SC}(k,k')
\hat{G}(k')\hat{G}(-k')\hat{\Delta}(k').
\label{Eliash}
\end{eqnarray}
In the Migdal approximation, the pairing interaction  $V^{\rm SC}$ is
simply given
by
\begin{eqnarray}
\hat{V}^{\rm SC}(k,k')=\frac{3}{2}\hat{U}^s\hat{\chi}^s(k-k')\hat{U}^s-\frac{1}{2}\hat{U}^c\hat{\chi}^c(k-k')\hat{U}^c.
\label{pairing}
\end{eqnarray}
Thus, $\hat{V}^{\rm SC}$ is composed of the
repulsive (positive) spin-fluctuation term and the
attractive (negative) charge-channel-fluctuation term.
The $s_{\pm}$-wave state with sign reversal in Fig. \ref{fig:S+-S++} (a) is mediated by the AF spin
fluctuations with peak at $\Q=(0,\pi)$, $(\pi,0)$
\cite{KKuroki-PRL2008,Mazin,maier-scalapino,Hirschfeld-s+-,Chubukov-s+-,Ikeda-s+-,Schmalian-s+-,DH-Lee} 
according to the gap equation (\ref{Eliash}).
Within the RPA, the $s_{\pm}$-wave state is uniquely obtained because of the relation 
${\hat \chi}^s(q)\gg {\hat \chi}^c(q)$ in the RPA.
However, high $T_{\rm c}$ ($\gtrsim 60$K) emerges in electron-doped FeSe without hole FS,
while spin fluctuations are weak\cite{Hrovat2015_eFeSe_NMR}.
Similarly, spin fluctuations are relatively weak in 
high-$T_{\rm c}$ $(\gtrsim 50{\rm K})$ 1111 compounds \cite{Fujiwara2017}. 
On the other hand,
the $s_{++}$-wave state without sign reversal in
Fig. \ref{fig:S+-S++} (a) is mediated by the ferro- and antiferro-orbital
fluctuations\cite{Kontani-RPA,Onari-SCVCS,Yanagi}. 
The smectic bond order discussed in Sect. \ref{sec:Kontani4} has been observed in 
several compounds \cite{Shimojima-hidden,Zhou-hidden,smectic-FeSe-STM}.
Therefore, novel spin + orbital fluctuation pairing mechanism 
has to be developed theoretically.

\subsubsection{Charge quadrupole interaction}
In the mean field approximation, the $s_{\pm}$-wave state is uniquely obtained.
In order to understand the $s_{++}$-wave state, beyond mean-field theories have to be developed.
As shown in Sect. \ref{sec:Kontani4}, the orbital susceptibilities
$\chi^c_{x^2-y^2}(\q)$ and $\chi^c_{xz(yz)}(\q)$ are enhanced by the paramagnon interference mechanism.
In this mechanism, $\chi^c_{x^2-y^2}(\q)$ has peak at $\q=\bm{0}$.
$\chi^c_{xz(yz)}(\q)$ has peak at the nesting vector $\q=\Q$, which acts as the
attractive pairing interaction between hole FSs and electron FSs. 
Both orbital fluctuations lead to the $s_{++}$-wave state.

For phenomenological analysis,
we introduce the following quadrupole interaction, which describes
the effective interaction due to the AL-VC:
\begin{equation}
{\cal H}_{\rm
 quad}=-g(\omega_l)\sum_i\sum_\Gamma^{xz,yz}\hat{O}^i_\Gamma\hat{O}^i_\Gamma,
\label{eq:quadrupole}
\end{equation}
where $\hat{O}_\Gamma$ is the quadrupole operator for channel $\Gamma$
introduced in Eqs. (\ref{eqn:O-x2y2}) and (\ref{eqn:O-xz}).
(Note that $\hat{O}_{\mu\nu}\propto\hat{l}_\mu\hat{l}_\nu+\hat{l}_\nu\hat{l}_\mu$,
where $\hat{l}$ is the angular momentum operator.)
$g(\omega_l)=g(0)\omega_c^2/(\omega_l^2+\omega_c^2)$ is
the quadrupole coupling with cutoff energy $\omega_c$.
By performing the RPA for ${\cal H}_{\rm Hub}+{\cal H}_{\rm quad}$,
the enhancement of  $\chi^c_{xz(yz)}(q)$ given by the SC-VC theory in
Fig. \ref{fig:SCVC-1111} is well reproduced \cite{Kontani-RPA}.

Here, we discuss the SC state and phase diagram, where the ratio $J/U=1/6$ is fixed.
Figure \ref{fig:S+-S++} (b) shows the $U$-$g(0)$ phase diagram for $n=6.1$
given by the RPA.
$\alpha_{s(c)}$ is the spin (charge) Stoner factor, which is given by the maximum eigenvalue of
$\hat{U}^{s(c)} \hat{\chi}^0 (\bm{q},0)$.
The transition line for the spin (orbital) order is given by the condition $\alpha_{s(c)}=1$. 
The orbital fluctuations are enhanced by the quadrupole interaction.
Near the orbital-order boundary, $s_{++}$-wave state is realized by the orbital fluctuations. 
In later sections, we develop beyond Migdal-Eliashberg (ME) gap equation,
and analyze the pairing state in BaFe$_2$As$_2$, LiFeAs, and electron-doped FeSe 
based on the Hubbard model,
without introducing phenomenological interaction in Eq. (\ref{eq:quadrupole}).

\subsection{Phase sensitive experiments}
In general, it is not easy to distinguish between $s_{\pm}$-wave and $s_{++}$-wave states, 
since both states belong to the same $A_{1g}$ symmetry.
Nonetheless of this difficulty, various phase sensitive experiments have been proposed and performed,
such as the impurity effect on $T_{\rm c}$
\cite{impurity1,Nakajima-imp,Li-imp,FeSe-Taen,Canfield2009,Ideta2013,imp-LiFeAs,Lee-LiFeAs-imp,Ghigo2018_s++_imp,Schilling2016_s++_imp}, 
the inelastic neutron scattering below $T_{\rm c}$
\cite{neutron-FeSe, keimer,neutron-NaFeAs,Christianson,Lumsden,neutron-112,neutron-FeSe,Lee2016}, and
the quasi-particle interference (QPI) in STM measurement
\cite{Hanaguri-QPI,Chi-QPI,Davis-QPI,Wen2018_LiFeOHFeSe_QPI,Gu2018_LiFeOHFeSe_QPI}.
However, 
to extract reliable conclusions from these experimental data,
we must develop the theories of phase sensitive phenomena 
by considering the multiorbital and the strong correlation effects accurately.
Below, we explain the important theoretical studies for phase sensitive experiments.

\subsubsection{Impurity effect}
First, we discuss the impurity effect on the superconductivity.
In the presence of impurity potentials,
a Cooper pair $(\k,-\k)$ on the FS with gap 
is scattered to $(\k',-\k')$ on the same or different FS.
Thus, the gap function $\Delta_\k$ on the FS
is averaged by the impurity scattering,
the and therefore $T_{\rm c}$ is reduced if the gap function $\Delta_k$
has sign reversal or strong anisotropy.
In contrast, simple $s$-wave gap without sign reversal is
robust against impurities, known as Anderson theorem \cite{Anderson}.

\begin{figure}[!htb]
\centering
\includegraphics[width=.8\linewidth]{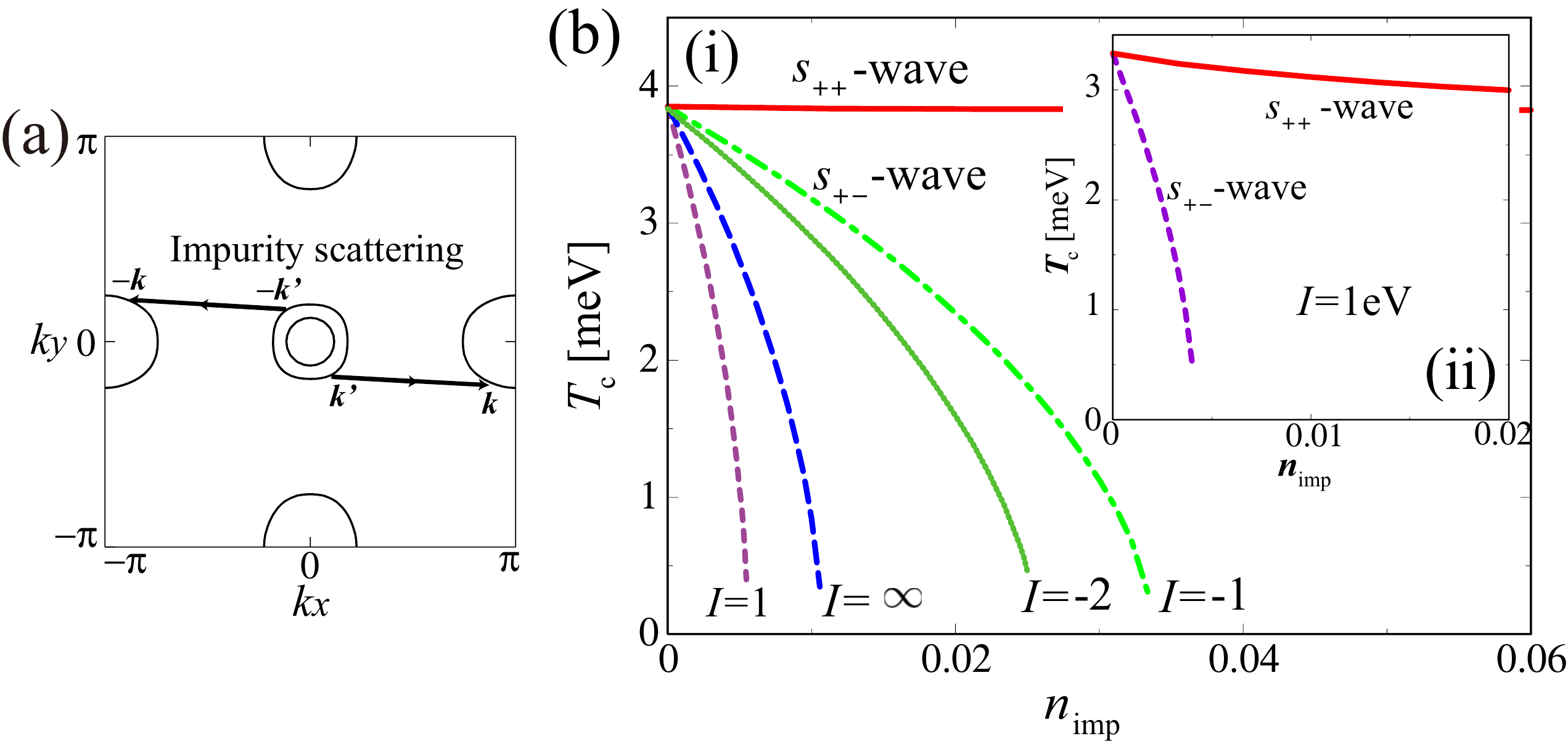}
\caption{(a) Schematic picture of interband impurity scattering.
(b) Obtained $T_{\rm c}$ for the $s_\pm$-wave and $s_{++}$-wave states
as functions of $n_{\rm imp}$,
in the case of
(i) isotropic $s_\pm$ wave gap ($T_{\rm c0}=46$ K) and 
(ii) anisotropic $s_\pm$ wave gap ($T_{\rm c0}=40$ K)
 \cite{Onari-impurity}. 
Cited from Ref. \cite{Onari-impurity}.
}
\label{fig:impurity-theory}
\end{figure}

Figure \ref{fig:impurity-theory} (a) shows the inter-FS
scattering in Fe-based superconductors.
In the case of $s_{++}$-wave state, $T_{\rm c}$ is essentially
insensitive to the impurity concentration, according to the Anderson theorem.
In the case of $s_{\pm}$-wave state, 
in contrast, superconductivity can be suppressed by the inter-FS
impurity scattering, after the impurity averaging of the electron- and hole-pockets 
gap functions with opposite signs.
Crudely speaking, the $T_{\rm c}$ of $s_{\pm}$-wave state state
is suppressed by the inter-FS impurity scattering $I_{\rm inter}$,
while it is not by the intra-FS scattering $I_{\rm intra}$.
Therefore, the ``reduction in the transition temperature 
$\Delta T_{c}\equiv T_{c0}-T_{\rm c} \ (>0)$'' in the $s_\pm$-wave state is expressed as
\begin{eqnarray}
\frac{\Delta T_{c}}{T_{c0}} \propto r_{\rm imp} (z \rho_{\rm imp}/T_{c0})
\label{eqn:r}
\end{eqnarray}
where $\rho_{\rm imp}$ is the impurity scattering residual resistivity that is proportional to 
$\tau_{\rm tr}^{-1}\propto n_{\rm imp}(|I_{\rm intra}|^2+ |I_{\rm inter}|^2)$,
and $z=m/m^*$ is the inverse of mass-enhancement factor. 
The coefficient $\displaystyle r_{\rm imp} \equiv \frac{2|I_{\rm inter}|^2}{ |I_{\rm intra}|^2+ |I_{\rm inter}|^2 }$
represents the weight of the inter-FS scattering among the total scattering.
This relation is driven by the Born approximation, 
which is valid for dilute and weak-potential impurities,
by following the Abrikosov-Gorkov theory \cite{AG-theory}.

As we shown in Fig. \ref{fig:CeCu2Si2-impurity},
the slope of $\Delta T_{c}/T_{c0}$ as function of $\hbar/\tau_{\rm imp}k_B T_{c0}$
in Ba122 compounds is comparable to that of BCS $s$-wave superconductors with anisotropic gap, 
while it is completely larger than that of $d$-wave superconductors.
This result puts significant constraint on the pairing state in Ba122.
In the case of $r_{\rm imp}\sim O(1)$, anisotropic $s$-wave state without sign reversal is safely concluded.
In the case of $r_{\rm imp}\ll 1$, both $s_{++}$- and $s_\pm$-wave states are allowed.
To find the value of $r_{\rm imp}$, one has to perform serious theoretical analyses
based on realistic multiorbital models.

Hereafter, we derive the value of $r_{\rm imp}$ in multiorbital
systems based on the theory in Ref. \cite{Onari-impurity}.
Here, we consider the impurity potential due to the Fe-site substitution or defect.
In this case, the impurity potential is diagonal with respect to $d$-orbital index: $I_{l,m}=I\delta _{l,m}$.
The Fe-site substitution impurities will be dominant in real compounds
because of the large scattering potential $I$.

We employ the following 
$10\times 10$ Nambu BCS Hamiltonian 
in the $d$-orbital basis:
\begin{eqnarray}
{\hat {\cal H}}_\k=\left(
\begin{array}{cc}
{\hat H}_\k^0 & {\hat \Delta}_\k \\
{\hat \Delta}_\k^\dagger & -{\hat H}_{-\k}^0 \\
\end{array}
\right) ,
 \label{eqn:H0}
\end{eqnarray}
where ${\hat H}_\k^0$ is the $5\times5$ hopping matrix of the five-orbital 
tight-binding model\cite{KKuroki-PRL2008}.
The Green function is given by
\begin{eqnarray}
{\hat {\cal G}}_\k(i\w_n) &\equiv&
\left(
\begin{array}{cc}
{\hat G}_\k(i\w_n) & {\hat F}_\k(i\w_n) \\
{\hat F}^\dagger_\k(i\w_n) & -{\hat G}_{\k}(-i\w_n) \\
\end{array}
\right)
 \nonumber \\
&=& (i\w_n{\hat 1}-{\hat \Sigma}_\k(i\w_n)-{\hat {\cal H}}_\k)^{-1} ,
 \label{eqn:G}
\end{eqnarray}
where $\w_n=\pi T(2n+1)$ is the fermion Matsubara frequency,
${\hat G}_\k$ (${\hat F}_\k$) is the $5\times5$ normal (anomalous)
Green function, and ${\hat \Sigma}_\k$ is the self-energy 
in the $d$-orbital basis.
We consider the local impurity potential 
due to the substitution of Fe by other 3$d$ elements
as a typical non-magnetic impurity potential \cite{Nakamura2011,Yamakawa-impurity}.
In the present $d$-orbital basis,
the impurity potential is momentum-independent and diagonal in the orbital-basis.

The impurity potential due to the Fe-site substitution, ${\hat {\cal I}}$, 
is simply given as ${\cal I}_{l,m}= I\delta_{l,m}$ for $1\le l,m \le5$, and 
${\cal I}_{l,m}= -I\delta_{l,m}$ for $6\le l,m \le10$.
Then, the $T$-matrix for a single impurity,
which is $\k$-independent in the $d$-orbital basis, is given as
\begin{eqnarray}
{\hat {\cal T}}(i\w_n)= ({\hat 1}-{\hat {\cal I}}
{\hat {\cal G}}_{\rm loc}(i\w_n))^{-1} {\hat {\cal I}},
 \label{eqn:Tmat}
\end{eqnarray}
where ${\hat {\cal G}}_{\rm loc}(i\w_n)
\equiv\frac1N \sum_\k {\hat {\cal G}}_\k(i\w_n)$.
In the $T$-matrix approximation, the self-energy matrix
in the $d$-orbital basis is $\k$-independent.
It is given as
\begin{eqnarray}
{\hat \Sigma}(i\w_n) &\equiv&
n_{\rm imp}{\hat {\cal T}}(i\w_n) .
 \label{eqn:Self}
\end{eqnarray}
%

The gap function ${\hat \Delta}_\k$ in eq. (\ref{eqn:H0}) 
is given by the solution of the Eliashberg equation:
\begin{eqnarray}
\Delta^{l,l'}(\k,\e_n)
= -\frac{T}{N}\sum_{\k',m} \sum_{m,m'}V_{\k,\k'}^{l,l';m,m'}(\e_n,\e_m)
F^{m,m'}(\k',\e_m) ,
 \label{eqn:Eliash1}
\end{eqnarray}
where $V_{\k,\k'}^{l,l';m,m'}$ is the pairing potential in the $d$-orbital basis.
In the fully self-consistent $T$-matrix approximation,
we solve Eqs. (\ref{eqn:G})-(\ref{eqn:Eliash1}) self-consistently.

We calculate the impurity effect on $T_{\rm c}$
by introducing interband pairing interaction
between the hole pockets and the electron pockets.
The fully-gapped $s_{++}$- ($s_\pm$-) wave state is realized
when the introduced inter-band pairing interaction is attractive (repulsive).
The obtained results are shown in Fig. \ref{fig:impurity-theory} (b) for 
(i) isotropic $s_\pm$ wave gap due to the band independent pairing ($T_{\rm c0}=46$ K at $n_{\rm imp}=0$) and 
(ii) anisotropic $s_\pm$ wave gap due to the band dependent pairing ($T_{\rm c0}=40$ K).
In both cases,
dilute impurities with $|I|\ge1$eV
induces sizable pair breaking for the $s_\pm$-wave state.
That is, the Anderson's theorem is completely violated 
in the $s_\pm$-wave state for Fe-site substitution impurities.
in many Fe-based superconductors.
Essentially the same results are obtained if we use the impurity potential 
based on first principles calculation \cite{Yamakawa-impurity}. 
(In Ref. \cite{Yamakawa-impurity}, we see that the critical residual resistivity 
$\rho_{\rm imp}^{\rm cr}$ for $s_\pm$-wave state is almost independent of $I$.)
These results mean that the coefficient $r_{\rm imp}$ in Eq. (\ref{eqn:r}) is of order unity,
because the e-pockets and h-pockets are composed of the same $d$-orbitals
Therefore, a special reason (about band and orbital structure) 
would be necessary for realizing $r_{\rm imp}\ll1$.

In the $s_{++}$-wave state, $T_{\rm c}$ is independent of impurity
effect in (i) due to the Anderson's theorem.
$T_{\rm c}$ in (ii) slowly decreases with $n_{\rm imp}$ with downward convex, 
since weak pair breaking occurs unless magnitude of all gap functions are the same.

\begin{figure}[!htb]
\centering
\includegraphics[width=.9\linewidth]{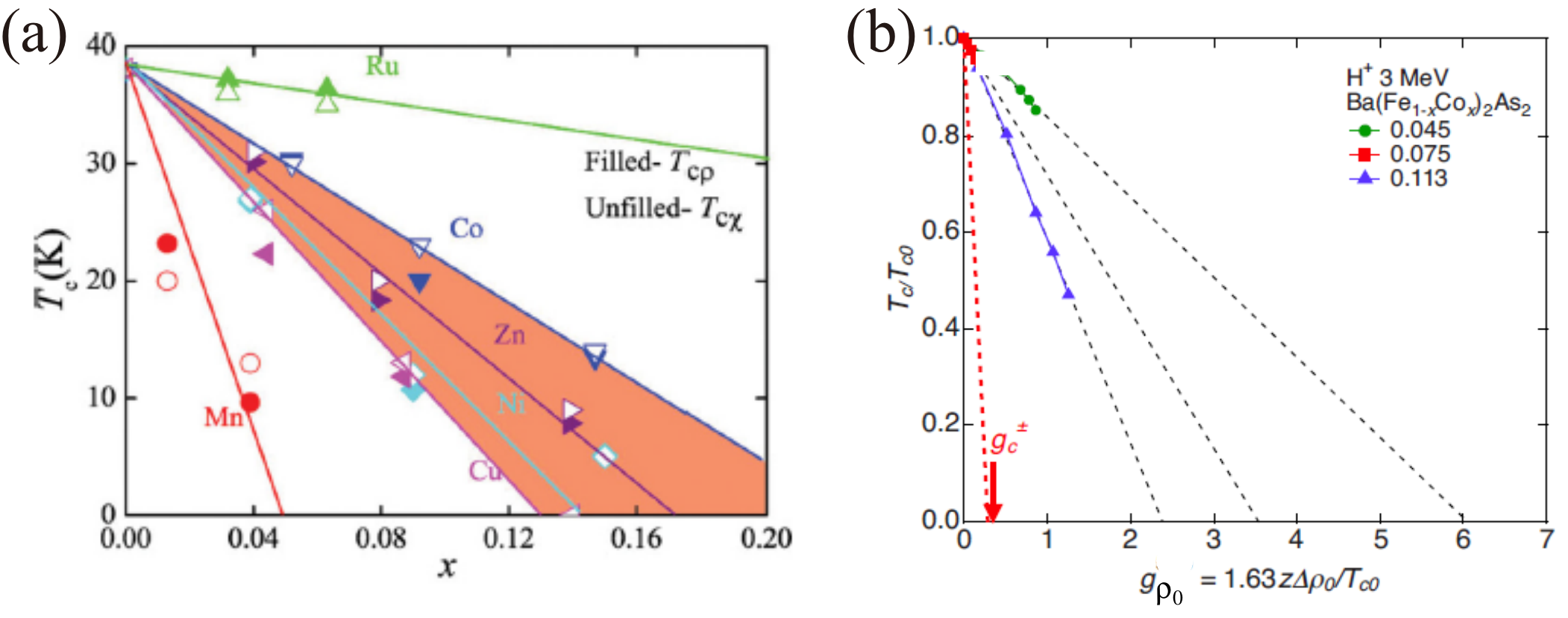}
\caption{
(a) $T_{c}$ as a function of $x$ for Ba$_{0.5}$K$_{0.5}$Fe$_{2-2x}$M$_{2x}$As$_2$ 
(M$=$Mn, Ru, Co, Ni, Cu, and Zn).
Reprinted with permission from \cite{Li-imp}. Copyright (2012) by the American Physical Society.
(b) $T_{c}/T_{c0}$ in Ba(Fe$_{1-x}$Co$_x$)$_2$As$_2$  
for $x=0.045$, 0.075, and 0.113.
Reprinted with permission from \cite{Nakajima-imp}. Copyright (2010) by the American Physical Society.
}
\label{fig:Muromachi-Nakajima-imp}
\end{figure}


Now, we introduce several important experiments on the impurity effect.
Figure \ref{fig:Muromachi-Nakajima-imp} (a) shows $T_{c}$ as a function of $x$ for
 Ba$_{0.5}$K$_{0.5}$Fe$_{2-2x}$M$_{2x}$As$_2$ (M$=$Mn, Ru, Co, Ni, Cu,
 and Zn) \cite{Li-imp}.
Applying a
linear function to $T_{\rm c}$ vs $x$, the suppression rates for Mn, Ru,
Co, Ni, and Cu are 6.98, 0.27, 1.73, 2.21, and 2.68 K/\%, respectively. 
Except for the nonmagnetic impurity Mn, the observed 
suppression of $T_{\rm c}$ is much weaker than that expected from the
$s_{\pm}$-wave model.
Similar impurity effect was also reported by proton irradiation study in
Ba(Fe$_{1-x}$Co$_x$)$_2$As$_2$ in Fig. \ref{fig:Muromachi-Nakajima-imp} (b),
where the normalized scattering rate is estimated as
$g_{\rho_0}=1.63z\rho_{\rm imp}/T_{c0}$ in the five-orbital model,
where  $\rho_{\rm imp}$ is the residual resistivity and
$z=m/m^*$ is the renormalization factor.
The proton irradiation
provides local nonmagnetic scattering centers without
changing electronic structure.
The obtained critical scattering rates are one-order of magnitude larger
than theoretically predicted critical value $g_c^{\pm}=0.23$ for the $s_\pm$-wave state.

Here, we calculated the $T$-matrix based on the $d$-orbital basis in Eq. (\ref{eqn:Tmat}).
Then, $|I_{\rm inter}|$ and $|I_{\rm intra}|$ are given by its unitary transformation,
and the relation $|I_{\rm inter}|\approx|I_{\rm intra}|$ is satisfied for any value of $I$.
In principle, we can also perform the same calculation based on the band-basis directly,
while it is difficult because the impurity potential possesses the momentum dependences
in the band-basis, $I_{b,b'}(\k,\k')^{\rm band}$.
If we neglect the $\k,\k'$-dependences of $I_{b,b'}(\k,\k')^{\rm band}$,
we met a serious artifact:
The $s_\pm$-wave state becomes very robust since $r_{\rm imp}$ is renormalized to zero 
in the unitary region \cite{Senga,Bang}.
However, this is an artifact due to the oversimplified impurity potential.
As we discussed here, the relation $r_{\rm imp}\sim O(1)$ has been confirmed based on 
realistic multiorbital and impurity models.

\begin{figure}[!htb]
\centering
\includegraphics[width=0.8\linewidth]{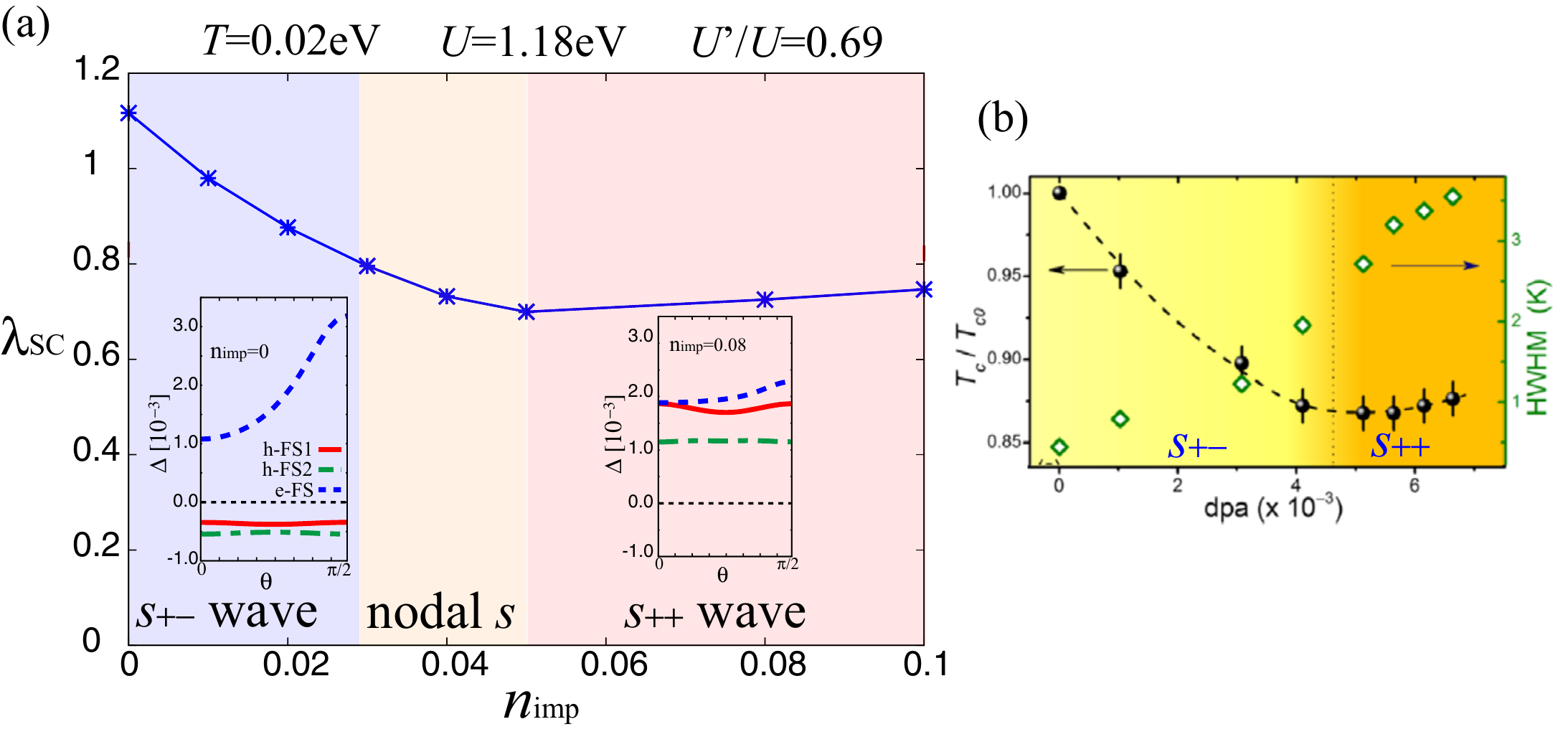}
\caption{
(a) Obtained $n_{\rm imp}$-dependence of $\lambda_{\rm SC}$ at $\a_{\rm c}=0.98$
\cite{Kontani-RPA}.
Blue-, orange-, and red-shaded areas represent $s_{\pm}$-wave state,
 nodal-$s$-wave state, and $s_{++}$-wave state, respectively. Insets
 show azimuthal angle $\theta$ dependences of $\Delta$ on FSs for
 $n_{\rm imp}=0$ and $n_{\rm imp}=0.08$.
Cited from Ref. \cite{Kontani-RPA}.
(b) $T_{\rm c}$ in Ba(Fe$_{1-x}$Rh$_x$)$_2$as a function
of irradiation-induced disorder.
Reprinted with permission from \cite{Ghigo2018_s++_imp}. Copyright (2018) by the American Physical Society.
}
\label{fig:lambda}
\end{figure}

In the next stage, we discuss the impurity induced $s_{\pm}$ to $s_{++}$ crossover.
It is natural to expect in many Fe-based superconductors,
in which both attractive and repulsive interaction coexist.
Figure \ref{fig:lambda} (a) shows the $n_{\rm imp}$-dependence of 
$\lambda_{\rm SC}$ at $\a_{\rm c}=0.98$ for $U=1.18$eV.
$s_\pm$-wave state is realized at $n_{\rm imp}=0$;
$\lambda_{\rm SC}$ decreases slowly as $n_{\rm imp}$ 
increases from zero, whereas it saturates for $n_{\rm imp}\ge0.05$,
indicating the smooth crossover from $s_\pm$-wave state to $s_{++}$-wave one
due to the interband impurity scattering. 
The nodal $s$-wave state 
emerges during the crossover $0.03\lesssim n_{\rm imp}\lesssim 0.05$.
These behavior is very similar to the experimental
irradiation-induced disorder effect of Ba(Fe$_{1-x}$Rh$_x$)$_2$as 
in Fig. \ref{fig:lambda} (b)
\cite{Ghigo2018_s++_imp,Schilling2016_s++_imp}. 
The impurity induced $s_{\pm} \rightarrow s_{++}$ crossover 
has also been discussed in Ref. \cite{Hirsch-cross}.

\begin{figure}[!htb]
\centering
\includegraphics[width=0.5\linewidth]{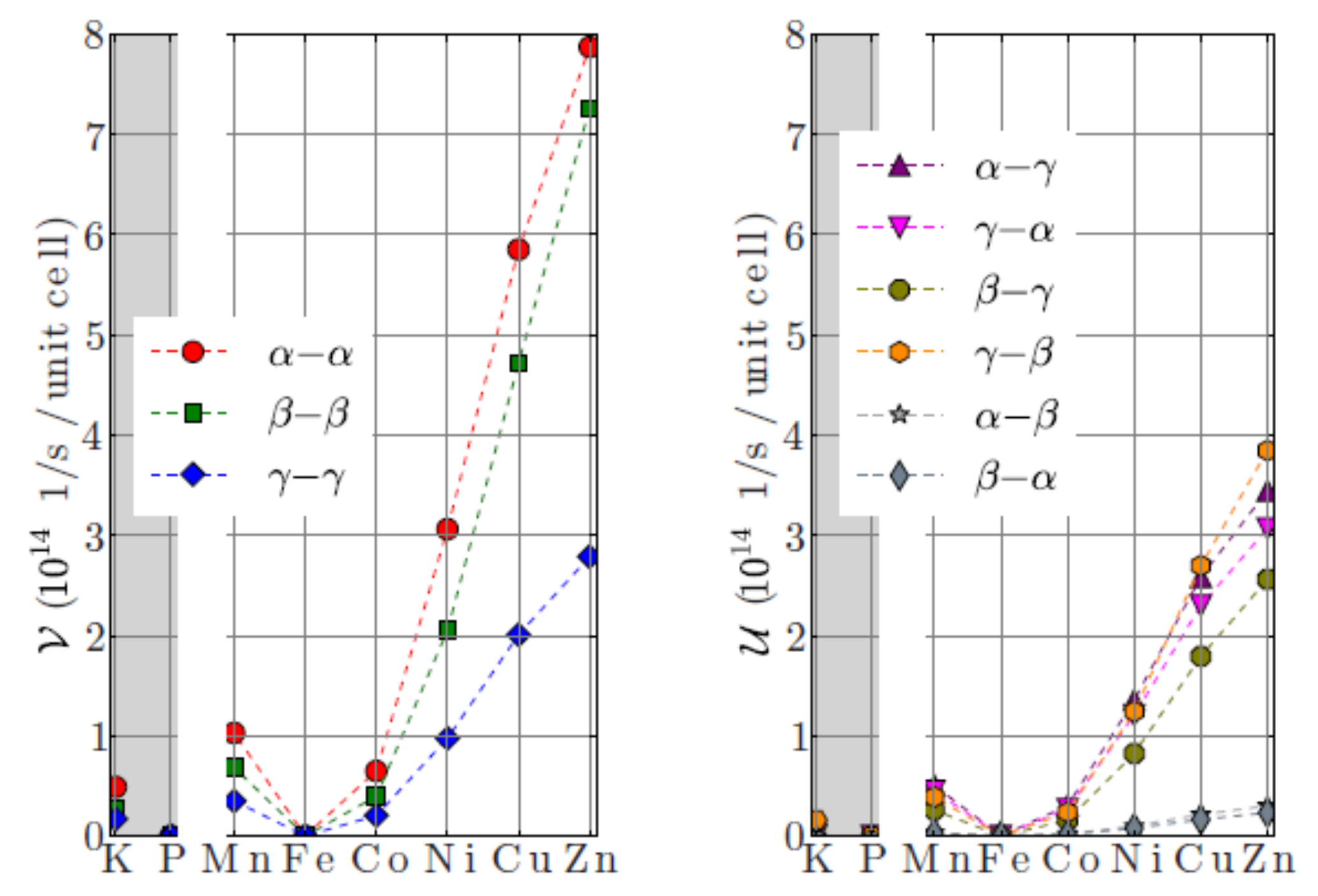}
\caption{Normalized intraband $\cal V$ and interband $\cal U$
scattering rates obtained from the single-impurity $T$ matrix in the
dilute limit of Ba122 compounds\cite{Schmalian2016}. $\alpha$ and
 $\beta$ denote the hole pockets, and $\gamma$ denotes the electron pockets.
Reprinted with permission from \cite{Schmalian2016}. Copyright (2016) by the American Physical Society.
}
\label{fig:CPA}
\end{figure}

Finally, we explain the first principles study on the impurity effect.
Figure \ref{fig:CPA} shows the interband impurity scattering rates $\cal U$ 
and the intraband ones $\cal V$ obtained by the systematic first principles
calculation in Ba122 compounds.
\cite{Schmalian2016}.
The magnitude of $\cal U$ between hole and electron pockets
is about a half of $\cal V$ within the hole pockets. 
In our study, we employed orbital-diagonal impurity potential.
This relation ${\cal U/\cal V}\sim 0.5$ ({\rm i.e.}, $r_{\rm imp}\sim 2/3$) 
is also satisfied in the present study, 
which is easily verified by performing the unitary transformation of 
the $T$-matrix Eq. (\ref{eqn:Tmat}) into the band-basis. 
To summarize, the relation $r_{\rm imp} \sim O(1)$ is concluded by several 
reliable theoretical methods.

\subsubsection{Resonance in inelastic neutron scattering}
Inelastic neutron scattering is another important phase-sensitive experiment.
In $d$-wave superconductors, 
such as cuprates \cite{iikubo-sato,ito-sato,keimer-highTc}
and CeCoIn$_5$ \cite{stock-CeCoIn5},
large and sharp resonance peak appears in the imaginary part of 
dynamical spin susceptibility below $T_{\rm c}$
\cite{Monthoux-Scalapino,pines,chubukov-resonance,takimoto-moriya}. 
It is widely accepted that the sharp resonance peak 
observed in cuprates and CeCoIn$_5$ originates from the positive 
coherence factor given by the sign-reversal of $d$-wave gap function.
The resonance peak energy $\w_{\rm res}$ satisfy the 
resonance condition $\w_{\rm res}<2\Delta^{\rm max}$,
where $\Delta^{\rm max}$ is the maximum value of the gap function.

\begin{figure}[!htb]
\centering
\includegraphics[width=0.99\linewidth]{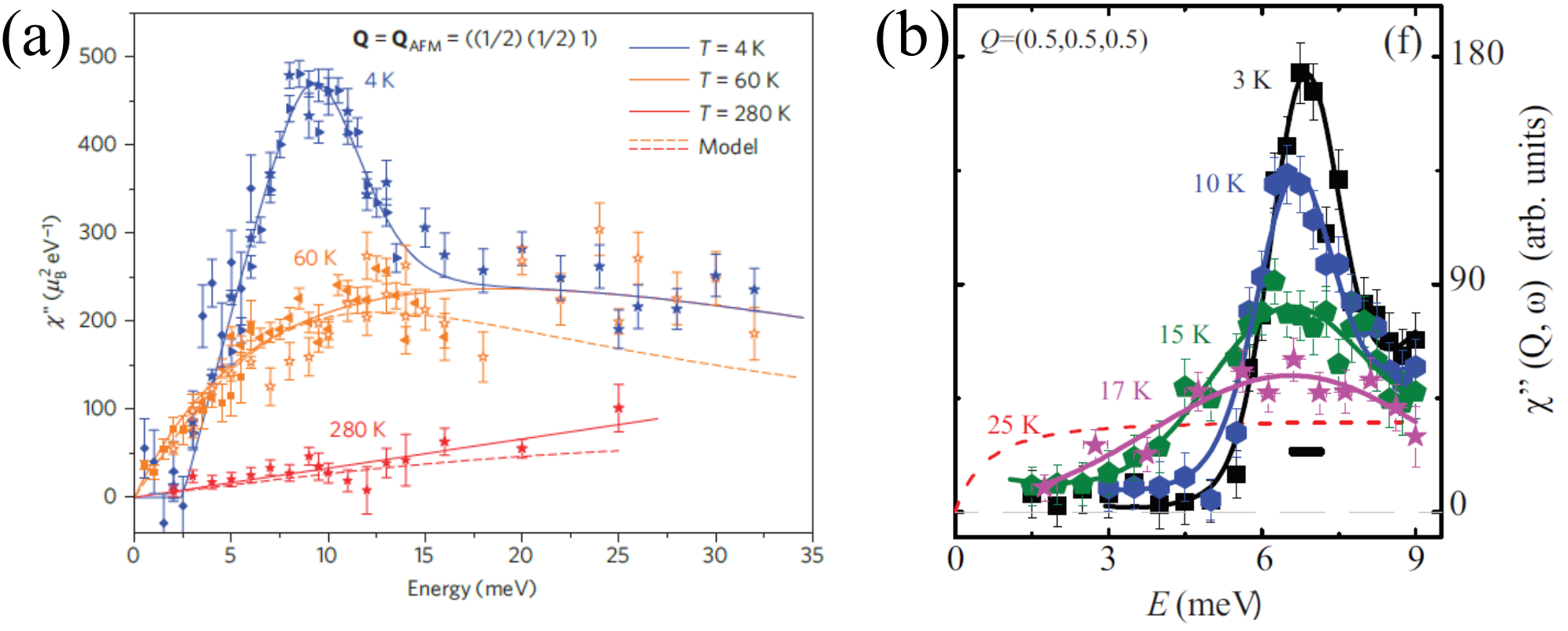}
\caption{
(a) Imaginary part of the spin susceptibility $\chi(\bm{Q},\omega)$
in BaFe$_{1.85}$Co$_{0.15}$As$_2$.
Reprinted by permission from Springer Nature:
Nat. Phys.  \cite{keimer}, copyright 2010.
(b) Im$\chi(\bm{Q},\omega)$ in NaFeAs.
Cited from Ref. \cite{neutron-NaFeAs}.
}
\label{fig:neu-keimer-Zhang}
\end{figure}

Inelastic neutron scattering studies have been performed 
in various Fe-based superconductor.
Figure \ref{fig:neu-keimer-Zhang} (a) exhibits the experimental data
for BaFe$_{1.85}$Co$_{0.15}$As$_2$ \cite{keimer}.
Below $T_{\rm c}=25$K,
resonance-like broad peak in ${\rm Im}\chi^s(\Q,\omega)$ emerges at $\w_{\rm res}\approx 8$meV.
However, the height of the resonance peak is 
just twice as large as the signal in normal state ($T=60$K). 
The resonance-like peak in NaFeAs \cite{neutron-NaFeAs} shown in 
Fig. \ref{fig:neu-keimer-Zhang} (b) is sharper than Ba122.
These results may indicate the sigh-reversing gap function.
However, the observed peak structure in these compounds
is much broader and smaller than the resonance peak in CeCoIn$_5$
\cite{stock-CeCoIn5}.

A key question is whether resonance-like peaks observed in 
Figs. \ref{fig:neu-keimer-Zhang} (a) and (b)
are understandable based on the $s_{++}$-wave state or not.
To answer this question, 
we focus on the nontrivial self-energy effect 
on the dynamical spin susceptibility 
\cite{onari-resonance,onari-resonance2,takeuchi-resonance}.
In Fe-based superconductors, 
the inelastic scattering rate $\gamma^* \equiv z{\rm Im}\Sigma$ 
is comparable to $T$ in the normal state,
while it is suddenly suppressed by the finite gap below $T_{\rm c}$.
Such drastic change in the self-energy due to the $s_{++}$-wave gap
gives rise to the resonance-like peak structure in ${\rm Im}\chi^s(\Q,\omega)$.

%
We study the five-orbital Hubbard model for Fe-based superconductors
to understand vast inelastic neutron scattering experiments 
\cite{keimer,neutron-NaFeAs,Christianson,Lumsden,neutron-FeSe,neutron-112}.
The bare susceptibilities in Matsubara frequency representation are written as
\cite{maier-scalapino,eremin,onari-resonance,onari-resonance2}
\begin{eqnarray}
  \chi^0_{ll'mm'} (q)
  &=& -\frac{T}{N}\sum_{k} G_{lm}(k+q) G_{m'l'}(k),\\
\label{chi0}
  \phi^0_{ll'mm'} (q)
  &=& -\frac{T}{N}\sum_{k} F_{lm'}(k+q) F^{\dagger}_{l'm}(k),
\label{phi0}
\end{eqnarray}
where $q=({\bm q},i\omega_l)$, $\omega_l=2\pi l T$, and $N$ is the number of $\k$-meshes. 


Here, we define $\hat{\Psi}(q)\equiv \hat{\chi}^{0}(q)+\hat{\phi}^{0}(q)$.
By calculating the retarded (advanced) function $\hat{\Psi}^{R(A)}(\q,\w)$ numerically
\cite{takimoto-moriya,onari-resonance2},
the retarded spin susceptibility $\chi^{s,R}$ is given as
\begin{eqnarray}
  \chi^{s,R}_{ll'mm'}(\q,\omega)
  = \left[\frac{\hat{\Psi}^{R}(\q,\omega)}
  {1-\hat{U}^s\hat{\Psi}^{R}(\q,\omega)}\right]_{ll'mm'},
\label{chiS}
\end{eqnarray}
Here, we introduce the dynamical Stoner factor $\a_S(\w)$ 
defined as the maximum eigenvalue of 
$\hat{U}^s(\hat{\Psi}^{R}(\q,\w)+\hat{\Psi}^{A}(\q,\w))/2$ for $\q=\Q$.

\begin{figure}[htb]
  \centering
  \includegraphics[width=0.99\linewidth]{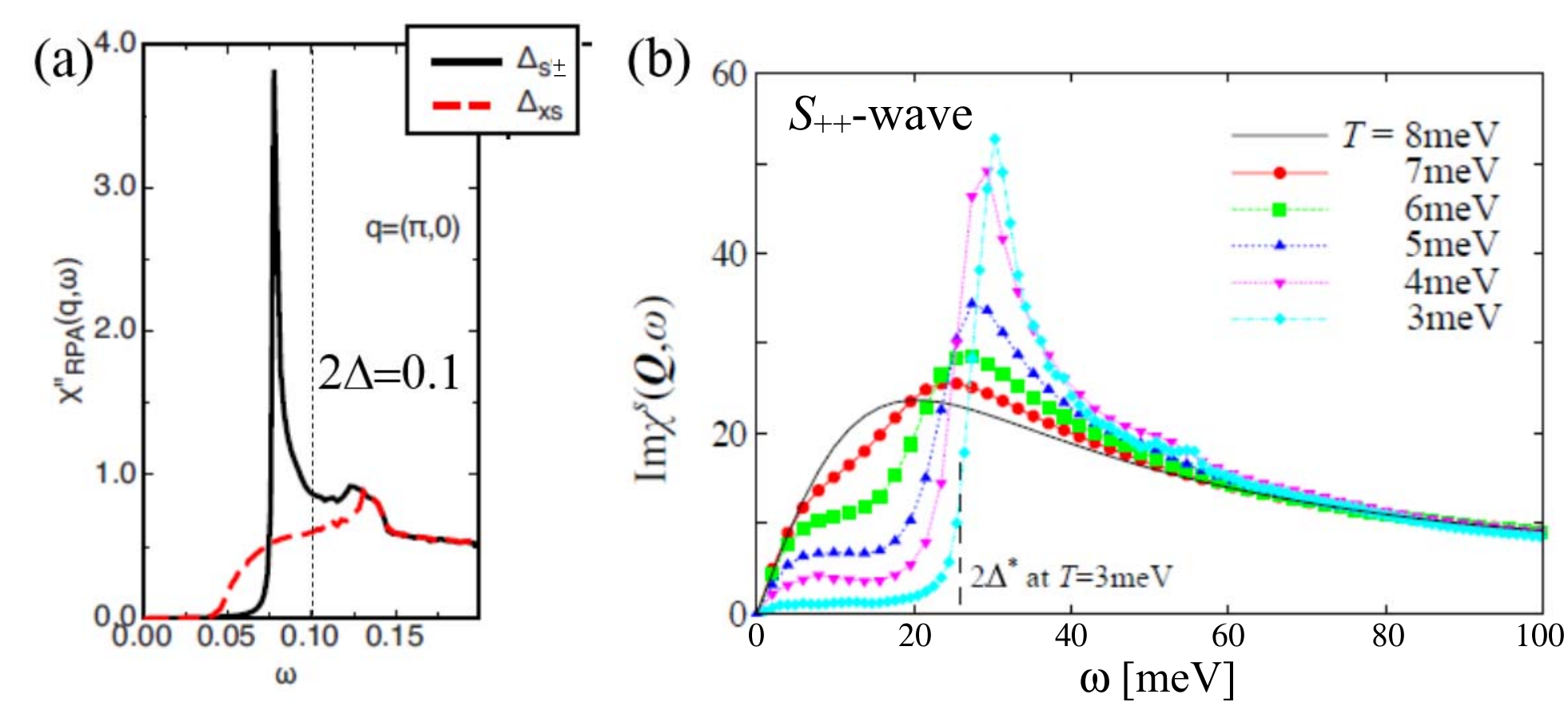}
  \caption{
(a) ${\rm Im}\chi^s(\Q,\w)$ given by the RPA for the $s_{\pm}$-wave state ($\Delta_{s\pm}$)
and the nodal $s$-wave state ($\Delta_{sx}$).
Reprinted with permission from \cite{maier-scalapino}. Copyright (2008) by the American Physical Society.
(b) ${\rm Im}\chi^s(\Q,\w)$ given by the multi-step FLEX approximation 
for the $s_{++}$-wave state \cite{takeuchi-resonance}:
Here, the normalized gap function is $\Delta^*\approx 13$meV
($2\Delta^*/T_{\rm c}\approx5.2$) for all FSs, 
and $\Delta^*$ follows the BCS-type $T$-dependence.
Black line ($T_{\rm c}=8$meV) is the normal state.
The obtained resonance-like peak 
is consistent with experiments shown in Fig. \ref{fig:neu-keimer-Zhang}.
Reprinted with permission from \cite{takeuchi-resonance}: Copyright (2018) by the American Physical Society.
}
  \label{resonance}
\end{figure}

First, we review the theory of resonance within the RPA by neglecting the self-energy effect.
In the SC state, ${\rm Im}\Psi(\Q,\w)$ is almost zero for $|\w|<2|\Delta|$
because the p-h excitation is prohibited for $E_{\rm p-h}<2|\Delta|$.
Then, the Kramers-Kronig relation leads that 
${\rm Re}\Psi(\Q,\w)$ is an increasing function of $\w^2$ for $|\w|<2|\Delta|$
for both $s_\pm$- and $s_{++}$-wave states.
In addition, in the $s_\pm$-wave state,
$\a_S(\w)$ is enlarged below $T_{\rm c}$ 
because ${\rm Re}\phi^0(\Q,\w)\propto -\Delta(\k)\Delta(\k+\Q)$ is positive.
For these reasons, 
$\a_S(\w)$ can reach unity at finite $\w_{\rm res} \ (<2|\Delta|)$
in the paramagnetic $s_\pm$-wave state with $\a_S(0)<1$. 
Then, $\chi^s(\Q,\w)$ exhibits sharp resonance peak at $\w=\w_{\rm res}$.
In the $s_{++}$-wave state, however,
$\a_S(\w)$ does not reach unity in the RPA because ${\rm Re}\phi^0(\Q,\w)<0$.
Thus, within the RPA, the resonance peak does not appear in the $s_{++}$-wave state.
Figure \ref{resonance} (a) shows the dynamical spin susceptibility 
${\rm Im}\chi^s(\Q,\w)$ obtained by the RPA \cite{maier-scalapino}.
In the $s_\pm$-wave state, sharp resonance peak appears at
$\w_{\rm res}=0.08$, which is smaller than $2\Delta^{\rm max}=0.1$.
In contrast, no resonance peak appears in the nodal $s_{++}$-wave state. 
Similar results have been published in Refs.
\cite{maier-scalapino,eremin,Das-res,Kuroki-res,Korshunov-res,Korshunov-res2}.

In the next stage,
we calculate ${\rm Im}\Psi(\Q,\w)$ by including the 
``self-energy due to inelastic scattering'' 
into Eqs. (\ref{chi0}) and (\ref{phi0}).
We apply the FLEX approximation for the self-energy.
In order to calculate the self-energy accurately at $T\sim1$meV,
we use the ``multi-step FLEX method'',
which enables us to use $128^2$ $\k$-meshes 
and $2^{16}$ Matsubara frequencies.
Figure \ref{resonance} (b) shows the obtained ${\rm Im}\chi^s(\Q,\w)$ 
in the $s_{++}$-wave state ($T_{\rm c}=8$meV).
Here, $\a_S(0)=0.95$ at $T=T_{\rm c}$.
The hump structure at $\w_{\rm res}\lesssim30$meV
becomes taller and sharper as $T$ is lowered.
The resonance energy $\w_{\rm res}$ slightly increases as
$T$ decreases, and $\w_{\rm res}$ is slightly above $2\Delta^{\ast}$.
The height of ${\rm Im}\chi^s(\Q,\omega)$ in the $s_{++}$ state
is just twice as large as the signal in normal state.
Thus, ${\rm Im}\chi^s(\Q,\omega)$ in the $s_{++}$ state 
obtained by the multi-step FLEX well explains experimental results
in Fig. \ref{fig:neu-keimer-Zhang}.

\begin{figure}[htb]
\centering
\includegraphics[width=0.8\linewidth]{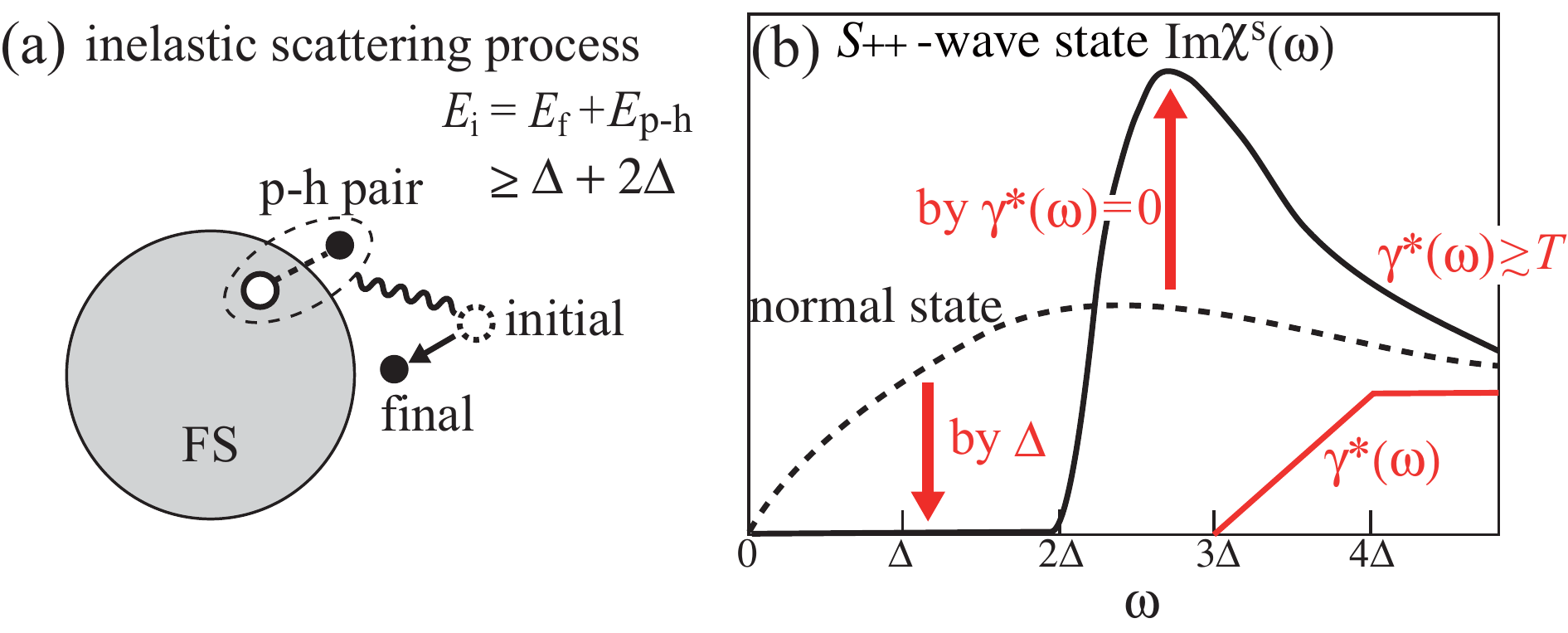}
\caption{(a) Inelastic scattering of the initial QP 
with energy $E_i \ (=E_\k)$
due to the creation of a particle-hole excitation $2\Delta$.
This process can occur when $E_i \ge 3\Delta$.
(b) Origin of the resonance-like peak in the $s_{++}$-wave state.
The relation $\gamma^*(\w)\approx 0$ for $\w<3\Delta$
magnifies Im$\chi^s(\w)$ at $\w\sim \w_{\rm res}$, where $\a_S(\w_{\rm res})\sim1$
\cite{onari-resonance}.
}
\label{Schema}
\end{figure}

Finally,  we explain a physical reason why resonance-like peak appears 
in the $s_{++}$-wave.
In Fig. \ref{Schema} (a), 
we show an inelastic scattering process, in which a quasi-particle (QP) 
with energy $E_\k=E_i$ is
scattered to the final state with energy $E_f$, 
with exciting a particle-hole (p-h) pair.
$E_f$ should be larger than $|\Delta|$,
and the energy of a particle-hole excitation $E_{\rm p-h}$ is larger than $2\Delta$. 
Therefore, the inelastic scattering is prohibited for $E\le 3|\Delta|$.
($E= E_{f} +E_{\rm p-h}\geq 3|\Delta|$).
That is, the QP inelastic scattering is absent 
for $|E|<3\Delta$ at $T\ll T_{\rm c}$.
Therefore, $\gamma^*(\w)$ is suddenly suppressed to become dissipationless 
in the SC state for $E\le 3|\Delta|$.

This fact drastically influences low-energy functional form of 
Im$\chi^s(\Q,\w)$,
as we illustrated in Fig. \ref{Schema} (b). 
In the normal state, Im$\chi^{\rm s}(\Q,\w)$ is strongly suppressed 
by large QP damping $\gamma^*\sim T$.
This suppression suddenly disappears in the SC state
because $\gamma^*(E)\sim0$.
In contrast, Im$\chi^{\rm s}(\Q,\w)$ for $\w \gtrsim 3|\Delta|$ 
is insensitive to the SC transition.
Owing to this ``dissipationless mechanism'', 
Im$\chi^s(\Q,\w)$ exhibit the resonance-like peak ($\a_S(\w_{\rm res})\sim1$) 
even in the $s_{++}$-wave state, as we show in Fig. \ref{resonance} (b).
This mechanism is different from the 
coherence factor mechanism in the sign-reversing superconductors.
The resonance-like peak in this mechanism becomes prominent near the magnetic QCP.
Thus, clear peak in optimally-doped compounds 
in Figs. \ref{fig:neu-keimer-Zhang} (a) and (b) are naturally explained 
based on this dissipationless mechanism.
In future, more detailed comparison between theatrical results and experimental ones 
would enable us to elucidate important information on the gap function.

\subsubsection{Nuclear relaxation rate $1/T_1$}
We also discuss the nuclear relaxation rate $1/T_1$,
which provides us important phase-sensitive information on 
the SC gap.
Here, we focus on the effect of the inelastic QP scattering 
on the size of the Hebel-Slichter peak (coherence peak) in $1/T_1$.

In the weak-coupling BCS theory without self-energy correction,
$1/T_1$ exhibit the Hebel-Slichter peak when the superconducting gap is sign-preserving,
reflecting the huge DOS for $|\w|\sim |\Delta|$
\cite{Hebel}.
The Hebel-Slichter peak is observed in many weak-coupling BCS $s$-wave superconductors.
However, in sign-reversing superconducting gap states,
the Hebel-Slichter peak is suppressed by the coherence factor mechanism
within the BCS theory.

In many Fe-based superconductors, the Hebel-Slichter peak is absent 
\cite{Nakai,Mukuda,Grafe,Yashima,impurity1,Nakai2}.
However, it is well-known that the Hebel-Slichter peak is easily
suppressed by finite QP damping rate,
because the inelastic scattering $\gamma^*$ is still large for $T\lesssim T_{\rm c}$.
Since $\gamma^*\propto T^2$ in Fermi liquids,
the Hebel-Slichter peak is not observed
in several high-$T_{\rm c}$ BCS $s$-wave superconductors ($T_{\rm c} > 15$ K), 
such as boron carbide YNi$_2$B$_2$C \cite{Kohara} 
and A-15 compounds V$_3$Si \cite{Kishimoto}.

\begin{figure}[htb]
\centering
\includegraphics[width=0.4\linewidth]{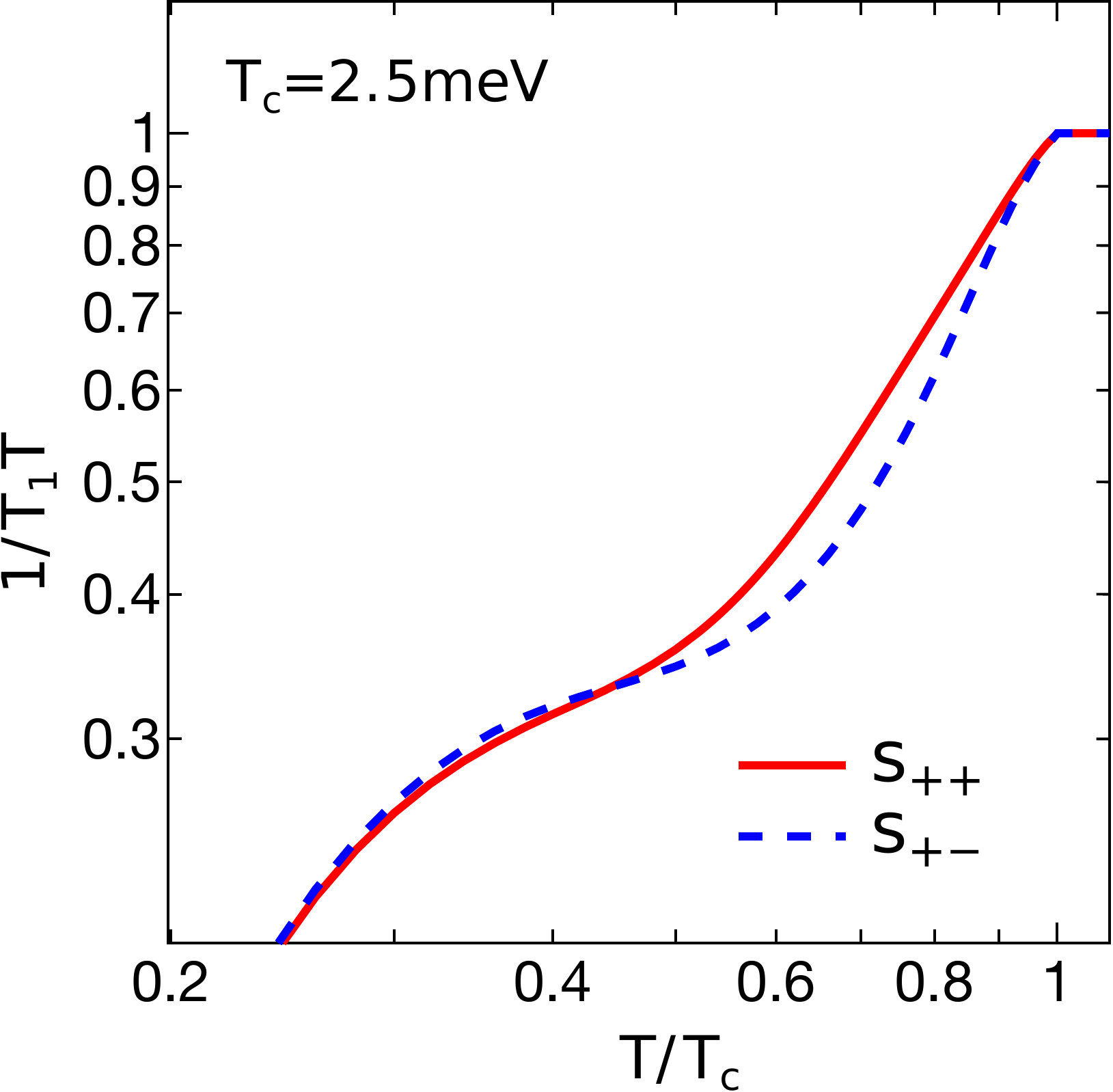}
\caption{Nuclear relaxation rate $1/T_1T$ normalized at 
$T=T_{\rm c}=2.5$meV in the $s_{++}$-wave and $s_{\pm}$-wave states
\cite{Yamakawa-Hebel}.
We set $2|\Delta_e/T_{\rm c}|=5$ and $|\Delta_e/\Delta_h|=3$,
and assume BCS-type $T$-dependence of $\Delta_{e,h}$.
Cited from Ref. \cite{Yamakawa-Hebel}.
}
\label{nmr}
\end{figure}

Here, we study the self-energy effect on 
the Hebel-Slichter peak in Fe-based superconductors,
like the analysis of the inelastic neutron scattering. 
For the quantitative analysis, 
we using the experimental QP inelastic scattering rate
that is accurately extracted from the resistivity above $T_{\rm c}$
based on the realistic five-orbital model \cite{Yamakawa-Hebel}.
Based on the Fermi liquid theory, the inelastic damping rate 
$\gamma^*(0)$ at $T=T_{\rm c}$ is represented as
\begin{equation}
\gamma^*(0)=z\frac{\pi^3}{2}V^2_{\rm eff}N(0)T_{\rm c}^2,
\end{equation}
where $z=m/m^* \ (\sim1/2)$ is the renormalization factor,
and $V_{\rm eff}$ is the effective electron-electron interaction
enhanced by the spin and orbital fluctuations.
Fortunately, $V_{\rm eff}$ is accurately derived from experimental 
$\rho(T)-\rho_{\rm imp}$ at $T=T_{\rm c}$ in several compounds 
\cite{Yamakawa-Hebel}.

We can calculate $\gamma^*(0)$ below $T_{\rm c}$ based on the Fermi liquid theory
\cite{Akis,Fujimoto,Yamakawa-Hebel}. 
Then, the dynamical spin susceptibility below $T_{\rm c}$ 
is obtained by using Eqs. (\ref{chi0})-(\ref{chiS}) with including $\gamma^*(0)$.
As a result, $1/T_1T$ in the SC state 
is derived from the standard formula:
\begin{equation}
\frac{1}{T_1T}\propto\frac{1}{N}\sum_{\q}\lim_{\omega\rightarrow 0}{\rm Im}\frac{\chi^s(\q,\omega)}{\omega}
\end{equation}
Figure \ref{nmr} shows the obtained $1/T_1T$, which is normalized at $T=T_{\rm c}=2.5$meV,
for both the $s_{++}$-wave and the $s_{\pm}$-wave states. 
Here, we set $V_{\rm eff}=17.7$eV and $z=1/2$ that correspond to the optimally doped Ba122.
The value of $V_{\rm eff}$ is derived from the experimental value
$\rho(T)-\rho_{\rm imp}=11.9\mu\Omega$cm at $T\gtrsim T_{\rm c}$
\cite{Yamakawa-Hebel}. 
In this case, $\gamma^*(0)/T_{\rm c}=1.4$.
Thus, the Hebel-Slichter peak is suppressed 
in both the $s_{++}$-wave and $s_{\pm}$-wave states
in the case of $T_{\rm c} \sim 30$K.

In overdoped Ba122 with $T_{\rm c}=11$K, $V_{\rm eff}$ is reduced to $12.7$eV,
and $\gamma^*(0)/T_{\rm c}=0.34$.
\cite{Yamakawa-Hebel}. 
Even in this case, the Hebel-Slichter peak is almost absent for the $s_{++}$-wave state.
We note that tiny Hebel-Slichter peak is expected to appear
when $\gamma^*(0)/T_{\rm c}\ll 1$
\cite{Yamakawa-Hebel}. 
Therefore, it is would not be easy to discriminate between
$s_{++}$-wave and $s_{\pm}$-wave states from the present NMR experimental data.

\subsection{Theory of superconductivity beyond-Migdal approximation}
\label{sect:Onari3}

\subsubsection{beyond-Migdal pairing interactions}
\label{sec:Onari3-bM}

\begin{figure}[t]
\centering
\includegraphics[width=0.7\linewidth]{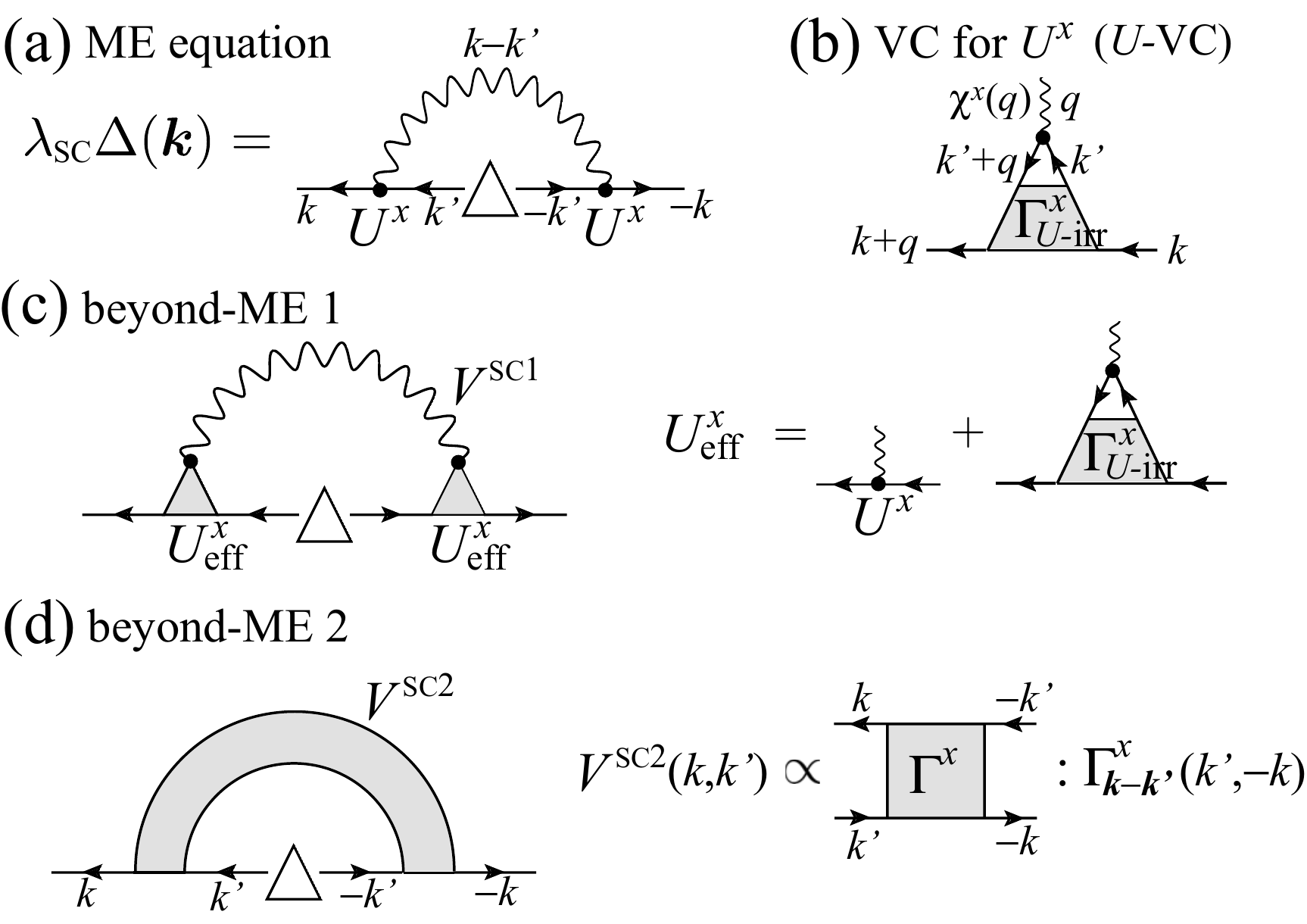}
\caption{
(a) SC gap equation within the ME theory.
(b) The VC that is neglected in the Migdal approximations.
(c) Beyond-Migdal interaction 1
\cite{Nakaoka}:
{\it U}-VC is dressed by $\Gamma^x_{U-{\rm irr}}$
that is irreducible with respect to $U^x$.
(d) Beyond-Migdal interaction 2
\cite{Kagome-Tazai}:
Here, the pairing interaction is directly given by the full vertex 
as $V^{{\rm SC2},x}(k,k')=\Gamma^x_{\k'-\k}(\k,\k')$, where $x=s,c$.
$\Gamma^x$ is expressed as infinite series of $I^x$'s shown in Fig. \ref{fig:DW-eq-diagram} (b),
and it includes irreducible terms with respect to bare $U^x$.
Then, the pairing interaction is expressed by the solution of the DW equation:
$\displaystyle V^{{\rm SC2},x}(k,k') \propto
 \frac{f_\q(k)f_\q^*(k')}{1-\lambda_\q}$ with $\q=\k'-\k$.
}
\label{fig:diagram}
\end{figure}

In the study of unconventional superconductors, 
the Migdal-Eliashberg gap equation in Fig. \ref{fig:diagram} (a)
(also in Eqs. (\ref{Eliash}) and (\ref{pairing}))
has been frequently used, by using the bare interaction $U^{s,c}$
for the electron-boson coupling constant.
However, the validity of the  Migdal approximation is guaranteed
only for weak or moderate $e$-ph interaction systems. 
In this section, we explain that the Migdal theorem is strongly violated
for the charge-channel fluctuation mediated superconductors,
because ``the VC for the electron-boson coupling constant'' is significant 
due to the quantum interference mechanism.

For preparation, we briefly review the Migdal's theorem for the $e$-ph interaction
with the BCS cut-off energy $\w_c \ (\ll E_F)$.
In this case, the $q$ dependence of the VC in Fig. \ref{fig:diagram} (b),
expressed as $\Gamma_{U-{\rm irr}}$, is moderate.
Here, we assume that both $\k$ and $\k+\q$ in Fig. \ref{fig:diagram} (b)
are on the FS, because we consider low-energy electrons with $|E|\lesssim \w_c$.
In taking the summation of $\k'$ in Fig. \ref{fig:diagram} (b),
both $\k'$ and $\k'+\q$ lay on the FS only when (i) $\k'\approx \k$
or (ii) $\k'\approx -\k-\q$, while such area in the momentum space
is very limited except for $\q\approx{\bm0}$.
Therefore, the VC due to the $e$-ph interaction becomes $O(\omega_c/E_F)\ll 1$,
and therefore it is negligible.

One may expect that the Migdal's theorem is applicable in metals with 
strong spin and/or orbital fluctuations, because their cutoff energy $\w_c$
is much smaller than $E_F$ near the QCP.
However, the Migdal's theorem is not satisfied in general.
For example, when nematic ($\q\approx{\bm0}$) fluctuations strongly develop,
the limitation on the $\k'$ summation in Fig. \ref{fig:diagram} (b) 
is not severe any more.
In a similar way, the Migdal's theorem is not satisfied 
when smectic fluctuations develop at the FS nesting vector $\q$.
Note that the VC has prominent momentum dependence because it is the function of 
$\chi^{s,c}(q)$, and some diagrams for $\Gamma_{U-{\rm irr}}(\k,\k',\q)$ take huge value 
($\gg U$) for special momenta.
(In contrast, the local vertex $\frac1{N^2} \sum_{\k,\k'}\Gamma_{U-{\rm irr}}(\k,\k',\q)$ is $O(U)$.)

For this reason, in Fe-based superconductors, 
the VCs for the electron-boson coupling is very important,
and therefore we construct ``the beyond-ME gap equation formalism''.
As we discussed in Sect. \ref{sec:Kontani2},
the VC for the charge-channel susceptibility, which we call the $\chi$-VC,
is significant near the magnetic QCP. 
Theoretically, the same VC should be important for the coupling constant ($U^{s,c}$)
in the gap equation, which we call the $U$-VC.
The gap equation with $U$-VC is shown in Fig. \ref{fig:diagram} (c).


The beyond-ME gap equation is given as
\begin{eqnarray}
\lambda_{\rm SC}\Delta_{l,l'}(k)&=& -T\sum_{k',m_i}
V^{\rm SC}_{l,m_1;m_4,l'}(k,k') G_{m_1,m_2}(k')
\nonumber \\
& &\times \Delta_{m_2,m_3}(k')G_{m_4,m_3}(-k'),
\label{eq:bME}
\end{eqnarray}
where $\lambda_{\rm SC}$ is the eigenvalue, 
$\Delta_{l,l'}(k)$ is the gap function, and
$G_{l,l'}(k)$ is the Green function.
The pairing interaction is no more a simple function of
$V^{\rm SC}(k,k')$ due to the $U$-VC.

First, we introduce ``the beyond-Migdal interaction 1'' shown in 
Fig. \ref{fig:diagram} (c)
\cite{Nakaoka}.
It is expressed as
\begin{eqnarray}
&&{\hat V}^{\rm SC1}(k,k')= \frac32 {\hat U}^s_{\rm eff}(k,k')
{\hat \chi}^s(k-k') {\hat U}'^s_{\rm eff}(-k,-k')
\nonumber \\
&&\ \ \ \ \ 
-\frac12 {\hat U}^c_{\rm eff}(k,k') {\hat \chi}^c(k-k')
{\hat U}'^c_{\rm eff}(-k,-k') +V^{(1)},
 \label{eqn:V} 
\end{eqnarray}
where $V^{(1)}=\frac12({\hat U}^s-{\hat U}^c) \sim U$.
Here, ${\hat U}^{x}_{\rm eff}(k,k')$ is the effective coupling
dressed by $\Gamma^x_{U-{\rm irr}}$,
which we replace with $I^x_{U-{\rm irr}} = I^x-U^x$ in the present numerical study.
Note that $({\hat U}'^{x}_{\rm eff})_{l,l';m,m'}(k,k')=({\hat U}^{x}_{\rm eff})_{m',m;l',l}(k,k')$

In the absence of $U$-VC, the spin fluctuation mediated pairing 
is expected because of the factor $3$ in the first term in Eq. (\ref{eqn:V}).
However, the $U$-VC enhances the charge channel pairing interaction, 
while it suppresses the spin channel pairing interaction.
As a result, the $s_{++}$-wave state can emerge in Fe-based superconductors.

In addition, 
we found that the following double fluctuation exchange pairing interaction $V^{(2)}$,
which is neglected in the ME approximation,
is important for some Fe-based superconductors:
\begin{eqnarray}
V^{(2)}_{l,l';m,m'}(k,k')&=&\frac{T}{4}\sum_{p}\sum_{a,b,c,d}G_{a,b}(k'-p)G_{c,d}(-k-p) \nonumber \\
&&\times \sum_{x,x'}^{x}b_{x,x'}V^{x'}_{l,a,m,d}(k-k'+p)V^{s}_{b,l',c,m'}(-p) 
\label{eq:Vcross}
\end{eqnarray}
where $b_{s,s}=b_{s,c}=b_{c,s}=3$ and $b_{c,c}=-1$.
Since $V^{(2)}$ gives attractive inter-pocket interaction ($\k-\k'\sim\Q$)
and repulsive intra-pocket interaction ($\q\sim{\bm0}$),
this term is favorable for the $s_{++}$-wave state.
In this article, $V^{(2)}$ is added to the pairing interaction ${\hat V}^{\rm SC1}(k,k')$
in the study of Ba122 in Sect. \ref{sec:Onari-Ba122}.

Next, we introduce ``the beyond-Migdal interaction 2'' shown in Fig. \ref{fig:diagram} (d)
\cite{Kagome-Tazai}.
It is directly given by 
the full four-point vertex $\Gamma^x_\q(k,k')$ ($x=s,c$).
It is expressed in Fig. \ref{fig:DW-eq-diagram} (b),
and is derived from the Bethe-Salpeter equation (\ref{eqn:BS1}).
Mathematically, the Bethe-Salpeter equation and the DW equation are essentially equivalent.
By applying the singular value decomposition, 
$\Gamma^x_\q(k,k')$ is well approximated as
$\Gamma^x_\q(k,k') ={\bar I}_\q \frac{f_\q(k)f_\q^*(k')}{1-\lambda_\q}$ with $\q=\k'-\k$
when $\lambda_\q$ is close to unity.
Here, $\lambda_\q$ and $f_\q(k)$ are the solution of the DW equation.
In Fe-based superconductor, the charge-channel eigenvalue strongly develops.
Therefore, the charge-channel pairing interaction is expressed with good accuracy as
\begin{eqnarray}
V^{{\rm SC2},c}(k,k')={\bar I}_\q \frac{f_\q(k)f_\q^*(k')}{1-\lambda_\q}
\label{eqn:Ibar}
\end{eqnarray}
with $\q=\k'-\k$,
At fixed $\q$, the coefficient ${\bar I}_\q$ is determined numerically
from the relation $I^c_{\q}(k,k+\q) ={\bar I}_\q f_\q(k)f_\q^*(k+\q)$,
where $k$ maximize the left-hand-side.
The $\k$-dependence of $f_\q(k)$ describes the beyond-Migdal pairing interaction.

Because the spin-channel eigenvalue is smaller in Fe-based superconductor, 
the total pairing interaction is expressed with good accuracy as
$\displaystyle V^{\rm SC2}(k,k')=V^{{\rm SC2},c}(k,k')- \frac32 (U^s)^2\chi^s(k-k') +V^{(1)}$.
Based on this beyond-ME equations 2,
we will analyze the pairing state in FeSe family in Sect. \ref{sec:Onari-FeSe}.
We note that the cross diagram $V^{(2)}$ in Eq. (\ref{eq:Vcross}) is included in $V^{{\rm SC2}}$,
and the pairing interaction 2 in Fig. \ref{fig:diagram} (d)
is a natural extension of the pairing interaction 1 in Fig. \ref{fig:diagram} (c).

\subsection{Ba122: Cooperation and Competition between spin and orbital fluctuations}
\label{sec:Onari-Ba122}

\begin{figure}[htbp]
\centering
\includegraphics[width=0.7\linewidth]{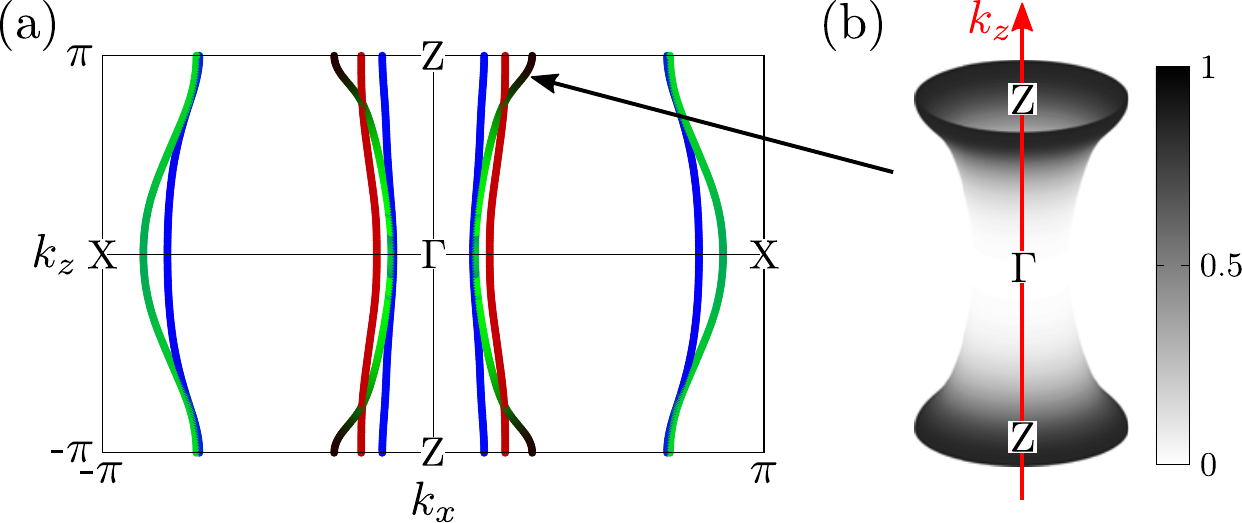}
\caption{
(a) FS in the $k_{y}=0$ plane in the optimally doped BaFe$_{2}$(As,P)$_{2}$. The solid lines show the Brillouin zone. Black, green, red and blue colors show the weight of the $z^{2}, {xz}, {yz}$ and ${xy}$ orbitals, respectively. (b) Schematic picture of the hole cylinder with the $k_{z}$ dependence of the weight of the $z^{2}$ orbital
\cite{Nakaoka}. 
}
\label{fig:model}
\end{figure}

Here, we analyze the 
orbital + spin fluctuation mediated superconductivity in Ba122 compounds
based on the beyond-ME gap equation (\ref{eq:bME}).
Ba122 is one of the most intensively studied family of Fe-based superconductors.
The SC gap structure has precisely determined by ARPES studies.
Figure \ref{fig:model} (a) shows the FSs of BaFe$_2$(As,P)$_2$ on the $k_y=0$ plane.
The hole-FSs around $\Gamma$ point and the electron-FSs around X point
are similar to those in Fig. \ref{fig:FeAs-layer} (c).
These FSs are composed of $t_{2g}$ orbitals ($xz,yz,xy$).
Interestingly, one of the hole-FSs around Z point is composed of $z^3$-orbital.
There is no $z^2$-orbital weight on electron-FSs.
Figure\,\ref{fig:model} (b) illustrates the weight of $z^{2}$-orbital
on the hole cylinder.

In the RPA analysis
\cite{Suzuki-Kuroki,Saito-loop},
spin fluctuations develop in all $t_{2g}$-orbitals,
whereas those in the $z^2$-orbital remain very small,
because spin fluctuations develop within the same $d$-orbital.
For this reason, spin fluctuation theories predict the 
horizontal node around the $z^2$-orbital hole-FS.
In the orbital-fluctuation theory in Refs. \cite{Saito-loop,Nakaoka},
in contrast, $z^2$-orbital hole-FS is fully-gapped
due to the intra-orbital pairing interaction.
Experimentally, the horizontal node was reported in Ref. \cite{Feng},
whereas other ARPES studies reported that
$z^2$-orbital hole-FS is fully-gapped 
\cite{Shimojima,Yoshida}.
Thus, presence or absence of the horizontal node in Ba122
is a significant key factor to distinguish the pairing mechanism
\cite{Suzuki-Kuroki,Shimojima,Yoshida,Feng,Saito-loop}.

\begin{figure*}[htbp]
\centering
\includegraphics[width=0.46\linewidth]{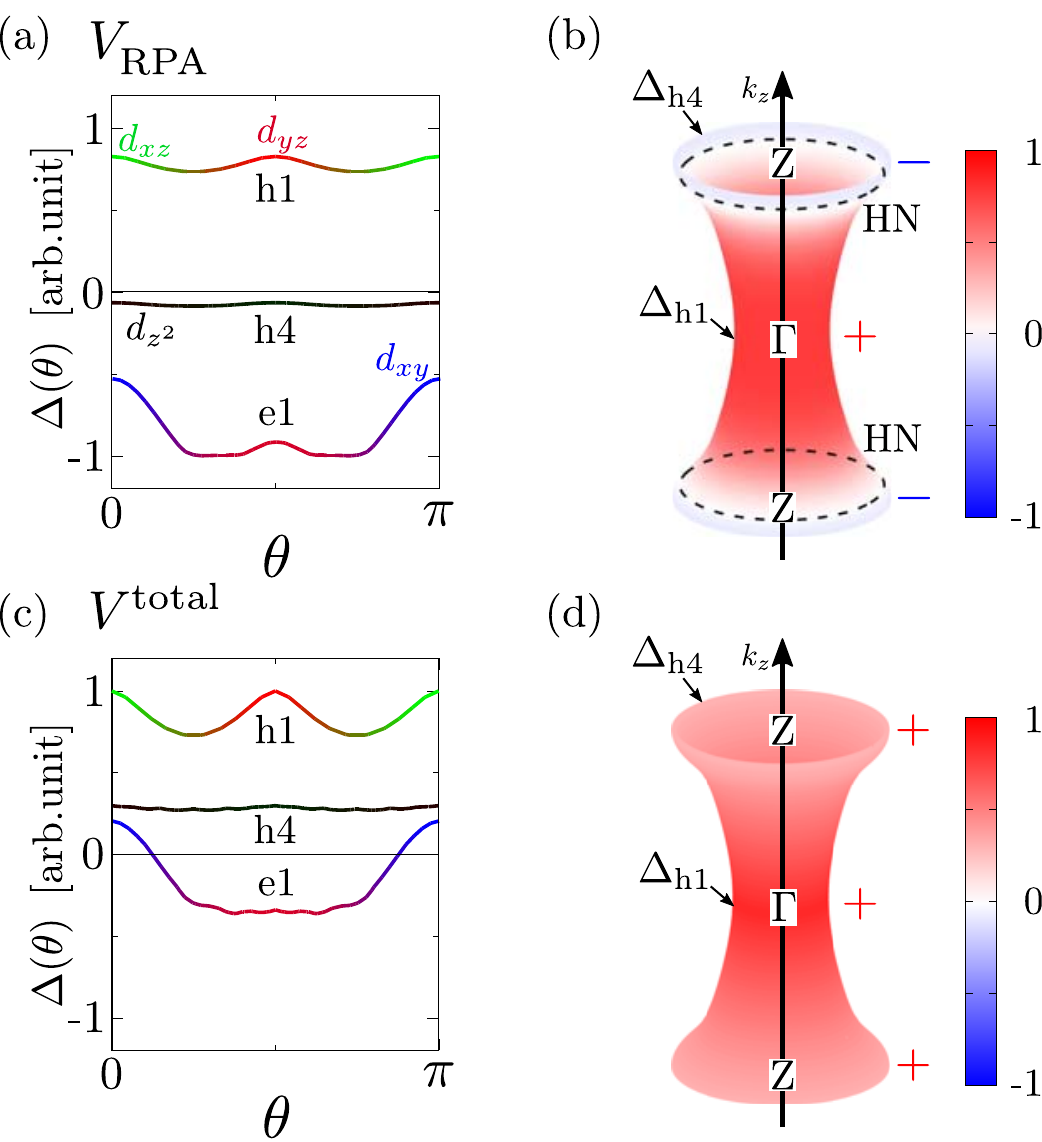}
\caption{
(a) Fully-gapped $s_{\pm}$ wave state obtained by $V_{{\rm RPA}}$,
where $\lambda_{\rm SC}=1.4$
\cite{Saito-loop}.
The obtained schematic horizontal node is shown in (b).
 The broken lines represent the expected horizontal node. 
(c) Nodal $s$ wave state obtained by the SC-VC theory,
where $\lambda_{\rm SC}=1.2$.
There is no sign reversal between h4 and h1, 
meaning the absence of horizontal node in (d). 
Instead, the loop-shaped nodes appear on electron-FSs
\cite{Saito-loop}.
Cited from Ref. \cite{Saito-loop}.
}
\label{fig:gap}
\end{figure*}

Figure\,\ref{fig:gap} (a) shows the obtained gap function in the RPA,
in the case of $\alpha_{s}=0.97$.
Here, both $\chi$-VC and $U$-VC are neglected.
The locations of hole-FSs $h1$ and $h4$ are shown in Fig. \ref{fig:gap} (b),
and $e1$ is the electron-FS around X point.
This is a $s_{\pm}$ wave state with horizontal node around the hole-FS h4.
The gap function on the hole cylinder is schematically shown in 
Fig.\,\ref{fig:gap} (b). 
This result is consistent with the previous RPA \cite{Suzuki-Kuroki}. 
However, small Volovik effect in the specific heat measurement 
\cite{Kim,Wang} indicate the absence of horizontal node.

Figure\,\ref{fig:gap} (c) is the gap function derived from 
${\hat V}^{{\rm SC1}}$ in Eq. (\ref{eqn:V}) based on the beyond-ME theory.
Here, $U=1.4$eV, $T=20$meV, and  $(\a_S,\a_C)=(0.97,0.88)$.
In this case, nodal $s$ wave state is obtained.
There is no sign reversal between h4 and h1, that is,
no horizontal node appear as illustrated in Fig.\,\ref{fig:gap} (d). 
In Fig.\,\ref{fig:gap}(c), nodes appear only on the electron-FSs,
at which the orbital character gradually changes between $xz(yz)$ and $xy$.
This result means the emergence of the loop-nodes on the electron-FSs,
consistently with the angle-resolved thermal conductivity measurement \cite{Yamashita}
and ARPES study \cite{Yoshida}.

The nodal $s$-wave gap structure in Figs.\ref{fig:gap} (c) and (d)
originates from the competition between attractive interaction 
and the repulsive interaction.
The former (latter) is induced by orbital (spin) fluctuations,
and the attractive interaction is magnified by the $U$-VC.
(The nematic fluctuations enlarge any SC states.)
In this situation, small number of impurities induce the crossover
from $s_{\pm}$-wave state to $s_{++}$-wave state 
shown in Fig. \ref{fig:lambda} (a).
The reduction in $T_{\rm c}$ during the crossover is expected to be small
\cite{Nakaoka}.
The impurity induced change in the gap function in Ba122
is reported by several experiments; see Fig. \ref{fig:lambda} (b).

\subsection{LiFeAs: Cooperation of $e$-ph and Coulomb interaction}

\begin{figure}[htbp]
\centering
\includegraphics[width=0.8\linewidth]{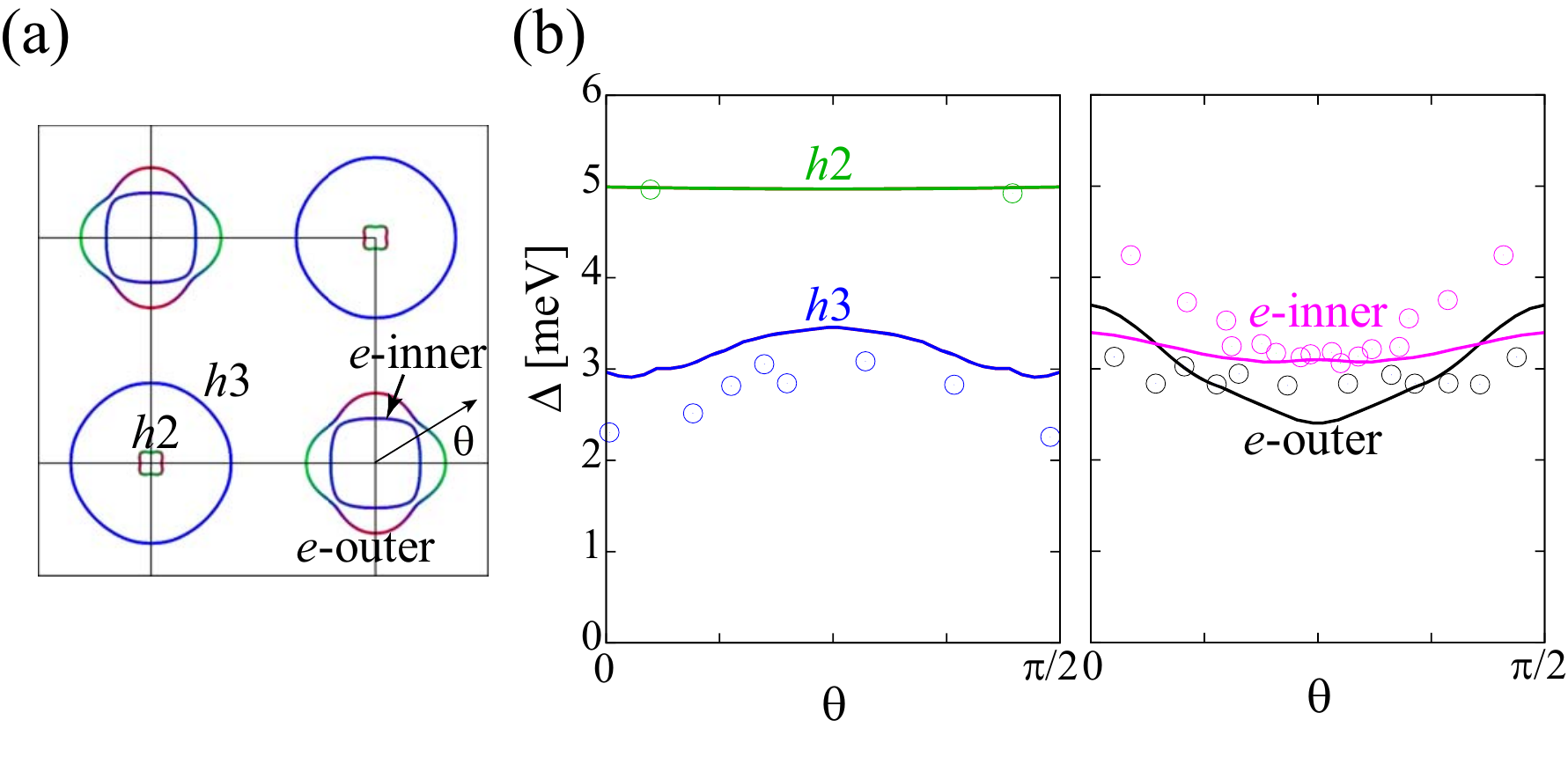}
\caption{
(a) FSs in 20 orbital model with SOI.
(b) Obtained gap function on each FS as a function of $\theta$ for
$\a_S=0.70$, and $\a_C=0.90$
\cite{Saito-LiFeAs2}.
The eigenvalue is $\lambda_{\rm SC}=0.38$ at $T=30$meV.
(Note that $\lambda_{\rm SC}\sim 0.1$ in the RPA).
Circles represent ARPES data in Ref. \cite{Umezawa-LiFeAs}.
Cited from Ref. \cite{Saito-LiFeAs2}.
}
\label{fig:LiFeAs-gap}
\end{figure}

In LiFeAs, neither AFM order nor nematic order exist.
Nonetheless of the weak correlation, its $T_{\rm c} \ (=17K)$ is relatively high.
The uniqueness of LiFeAs ($T_{\rm c}=17$K) lies in its peculiar FSs that are very 
different from other Fe-based superconductors.
In LiFeAs, $xz,yz$-orbital hole-FSs are very tiny, while
$xy$-orbital hole-FS and two electron-FSs are very large,
as shown in Fig. \ref{fig:LiFeAs-gap} (a).
Reflecting the wrong nesting of the FSs,
the observed spin fluctuations are very weak.
The magnitude of spin susceptibility observed in neutron inelastic scattering study
is about $1/8$ of that in optimally-doped Ba(Fe,Co)$_2$As$_2$ \cite{Li111-NMR,Qureshi}.
In this respect, the pairing mechanism in LiFeAs is very mysterious.

In LiFeAs,
precise SC gap structure has been determined by ARPES measurements
\cite{Borisenko-LiFeAs,Umezawa-LiFeAs}.
The experimental anisotropic gap structure is shown by circles
in Fig. \ref{fig:LiFeAs-gap} (b) \cite{Umezawa-LiFeAs}.
Theoretical analyses of gap structure have been performed in Refs. 
\cite{Hirschfeld-LiFeAs,Saito-LiFeAs1,Saito-LiFeAs2,Thomale-LiFeAs}
based on the RPA and the fRG.
However, the eigenvalue $\lambda_{\rm SC}$ is small
when spin fluctuations are weak as observed in LiFeAs.
In addition, the impurity effect on $T_{\rm c}$ is very weak in LiFeAs.
In fact, $T_{\rm c}$ in LiFeAs vanishes by introducing Co-impurities
when $\rho_{\rm imp}$ reaches $130\mu\Omega$ \cite{Lee-LiFeAs-imp},
which is one order of magnitude larger than the theoretical 
critical $\rho_{\rm imp}$ for the $s_\pm$-wave state.
In addition, $T_{\rm c}(x;{\rm Co})$ in LiFe$_{1-x}$Co$_x$As and 
$T_{\rm c}(y;{\rm Na}$ in LiFe$_{1-y}$Na$_y$As satisfy the relation 
$T_{\rm c}(x;{\rm Co})\approx T_{\rm c}(y/2;{\rm Na})$
\cite{imp-LiFeAs}.
Thus, $T_{\rm c}$ is scaled by the carrier concentration,
and therefore $T_{\rm c}$ is robust against impurity potential.


To understand the superconductivity ($T_{\rm c}=17$K)
in LiFeAs under small spin and nematic fluctuations,
we focus on the cooperation between $e$-ph and Coulomb interaction.
This idea is supported by the observed strong spin-lattice coupling
\cite{Egami-spin-lattice}.
This idea also is supported by the observation of large Fano effect for 
the in-plane Fe-As stretching mode in Ba$_{1-x}$K$_x$Fe$_2$As$_2$
by the infrared spectroscopy \cite{Fano}.
The fact that the Fano parameter $1/q^2$ is well scaled by the $T_{\rm c}$
means the importance of the Fe-As stretching mode for the SC pairing 
\cite{Fano}.

The in-plane Fe-As stretching mode indices the 
quadrupole interaction given in Eq. (\ref{eq:quadrupole})
\cite{Kontani-RPA}.
Here, we introduce the quadrupole interaction in Eq. (\ref{eq:quadrupole})
into the multiorbital Hubbard model,
and solve the beyond-ME gap equation based on the SC-VC theory.
(We use the LiFeAs model in Ref. \cite{Hirschfeld-LiFeAs} with a tripled energy scale.)
It is found that the quadrupole interaction and the $\chi$-VC
induce the smectic inter-orbital fluctuations cooperatively
\cite{Saito-LiFeAs1,Saito-LiFeAs2},
and the attractive paring interaction is magnified by the $U$-VC.
Figure \ref{fig:LiFeAs-gap} (b) shows obtained $s_{++}$-wave state
for $(\a_S,\a_C)=(0.70,0.90)$,
for $U=0.96$eV and $g(0)=0.29$eV at $T=20$meV ($\w_c=0.06$eV).
Here, we solve the gap equation in the 10-orbital model 
to introduce the atomic spin-orbit interaction (SOI) in Fe-ion
 \cite{Saito-LiFeAs1,Saito-LiFeAs2}.
The obtained gap structure well-reproduces ARPES results 
depicted by circles\cite{Umezawa-LiFeAs}. 
The largest magnitude of gap on the $h2$ FS is induced
by the attractive inter-orbital pairing interaction 
between the hole-pockets and the electron-pockets.
We note that similar gap structure is obtained even if $U$-VC is dropped,
when the system is close to the orbital-order QCP ($\a_C\sim 0.98$)
\cite{Saito-LiFeAs1,Saito-LiFeAs2}.
Thanks to the $U$-VC, fully-gapped $s_{++}$-wave state with large $\lambda_{\rm SC}$
is realized even for $\a_C=0.8\sim0.9$.

We comment that 
the pairing state and the mechanism of LiFeAs are still open problems.
For instance, orbital-antiphase $s_{\pm}$-wave state\cite{Kotliar2014} 
and the fully gapped $s_{\pm}$-wave state \cite{Kreisel2017_FeSe_gap}
has been proposed.
Impurity-induced in-gap state has been discussed in Ref.
\cite{LiFeAs-STM}.

\subsection{La1111: Double-dome superconducting phase}

The electron-doped LaFeAsO is the earliest Fe-based superconductor discovered in 2008.
As shown in Fig. \ref{fig:phase-1111} (a), 
interesting double-dome superconducting phase has been reported
in LaFeAsO$_{1-x}$H$_x$ \cite{Hosono-phase1111}.
To discuss $x$ dependence of $T_{\rm c}$, 
we analyzed the superconducting states of LaFeAsO$_{1-x}$H$_x$
by means of the beyond-ME gap equation,
based on the SC-VC theory together with the self-energy correction
\cite{Onari-SCVCS}.

In the phase diagram of LaFeAsO$_{1-x}$H$_x$ 
at ambient pressure, 
$T_{\rm c}\approx 30K$ at $x\approx 0.1$ 
gradually decreases with increasing $x$, 
while $T_{\rm c}$ starts to increase again and exceeds 30K around $x\sim0.35$.
This interesting double-dome structure in $T_{\rm c}$
has been discussed based on the spin-fluctuation mechanism
\cite{Kuroki-PRL2014}.
However, low-energy spin fluctuations observed in 
LaFeAsO$_{1-x}$H$_x$ is rather small, and the correlation between
$T_{\rm c}$ and the spin fluctuation strength is unclear
\cite{Fujiwara-NMR-1111}.
The phase diagram of LaFeAsO$_{1-x}$H$_x$ has also been discussed 
based on the orbital-fluctuation mechanism
\cite{Onari-SCVCS}.
In this theory,
the second-SC dome is caused by the inter-(xz,yz) and xy orbital fluctuations.

Interestingly, $T_{\rm c}$ in LaFeAsO$_{1-x}$H$_x$ increases further 
under pressure $P=2$GPa, as shown in Fig. \ref{fig:phase-1111} (a).
This phase diagram is similar to that in SmFeAsO$_{1-x}$H$_x$ at ambient pressure
 \cite{Hosono-phase1111}.
In the orbital fluctuation scenario,
the increment of $T_{\rm c}$ in SmFeAsO$_{1-x}$H$_x$ is ascribed to 
the large $d_{xy}$-orbital hole-FS \cite{Onari-SCVCS}.

\subsection{FeSe: High-$T_{\rm c}$ superconductivity without hole-pocket}
\label{sec:Onari-FeSe}

FeSe family is the most intensively studied Fe-based superconductors
in the last five years.
Rich phase diagrams of bulk FeSe are shown in 
Figs. \ref{fig:phase-11} and \ref{fig:phase-11-PT},
and the normal-state electronic states of FeSe has been discussed
in Sect. \ref{sec:Kontani4}.
The SC state has been studied intensively by both theorists
\cite{Yamakawa2017_FeSe_pressure,Chubukov-nematic-rev,Kreisel2017_FeSe_gap,Fanfarillo-FeSe-SC,Andersen-FeSe-SC}
and experimentalists
\cite{Bohmer2017_FeSe_review,Davis-QPI, neutron-FeSe, FeSe-Taen, Baek2020_FeSe_NMR}.
Very anisotropic orbital-selective $s$-wave gap function has been 
observed by ARPES measurement
\cite{Feng2016,Hashimoto2018,Borisenko2018}. 
The $s_{\pm}$-wave state was proposed by the QPI measurement for FeSe \cite{Davis-QPI}.
On the other hand, $T_{\rm c}$ in bulk FeSe seems to be robust against impurities
in Refs. \cite{FeSe-Urata,FeSe-Taen}.

Here, we discuss the 
high-$T_{\rm c}$ SC state with $T_{\rm c} = 40$--$100$~K in heavily
electron-doped (e-doped) FeSe systems.
By introducing only a few-percent e-doping, 
the orbital order is suppressed, and instead, 
a high-$T_{\rm c}$ SC phase with $T_{\rm c} \ge 40$~K appears 
for a wide doping range.
High-$T_{\rm c}$ state universally emerges in many e-doped FeSe compounds,
such as an ultra-thin FeSe layer on SrTiO$_3$ ($T_{\rm c} = 40$--$100$~K) 
\cite{Wang2012_FeSeSTO_1st, Lee2014_FeSeSTO_ARPES, Fan2015_FeSe_STM, Zhang2016_FeSeSTO_ARPES, Shi2017_FeSeSTO_phase}, 
K-dosed FeSe ($T_{\rm c} \sim 40$~K) 
\cite{Miyata2015_KdoseFeSe, Wen2016_KdosedSTO_phase}, 
and Li-intercalated superconductors ($T_{\rm c} \sim 40$~K) 
\cite{Lu_LiFeOHFeSe_crystal, Du2016_LiFeOHFeSe_STM, Noji2014_FeSe_intercalate, Yan2016_LiFeOHFeSe_QPI, Niu2015_LiFeOHFeSe_ARPES, Ren2017_LiFeOHFeSe_ARPES, Gu2018_LiFeOHFeSe_QPI}.

\begin{figure}[!htb]
\centering
\includegraphics[width=.99\linewidth]{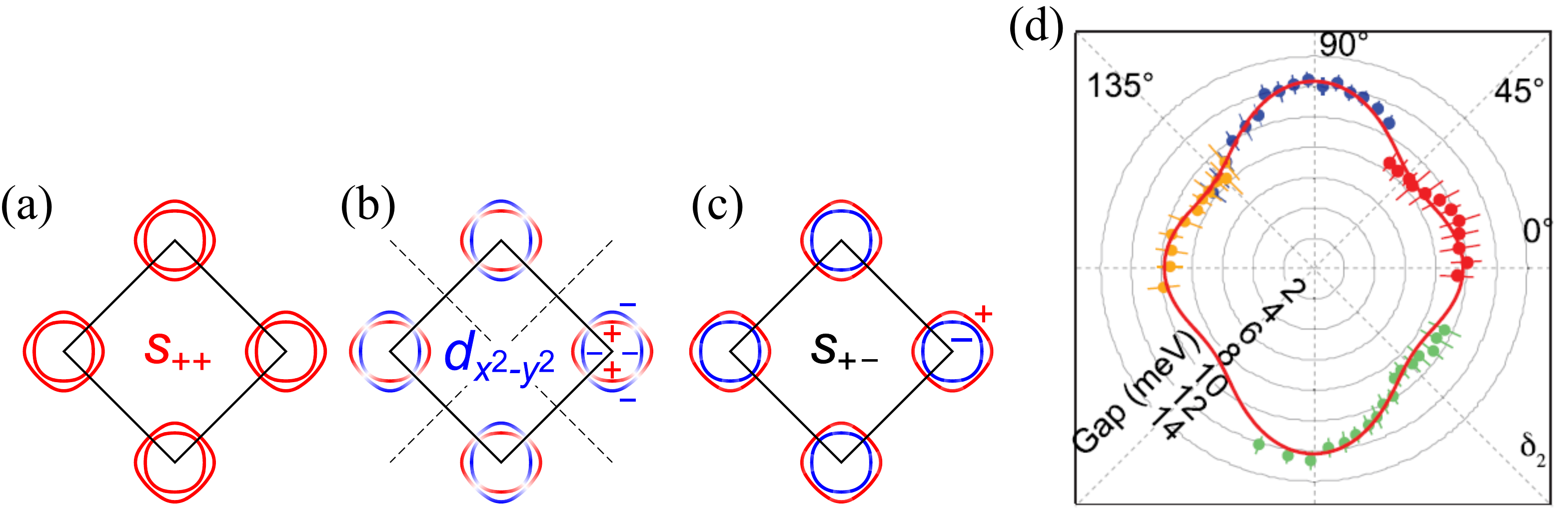}
\caption{
(a) The $s_{++}$-wave state and (b) $d$-wave state, and (c)$s_\pm$-wave
 state in the 2Fe unit-cell Brillouin zone.
The inner FS and outer FS are formed due to the 
SOI-induced band hybridization.
(d) Angular dependence of gap structure observed by ARPES measurement 
on the unfolded FS in monolayer FeSe.
Reprinted with permission from \cite{Zhang2016_FeSeSTO_ARPES}. Copyright (2016) by the American Physical Society.
}
\label{fig:FeSe-fig3}
\end{figure}

\color{black}
At the present stage,
the origin and the mechanism of high-$T_{\rm c}$ state in e-doped FeSe 
is a very important open problem.
In analogy with the $s_\pm$ wave state in Fig. \ref{fig:S+-S++} (a),
the incipient $s_\pm$-wave state between the missing hole-pocket 
and the electron-pocket has been proposed
\cite{Saito2011_KFeSe, Chen2015_KFeSe_s+-, Mishra2016_eFeSe_s+-,FeSe-incipient-Hirsdhfeld},
and high-$T_c$ state is obtained by the FLEX approximation
even when the top of the hole-band in FeSe is about $-0.1$eV below $E_F$.
Other possible pairing states on two electro-FSs are illustrated
in Figs. \ref{fig:FeSe-fig3} (a)-(c) in the folded BZ with finite SOI.
The gap structure in (a) $s_{++}$-wave state 
\cite{Yamakawa2020_eFeSe}
and that in (c) $s_{+-}$-wave state
\cite{Chubukov-KSe} are fully-gapped.
On the other hand, nodal gap appears in (b) $d_{x^2-y^2}$-wave state
due to the SOI-induced pair breaking \cite{Lee2015_eFeSe_SOI},
while nodeless $d$-wave state is realized when the SC gap
is larger than $\lambda_{\rm SOI}$ 
\cite{Agterberg2017_full_dwave}.
Experimentally,
fully gapped superconducting state has been observed 
by ARPES and STM measurements
\cite{Fan2015_FeSe_STM, Zhang2016_FeSeSTO_ARPES, Du2016_LiFeOHFeSe_STM,Lee2014_FeSeSTO_ARPES, Yan2016_LiFeOHFeSe_QPI,Niu2015_LiFeOHFeSe_ARPES,  Ren2017_LiFeOHFeSe_ARPES,Gu2018_LiFeOHFeSe_QPI},
as we show in Fig. \ref{fig:FeSe-fig3} (d).
It is notable that the observed spin fluctuations above $T_{\rm c}$ 
in $e$-doped FeSe by NMR measurements are very weak 
\cite{Hrovat2015_eFeSe_NMR}. 
Recent STM/STS study on single-layer FeSe/STO 
\cite{FeSe-ABS} reports 
the absence of topologically non-trivial edge/corner modes
that is consistent with the $s_{++}$ wave state.
\color{black}


In FeSe/SrTiO$_3$, strong $A_{1g}$ interfacial $e$-ph coupling 
is observed as the replica bands in ARPES measurements
\cite{Lee2015_eFeSe_SOI, Rademaker2016_eFeSe_Phonon,
 Wang2012_FeSeSTO_1st, Lee2014_FeSeSTO_ARPES}. 
The $e$-ph coupling has been expected to increase $T_{\rm c}$ up to $\sim 60$~K.
On the other hand, $T_{\rm c}\sim 40$~K is realized in (Li,Fe)OHFeSe 
even in the absence of strong interfacial $e$-ph interaction \cite{Lu_LiFeOHFeSe_crystal}.
This fact indicates that the main pairing glue originates from electron correlations.

\begin{figure}[!tb]
\centering
\includegraphics[width=0.99\linewidth]{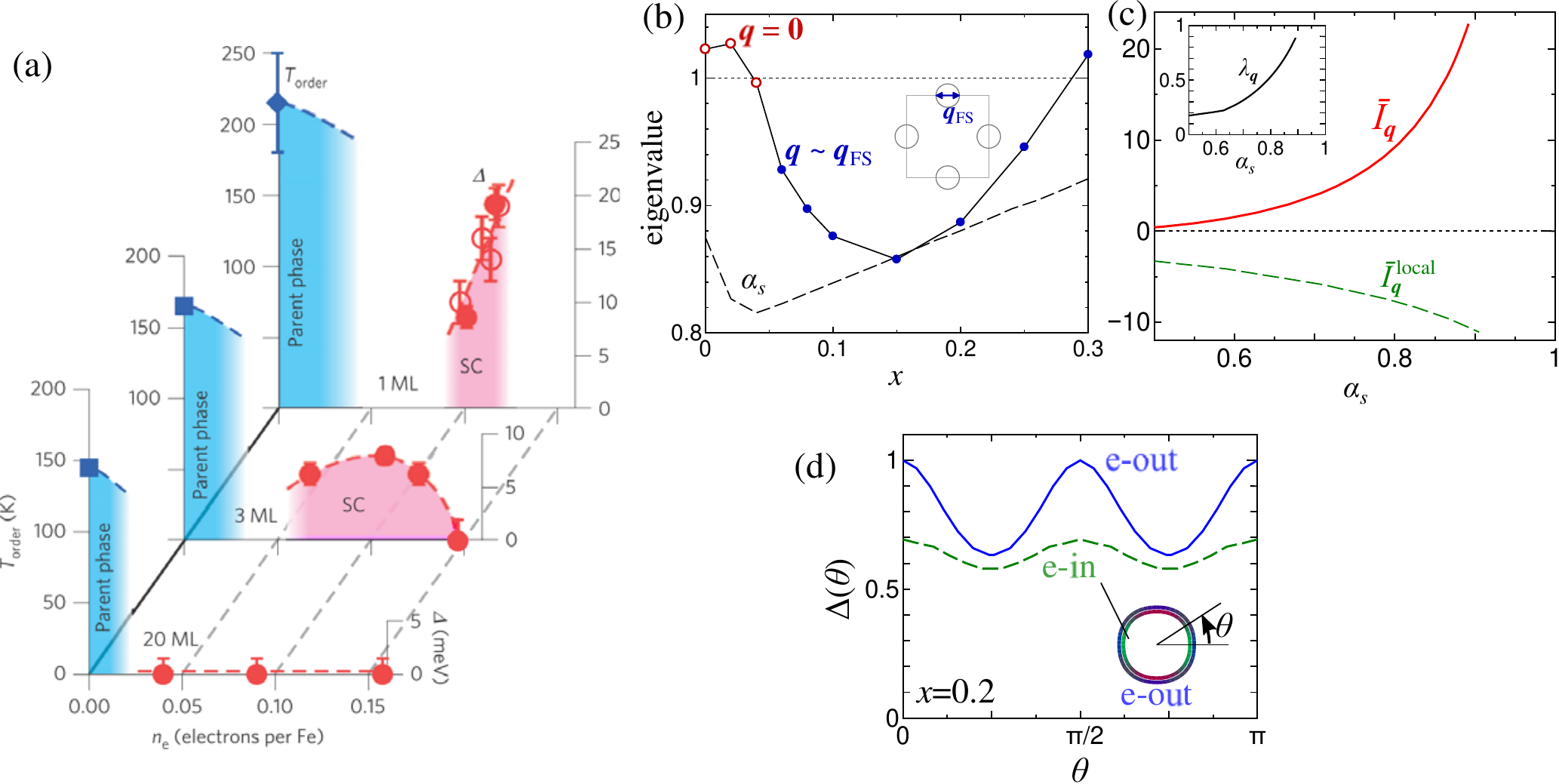}
\caption{
(a) Phase diagram of thin layer FeSe.
Reprinted by permission from Springer Nature:
Nat. Mater.  \cite{Miyata2015_KdoseFeSe}, copyright 2015.
Theoretically obtained 
(b) $x$ dependences of the DW equation eigenvalue and $\a_S$, and
(c) $\a_S$-dependences of 
the DW equation eigenvalue $\lambda_{\q}$ ($\q\approx \q_{\rm FS}$) and 
the coupling constant ${\bar I}_\q$ in Eq. (\ref{eqn:Ibar}).
In (c), the local coupling constant ${\bar I}_\q^{\rm local}$ 
is also shown for comparison.
(d) $s_{++}$-wave gap function at $x=0.2$ obtained by the beyond-ME2 theory.
The fully-gapped state is mediated by the smectic nematic fluctuations.
The FSs with SOI are shown in the inset.
((b)-(d): Y. Yamakawa {\it et al.}, unpublished.)
}
\label{FeSe-edope2}
\end{figure}

Figure \ref{FeSe-edope2} (a) is a typical phase diagram of 
electron-doped thin layer FeSe \cite{Miyata2015_KdoseFeSe}.
The highest-$T_{\rm c}$ (or largest SC gap)
is observed for single monolayer (ML) sample.
Next to the high-$T_{\rm c}$ phase,
smectic order at $\q\approx(\pi/4,0)$ has been discovered
by recent STM/STS measurements \cite{smectic-FeSe-STM}.
This finding provides evidence that 1ML FeSe is in close proximity 
to the smectic phase, and its superconductivity is likely 
enhanced by this electronic instability
\cite{smectic-FeSe-STM}.

In order to study e-doped FeSe systems, 
we construct Li$_{1-x}$Fe$_x$OHFeSe models based on the
first principles calculation, where $x$ dependence is given by
using the virtual crystal approximation\cite{Yamakawa2020_eFeSe}. 
The RPA spin Stoner factor $\a_S$ as function of $x$
at $U=1.6$eV and $T=20$meV is shown in Fig. \ref{FeSe-edope2} (b).
Here, $\a_S$ decreases 
when electron-FS disappears at $x\sim x_c\approx 0.05$.
However, with increasing $x$ further, $\a_S$ gradually increases 
with the aid of the non-rigid band deformation
\cite{Yamakawa2020_eFeSe}.
Thus, electron correlation increases for $x\gtrsim0.1$.
Based on this model,
we perform the DW equation analysis, 
and the obtained eigenvalues are shown in Fig. \ref{FeSe-edope2} (b).
For $x\sim0$, the eigenvalue of the nematic ($\q={\bm 0}$) state is the largest.
However, the smectic ($\q\approx(\pi/4,0)$) eigenvalue becomes the largest 
when electron-FS disappears ($x>x_c$).
The obtained strong smectic DW instability is consistent 
with the recent STM/STS measurement
\cite{smectic-FeSe-STM}.


Next, we explain that the smectic fluctuations induce
sizable pairing interaction. 
Figure \ref{FeSe-edope2} (c) shows the obtained DW equation eigenvalue 
$\lambda_\q$ and the coupling constant ${\bar I}_\q$ in Eq. (\ref{eqn:Ibar}),
where $\q\approx \q_{\rm FS}$.
Both quantities drastically increase under moderate spin fluctuations:
When $\a_S\approx 0.9$, $\lambda_\q$ is almost unity,
and ${\bar I}_\q$ is much larger than $U \approx 2$eV.
Since the pairing attraction is proportional to 
${\bar I}_\q/(1-\lambda_\q)$,
moderate smectic fluctuations cause sizable attraction for $\a_S\gtrsim0.85$.
We stress that the irreducible four-point vertex $I_\q(k,k')$
is large only when four momenta $\k,\k',\k+\q,\k'+\q$ are near the FS,
and the outer Matsubara frequencies are small.
For comparison, we show the local four-point vertex 
$I_\q^{\rm local}(k,k') \equiv \frac{1}{N^2}\sum_{\k,\k'} I_\q(k,\k')$
with $k=(\k,\pi T)$ in Fig. \ref{FeSe-edope2} (c).
We see that $I_\q^{\rm local}(k,k')$ takes a negative value,
and its magnitude is much smaller than ${\bar I}_\q$.
(Note that $I_\q^{\rm local}(k,k')\approx -2U$ 
in the mean-field approximation.)
Thus, the AL-VC due to moderate spin fluctuations is important in FeSe.

We study the SC state in e-doped FeSe
by means of the beyond-ME gap equation 2 given in Fig. \ref{fig:diagram} (d).
In this method, 
the charge-channel interaction is expressed with good accuracy 
in Eq. (\ref{eqn:Ibar}),
and the total pairing interaction is given as
$\displaystyle V^{\rm SC2}(k,k')=V^{{\rm SC2},c}(k,k')- \frac32 (U^s)^2\chi^s(k-k') +V^{(1)}$.
Figure \ref{FeSe-edope2} (d)
shows the obtained $s_{++}$-wave gap function on the FS
mediated by the smectic fluctuations.
The FS is illustrated in the inset.
The obtained gap anisotropy is well consistent with ARPES result
in Fig. \ref{fig:FeSe-fig3} (b).
This result is also consistent with the recent QPI measurement reported in Ref. \cite{Fan2015_FeSe_STM}, 
while another QPI study indicates sign reversal between inner- and outer-electron FSs 
\cite{Du2016_LiFeOHFeSe_STM, Gu2018_LiFeOHFeSe_QPI}. 
In this study, $\lambda_{\rm SC}$ for the $d$-wave state is strongly suppressed by
the SOI-induced pair breaking.

\color{black}
Finally, we briefly discuss the important role of 
the $\q\approx{\bm0}$ $A_{1g}$ interfacial $e$-ph interaction on $T_{\rm c}$
\cite{Lee2015_eFeSe_SOI, Rademaker2016_eFeSe_Phonon,Wang2012_FeSeSTO_1st,Lee2014_FeSeSTO_ARPES}.
The nematic/smectic fluctuations have $d$-wave form factor,
while the interfacial phonon mode has $A_{1g}$ form factor.
Because they are orthogonal,
the correlation-driven nematic/smectic fluctuations
are unchanged by the interfacial phonons.
Therefore, both nematic/smectic fluctuations
and $\q\approx{\bm0}$ interfacial phonons
contribute to the $s$-wave pairing just additively.
Then, the eigenvalue of gap equation is simply given as
$\lambda_{\rm SC}^{\rm tot}= \lambda_{\rm SC}+{\bar \lambda}_{e-{\rm ph}}$,
where ${\bar \lambda}_{e-{\rm ph}}$ is the $e$-ph coupling constant averaged on the FS.
More quantitative theoretical study on
the high-$T_{\rm c}$ mechanism of monolayer FeSe/STO
would be desired.
\color{black}

\section{Unconventional superconductivity and multipole orders in heavy fermions}
\label{sec:Tazai}
\subsection{Introduction of heavy fermions}
\label{sec:Tazai1}


In this section, we discuss interesting exotic metallic states 
in Ce-, Yb-, U-based compounds.
They are called the ``heavy fermions'' because effective mass
of conduction electrons $m^*$ is largely magnified by the strong 
electron correlation on $f$-electron ions.
The mass enhancement factor 
$z^{-1}= 1-{\rm Re}\d\Sigma(\e)/\d\e|_{\e=0} \ (\approx m^*/m_0)$
reaches $O(100)$ in typical heavy fermions at low temperatures.
Here, $\Sigma(\e)$ is the $f$-electron self-energy.
In addition, strong $f$-electron correlation gives rise to 
exotic quantum phase transitions and the SC states.
For example, nodal $d$-wave superconductivity is realized in 
CeMIn$_5$ (M=Co,Rh,Ir), and interesting spin triplet states 
appear in several U-based heavy fermions. 
Surprisingly, fully-gapped $s$-wave superconductivity appears
in CeCu$_2$Si$_2$,
irrespective of the presence of strong magnetic fluctuations.
In fact, the nodeless SC state has been confirmed  
by the specific heat, thermal conductivity,
and penetration depth measurements \cite{Kit1-122,Kit2-122,Yama-122,Steglich},
as explained in Sect. \ref{sec:Kontani1-4}.
Furthermore, $T_{\rm c}$ is quite robust against randomness,
comparable to other $s$-wave superconductors as shown in 
Fig. \ref{fig:CeCu2Si2-impurity}.
Thus, many-body electronic states in heavy fermions 
are not fully understood, and there are many unsolved problems.

A remarkable characteristic of heavy fermions 
is the strong spin-orbit interaction (SOI) of $f$-electrons.
The strong SOI in heavy fermions induces drastic change 
in the electronic states that cannot be treated perturbatively,
in contrast to usual 3$d$-electron systems.
In addition, the crystalline electric field (CEF)
is small because of the small radius of $f$-orbitals.
For these reasons, higher-order multipole degrees of freedom,
such as the quadrupole, octupole, and hexadecapole states,
become active in various heavy fermions. 
Due to the combination of the strong correlation and the
multipole degrees of freedom,
rich electronic states are realized in heavy fermion systems.
This issue will be discussed for Ce-compounds
in Sect. \ref{sec:Tazai2}.

In heavy fermion metals,
$f$-electrons are localized on $f$-ions when the temperature is 
higher than the Kondo temperature $T_{\rm K}$.
($T_{\rm K}$ describes the renormalized Fermi energy,
and it is proportional to $z$.)
Below, $T_{\rm K}$, on the other hand,
$f$-electrons start to hybridize with conduction electrons.
Due to this $f$-$c$ hybridization process,
$f$-electrons contribute to the formation of the itinerant heavy quasiparticles,
and therefore the realized FS is ``large'' in volume.
This ``itinerant picture'' is plausible for $f^1$ (Ce-ion)
and $f^{13}$ (Yb-ion) compounds based on the Fermi liquid theory,
by considering the adiabatic continuity from $U=0$
\cite{Yamada-text,Ueda-RMP}.
The correlation effects in heavy fermions have been intensively 
studied by applying various theoretical methods.
\cite{Ueda-RMP}.
Recently, the dynamical-mean-field-theory (DMFT)
has been successfully applied to various heavy fermions.
In this article,
we study the roles of the quantum interference mechanism 
in heavy fermion systems based on the itinerant picture, 
by focusing on the significance of the ``nonlocality of electron correlations''.
The present mechanism gives rise to various multipole orders, 
which are classified as the unconventional orders 
that are traceless in the $f$-orbital basis.
The strong quantum fluctuations of multipole orders 
mediate exotic  superconductivity.

\subsection{Multipole degrees of freedom by spin-orbit coupling}
\label{sec:Tazai2}
In the previous sections, we discussed the various charge-channel unconventional orders in $3d$-electron systems. They
originate from the strong coupling between spin- and charge-channel 
fluctuations due to the many-body quantum interference. 
However, it is well known that the spin and orbital degrees of freedom are coupled via the
spin-orbital interaction (SOI), independently of the Coulomb interaction $U$. 
This fact indicates that more exotic
unconventional orders emerge in $5d$- and $f$-electron systems with strong atomic SOI. Motivated 
by this naive expectation, here we analyze the Ce-based ($4f^{1}$) heavy fermion systems, by taking the strong coupling
limit of the SOI. We discuss the emergence of the exotic multipole order and the multipole-fluctuation-mediated superconductivity.

Heavy fermion systems are interesting platform for exotic phenomena 
owing to the combination of strong SOI and strong Coulomb repulsion 
in $4f$ and $5f$ electrons.
Due to the strong SOI ($\Delta E \sim 1000$ [K]), 
total angular momentum $J=L+S$
becomes a good quantum number for describing electric ground states.
Then, the rank $k$ of  multipole degrees of freedoms described by
the linear combinations of the $k$-th powers of
$J_{\mu(=x,y,z)}$ comes to be active, which cause
unconventional superconductivity and multipole order so called ''hidden-order''. 
For instance, quadrupole (rank 2) 
and octupole (rank 3)'phase transitions were reported in CeB$_6$ \cite{Shiina1,Shiina4}.
Also, hexadecapole (rank 4) and dotriacontapole (rank 5) ordering
were predicted in PrRu$_4$P$_{12}$ \cite{Takimoto} and URu$_2$Si$_2$ 
\cite{Haule,Oka,Ikeda-Uru2}.
A rank-$k$ multipole order is described by the linear combination of the
spherical tensor operator
$J^{(k)}_{q} (q=-k\sim k)$\cite{Shiina1,Shiina4,Springer} :
 \begin{eqnarray}
 [J_{\pm},J^{(k)}_{q}]=\sqrt{(k\mp q)(k\pm q+1)}J^{(k)}_{q\pm1}, \hspace{15pt}
J_{k}^{(k)}=(-1)^{k}\sqrt{\frac{(2k-1)!!}{(2k)!!}}(J_{+})^{k}.
 \label{eqn:J3}
\end{eqnarray}

Here, we consider the $4f^{1}$ states in Ce ion. Due to the strong SOI, $4f$ (14 folded) states are split into $J=3/2$ (8 folded) and $J=5/2$ (6 folded) states.
In addition,  considering finite CEF,
$J=5/2$ states split into three Kramers doublets $|f_{l}\Uparrow \rangle,  |f_{l}\Downarrow \rangle $ $(l=1\sim 3)$. When two or three Kramers pairs
are nearly degenerated and hybridize with conduction electrons,
then higher-order multipoles ($k\geq 2$) become active. 
Typical examples of this situation are CeB$_6$ and CeCu$_2$Si$_2$. Active multipoles of the effective 2-orbital systems for 
CeB$_6$ and CeCu$_2$Si$_2$ are given in the Table \ref{tab:multipole} and \ref{tab:multipole2}, respectively.
Even (odd)-rank operators correspond to electric (magnetic) channel
in the presence of space inversion symmetry.
\begin{table}[htb]
\begin{minipage}{.45\textwidth}
\begin{center}
  \begin{tabular}{|c|c|c|} \hline
    \hspace{3mm}IR ($\Gamma$) \hspace{3mm} & rank (k) & multipole (Q)  \\ \hhline{|=|=|=|}
     & $0$  & $\hat{1}$ \\ 
     $A_{1}^{+}$& $2$  & $\hat{O}_{20}$  \\ 
      & $4$  & $\hat{H}_{0}$ \\ \hline
    $A_{2}^{+}$ & $4$  & $\hat{H}_{z}$ \\  \hline
    $E^{+}$ & $2$  & $\hat{O}_{yz(zx)}$   \\  \hhline{|=|=|=|}
    $A_{1}^{-}$ & $5$  & $\hat{D}_{4}$ \\   \hline
     & $1$  & $\hat{J}_{z}$  \\  
    $A_{2}^{-}$  & $3$ & $\hat{T}_{z}$\\ 
      & $5$  & $\hat{D}_{z}$ \\ \hline
       & $1$  & $\hat{J}_{x(y)}$ \\ 
      $E^{-}$ & $3$  & $\hat{T}_{x(y)}$ \\  
      & $5$  & $\hat{D}_{x(y)}$  \\ \hline
    \end{tabular}
\end{center}
    \caption{16-type active multipoles for the CeCu$_2$Si$_2$. }
    \label{tab:multipole}
\end{minipage}
\hfill
\begin{minipage}{.45\textwidth}
\begin{center}
  \begin{tabular}{|c|c|c|} \hline
    \hspace{3mm}IR ($\Gamma$) \hspace{3mm} & rank (k) & multipole (Q) 
     \\ \hhline{|=|=|=|}
     $\Gamma_{1}^{+}$ & $0$  & $\hat{1}$ 
  \\ \cline{2-3}
                             & $2$  & $\hat{O}_{20}$ 
 \\ \hline
     $\Gamma_{3}^{+}$ & $2$  & $\hat{O}_{22}$ 
 \\ \hline
     $\Gamma_{4}^{+}$ & $2$  & $\hat{O}_{xy}$ 
\\   \hline
     $\Gamma_{5}^{+}$ & $2$  & $\hat{O}_{yz(zx)}$ 
\\  \hhline{|=|=|=|}
     $\Gamma_{2}^{-}$ & $1$  & $\hat{J}_{z}$
  \\  \cline{2-3}
                             & $3$  & $\hat{T}_{z\a}$ 
 \\   \hline
     $\Gamma_{3}^{-}$ & $3$  & $\hat{T}_{xyz}$
 \\  \hline
     $\Gamma_{4}^{-}$ & $3$  & $\hat{T}_{z\b}$ 
\\ \hline
     $\Gamma_{5}^{-}$ & $1$  & $\hat{J}_{x(y)}$
\\  \cline{2-3}
                             & $3$  & $\hat{T}_{x\a(y\a)}$  \\  \cline{2-3}
                   & $3$  & $\hat{T}_{x\b(y\b)}$ 
  \\ \hline
    \end{tabular}
\end{center}
\caption{16-type active multipoles for the CeB$_6$.}
    \label{tab:multipole2}
\end{minipage}

\end{table}
The characteristic feature of the $f$-electrons change across the Kondo temperature 
$T_K$. Especially at $T<T_K$, 
they are described as itinerant quasi particles with large effective mass $m^{*} \gg 100m_{e}$ based on the Fermi liquid theory.
Here, we introduce the Periodic Anderson Model (PAM). The kinetic term is given by
\begin{eqnarray}
\hat{H}_{0}=\sum_{\k\sigma}\epsilon_{\k}c^{\dagger}_{\k\sigma}c_{\k\sigma}+\sum_{\k l\Sigma}E_{l\k}f^{\dagger}_{\k l\Sigma}f_{\k l\Sigma} 
+\sum_{\k l\sigma\Sigma}\left(V^{*}_{\k l\sigma\Sigma}f^{\dagger}_{\k l\Sigma}c_{\k \sigma}
+V_{\k l\sigma\Sigma}c^{\dagger}_{\k\sigma}f_{\k l\Sigma}\right),
\label{eqn:hamiltonian}
\end{eqnarray}
where $f^{\dagger}_{\k l \Sigma}$ is a creation operator for 
$f$-electron with $\k$, orbital $l=1,2$, pseudo-spin $\Sigma (=\Uparrow,\Downarrow)$,
and energy $E_{l\k}$.
$\sigma(=\uparrow,\downarrow)$ is real spin of conductive electron.
In general, we can set $V^{*}_{\k l\uparrow\Downarrow}=V^{*}_{\k l\downarrow\Uparrow}=0$ by choosing appropriate Kramers pair.
In this case, however, the definition of  Kramers pair depends on $\k$,
which gives serious problem in the theoretical treatment.
This difficulty is removed by considering two-dimensional systems
as shown in Refs. \cite{Tazai-HF}, as we analyze in later subsections.
In this case, the relations $\sigma=\Sigma$ and 
$V^{*}_{\k l\sigma\Sigma}=V^{*}_{\k l\sigma\Sigma}$ are satisfied.

\subsection{$S$-wave superconductivity in $\rm{CeCu}_{2}\rm{Si}_{2}$}
\label{sec:Tazai3}

CeCu$_2$Si$_2$ is the first discovered superconductor in heavy fermion systems \cite{Ste-122,Yuan-122,Peid}.
Superconducting (SC) transition occurs at $T_{\rm c}\approx0.6$ [K]
near the AFM quantum critical point at ambient pressure \cite{Ishida},
while it goes up to $1.5$ [K] around $P_c\approx4.5$ [GPa].
Historically, it was believed as a typical nodal $d$-wave superconductor
in accordance with previous NMR and specific heat measurements. 
Theoretically, $d$-wave SC  was considered as a reasonable scenario since it can avoid the energy loss due to the strong on-site Coulomb repulsion among $f$-electrons.

\begin{figure}[htb]
\centering
\includegraphics[width=.8\linewidth]{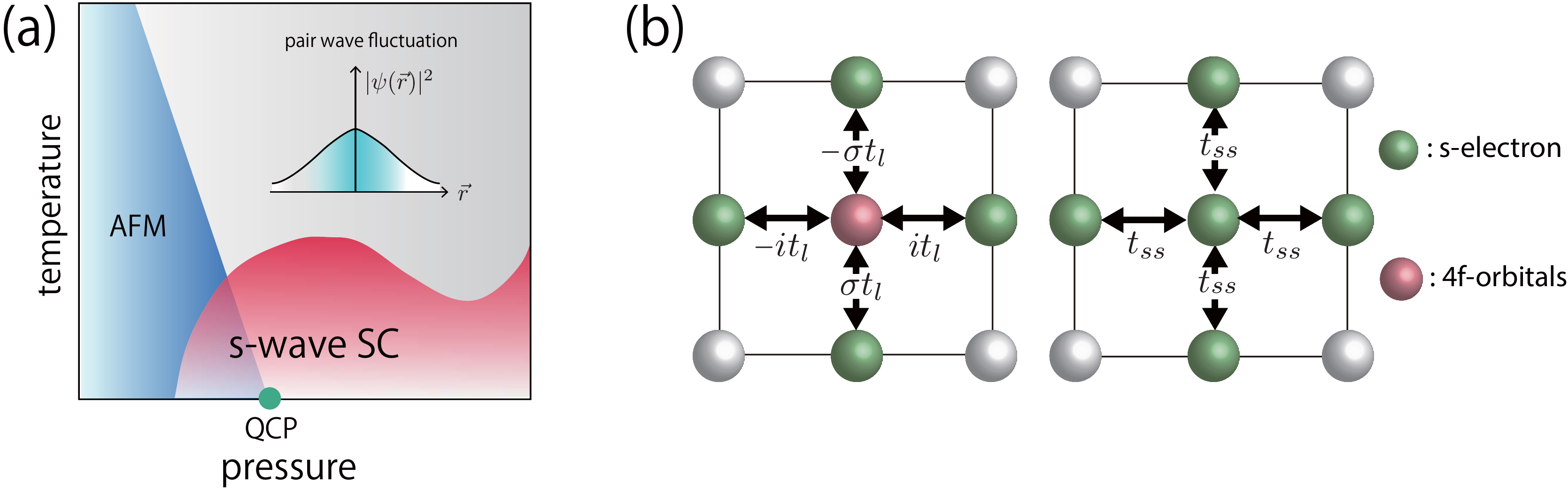}
\caption{(a) Schematic phase diagram of $\rm{CeCu_{2}Si_{2}}$.
(b) The square lattice model with nearest neighbor hoppings $s$-$s$ and $s$-$f$.
$\sigma=1(-1)$ is pseudo-spin up (down) and $ t_{l} \equiv (-1)^{l-1}t_{sf}^{l}$.
Cited from Ref. \cite{Tazai-HF}.
}
\label{fig:band}
\end{figure}

However, this historical belief was broken by
recent experiments based on specific heat, thermal conductivity and penetration depth measurements \cite{Kit1-122,Kit2-122,Yama-122,Steglich}.
Surprisingly, they revealed that fully gapped $s$-wave SC is realized against the 
strong Coulomb repulsion in CeCu$_2$Si$_2$  as shown in the phase diagram of Fig. \ref{fig:band} (a).
Furthermore, $T_{\rm c}$ is quite robust against randomness,
comparable to other $s$-wave superconductor as shown in Fig.\ref{fig:CeCu2Si2-impurity} in Sect.\ref{sec:Kontani1}.
Therefore, $s$-wave SC state
without any sign-reversal emerges in CeCu$_2$Si$_2$
\cite{Yama-122}, which brings a paradigm shift in
the long history of study of heavy fermion superconductor.
To understand the origin of the $s$-wave SC states require us to 
face the following fundamental issue: Why attractive pairing interaction
overcomes strong Coulomb repulsion in heavy fermion system?
To attack this issue, we have to go beyond the mean-field approximation.

The superconductivity in $\rm{CeCu_{2}Si_{2}}$ originates from
$4f^{1}$ ($L=3,S=1/2$) electrons on Ce-ion. 
The $J=5/2$ ground states split into 3 Kramers doublets due to 
the CEF and the following two Kramers doublets give large density of states around the Fermi energy.
They are expressed in the $J_{z}$ basis as
\begin{eqnarray}
&&|f_{1}\Downarrow \rangle=a|+ \frac{5}{2}\rangle +b|-\frac{3}{2}\rangle, \,\,\,\,
|f_{1}\Uparrow \rangle=a|- \frac{5}{2}\rangle +b|+\frac{3}{2}\rangle ,\nonumber \\
&&|f_{2}\Uparrow \rangle=-a|+ \frac{3}{2}\rangle+b|- \frac{5}{2}\rangle, \,\,\,\,
|f_{2}\Downarrow \rangle=-a|- \frac{3}{2}\rangle +b|+ \frac{5}{2}\rangle,
\label{ground}
\end{eqnarray}
where $\Downarrow (\Uparrow)$ represents pseudo spin up (down).
$a$ and $b(=\sqrt{1-a^2})$ are coefficient parameter determined by CEF.
The present multi orbital model is consistent with  the LDA+DMFT study \cite{LDADMFT_multiporbital} at ambient pressure.

\begin{figure}[t]\centering
\includegraphics[width=.99\linewidth]{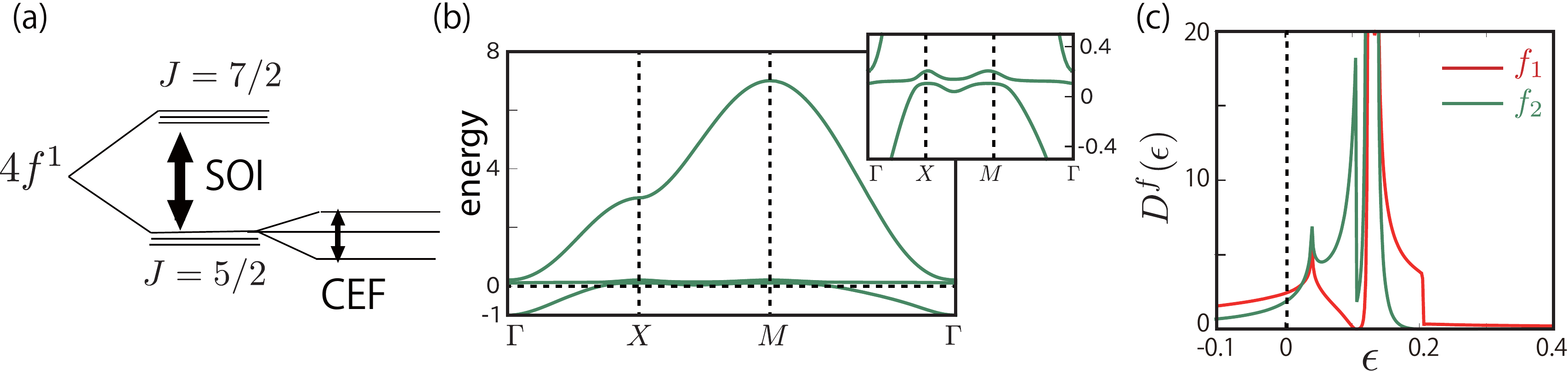}
\caption{(a) Splitting of the $4f^1$ states due to spin-orbit coupling (SOI) and crystalline electric field.
(b) Band dispersion along high-symmetry line.
(c) Partial density of states of $f_{l}$-electrons. 
The red (green) line corresponds to $f_{1}$($f_{2}$)-orbital.
}
\label{fig:band1}
\end{figure}

Here, we set the energy dispersion of conductive $s$-electron  as
$\epsilon_{\k}=2t_{ss}(\cos k_{x}+\cos k_{y})+\epsilon_{0}$
with $t_{ss}=-1$.
$V_{\k l \Sigma \sigma}$ is the 
hybridization term between $f$- and $s$-electron.
In this study, we consider a two-dimensional square lattice model in Fig.\ref{fig:band} (b). 
Both $f$- and $s$-orbital are on Ce-ion.
$V_{\k l\Sigma \sigma}$  is calculated
by using Slater-Koster table \cite{Slater-U}.
To simplify the analysis, we put $a=1,b=0$ and obtain
\begin{eqnarray}
V_{\k l \Sigma \sigma}=\sigma(-1)^{l} t_{sf}^{l} (\sin k_{y}-i\sigma \sin k_{x})\delta_{\sigma,\Sigma},
\label{eqn:hopping}
\end{eqnarray}
where $\delta_{\sigma, \Sigma}$ is Kronecker delta function.
The imaginary part comes from strong SOI considered in the ground states of Eq.(\ref{ground}).
We put the chemical potential $\mu=-5.52\times 10^{-3}$ and temperature $T=0.02$. 
The $f$- and $s$-electron numbers are $n_{f}=0.9$ and $n_{s}=0.3$, respectively.
In this case, $t_{sf}^{f_{1}}/t_{sf}^{f_{2}}\simeq 7/3$ is obtained, which means that the 2-orbitals have different itineracy.
In addition, we set $E_{1\k}=0.2$ and $E_{2\k}=0.1$ by considering finite CEF splitting described in Fig.\ref{fig:band1}(a).
We show the obtained band structure and density of states $D^{f_{l}}(\epsilon)$ in Fig.\ref{fig:band1} (b) and (c), respectively.
The relation $D^{f_{1}}(0)\simeq D^{f_{2}}(0)$ is satisfied.
$|t_{ss}|$ is of order $1$ [eV] since $W_{D}\sim 10$ [eV] holds in CeCu$_2$Si$_2$ \cite{Ikeda-122}. 

In addition, we introduce on-site Coulomb interaction among $f$-electrons:
\begin{eqnarray}
\hat{\rm{H}}_{U}=u\cdot\frac{1}{4}\sum_{i}\sum_{l l' m m'}\sum_{\sigma \sigma' \rho \rho'}
{\bar U}_{l l';m m'}^{\sigma \sigma';\rho \rho'}
f^{\dagger}_{il\sigma}
f_{i l'\sigma'}
f_{i m\rho} 
f^{\dagger}_{i m'\rho'} \label{eqn:U0}
\end{eqnarray}
where $i$ is site index.
$\hat{\bar U}$ is the interaction matrix normalized as ${\bar U}^{\Uparrow \Downarrow; \Uparrow \Downarrow}_{11;11}=1$.
Note that $\hat{\bar U}$ is the antisymmetrized 4-point vertex function, which is obtained by Slater-Condon parameter
$F^{p}$ \cite{Slater-U}.
We put $(F^{0},F^{2},F^{4},F^{6})\propto (5.3,9.09,6.927,4.756)$ in unit eV by referring Ref.\cite{F0F2F4F6}.

In the present model, there are 
16-type active multipoles labeled by $Q \equiv(\Gamma, \phi)$, 
where $\Gamma$ is the index of the irreducible representation (IR)
 $(\Gamma=A^{+}_{1},A^{+}_{2},E^{+},A^{-}_{1},A^{-}_{2},E^{-})$ and $\phi$ 
 is the index of independent multipole operator $(\phi=1\sim  N_{\Gamma})$. 
They are expressed as monopole (rank 0), dipole (rank 1), quadrupole (rank 2), octupole (rank 3), hexadecapole (rank 4) and dotriacontapole (rank 5) as shown in TABLE {\ref{tab:multipole} \cite{Ikeda-Uru2}.
The on-site Coulomb interaction $U$ is decomposed into the multipole channel.
\begin{eqnarray}
\hat{\rm{H}}_{U}=u\cdot \frac{1}{4}\sum_{LL'MM'}\sum_{QQ'}\bar{U}^{QQ'}(f^{\dagger}_{i L} O^{Q}_{L,L'} f_{i L'})(f^{\dagger}_{iM}O^{Q'}_{M,M'}f_{iM'})
\end{eqnarray}
where $O^{Q}_{L,M}$ is $4 \times 4$ matrix expression of multipole operator $Q$
with $L=(l,\sigma)$ and $M=(m,\rho)$. 
We verified that the magnetic Coulomb interaction ${\bar U}^{QQ} (Q=J,T,D)$ is larger than electric ones $(Q=C,O,H)$ \cite{Tazai-HF}.
Thus, magnetic fluctuations always dominate over the electric fluctuations
within the RPA. 
The particle-hole susceptibility in the multipole channel is written as
\begin{eqnarray}
\chi^{Q,Q'}(q)&=&\int^{\beta}_{0} d\tau \left\langle {\cal O}^{Q}({\bm q},\tau){\cal O}^{Q'}({\bm -\q},\tau)\right\rangle e^{i\w_j\tau}, \\
{\cal O}^{Q}(\q, \tau)&=&\sum_{L,M,\k} O^{Q}_{L,M} 
f^{\dagger}_{\k M}(\tau) f_{\k +\q L}(\tau),
\label{eq:suscep}
\end{eqnarray}
A compact expression of $O^{Q}_{L,M}$ using pseudo-spin Pauli matrices is given in Refs. \cite{Tazai-HF}.
Note that $\chi^{(\Gamma,\phi),(\Gamma',\phi')}(q)=0$ holds for $\Gamma \neq \Gamma'$, which is a great merit in analysis.

Considering the multipole fluctuations, 
we solve the linearized SC gap equation for spin singlet pairing given by
\begin{eqnarray}
\lambda \Delta(\k,\epsilon_{n})=
-\frac{\pi T}{(2\pi)^2}\sum_{\epsilon_{m}}
\oint \frac{d\k'}{v_{\k'}} 
\frac{\Delta(\k',\epsilon_{m})}{|\epsilon_{m}|} V^{\rm{sing}}(k,k') ,\label{eqn:linear}
\end{eqnarray}
where $\Delta(\k,\epsilon_{n})$ is the gap function on FS and
$v_{\k}$ is Fermi velocity. 
$V^{\rm{sing}}(k,k')$ is the SC pairing interaction with $U$-VC
expressed in Fig. \ref{fig:diagram} (c).
Here, we introduce the phonon-mediated electron interaction.
Then, the effective Coulomb interaction matrix is given by
\begin{eqnarray}
\hat{U}^{\rm{eff}}&=&\hat{U}+ {\hat I}_{\rm ph},
\\
{\hat I}_{\rm ph}&=&2g(0) \vec{C}_{A^{+}_{1}} \vec{C}_{A^{+}_{1}}^{\dagger},
 \label{eqn:retph}
\end{eqnarray}
where 
$\hat{U}\equiv u\cdot \hat{\bar U}$,
$g(\omega_{j})\equiv \tilde{g}\frac{\omega_{D}^2}{\omega_{D}^{2}+\omega_{j}^{2}}$ 
and $\tilde{g}=\frac{2\eta^{2}}{\omega_{D}}$. 
Here, $\omega_{D}$ is the frequency of the $A_{1g}$ phonon
induced by oscillation of $c$-axis length \cite{Kontani-RPA},
and $\eta$ is the electron-phonon coupling constant.
$\hat{C}_{A^{+}_{1}}$ is a linear combination of multipole operators in
$A^{+}_{1}$ symmetry.

Figures \ref{fig:SCphase} (a) shows the obtained phase diagram
by solving the gap equation with $U$-VC,  
which magnifies the electric-channel pairing interaction
near the magnetic QCP as we explained in Fig. \ref{fig:diagram} (c)
in Sect. \ref{sec:Onari}.
Fully gapped $s$-wave state without any sign reversal
emerges as plotted in Fig.\ref{fig:SCphase} (b).
Moreover, the region of $s$-wave phase gets wider as the 
magnetic multipole (odd-rank) fluctuations develop.
This counterintuitive result originates from the fact that
the SC pairing attraction due to the electron-phonon interaction is strongly enhanced by magnetic multipole fluctuations, which is realized by the $U$-VC.
This interesting cooperation mechanism 
is illustrated in Fig.\ref{fig:SCphase} (c). 
Then, a quite small $g$ is enough for realizing the $s$-wave SC state.
In fact, $s$-wave state emerges even at $g=0.025$, which
is much smaller than $u=0.31$ at $\alpha_{S}=0.9$.
Therefore, we reveal that
fully gapped $s$-wave SC state is
strongly stabilized by AL-type $U$-VC near the magnetic QCP in the presence of small electron-phonon interaction.
It is noteworthy that phonon-mediated $s$-wave SC states 
in heavy fermion systems have been discussed in Refs.\cite{Razaf,Ohkawa,Nagaoka}
by focusing on the large Gruneisen parameter
($\eta\equiv -d{\rm log}T_K/d{\rm log}\Omega \sim 30-80$) \cite{Razaf}.
Now, this scenario becomes more realistic by considering
AL-type $U$-VC.
\begin{figure}[t]\centering
\includegraphics[width=.99\linewidth]{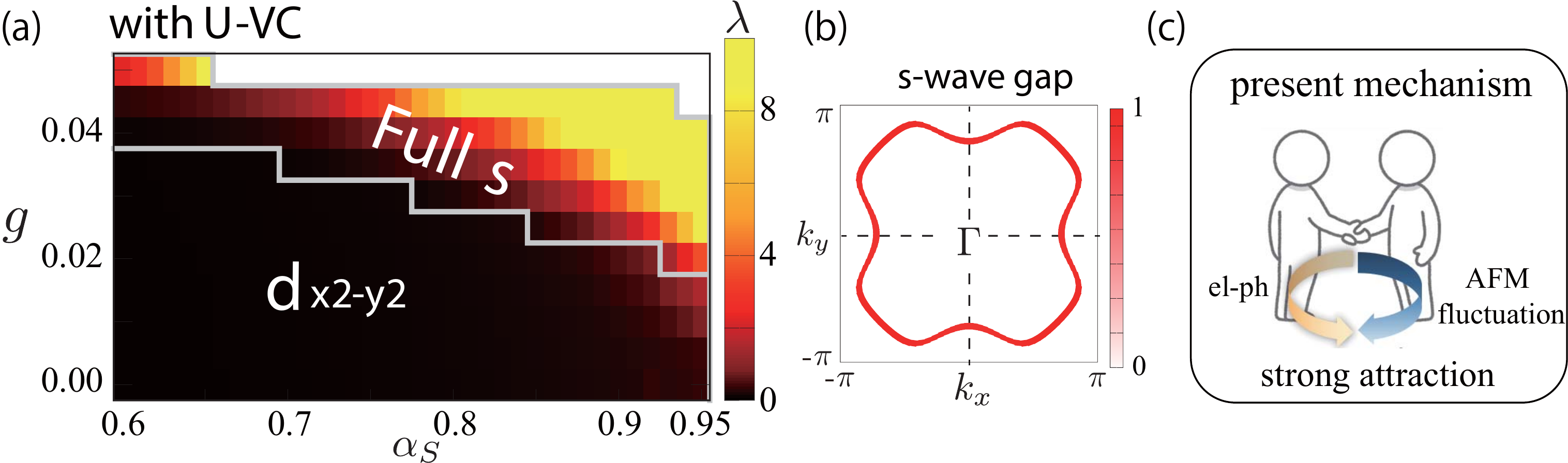}
\caption{(a) Phase diagram in the presence of $U$-VC.
The $s$-wave state emerges due to the significant contribution of $U$-VC. 
The white region corresponds to $\alpha_{C}>1$.
(b) The $s$-wave gap function on the FS.
(c) Present proposed mechanism for realizing the $s$-wave SC.
Cited from Ref. \cite{Tazai-HF}.
}
\label{fig:SCphase}
\end{figure}

In the next stage, we reveal that the $s$-wave SC phase can
appear even in the absence of electron-phonon interaction $(g=0)$.
In this case, the AL-type $\chi$-VC for the charge-channel 
irreducible susceptibility gives rise to the attractive force.
It is schematically shown in Fig.\ref{fig:resultmodel2} (a).
Its analytic expression is 
\begin{eqnarray}
\hat{X}^{\rm{AL}}_{QQ'}\propto \sum_{Q_1\sim Q_4}R ^{Q}_{Q_1,Q_4}R ^{Q'}_{Q_2,Q_3}
\hat{V}^{Q_1,Q_2}\hat{V}^{Q_3,Q_4},
\label{eq:chiVC}
\end{eqnarray}
where $R^{Q}_{Q_1,Q_2}$ is three point vertex function made of 
three multipole operators $(Q,Q_1,Q_2)$ and three Green functions.
Also, $\hat{V}^{Q,Q'}={\hat U}^{Q,Q'}+ \sum_{Q_1,Q_2}{\hat U}^{Q,Q_1}\chi^{Q_1,Q_2}(q){\hat U}^{Q_2,Q'}$ is the RPA interaction,
which is large only for the magnetic channels.
In Fig.\ref{fig:resultmodel2} (b), we show obtained multipole susceptibilities
by considering the $\chi$-VC.
With increasing $u$, all the electric fluctuations 
strongly develop thanks to the AL-type $\chi$-VC.
Thus, large electric susceptibilities originate 
from the interference among magnetic fluctuations.
By considering both the $\chi$-VC and the $U$-VC, 
we obtain the eigenvalue of the SC 
gap equation in Fig.\ref{fig:resultmodel2} (c).
$s$-wave SC state appears at $u>0.55$ reflecting the strong electric fluctuations
due to the $\chi$-VC.
Especially, the obtained $s$-wave state is mainly caused by the 
hexadecapole (rank 4) fluctuations as well as quadrupole and monopole ones.
Thus, we discover the mechanism of
multipole-fluctuation-mediated $s$-wave SC pairing 
even in the absence of the electron-phonon coupling.
This result is consistent with the ``$s$-wave SC phase near the magnetic QCP'' 
in CeCu$_2$Si$_2$.

Here, we discuss important roles of retardation effects.
In Fig.\ref{fig:resultmodel2} (d), we show the energy-dependence 
of the SC pairing interaction, 
which is attractive (positive) at $\omega_{j}=0$,
whereas it becomes to be repulsive for $|\omega_{j}|>0$.
This is a hallmark of the retardation effects 
due to the strong $\omega_{j}$-dependence of the 
electric (even-rank) fluctuation.
The direct Coulomb depairing potential
is reduced as
\begin{eqnarray}
U^* \sim \frac{U}{1+UD(0){\rm ln}(E_F/\w_0)},
\label{eqn:Coulomb_ret}
\end{eqnarray}
where $\w_0$ is the energy cutoff due to the electric fluctuations due to 
the $\chi$-VC.
Since $\w_0\ll E_F$, 
the fully-gapped $s$-wave SC phase is stabilized.
\begin{figure}[t]\centering
\includegraphics[width=.9\linewidth]{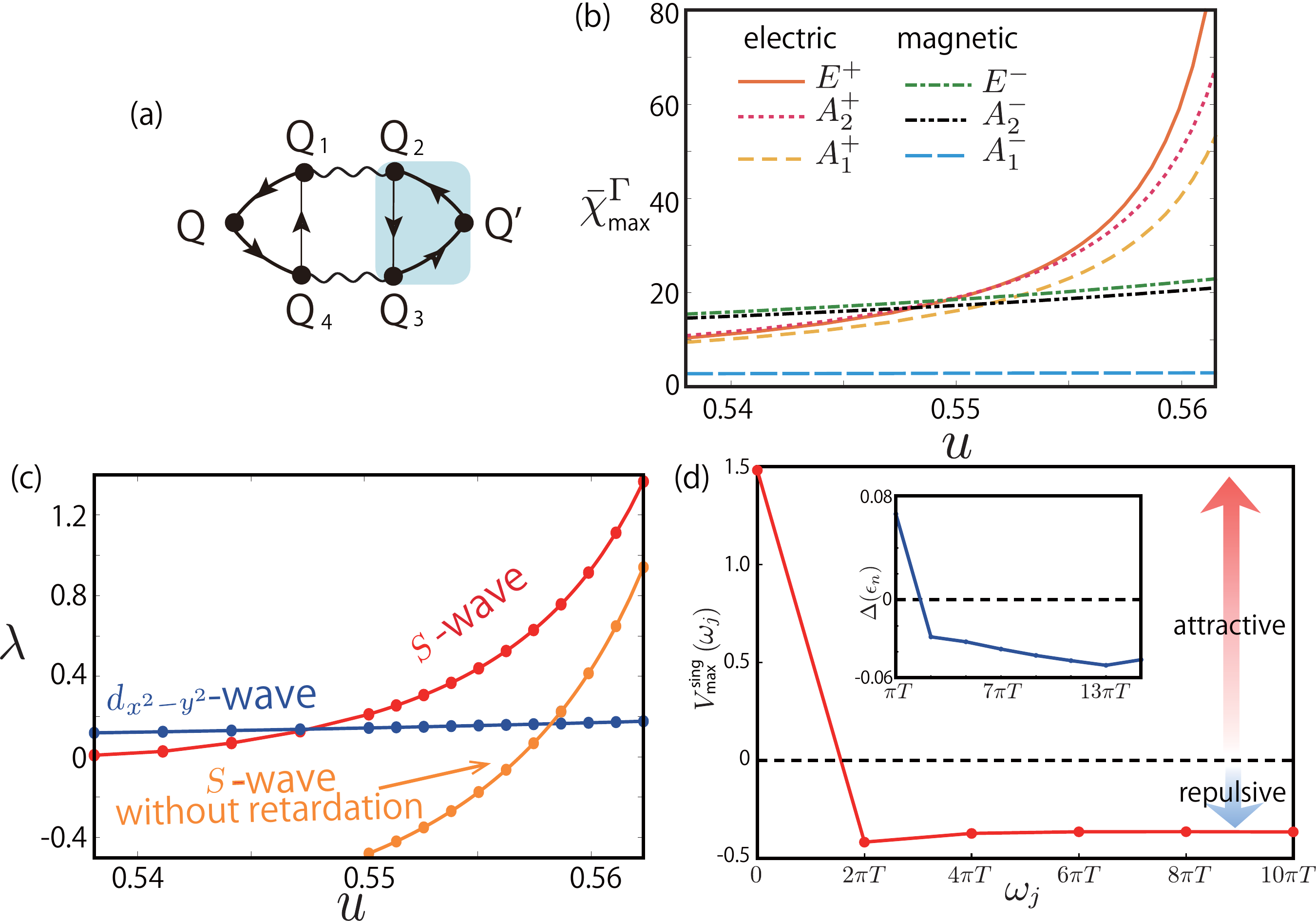}
\caption{
(a) Diagrammatic expression of the AL-VC in $f$-electron systems.
(b) Obtained susceptibility for each IR
\cite{tazaijp2}. 
Electric susceptibilities ($\Gamma=E^{+},A^{+}_{2},A^{+}_{1}$) 
develop due to the AL-type $\chi$-VC. 
(c) Obtained eigenvalue as a function of Coulomb interaction $u$
\cite{tazaijp2}. 
Fully gapped $s$-wave state appears at $u>0.55$.
(d) Obtained pairing interaction $V_{\rm max}^{\rm sing}(\omega_{j})$ 
and gap function $\Delta(\epsilon_{j})$ (inset) 
as the function of Matsubara frequency. 
Strong retardation effect is recognized
\cite{tazaijp2}.
Cited from Ref. \cite{tazaijp2}.
}
\label{fig:resultmodel2}
\end{figure}

In summary,
we proposed a microscopic origin of fully gapped $s$-wave superconductivity
in multi-orbital heavy fermion systems beyond Migdal-Eliashberg formalism,
expressed in Fig. \ref{fig:diagram} (c).
In the present system,
various magnetic multipole fluctuations develop
due to the cooperation between strong SOI and Coulomb interaction.
These multipole fluctuations mediate unconventional 
attractive pairing interaction, and its strength is magnified 
AL-type $U$-VC for the electron-boson coupling.
This mechanism gives rise to the $s$-wave SC state
when the system approaches to the magnetic QCP.
The present mechanism may be responsible for the
fully gapped $s$-wave superconducting state
realized in CeCu$_2$Si$_2$.

In addition, electric multipole fluctuations strongly develop 
due to the many body effects beyond RPA ($=\chi$-VC). 
Owing to the $\chi$-VC, electric multipole fluctuations induce 
the $s$-wave SC state even in the absence of the electron-phonon coupling.

There remain a lot of important unsolved future issues.
For instance, renormalization effect due to self-energy,
which brings the strong mass enhancement, is one of the important issues.
Also, pressure induced second SC dome observed in CeCu$_2$Si$_2$
is uncovered problem \cite{Hol-122}.

\subsection{Quadrupole order in CeB$_6$}
\label{sec:Tazai4}
In heave fermion systems, various multipole orders appear
due to the strong SOI and Coulomb repulsion,
which is absent in transition metal oxides.
Here, we study the microscopic origin of quadrupole ordering in CeB$_6$
It is known that antiferro-quadrupole order appear at $T_Q=3.2$ [K] with $\q=(\pi,\pi,\pi)$
and magnetic dipole order appears at $T_N=2.4$ [K] \cite{Kasuya,Inosov-review}.
Moreover, antiferro-octupole order is induced under the small magnetic field
\cite{Shiina1,Shiina4}. 
Until now, CeB$_6$ has been studied 
intensively mainly based on localized $f$-electron picture 
\cite{Shiina1,Shiina4,Hanzawa}.
However, recent ARPES \cite{ARPES1,ARPES2} and inelastic neutron scattering 
\cite{Inosov1,Inosov2} for CeB$_{6}$,
as well as dHvA for Ce$_x$La$_{1-x}$B$_{6}$ \cite{Endo}, uncovered the itinerant nature of
the $f$-electron system above $T\sim T_Q$. 
These findings indicate that itinerant picture provides a reasonable starting point
to study the multipole physics of CeB$_{6}$.
Thus, we study the itinerant $f$-electron periodic Anderson model based on Fermi liquid theory.
Up to now, Fermi Liquid approach has been succeeded in heavy fermion materials, 
such as CeB$_{6}$ \cite{Thal}, URu$_2$Si$_2$  \cite{Ikeda-Uru2}, and CeCu$_{2}$Si$_{2}$ \cite{Tazai-HF}. 
Since large Coulomb interaction is renormalized to $\sim zU$,
Fermi liquid theory is applicable for heavy fermion systems with $z \ll 1 $.

First, we introduce $J=5/2$ PAM describing CeB$_6$
with $\Gamma_{8}$ quartet \cite{Shiina1} as follows
\begin{eqnarray}
|f_{1}\Downarrow (\Uparrow) \rangle &=&\sqrt{\frac{5}{6}}|+(-)\frac{5}{2}\rangle+
\sqrt{\frac{1}{6}}|-(+)\frac{3}{2}\rangle, \nonumber \\
|f_{2}\Downarrow (\Uparrow) \rangle &=&|+(-)\frac{1}{2}\rangle,  
\label{eqn:wavefunc}
\end{eqnarray}
where $\Downarrow (\Uparrow) $ is the pseudo-spin up (down).
Using the Slate-Koster method \cite{Takegahara}, 
$V_{\k l\Sigma \sigma}$ is given as
\begin{eqnarray}
V_{\k l\Sigma \sigma}=-\sigma t_{sf}(\sin k_{y} +(-1)^{l}\sigma i\sin k_{x}) 
\delta_{\sigma, \Sigma}. \label{eqn:hybri}
\end{eqnarray}
Hereafter, we set $2|t_{ss}^{1}|=1$ as energy unit,
and put $t_{sf}=\sqrt{{18}/{14}}\times(0.78)$, $E_1=-2.0$, $T=0.01$, and $\mu=-2.45$.
Then, $f(s)$-electron number is $n_{f}=0.58$ ($n_{s}=0.69$).
We comment that $n_{f}$ increases if we put the level of $E_{l}$ lower under the condition
$n_{f}+n_{c}=$const. By this procedure, our main results will not change since the shape of the FS 
is essentially unchanged.
Figure \ref{fig:modelceb6} (a) shows the band structure of the present PAM.
The lowest band crosses the Fermi level ($\epsilon=0$).
Since $W_{D}\sim 5$eV \cite{ARPES1,ARPES2,LDA1,LDA2} in CeB$_{6}$,
$2|t_{ss}^{1}|$ corresponds to $\sim0.5$eV.  
The bandwidth of itinerant $f$-electron is 
$W_{D}^{qp} \sim |V_{\k l\Sigma \sigma}| \sim 1$.
The FSs shown in Fig.\ref{fig:modelceb6} (b) 
are composed of large ellipsoid electron pockets around X,Y points, 
consistently with recent ARPES studies
\cite{ARPES1,ARPES2}.
We also consider the Coulomb interaction introduced
in Eq.(\ref{eqn:U0}).
The maximum element of $\hat{U}$ of Eq. (\ref{eqn:U0}) is set to unity. 
In the $\Gamma_{8}$ quartet model, there are 
16-type active multipole operators up to rank 3; monopole, 
dipole (rank 1), quadrupole (rank 2), octupole (rank 3)
as summarized in TABLE \ref{tab:multipole2}.
\begin{figure}[t]\centering
\includegraphics[width=.7\linewidth]{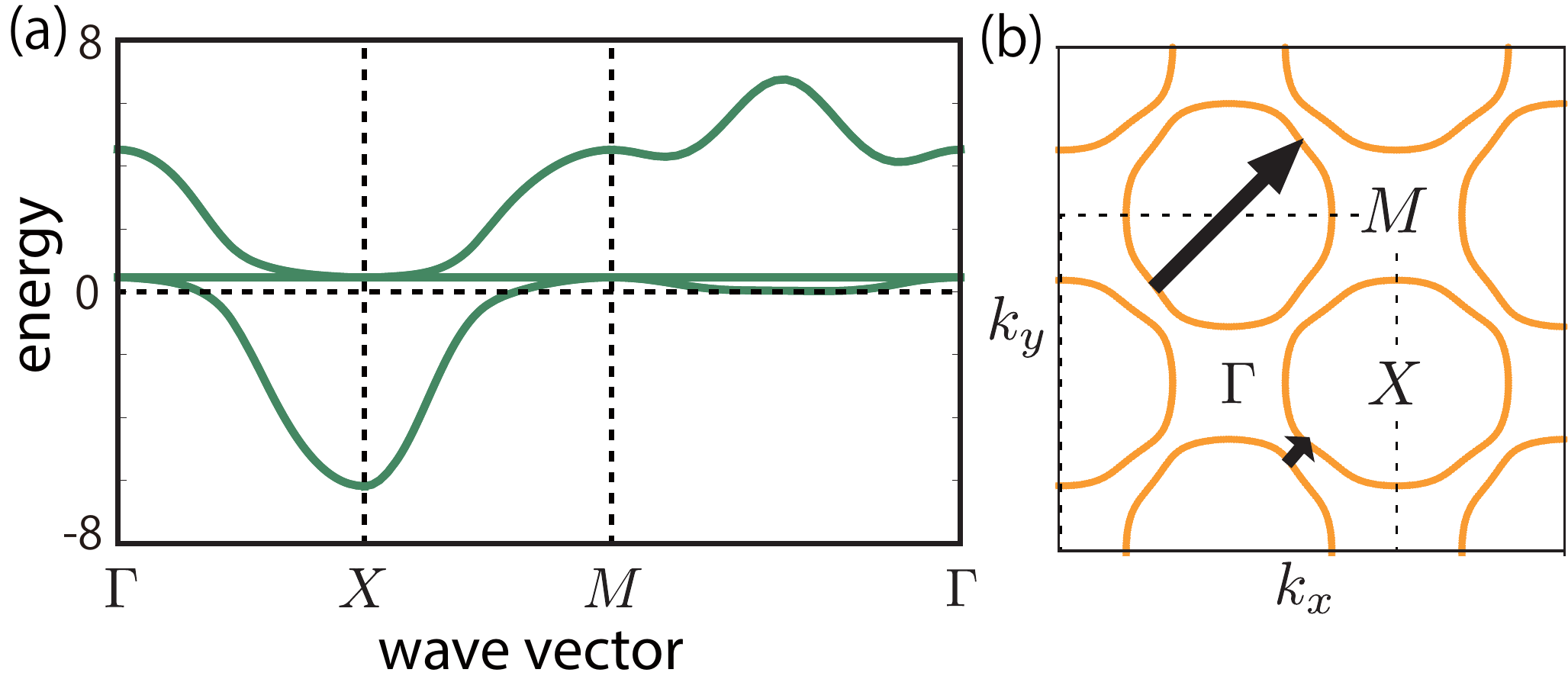}
\caption{(a) Band dispersion and (b) FSs of the present 
periodic Anderson model for CeB$_6$ studied in Ref. \cite{tazai-last}.
Black vector represents major nesting vectors.
}
\label{fig:modelceb6}
\end{figure}

First, we explain that the quadrupole phase cannot be explained within the RPA. 
Within the RPA, in $f$-electron systems, 
odd-rank (=magnetic) multipole fluctuations are enlarged 
by the Coulomb interaction,
while even-rank (=electric) ones remain small 
\cite{Thal,Ikeda-Uru2,Tazai-HF}.
This RPA result is naturally understood by considering
the multipole-dependence of the Coulomb interaction $\bar{U}^{Q}$.
As shown in TABLE \ref{tab:multipole3},  
on-site quadrupole ($O_{xy}$) interaction 
is about 60\% of dipole ($J$) and octupole ($T$) one.
Thus, only magnetic multipole fluctuations develop in the RPA.
In particular, both ferro- and antiferro-magnetic multipole fluctuations 
are induced around nesting vector of the FS, which 
is consistent with the recent neutron experiments \cite{Inosov1,Inosov2}.

Recently, important role of the AL-VC on the quantum phase transition
has been revealed in various $d$-electron systems. 
For example, AL-VC works a trigger for realizing 
the nematic order in Fe-based superconductors
\cite{SOnari-PRL2012,Onari-FeSe,YYamakawa-PRX2016}.
Analytically, AL-VC is related to the magnetic correlation length $\xi$ as $\xi^{4-d}$ in 
$d$-dimension systems.
Thus, AL-VC plays significant roles near the AFM-QCP, which is verified
by fRG study with higher-order VC in an unbiased way
\cite{Tsuchiizu-PRL,Tsuchiizu2,Tsuchiizu2016,Tazai-FRG,Tsuchiizu2018,Chubukov-RG1}.
In the present study, it is verified that the enhancement of 
$O_{xy}$ quadrupole fluctuations originates from the significant roles of AL-VC.

Now, we perform the beyond-RPA analysis by including the $\chi$-VC 
due to MT- and AL-type vertex corrections  \cite{tazai-last}.
The diagrammatic expression of the AL-VC is shown in 
Fig.\ref{fig:resultmodel2} (a).
The obtained quadrupole susceptibility $\chi^{O_{xy}}(\q,0)$
in Fig. \ref{fig:chiceb6} (a) is strongly enhanced 
at $\q=\bm{Q}$ and $\q=\bm{0}$.
The highest peak at $\q=\bm{Q}$ is consistent with 
the antiferro-$O_{xy}$ order in CeB$_6$.
Moreover, the second highest peak of $\chi^{O_{xy}}(\q,0)$ 
at $\q=\bm{0}$ explains the softening of shear modulus $C_{44}$ 
in CeB$_6$ \cite{Goto-CeB6}.
In Fig. \ref{fig:chiceb6} (b), we plot the $u$-dependence of the quadrupole susceptibility.
We find that $\chi^{O_{xy}}(\q,0)$ strongly increases with $u$ due to the AL-type VC.
In the previous study, 
MT-term was investigated as the origin of the rank-5 multipole ordered state in URu$_2$Si$_2$ \cite{Ikeda-Uru2}.
On the other hand, the MT-VC does not enhance even-rank multipole fluctuations.

\begin{figure}[t]\centering
\includegraphics[width=.99\linewidth]{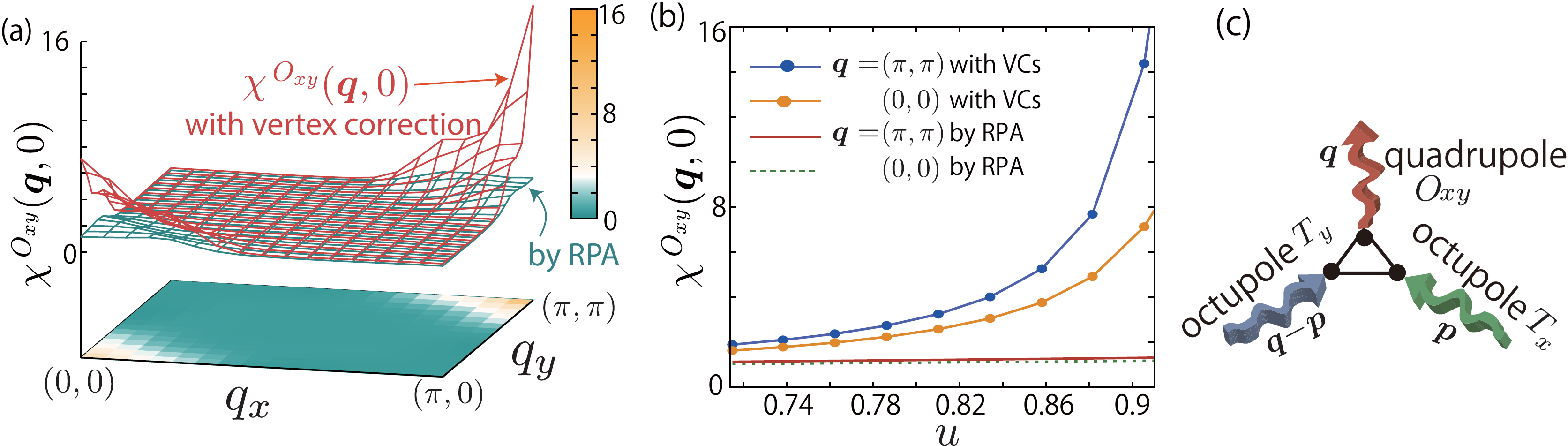}
\caption{
(a) $\q$-dependence of $\chi^{O_{xy}}(\q,0)$ at
$\a^{\Gamma_4^{+}}=0.94$ with AL-VC$+$MT-VC.
(b) $u$-dependence of  $\chi^{O_{xy}}(\q,0)$ at $\q=\bm{Q}$ and $\q=\bm{0}$.
(c) Quantum interference process for $O_{xy}$ quadrupole order.
Cited from Ref. \cite{tazai-last}. 
}
\label{fig:chiceb6}
\end{figure}

Figure \ref{fig:chiceb6} (c) presents the quantum process for
$O_{xy}$ quadrupole order, which is driven by 
the interference between $(T_{x},T_{y})$ fluctuations.
This process is realized because the following coupling constant 
among ($Q_{xz}$, $Q$, $Q'$)-channel fluctuations,
\begin{eqnarray}
\Lambda^{O_{zx}QQ'}\propto
{\rm Tr}\{ \hat{O}_{xy}\cdot \hat{Q}\cdot \hat{Q}' \}
\label{eqn:qdep}
\end{eqnarray}
is finite for $(Q,Q')=(T_{x},T_{y})$ because of the symmetry of the model.
This fact is easily understood based on the pseudo-spin representation 
for multipole operators in Table \ref{tab:multipole2} 
by using two-types of Pauli matrices ($\sigma^{\mu},\tau^{\nu}$)
\cite{Shiina1,Shiina4,tazaijp2}.
In contrast, $\Lambda^{QT_{x}T_{y}}\propto
{\rm Tr}\{ \hat{Q} \cdot \hat{T}_{x}\cdot \hat{T}_{y} \}=0$ for odd-rank $Q$.
For this reason, the dipole ($J$) and the octupole ($Q$)
fluctuations remain small even when the AL-VC is taken into consideration.
Note that the field-induced octupole order in CeB$_6$
is naturally understood based on the present AL-VC mechanism
\cite{tazai-last}.

\begin{table}
\begin{center}
  \begin{tabular}{|c|c|c|c|c|c|c|c|} \hline
  Q &   1 & $O_{20(22)}$  & $O_{xy(yz,zx)}$ & $T_{xyz}$ & $J_{z(x,y)}$ &  $T^{\alpha}_{z(x,y)}$ &$T^{\beta}_{z(x,y)}$
     \\ \hhline{|=|=|=|=|=|=|=|=|}
 ${\bar U}^{Q}$ &   -2.4 & 0.50  & 0.63 & 0.81 & 1.03 & 0.94 & 0.94 \\ \hline
   \end{tabular}
\end{center}
\caption{Normalized Coulomb interaction $\bar{U}^{Q}$
\cite{tazai-last}.
The relation $\bar{U}^{Q,Q'}=0$ holds for $Q\ne Q'$ except for $\bar{U}^{J_{\mu},T^{\a}_{\mu}}=0.58$ where $\mu=x,y,z$. 
}
\label{tab:multipole3}
\end{table}

In summary,
we proposed multipole fluctuation mechanism to 
explain the quadrupole ordering in CeB$_6$ by 
considering AL-type VC.
Near the AFM-QCP,
several multipole fluctuations strongly develop, simultaneously
including higher-rank (octupole $T$) fluctuations.
Development of magnetic multiple multipole fluctuations gives
large AL-type $\chi$-VC  for electric multipole fluctuations, which cause violation of RPA.
Owing to AL-VC,  AF-quadrupole fluctuations 
$\chi^{O_{xy}}(\q)$ at $\q=(\pi,\pi)$ develop due to the
 interference between magnetic octupole fluctuations for ($T_{x},T_{y}$).
The inter-multipole coupling mechanism
will be important even in other heavy fermion systems \cite{Thal-Yb,Kubo-Np} as well as
4$d$, 5$d$ transition metal system.
Therefore, it is an important future problem to analyze the AL-VC 
in these systems.


\section{Summary}


We reviewed the recent progress on the 
theoretical studies of (i) unconventional superconductivity
and (ii) exotic normal state order parameters
in unconventional superconductors,
mainly in Fe-based and cuprate superconductors.
They are important open problems in condensed matter physics.
In this article, we discussed the topics (i) and (ii) in a parallel way,
because these two issues are closely related to each other.
For example, quantum fluctuations of exotic order parameters
can mediate exotic pairing states.

To understand the topic (ii), 
we introduced the significant developments of  
the theory of spontaneous symmetry breaking phenomena in metals.
In Sect. \ref{sec:Kontani2}, we introduced the form factor $f_\q(\k)$,
in order to discuss various exotic order parameters,
such as the bond-order, current order, and spin-current order,
in a unified way.
In Sect. \ref{sec:Kontani3}, 
we explained the microscopic mechanism of exotic order parameters,
which are expressed as non-$s$-wave form factors,
by going beyond the mean-field-level approximations.
The ``paramagnon interference'' in Fig. \ref{fig:phase-AFM} (b)
is a key mechanism of exotic phase transitions.
In Sect. \ref{sec:Kontani5} and Sect. \ref{sec:Kontani4},
We discussed various nematic and smectic bond orders 
(=even-parity $f_\q(\k)$) in cuprate and Fe-based superconductors
based on the paramagnon interference mechanism.
Exotic current orders (=odd-parity $f_\q(\k)$) are also analyzed.
We hope the present theory would contribute 
in understanding the pseudogap mechanism,
which is one of the most important open issues at present.

Next, we discussed the topic (i) 
based on the recently achieved knowledge on 
the correlation-driven spontaneous symmetry breaking in metals.
In Sect. \ref{sec:Onari},
we discussed the mechanism of superconductivity
mediated by the quantum fluctuations of exotic order parameters
with non-$s$-wave form factors.
For this purpose, we constructed beyond-Migdal-Eliashberg gap equation.
Based on the nematic/smectic charge-channel fluctuation mechanism,
we explained the SC states in Ba122, LiFeAs, and FeSe families.
High-$T_{\rm c}$ SC state in monolayer FeSe without hole-FS
is naturally explained by means of the smectic fluctuation mechanism.
The theory of superconductivity in Fe-based superconductors
is still developing, and the present theoretical study 
would be useful for future progress.

In Sect. \ref{sec:Tazai},
we introduced interesting multipolar physics in $f$-electron systems
due to the strong SOI and strong electron correlation.
We discussed exotic multipole fluctuation pairing mechanism,
and explained the fully-gapped $s$-wave superconductivity in CeCu$_2$Si$_2$.
We also discussed the exotic multipole order in $f$-electron systems.

\color{black}
In the present article,
we discuss the origin of the unconventional density-waves
based on the DW equation method and the fRG method.
The solution of the former method satisfies the 
stationary condition of the Luttinger-Ward free energy,
so the macroscopic conservation laws are satisfied 
\cite{RTazai-arXiv2021}.
In the latter methods, all the parquet diagrams
for the four-point vertex are calculated in an unbiased way.
However, these methods are classified as weak-coupling theories
based on the Fermi liquid picture.
On the other hand,
remarkable progress in the numerically exact studies
for the single-orbital Hubbard model
has been achieved recently,
such as several quantum Monte Carlo methods
and the density matrix renormalization group method
\cite{various1,various2,exact3,exact4}.
The formation of nontrivial charge/spin stripe orders is found
in large cluster Hubbard models in 
Refs. \cite{various1,various2,exact3,exact4}.
It is a very interesting future issue to
make comparison between the present weak-coupling theories
and the numerically exact studies to understand 
the physical origin of unconventional DWs in cuprates.
\color{black}

\begin{figure}[t]\centering
\includegraphics[width=0.8\linewidth]{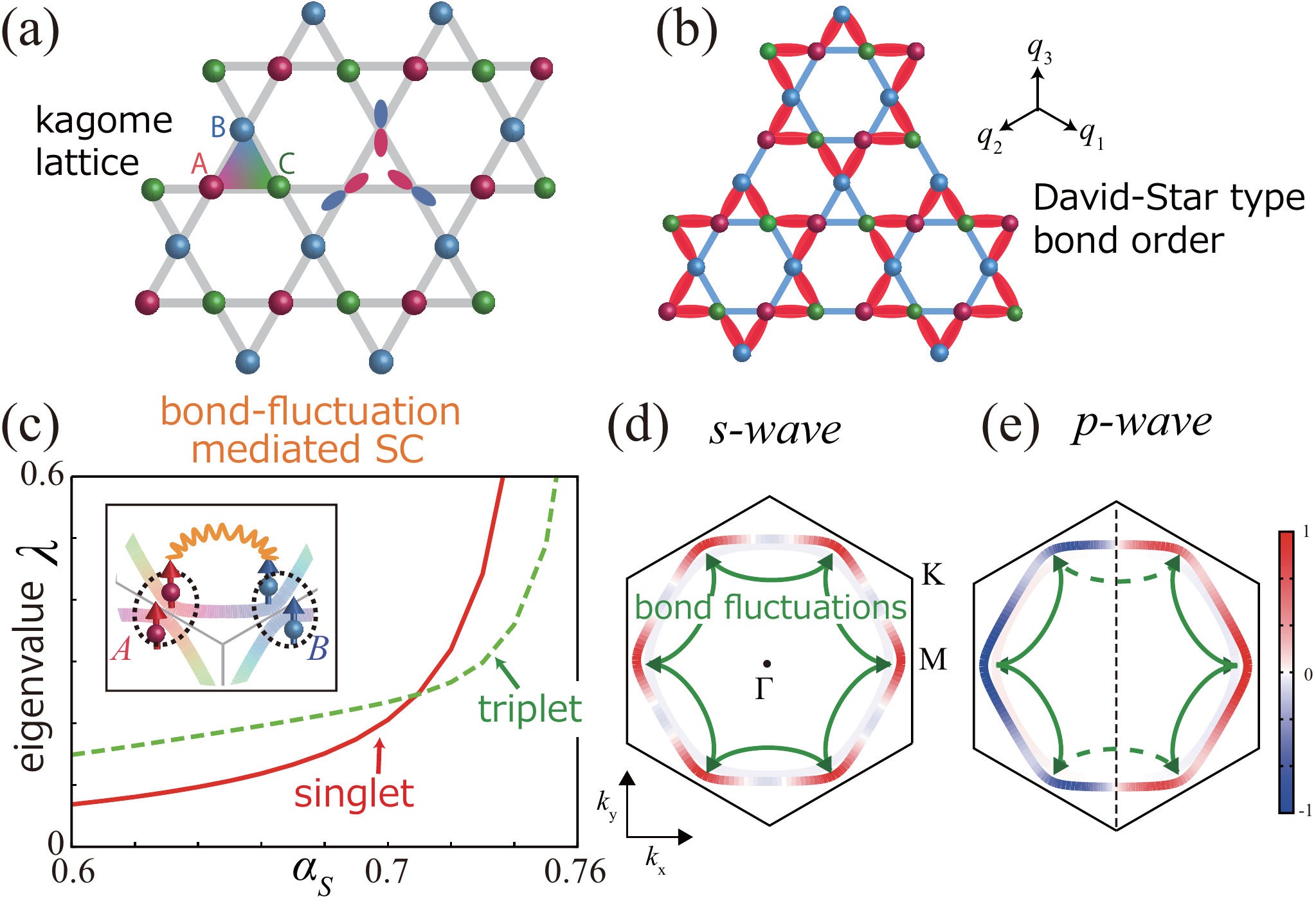}
\caption{
(a) Lattice structure of kagome lattice in AV$_3$Sb$_5$.
Each unit cell is composed three V-ion sites (A,B,C).
(b) Star of David bond-order derived by DW equation analysis
in Ref. \cite{Kagome-Tazai}.
It is given by the combination of three bond orders 
($\q=\q_1,\q_2,\q_3$).
(c) Eigenvalue of SC gap equation as function of $\a_S \ (\propto U)$
 \cite{Kagome-Tazai}.
Both singlet SC state and triplet SC state are mediarted
by the star-of-David BO fluctuations.
(inset) Beyond-Mgidal SC gap equation:
Wavy line and green circles represent the BO fluctuations
and the BO form factor, respectively.
(d) Singlet anisotropic $s$-wave SC gap function.
(e) Triplet two-dimensional ($p_x,p_y$)-wave SC gap function.
Cited from Ref. \cite{Kagome-Tazai}.
}
\label{fig:kagome}
\end{figure}

Finally, we shortly review very recent remarkable progress.
In 2019, interesting density-wave formation
and unconventional superconductivity have been discovered 
in kagome lattice metal AV$_3$Sb$_5$ (A=Cs,Rb,K).
Here, the cooperation between the geometrical frustration and 
strong electron correlation leads to verious exotic phase transitions. 
Figure \ref{fig:kagome} (a) shows the lattice structure of kagome lattice.
Each unit cell is composed of three V-ion sites (A,B,C).
Here, we analyze this multiorbital kagome lattice Hubbard model 
based on the DW equation \cite{Kagome-Tazai}, and obtain the 
star of David bond-order shown in Fig. \ref{fig:kagome} (b).
It is given by the combination of three bond orders ($\q=\q_1,\q_2,\q_3$).
The obtained bond-order is consistent with experimental reports.
In the next stage, we study unconventional superconductivity
by means of the bond-order fluctuation mechanism.
For this purpose, we solve the beyond-ME gap equation
introduced in Sect. \ref{sec:Onari-FeSe},
using the form factor obtained by the DW equation.
Figure \ref{fig:kagome} (c) exhibits the eigenvalue of SC gap equation 
as function of $\a_S \ (\propto U)$.
Here, both singlet SC state and triplet SC state can emerge.
They are mediated by the fluctuation of star-of-David bond-order.
The inset exhibits the Beyond-Mgidal SC gap equation:
Wavy line represents the BO fluctuations,
and green circles are the BO form factor derived from the DW equation
\cite{Kagome-Tazai}.
The obtained singlet anisotropic $s$-wave SC gap function
and triplet two-dimensional ($p_x,p_y$)-wave SC gap function
are shown in Figs. \ref{fig:kagome} (d) and (e), respectively.
In Kagome metals, the charge-loop-current (cLC) state has been 
observed by several experimental methods,
and the mechanism of the cLC has been intensively studied recently
\cite{Kagome-Tazai-cLC}.

\vspace{5mm}
{\large \bf acknowledgements}

This study has been supported by Grants-in-Aid for Scientific
Research from MEXT of Japan (JP18H01175, JP17K05543, JP20K03858, JP20K22328),
and by the Quantum Liquid Crystal
No. JP19H05825 KAKENHI on Innovative Areas from JSPS of Japan.


\end{document}